\RequirePackage[l2tabu, orthodox]{nag}		
\documentclass[12pt]{book}

\usepackage[T1]{fontenc}										
\usepackage{lmodern}												
\usepackage{microtype}											
\usepackage[frenchmath,LGRgreek]{mathastext}
\MTgreekfont{lmtt} 													
\Mathastext
\let\varepsilon\epsilon	
\let\varphi\phi		
\let\vartheta\theta											
\usepackage[mathscr]{eucal}                 
\usepackage[sans]{dsfont}										
\usepackage{amsbsy}													
\usepackage{natbib}       									
\usepackage{amsmath}                        
\numberwithin{equation}{chapter}
\usepackage{graphicx}                       
\usepackage{mdwlist}												
\usepackage[dvipsnames]{xcolor}							
\usepackage[plainpages=false, pdfpagelabels]{hyperref} 
\usepackage{xurl}                           
	\hypersetup{
		colorlinks   = true,
		citecolor    = RoyalBlue,
		linkcolor    = RubineRed,
		urlcolor     = Turquoise
		}
\usepackage[paperwidth=8.5in,paperheight=11in,top=1.00in, bottom=1.00in, left=0.50in, right=0.50in]{geometry}
\usepackage{mathtools}                      
\mathtoolsset{showonlyrefs=true}            
\usepackage{amsthm}                         
\allowdisplaybreaks                         
\newtheoremstyle{plain}
  {}   
  {}   
  {\itshape}  
  {}       
  {\mdseries\scshape} 
  {.}         
  { } 
  {\thmname{#1}\thmnumber{ #2}\ifx#3\empty\else\ (#3)\fi}
\theoremstyle{plain}
\newtheorem{theorem}{\underline{Theorem}}
\numberwithin{theorem}{section}
\newtheorem{lemma}[theorem]{\underline{Lemma}}   		

\newtheoremstyle{definition}
  {}   
  {}   
  {}  
  {}       
  {\mdseries\scshape} 
  {.}         
  { } 
  {\thmname{#1}\thmnumber{ #2}\ifx#3\empty\else\ (#3)\fi}
\theoremstyle{definition}
\newtheorem{definition}[theorem]{\underline{Definition}}
\newtheorem{example}[theorem]{\underline{Example}}
\newtheorem{remark}[theorem]{\underline{Remark}}
\newtheorem{assumption}[theorem]{\underline{Assumption}}
\newtheorem{exercise}{\underline{Exercise}}
\numberwithin{exercise}{chapter}

\expandafter\let\expandafter\oldproof\csname\string\proof\endcsname
\let\oldendproof\endproof
\renewenvironment{proof}[1][\proofname]{%
  \oldproof[\scshape\underline{#1}]%
}{\oldendproof}

\usepackage{sectsty}
\allsectionsfont{\mdseries\scshape}

\usepackage{pdfrender}
\renewcommand{\textbf}[1]{\textpdfrender{
  TextRenderingMode=FillStroke,
  LineWidth=.3pt,
}{#1}}

\renewcommand{\(}{\left(}
\renewcommand{\)}{\right)}
\renewcommand{\[}{\left[}
\renewcommand{\]}{\right]}

\newcommand\Eb{\mathds{E}}
\newcommand\Pb{\mathds{P}}

\newcommand\Rb{\mathds{R}}
\newcommand\Ib{\mathds{1}}
\newcommand\Nb{\mathds{N}}

\newcommand\Vb{\mathds{V}}

\newcommand\Ac{\mathscr{A}}

\newcommand\Dc{\mathscr{D}}

\newcommand\Fc{\mathscr{F}}

\newcommand\Mc{\mathscr{M}}

\newcommand\Oc{\mathscr{O}}

\newcommand\Vc{\mathscr{V}}

\newcommand\om{\omega}
\newcommand\Om{\Omega}
\newcommand\sig{\sigma}

\newcommand\Lam{\Lambda}
\newcommand\gam{\gamma}
\newcommand\Gam{\Gamma}
\newcommand\lam{\lambda}
\newcommand\del{\delta}
\newcommand\Del{\Delta}

\newcommand\Tb{\overline{T}}

\newcommand\rhob{\bar{\rho}}

\newcommand\Pbh{\widehat{\Pb}}
\newcommand\Ebh{\widehat{\Eb}}

\newcommand\Wh{\widehat{W}}
\newcommand\Fh{\widehat{F}}

\newcommand\Xh{\widehat{X}}

\newcommand\uh{\widehat{u}}

\newcommand\fh{\widehat{f}}

\newcommand\ph{\widehat{p}}
\newcommand\qh{\widehat{q}}

\newcommand\Ebt{\widetilde{\Eb}}
\newcommand\Pbt{\widetilde{\Pb}}
\newcommand\Act{\widetilde{\Ac}}

\newcommand\Wt{\widetilde{W}}

\newcommand\qt{\widetilde{q}}
\newcommand\rt{\widetilde{r}}

\newcommand\ft{\widetilde{f}}
\newcommand\Ft{\widetilde{F}}

\renewcommand\d{\partial}
\newcommand{\ind}{\perp \! \! \! \perp}
\newcommand\ii{\mathtt{i}}

\newcommand\dd{\mathrm{d}}
\newcommand\ee{\mathrm{e}}
\newcommand\BS{\textrm{BS}}

\newcommand{\cf}{\text{CF}}

\usepackage{booktabs} 
\usepackage{placeins} 
\usepackage{tikz} 
\usetikzlibrary{matrix}
\usepackage{pgfplots}
\pgfplotsset{compat=1.18}
\pgfplotsset{
  payoff/.style={                         
    width=5.5cm, height=4.5cm,
    axis lines=left,
    xmin=0, xmax=2, ymin=-1, ymax=1,
    xtick=\empty, ytick=\empty,
    clip=false,
    every axis plot/.append style={blue, thick, no marks},
  },
  payoff3/.style={payoff, width=4.6cm},   
  boxpayoff/.style={                      
    width=5.5cm, height=3.8cm,
    xmin=0, xmax=2, ymin=0, ymax=1.05,
    xtick={0}, ytick={0}, xlabel={$S_T$},
    every axis plot/.append style={blue, thick, no marks},
  },
  greek/.style={                          
    width=11cm, height=7cm,
    xmin=0.7, xmax=1.3, xlabel={$s$},
    domain=0.7:1.3, samples=121,
    legend cell align=left,
    every axis plot/.append style={thick, no marks},
  },
}
\tikzset{zeroline/.style={dotted, thin, black}}
\newcommand{\zeroline}[1]{%
  \draw[zeroline] (axis cs:\pgfkeysvalueof{/pgfplots/xmin},0) -- (axis cs:\pgfkeysvalueof{/pgfplots/xmax},0) node[right] {#1};
  \node[left] at (axis cs:\pgfkeysvalueof{/pgfplots/xmin},0) {$0$};}
\pgfmathdeclarefunction{normpdf}{1}{\pgfmathparse{exp(-0.5*(#1)^2)/sqrt(2*pi)}}
\pgfmathdeclarefunction{normcdf}{1}{%
  \pgfmathparse{1/(1+0.2316419*abs(#1))}\let\tTmp\pgfmathresult
  \pgfmathparse{1-normpdf(abs(#1))*\tTmp*(0.319381530+\tTmp*(-0.356563782+\tTmp*(1.781477937+\tTmp*(-1.821255978+\tTmp*1.330274429))))}\let\pTmp\pgfmathresult
  \pgfmathparse{(#1>=0)*\pTmp + (#1<0)*(1-\pTmp)}}
\pgfmathdeclarefunction{dplus}{4}{\pgfmathparse{(ln(#1/#4)+0.5*(#3)^2*(#2))/((#3)*sqrt(#2))}}
\pgfmathdeclarefunction{dminus}{4}{\pgfmathparse{(ln(#1/#4)-0.5*(#3)^2*(#2))/((#3)*sqrt(#2))}}
\usepackage{mathtools} 

\usepackage{listings}
\usepackage{textcomp}   

\lstdefinestyle{pythonstyle}{
  language=Python,
  basicstyle=\ttfamily\small,
  keywordstyle=\bfseries,
  commentstyle=\itshape,
  stringstyle=\ttfamily,
  showstringspaces=false,
  upquote=true,
  frame=single,
  framerule=0.4pt,
  xleftmargin=1em,
  xrightmargin=1em,
  breaklines=true,
  breakatwhitespace=false,
  numbers=left,
  numberstyle=\tiny,
  numbersep=6pt,
  tabsize=4,
  columns=flexible,
}
\theoremstyle{definition}

\numberwithin{question}{section}

\newcommand\pt{\widetilde{p}}

\tikzstyle{bag} = [text width=2em, text centered]
\tikzstyle{end} = []

\makeatletter
\AtBeginDocument{%
  \begingroup\lccode`~=`'\lowercase{%
    \endgroup\def~}{\ifmmode\expandafter\mst@active@math@prime \else '\fi}%
}
\makeatother

\begin{document}

\frontmatter

\title{An Introduction to Financial Derivatives\\[0.5em] \large Open Lecture Notes}

\author{Matthew Lorig \thanks{Department of Applied Mathematics, University of Washington, Seattle, WA, USA.  \textbf{e-mail}: \url{mlorig@uw.edu}}}
\date{This version: \today}
\maketitle

\thispagestyle{empty}
\vspace*{\fill}
\noindent
\copyright\ 2026 Matthew Lorig.
\\[0.5em]
\noindent
This work is licensed under the Creative Commons Attribution-ShareAlike 4.0 International License (CC BY-SA 4.0), \url{https://creativecommons.org/licenses/by-sa/4.0/}.  You are free to share (copy and redistribute) and adapt (remix, transform and build upon) this material for any purpose, including commercially, provided that you (i) give appropriate credit to the original author, (ii) indicate if changes were made, and (iii) distribute your contributions under the same license.  Code listings in these notes are additionally released under the MIT License.
\\[0.5em]
\noindent
The LaTeX source of these notes is available at\\
\url{https://github.com/mattlorig/intro-financial-derivatives}.
\\[0.5em]
\noindent
\textit{Please cite as}: M.~Lorig, \textit{An Introduction to Financial Derivatives: Open Lecture Notes}, \url{https://github.com/mattlorig/intro-financial-derivatives}, 2026.
\clearpage
\tableofcontents
\chapter{Preface} 
These notes give a broad introduction to financial derivatives.  They were developed for CFRM 504, a course for Masters students in Computational Finance and Risk Management at the University of Washington, and they assume a working knowledge of calculus and basic probability.  Because the focus of this course is on the \textit{applied} aspects of this topic, we will sometimes forgo mathematical rigor, favoring instead a heuristic development.  The mathematical statements in these notes should be taken as ``true in spirit,'' but perhaps not always rigorously true mathematically.  The hope is that, what the notes lack in rigor, they make up in clarity.
\\[0.5em]
These notes are very much a work in progress.

\subsection*{How to contribute}
These notes are open source.  The LaTeX source is available at \url{https://github.com/mattlorig/intro-financial-derivatives}, and the entire document compiles from that single file, with no external figures or bibliography files.  Anyone may download it, correct errors, add examples, exercises or sections, and adapt the notes for their own courses, subject to the license on the reverse of the title page.  To report an error, open an issue on the repository; to propose a change, submit a pull request.  Contributors whose changes are accepted will be listed below.

\subsection*{Contributors}
The author of these notes wishes to express his sincere thanks to the following people for their contributions.
\begin{itemize}
\item Weston Barger: checking and writing homework solutions, as well as making corrections and improvements to the text.
\end{itemize}

\subsection*{Donations}
If you find these notes useful and would like to make a donation to help me develop them further you can donate ETH or ERC20 tokens to the address below.\\[0.25em]
\underline{ETH address}: 0x3D6042D6a6f39735bbAd43ad1Aa676113D55BaAC

\mainmatter





\chapter{Review of Probability}

In this chapter, we will \textit{very briefly} review some concepts from probability theory.  This material is \textit{essential} for understanding the subsequent chapters.


\section{Sample spaces, events, and probability measures}

\noindent
We begin this section with some important definitions.

\begin{definition}
The set of all possible outcomes of a random experiment is called the \textit{sample space} and is denoted by $\Om$.
\end{definition}

\noindent An \textit{outcome} of a random experiment is an \textit{element} of $\Om$ and is denoted by $\om$.

\begin{definition}
An \textit{event} is any subset of the sample space $\Om$.  We denote events by capital roman letters $A, B, C, \ldots$.
\end{definition}

\noindent
Finally, we will denote by $\Fc$ all of the subsets of a sample space $\Om$.
\footnote{Rigorously, $\Fc$ should be a \textit{$\sig$-algebra} of $\Om$.  But, we need not get into such detail in this text.}
Note that a single element $\om$ of $\Om$ is also a subset of $\Om$, as are the empty set $\emptyset$ and $\Om$ itself.  Thus, we have $\{\om\}, \emptyset, \Om \in \Fc$.

\begin{example}[Toss a coin]
When one tosses a coin, there are two possible outcomes: heads (denoted by $H$) and tails (denoted by $T$).  Thus, our sample space is
$\Om = \{ H, T \}$.  One possible element or outcome of $\Om$ is to ``toss a heads'' $\om = H$.  Two possible events are ``toss a heads'' $A = \{H\}$ and ``toss a heads or a tails'' $B = \{ H , T \}$.
\end{example}

\begin{example}[Roll a die]
A standard die has six sides, labeled one through six.  Thus, our sample space is $\Om = \{ 1, 2, 3, 4, 5, 6 \}$.  One possible element or outcome of $\Om$ is ``roll a two'' $\om = 2$.  Two possible events are ``roll an odd number'' $A = \{ 1, 3, 5 \}$ and ``roll a one or two'' $B = \{ 1, 2 \}$.
\end{example}

\noindent
So far, we have not yet talked about probabilities -- only outcomes of a random experiment (elements $\om \in \Om$) and events (subsets $A \in \Fc$).  We now introduce the concept of a probability measure.

\begin{definition}
\label{def:P}
A \textit{probability measure} $\Pb$ assigns probabilities to events $\Pb : \Fc \to [0,1]$ and satisfies
\begin{enumerate}
\item $\Pb( \Om ) = 1$, and
\item If $(A_i)$ is a countable sequence of disjoint subsets of $\Om$, (i.e., $A_i \cap A_j = \emptyset$ if $i \neq j$), then
\begin{align}
\Pb \Big( \bigcup_i A_i \Big)
	&=	\sum_i \Pb( A_i) .
\end{align}
\end{enumerate}
\end{definition}

\noindent
The first property in Definition \ref{def:P} is easy to understand.
When we perform a random experiment, the probability that \textit{something} happens is one.
Thus, in order for a probability measure $\Pb$ to make any sense at all, we must have $\Pb(\Om) = 1$.  
The second property in Definition \ref{def:P}, known as \textit{countable additivity}, ensures that we do not double-count the probability of events.
For example, if we toss a coin, the probability of getting a heads or tails should equal the probability of getting a heads plus the probability of getting a tails
\begin{align}
\Pb( H \cup T )
	&=	\Pb(H) + \Pb(T) .
\end{align}
Note that, together, the two properties in Definition \ref{def:P} imply that $\Pb(\emptyset) = 0$ as
\begin{align}
1 
	&= \Pb(\Om) = \Pb (\Om \cup \emptyset) = \Pb(\Om) + \Pb(\emptyset) = 1 + \Pb(\emptyset) , &
	&\Rightarrow&
\Pb(\emptyset)
	&=	0 ,
\end{align}
where we have used the fact that $\Om$ and $\emptyset$ are disjoint.
As we shall see with the next few examples, a probability measure $\Pb$ need not assign probabilities to events in a way that is consistent with our real-world experience.

\begin{example}[Toss a coin]
When we toss a coin, our experience tells us that the probability of getting a heads is $1/2$ and the probability of getting tails is $1/2$.  Thus, if we want a probability measure $\Pb$ to reflect our experience, we should have $\Pb(H) = \Pb(T) = 1/2$.  We can think of $\Pb$ as the ``real world'' or ``statistical'' probability measure.  However, we could consider another probability measure $\Pbt$ that assigns different probabilities to heads and tails.  For example, we could have $\Pbt(H) = 3/4$ and $\Pbt(T) = 1/4$.  The measure $\Pbt$ does not capture what goes on in the real world.  But, it is nonetheless a valid probability measure for the random experiment of tossing a coin.  You should check for yourself that $\Pbt$ satisfies the conditions of Definition \ref{def:P}.
\end{example}

\begin{example}[Roll a die]
When we roll a die, our experience tells us that the probability of getting any number $\om \in \{1,2,3,4,5,6\}$ is $1/6$.  Thus, if we want a probability measure $\Pb$ to reflect our experience, we should have $\Pb(\om) = 1/6$ for all $\om \in \{1,2,3,4,5,6\}$.  As in the coin toss example, we can think of $\Pb$ as the ``real world'' or ``statistical'' probability measure.  However, we could consider another probability measure $\Pbt$ that assigns different probabilities to each outcome.  For example, we could have $\Pbt(1) = 1/2$ and $\Pbt(\om) = 1/10$ for $\om \in \{ 2, 3, 4, 5, 6 \}$.  The measure $\Pbt$ does not capture what goes on in the real world.  But, it is nonetheless a valid probability measure for the random experiment of rolling a die.  You should check for yourself that $\Pbt$ satisfies the conditions of Definition \ref{def:P}.
\end{example}

\noindent
It may not be clear at this time why one would ever consider a probability measure $\Pbt$ that does not reflect our real-world experience.  However, when we discuss option pricing, it will become apparent how such a measure can be useful.

\begin{definition}
\label{def:equivalent}
Two probability measures $\Pb$ and $\Pbt$ are said to be \textit{equivalent}, denoted $\Pb \sim \Pbt$, if they agree on events that occur with probability one.  That is, if
\begin{align}
\Pb(A)
	&=	1  &
	&\Leftrightarrow&
\Pbt(A)
	&=	1 .
\end{align}
\end{definition}

\noindent
Note that if $\Pb \sim \Pbt$ we also have
\begin{align}
\Pb(A)
	&=	0  &
	&\Leftrightarrow&
\Pbt(A)
	&=	0 .
\end{align}
In other words, equivalent measures agree on events that occur with probability one and on events that occur with probability zero.

\begin{example}
Let us return to the coin toss example.  Consider two probability measures $\Pb$ and $\Pbt$.  Suppose that $\Pb(H) = 1/2$ and $\Pbt(H) = 3/4$.  Then $\Pb \sim \Pbt$ because
\begin{align}
\Pb(H \cup T)
	&=	\Pb(H) + \Pb(T) 
	 =	1/2 + 1/2
	 =	1 , \\
\Pbt(H \cup T)
	&=	\Pbt(H) + \Pbt(T)
	 =	3/4 + 1/4
	 =	1 .
\end{align}
Now, consider a third probability measure $\Pbh$ that satisfies $\Pbh(H) = 1$.  Then $\Pbh$ is \textit{not} equivalent to $\Pb$ (or to $\Pbt$) because $\Pb(H) = 1/2 \neq 1$.
\end{example}


\section{Random variables}
\label{sec:rvs}

\noindent
Intuitively, we think of a random variables as a quantity whose value is uncertain.  For example, the value of a stock 10 days from now is uncertain and, as such, we think of it as being random.  We give a precise definition of a random variable below.

\begin{definition}
A random variable $X$ is a function that maps every outcome $\om$ of a random experiment to a number in $\Rb$
\begin{align}
X
	&: \Om \to \Rb .
\end{align}
The set $S$ of values that $X$ can take is called the \textit{state space of $X$}.
\end{definition}

\begin{example}[Toss a coin]
\label{ex:coin.toss.rvs}
If we toss a coin, we have $\Om = \{H,T\}$.  Two possible random variables on this sample space are
\begin{align}
X(\om)
	&=	\left\{ \begin{aligned}
			1 , & & \text{if $\om = H$} , \\
			0 , & & \text{if $\om = T$} ,
			\end{aligned} \right. &
Y(\om)
	&=	\left\{ \begin{aligned}
			10 , & & \text{if $\om = H$} , \\
			20 , & & \text{if $\om = T$} .
			\end{aligned} \right. &
\end{align}
Thus, if we toss a heads $\om = H$, we have $X = 1$ and $Y = 10$.
The state space of $X$ is $S = \{0,1\}$.  The state space of $Y$ is $S = \{10,20\}$.
\end{example}

\begin{example}[Roll a die]
\label{ex:dice-roll}
If we roll a die, we have $\Om = \{1,2,3,4,5,6\}$.  Two possible random variables on this sample space are
\begin{align}
X(\om)
	&=	\om , &
Y(\om)
	&=	\om^2 .
\end{align}
Thus, if we roll a three $\om = 3$, we have $X = 3$ and $Y = 3^2 = 9$.
The state space of $X$ is $S = \{1,2,3,4,5,6\}$.  The state space of $Y$ is $S = \{1,4,9,16,25,36\}$.
\end{example}

\noindent
Random variables can be described by their \textit{cumulative distribution function}.

\begin{definition}
\label{def:cdf}
The \textit{cumulative distribution function} (CDF) of a random variable $X$ is a function $F_X : \Rb \to [0,1]$ defined by
\begin{align}
F_X(x)
	&:= \Pb( X \leq x ) , \label{eq:cdf}
\end{align}
where $x \in \Rb$.
\end{definition}

\noindent
Here, we have used the short-hand notation
\begin{align}
\{ X \leq x \}
	&=	\{ \om \in \Om : X(\om) \leq x \} .
\end{align}
Now, we make an important note.  A random variable $X$ is defined \textit{independently} of a probability measure.  However, we see from \eqref{eq:cdf} that the CDF $F_X$ is defined with respect to a probability measure $\Pb$.  Thus a random variable $X$ can have a CDF $F_X$ under $\Pb$ and have an entirely different CDF $\Ft_X$ under $\Pbt$.  We will (hopefully) make this concept clear through some examples in the coming sections.
\\[0.5em]
Sometimes, the value of one random variable $X$ may affect the value of another random variable $Y$.  
For example, suppose you roll two 6-sided dice.  Let $X$ be the result of the first roll, let $Y$ be the result of the second and let $Z = X+Y$.  If I tell you that $Z = 2$, then you know that both $X = 1$ and $Y = 1$.
In cases such as these, in order to describe the relationship between two random variables, we will need to consider \textit{joint cumulative distribution function}.

\begin{definition}
\label{def:joint.cdf}
The \textit{joint cumulative distribution function} (joint CDF) of two random variables $X$ and $Y$ is a function $F_{X,Y} : \Rb^2 \to [0,1]$ defined by
\begin{align}
F_{X,Y}(x,y)
	&:= \Pb( X \leq x, Y \leq y ) , \label{eq:joint.cdf}
\end{align}
where $x,y \in \Rb$.
\end{definition}

\noindent
Here, we have used the short-hand notation
\begin{align}
\{ X \leq x , Y \leq y\}
	&=	\{ \om \in \Om : X(\om) \leq x, Y(\om) \leq y \} .
\end{align}
As in the univariate case, the joint CDF $F_{X,Y}$ is defined with respect to a particular probability measure (in the case of equation \eqref{eq:joint.cdf}, the probability measure is $\Pb$).
\\[0.5em]
Definition \ref{def:joint.cdf} can easily be extended to any finite number of random variables.  For example, the joint CDF of three random variables $(X,Y,Z)$ would be defined as follows
\begin{align}
F_{X,Y,Z}(x,y,z) 
	&:= \Pb( X \leq x, Y \leq y, Z \leq z) ,
\end{align}
where, once again, $F_{X,Y,Z}$ is defined relative to the probability measure $\Pb$.
\\[0.5em]
Many (but not all) random variables fall in to one of two categories: \textit{discrete} or \textit{continuous}.  
We will discuss discrete random variables first.  Once we understand discrete random variables, we will be able to understand continuous random variables by making some obvious modifications.


\subsection{Discrete random variables}
\label{sec:disc.rv}

\begin{definition}
\label{def:disc.rv}
A \textit{discrete random variable} $X$ is any random variable whose state space consists of at most countably many values $S := \{ x_1, x_2, \ldots \}$.
\end{definition}

\begin{example}
Let us return to the example of rolling a single die.  Our sample space is $\Om = \{1,2,3,4,5,6\}$.
Consider a random variable $Z$ defined as follows
\begin{align}
Z(\om)
	&=	\om/10 .
\end{align}
The state space of $Z$ is $S = \{0.1, 0.2, 0.3, 0.4, 0.5, 0.6\}$.  As the state space $S$ contains only six values, it follows that $Z$ is a discrete random variable.
\end{example}

\noindent
A discrete random variable can be described by its \textit{probability mass function}.

\begin{definition}
\label{def:pmf}
The \textit{probability mass function} (PMF) of a discrete random variable $X$ is a function $f_X : S \to [0,1]$ is defined by
\begin{align}
f_X(x_i)
	&:=	\Pb( X = x_i ) . \label{eq:pmf}
\end{align}
\end{definition}

\noindent
Here, as in the previous section, we have introduced short-hand notation
\begin{align}
\{ X = x_i\}
	&=	\{ \om \in \Om : X(\om) = x_i \} .
\end{align}
Just as the CDF $F_X$ of a random variable $X$ is defined with respect to a particular probability measure $\Pb$, the PMF $f_X$ is also defined with respect to a particular probability measure $\Pb$.  If we wish to speak about the PMF of a discrete random variable $X$ under a probability measure $\Pbt$ we will write $\ft_X$.

\begin{example}[Roll a die]
\label{ex:die.roll.rv}
Let us return to the example of rolling a single die.  Our sample space is $\Om = \{1,2,3,4,5,6\}$.  We will consider the following random variable on this space
\begin{align}
X(\om)
	&=	\left\{ \begin{aligned}
			&1 , & & \text{if $\om$ is odd} , \\
			&0 , & & \text{if $\om$ is even} .
			\end{aligned} \right.
\end{align}
Now, consider two probability measures $\Pb$ and $\Pbt$ defined as follows
\begin{align}
\Pb(\om)
	&=	\frac{1}{6} , & 
\Pbt(\om)
	&=	\frac{\om}{21} , &
\om 
	&\in \{ 1, 2, 3, 4, 5, 6 \} .
\end{align}
First, we note that both $\Pb$ and $\Pbt$ satisfy Definition \ref{def:P}, as they must.
Let $f_X$ and $\ft_X$ denote the probability mass functions of $X$ under $\Pb$ and $\Pbt$, respectively.  
As the state space of $X$ is $S = \{ 0 , 1 \}$ we need only to compute $f_X(x_i)$ and $\ft_X(x_i)$ for $x_i = 0$ and $x_i = 1$.
We have
\begin{align}
f_X(1)
	&=	\Pb( X = 1 )
	=		\Pb(  \text{$\om$ is odd}  )
	=		1/6 + 1/6 + 1/6 
	= 	1/2 , \\
\widetilde{f}_X(1)
	&=	\Pbt( X = 1 )
	=		\Pbt( \text{$\om$ is odd}  )
	=		1/21 + 3/21 + 5/21 
	= 	9/21 .
\end{align}
As the complement of ``$\om$ is odd'' is ``$\om$ is even,'' we have that $f_X(0) = 1 - f_X(1) = 1/2$ and $\ft_X(0) = 1 - \ft_X(1) = 12/21$.
\end{example}

\noindent
Using the PMF $f_X$ we can compute $\Pb(X \in A)$ where $A \subseteq \Rb$ and the \textit{expectation} $\Eb g(X)$ as follows
\begin{align}
\Pb( X \in A)
	&=	\sum_{\{i : x_i \in A\}} f_X(x_i) , &
\Eb g(X)
	&=	\sum_i f_X(x_i) g(x_i) . \label{eq:E.disc}
\end{align}
Note that, as the expectation $\Eb g(X)$ in \eqref{eq:E.disc} depends on PMF $f_X$, which depends on a probability measure $\Pb$, it follows that the expectation $\Eb g(X)$ also depends on the probability measure $\Pb$.  Here and throughout the text we will use $\Eb$ to indicate an expectation taken with respect to $\Pb$.
If we wish to take an expectation under a different probability measure such as $\Pbt$ or $\Pbh$ we will use the notation $\Ebt$ or $\Ebh$.  For example
\begin{align}
\Pbt( X \in A)
	&=	\sum_{\{i : x_i \in A\}} \ft_X(x_i) , &
\Ebt g(X)
	&=	\sum_i \ft_X(x_i) g(x_i) ,
\end{align}
where $\ft_X(x_i) = \Pbt(X = x_i)$.

\begin{example}
A \textit{binomial random variable} $\text{Bin}(n,p)$ is a discrete random variable $X$ whose probability mass function $f_X$ is given by
\begin{align}
f_X(k)
	&=	\frac{n!}{k!(n-k)!} p^k q^{n-k} , &
q
	&=	1-p , &
k
	&\in \{ 0, 1, \ldots, n \} , 
\end{align}
where $n \in \Nb$ and $p \in (0,1)$.  Let us compute $\Pb(X \in \{0,n\})$ and $\Eb X$.  Using \eqref{eq:E.disc} we have
\begin{align}
\Pb(X \in \{0,n\} )
	&=	f_X(0) + f_X(n) 
	=		p^n + q^n , &
\Eb X
	&=	\sum_{k=0}^n f_X(k) k
	=		n p .
\end{align}
In fact, showing that $\Eb X = n p$ takes a bit of work.  If you cannot show this analytically for general $(n,p)$, check a few simple cases (e.g., $n=2$ and $p = 1/3$).
\end{example}

\begin{example}
Let us return to Example \ref{ex:die.roll.rv}.  The state space of $X$ was $S = \{ 0 ,1 \}$ and we computed
\begin{align}
f_X(x_i)
	&=	\left \{ \begin{aligned}
			1/2 && \text{if $x_i = 0$}, \\
			1/2 && \text{if $x_i = 1$},
			\end{aligned} \right. &
\ft_X(x_i)
	&=	\left \{ \begin{aligned}
			12/21 && \text{if $x_i = 0$}, \\
			9/21 && \text{if $x_i = 1$},
			\end{aligned} \right. &
\end{align}
Let us compute $\Eb (X+3)^2$ and $\Ebt (X+3)^2$.  Using \eqref{eq:E.disc}, we have
\begin{align}
\Eb (X+3)^2
	&=	\sum_{x_i=\{0,1\}} f_X(x_i) (x_i+3)^2
	=		f_X(0) (0+3)^2 + f_X(1) (1+3)^2 \\
	&=		1/2 \cdot 9 + 1/2 \cdot 16
	=		12.5 , \\
\Ebt (X+3)^2
	&=	\sum_{x_i=\{0,1\}} \ft_X(x_i) (x_i+3)^2
	=		\ft_X(0) (0+3)^2 + \ft_X(1) (1+3)^2 \\
	&=		12/21 \cdot 9 + 9/21 \cdot 16
	=		12.
\end{align}
\end{example}

\noindent
When the value of one discrete random variable $X$ affects the value of another discrete random variable $Y$, in order to describe how these two random variables interact, we need to consider the \textit{joint probability mass function}.

\begin{definition}
\label{def:jpmf}
Let $X$ and $Y$ be discrete random variables.  The \textit{joint probability mass function of $X$ and $Y$}, written $f_{X,Y}$ is defined by
\begin{align}
f_{X,Y}(x_i,y_j)
	&:=	\Pb( X = x_i, Y = y_j ) .
\end{align}
\end{definition}

\noindent
Using the joint probability mass function $f_{X,Y}$ we can compute $\Pb((X,Y) \in A)$ where $A \subseteq \Rb^2$ and $\Eb g(X,Y)$ as follows
\begin{align}
\Pb( (X,Y) \in A)
	&=	\sum_{\{i,j : (x_i,y_j) \in A \}} f_{X,Y}(x_i,y_j) , &
\Eb g(X,Y)
	&=	\sum_{i,j} f_{X,Y}(x_i,y_j) g(x_i,y_j) .
\end{align}
We can obtain the probability mass function $f_X$ from the joint probability mass function $f_{X,Y}$ as follows
\begin{align}
f_X(x_i)
	&=	\sum_j f_{X,Y}(x_i,y_j) .
\end{align}
The \textit{conditional probability mass function of $X$ given $Y$}, written $f_{X|Y}$ is defined by
\begin{align}
f_{X|Y}(x_i , y_j)
	&:=	 \Pb( X = x_i | Y = y_j ) .
\end{align}
We can compute the conditional probability mass function $f_{X|Y}$ from the joint probability mass function $f_{X,Y}$ like so
\begin{align}
f_{X|Y}(x_i , y_j)
	&=	\frac{f_{X,Y}(x_i,y_j) }{ f_Y(y_j)} . \label{eq:fXgivenY}
\end{align}
Note that \eqref{eq:fXgivenY} follows from the definition of \textit{conditional probability}
\begin{align}
\Pb( A | B )
	&= \frac{\Pb( A \cap B )}{\Pb(B)} , &
\Pb(B)
	&> 0 ,
\end{align}
by taking $A = \{ X = x_i \}$ and $B = \{ Y = y_j \}$.
We can compute the conditional probability $\Pb(X \in A | Y = y_j)$ where $A \subseteq \Rb$ and the conditional expectation $\Eb[ g(X,Y) | Y = y_j ]$ as follows
\begin{align}
\Pb( X \in A | Y = y_j )
	&=	\sum_{ \{i : x_i \in A \}} f_{X|Y}(x_i,y_j) , &
\Eb[ g(X, Y ) | Y = y_j ]
	&=	\sum_i f_{X|Y}(x_i,y_j) g(x_i,y_j) . \label{eq:cond.disc}
\end{align}

\begin{example}
Suppose $X$ and $Y$ have a joint probability mass function given by
\begin{align}
f_{X,Y}(k,n)
	&=	q (1-q)^{n-1} \binom{n}{k} p^k (1-p)^{n-k} , &
k
	&\in \{ 0, 1, 2, \ldots, n \} , &
n
	&\in \{ 1, 2, 3, \ldots \} ,
\end{align}
where $p,q \in (0,1)$ and we have introduced the binomial coefficient
\begin{align}
\binom{n}{k}
	&:=	\frac{n!}{k!(n-k)!} .
\end{align}
First, let us compute $f_Y(n)$, the probability mass function of $Y$.  We have
\begin{align}
f_Y(n)
	&=	\sum_{k=0}^n f_{X,Y}(k,n)
	=		\sum_{k=0}^n q (1-q)^{n-1} \binom{n}{k} p^k (1-p)^{n-k} \\
	&=	q (1-q)^{n-1}  \sum_{k=0}^n \binom{n}{k} p^k (1-p)^{n-k}
	=   q (1-q)^{n-1} .
\end{align}
Using the expression for $f_Y$ we can obtain $f_{X|Y}$, the conditional probability mass function of $X$ given $Y = n$.  We compute
\begin{align}
f_{X|Y}(k,n)
	&=	\frac{ f_{X,Y}(k,n) }{ f_Y(n)}
	=		\frac{ q (1-q)^{n-1} \binom{n}{k} p^k (1-p)^{n-k} }{q (1-q)^{n-1}}
	=		 \binom{n}{k} p^k (1-p)^{n-k} .
\end{align}
Finally, let us compute $\Eb[ X | Y = n ]$, the conditional expectation of $X$ given $Y = n$.  We have
\begin{align}
\Eb[ X | Y = n ]
	&=	\sum_{k=0}^n k \, f_{X|Y}(k,n) 
	=		\sum_{k=0}^n k \, \binom{n}{k} p^k (1-p)^{n-k}
	=		n p .
\end{align}
Again, if you cannot follow all of the algebra above, check that the above equations are true for some simple cases.
\end{example}

\noindent
All of the above can be generalized to $n$ discrete random variables.


\subsection{Continuous random variables}
\label{sec:cont.rv}

\noindent
Now that we understand discrete random variables, it will be quite easy to understand continuous random variables.  The main difference between discrete and continuous random variables is simply that, where we saw sums $\sum$ in Section \ref{sec:disc.rv}, these will become integrals $\int$ in Section \ref{sec:cont.rv}.

\begin{definition}
\label{def:cont.rv}
A \textit{continuous random variable} $X$ is a random variable whose CDF $F_X$ can be written as
\begin{align}
F_X(x)
	&=	\int_{-\infty}^x f_X(y) \dd y , \label{eq:CDF2}
\end{align}
for some non-negative function $f_X : \Rb \to \Rb_+$ called the \textit{probability density function} (PDF).
\end{definition}

\noindent
It follows from \eqref{eq:CDF2} that
\begin{align}
f_X(x)
	&=	\frac{\dd F_X}{\dd x}(x) . \label{eq:fX=dFX}
\end{align}
We can think of \eqref{eq:fX=dFX} as a definition of the PDF $f_X$.  Alternatively, we can think of the PDF $f_X$ as describing the probability that $X$ is in some small interval $\dd x$, i.e.,
\begin{align}
f_X(x) \dd x
	&:=	\Pb( X \in \dd x ) . \label{eq:pdf}
\end{align}
Note once again that the density $f_X$ is defined with respect to a particular probability measure $\Pb$.
If we want to express the density of $X$ under a different probability measure $\Pbt$, we would have $\ft_X(x) = \dd \Ft_X(x) / \dd x$ and $\ft_X(x) \dd x = \Pbt( X \in \dd x)$. 

\begin{example}
Suppose $\Om = [0,1]$.  Let us define a random variable $X$ by $X(\om) := \om^2$.  Consider two probability measures $\Pb$ and $\Pbt$, defined as follows
\begin{align}
\Pb (\dd \om )
	&=	\dd \om , &
\Pbt (\dd \om )
	&=	2 \om \dd \om .
\end{align}
Note that both $\Pb(\Om) = 1$ and $\Pbt(\Om) = 1$, as they must, because
\begin{align}
\Pb(\Om)
	&=	\int_0^1 \Pb(\dd \om) = \int_0^1 \dd \om = \om \Big|_0^1 = 1 , \\
\Pbt(\Om)
	&=	\int_0^1 \Pbt(\dd \om) = \int_0^1 2 \om \dd \om = \om^2 \Big|_0^1 = 1 .
\end{align}
Let us find the density of $X$ under both $\Pb$ and $\Pbt$.  First, under $\Pb$ for any $x \in [0,1]$ we have
\begin{align}
F_X(x)
	&=	\Pb( X(\om) \leq x )
	=		\Pb( \om^2 \leq x ) 
	=		\Pb( \om \leq \sqrt{x} ) \\
	&=		\int_0^{\sqrt{x}} \Pb(\dd \om ) 
	=		\int_0^{\sqrt{x}} \dd \om
	=		\sqrt{x} , \\
f_X(x)
	&=	\frac{\dd F_X}{\dd x}(x)
	=		\frac{1}{2\sqrt{x}} .
\end{align}
Now, under $\Pbt$ for any $x \in [0,1]$ we have
\begin{align}
\Ft_X(x)
	&=	\Pbt( X(\om) \leq x )
	=		\Pbt( \om^2 \leq x ) 
	=		\Pbt( \om \leq \sqrt{x} ) \\
	&=	\int_0^{\sqrt{x}} \Pbt(\dd \om ) 
	=		\int_0^{\sqrt{x}} 2 \om \dd \om
	=		x , \\
\ft_X(x)
	&=	\frac{\dd \Ft_X}{\dd x}(x)
	=		1 .
\end{align}
\end{example}

\noindent
Using the probability density function $f_X$ we can compute $\Pb(X \in A)$ where $A \subseteq \Rb$ and the \textit{expectation} $\Eb g(X)$ as follows
\begin{align}
\Pb( X \in A)
	&=	\int_A f_X(x) \dd x, &
\Eb g(X)
	&=	\int f_X(x) g(x) \dd x . \label{eq:E.disc2}
\end{align}
Compare \eqref{eq:E.disc2} to \eqref{eq:E.disc} and note the similarity.  Observe once again that, as $\Eb g(X)$ depends on $f_X$, which depends on $\Pb$, we have that $\Eb g(X)$ also depends on the probability measure $\Pb$.  
If we wish to take an expectation under a different probability measure such as $\Pbt$, we will use the notation $\Ebt$.  For example
\begin{align}
\Pbt( X \in A)
	&=	\int_A \ft_X(x) \dd x, &
\Ebt g(X)
	&=	\int \ft_X(x) g(x) \dd x .
\end{align}
where $\ft_X(x) \dd x = \Pbt(X \in \dd x)$.

\begin{example}
An \textit{exponential random variable} $\text{Exp}(\lam)$ is a continuous random variable $X$ whose probability density function $f_X$ is given by
\begin{align}
f_X(x)
	&=	\Ib_{\Rb_+}(x) \lam \ee^{- \lam x} , 
\end{align}
where $\lam > 0$.  Here, we have introduced the \textit{indicator function} $\Ib_A$, which is defined as follows
\begin{align}
\Ib_A(x)
	&:=	\left\{ \begin{aligned}
			1 && x \in A , \\
			0 && x \notin A .
			\end{aligned} \right.
\end{align}
Note that, due to the presence of the indicator function, we have that $f_X(x) = 0$ for all $x < 0$.
Let us compute $\Pb(X \in (a,b))$ with $0 \leq a \leq b$ and $\Eb X^2$.  Using \eqref{eq:E.disc2} we have
\begin{align}
\Pb(X \in (a,b))
	&=	\int_a^b \lam \ee^{- \lam x} \dd x
	=		- \ee^{- \lam b} + \ee^{- \lam a}, \\
\Eb X^2
	&=	\int_0^\infty \lam \ee^{- \lam x} x^2 \dd x 
	=		\frac{-\ee^{-\lam x}}{\lam^2 } \Big( \lam^2 x^2+2 \lam  x+2 \Big) \Big|_{0}^{\infty}
	=		\frac{2}{\lam^2 } .
\end{align}
As a check of your understanding, see if you can compute $F_X(x)$ and $\Eb X^3$ (or at least write the integrals that would need to be computed).
\end{example}

\noindent
When the value of one continuous random variable $X$ affects the value of another continuous random variable $Y$, in order to describe this interaction, we will need to consider the \textit{joint probability density function}.

\begin{definition}
\label{def:joint.cont.rv}
Two random variables $X$ and $Y$ are \textit{jointly continuous} if their joint CDF $F_{X,Y}$ can be written as
\begin{align}
F_{X,Y}(x,y)
	&=	\int_{-\infty}^x \int_{-\infty}^y f_{X,Y}(u,v) \dd u \dd v, \label{eq:joint.CDF2}
\end{align}
for some non-negative function $f_{X,Y} : \Rb^2 \to \Rb_+$ called the \textit{joint probability density function} (PDF).
\end{definition}

\noindent
It follows from \eqref{eq:joint.CDF2} that
\begin{align}
f_{X,Y}(x,y)
	&=	\frac{\d^2 F_{X,Y}}{\d x \d y}(x,y) . \label{eq:joint.fX=dFX}
\end{align}
We can think of \eqref{eq:joint.fX=dFX} as a definition of the joint PDF $f_{X,Y}$.  Alternatively, we can think of the joint PDF $f_{X,Y}$ as describing the probability that $X$ is in some small interval $\dd x$ and $Y$ is in some interval $\dd y$, i.e.,
\begin{align}
f_{X,Y}(x,y) \dd x \dd y
	&:=	\Pb( X \in \dd x , Y \in \dd y) . \label{eq:joint.pdf}
\end{align}

\noindent
Using the joint probability density function $f_{X,Y}$ we can compute $\Pb((X,Y) \in A)$ where $A \subseteq \Rb^2$ and $\Eb g(X,Y)$ as follows
\begin{align}
\Pb( (X,Y) \in A)
	&=	\int_A f_{X,Y}(x,y) \dd x \dd y, &
\Eb g(X,Y)
	&=	\int f_{X,Y}(x,y) g(x,y) \dd x \dd y .
\end{align}
We can obtain the probability density function $f_X$ from the joint probability density function $f_{X,Y}$ as follows
\begin{align}
f_X(x)
	&=	\int f_{X,Y}(x,y) \dd y .
\end{align}
The \textit{conditional probability density function} $f_{X|Y}$ is defined in terms of infinitesimal probabilities as follows
\begin{align}
f_{X|Y}(x,y) \dd x
	&:= \Pb( X \in \dd x | Y = y ) .
\end{align}
From the joint probability density function $f_{X,Y}$ we can compute the conditional probability density function $f_{X|Y}$ as follows
\begin{align}
f_{X|Y}(x , y)
	&=	\frac{f_{X,Y}(x,y) }{ f_Y(y)} . \label{eq:fXgivenY2}
\end{align}
To compute the conditional probability $\Pb(X \in A | Y = y)$ where $A \subseteq \Rb$ and the conditional expectation $\Eb[ g(X,Y) | Y = y ]$ we use
\begin{align}
\Pb( X \in A | Y = y )
	&=	\int_A f_{X|Y}(x,y) \dd x, &
\Eb[ g(X, Y ) | Y = y ]
	&=	\int f_{X|Y}(x,y) g(x,y) \dd x. \label{eq:cond.cont}
\end{align}

\begin{example}
Suppose $X$ and $Y$ have a joint density given by
\begin{align}
f_{X,Y}(x,y)
	&=	\Ib_{\Rb_+^2}(x,y) \frac{1}{y} \exp \Big( - y - \frac{x}{y} \Big) , 
\end{align}
where the indicator function $\Ib_{\Rb_+^2}$ indicates that $f_{X,Y}(x,y) = 0$ if either $x<0$ or $y<0$.
Let us compute the probability density $f_Y$.  We have
\begin{align}
f_Y(y)
	&=	\int_0^\infty f_{X,Y}(x,y) \dd x 
	=	\int_0^\infty \frac{1}{y} \exp \Big( - y - \frac{x}{y} \Big) \dd x \\
	&=	-\exp \Big( - y - \frac{x}{y} \Big) \Big|_{x=0}^{x = \infty} 
	=	\ee^{-y} .
\end{align}
Next, let us compute $f_{X|Y}(x,y)$, the conditional density of $X$ given $Y = y$.  We have 
\begin{align}
f_{X|Y}(x,y)
	&=	\frac{ f_{X,Y}(x,y) }{f_Y(y)}
	=		\frac{\frac{1}{y} \exp \Big( - y - \frac{x}{y} \Big)}{\ee^{-y}}
	=		\frac{1}{y} \exp \Big( - \frac{x}{y} \Big) .
\end{align}
We can compute the conditional probability $\Pb( x_1 < X < x_2 | Y = y )$ where $0 < x_1 < x_2 < \infty$ as follows
\begin{align}
\Pb( x_1 < X < x_2 | Y = y )
	&=	\int_{x_1}^{x_2} f_{X|Y}(x,y) \dd x
	=		\int_{x_1}^{x_2} \frac{1}{y} \exp \Big( - \frac{x}{y} \Big)  \dd x \\
	&=	- \exp \Big( - \frac{x}{y} \Big)  \Big|_{x = x_1}^{x = x_2} 
	=		\exp \Big( - \frac{x_1}{y} \Big) - \exp \Big( - \frac{x_2}{y} \Big) .
\end{align}
Lastly, we compute  $\Eb( X | Y = y )$, the conditional expectation of $X$ given $Y=y$.  We have
\begin{align}
\Eb[ X | Y = y ]
	&=	\int_0^\infty x \, f_{X|Y}(x,y) \dd x
	=		\int_0^\infty \frac{x}{y} \exp \Big( - \frac{x}{y} \Big) \dd x \\
	&=	-( x + y ) \exp \Big( - \frac{x}{y} \Big) \Big|_{x=0}^{x = \infty} 
	=		y .
\end{align}
\end{example}

\noindent
All of the above can be generalized to $n$ continuous random variables.


\section{Independence}
\label{sec:independence}
Intuitively, two events $A$ and $B$ are independent if the outcome of $A$ has no impact on the outcome of $B$, and vice versa.  We can define independence mathematically as follows:

\begin{definition}
\label{def:ind.sets}
Two sets $A$ and $B$ are \textit{independent} if
\begin{align}
\Pb(A \cap B)
	&=	\Pb(A) \cdot \Pb(B) ,
\end{align}
and we write $A \ind B$.
\end{definition}

\noindent
Independent random variables are defined in a similar fashion.

\begin{definition}
\label{def:ind.rvs}
Two random variables $X$ and $Y$ are said to be \textit{independent} if
\begin{align}
\Pb( X \in A, Y \in B )
	&=	\Pb(X \in A) \cdot \Pb( Y \in B ) ,
\end{align}
for all sets $A,B \subseteq \Rb$, and we write $X \ind Y$.
\end{definition}

\noindent
It follows from Definition \ref{def:ind.rvs} that
\begin{align}
f_{X,Y}(x,y)
	&=	f_X(x) f_Y(y) , &
	&\text{and}&
\Eb g(X) h(Y) 
	&=	\Eb g(X) \cdot \Eb h(Y) , &
	&\text{if} &
X 
	&\ind Y ,
\end{align}
where $f_{X,Y}$, $f_X$ and $f_Y$ may be probability mass functions (if $X$ and $Y$ are discrete random variables) or probability density functions (if $X$ and $Y$ are continuous random variables).


\section{Conditional Expectation revisited}
\label{sec:expect}
At times, we may wish to compute the conditional expectation of $g(X,Y)$ given $Y$ without specifying the value of $Y$.  For this, we define
\begin{align}
\Eb[ g(X,Y) | Y ]
	&:=	\psi(Y) , &
	&\text{with}&
\psi(y)
	&:=	\Eb[ g(X, Y) | Y = y ] , \label{eq:cond.gen}
\end{align}
where expressions for $\Eb[ g(X, Y) | Y = y ]$ are given in \eqref{eq:cond.disc} for discrete random variables and \eqref{eq:cond.cont} for continuous random variables, respectively.  Note that $\Eb[ g(X,Y) | Y ]$ is a random variable $\psi(Y)$ whereas $\Eb[ g(X, Y) | Y = y ]$ is simply a number $\psi(y)$.  Conditional expectation, as we have defined it in \eqref{eq:cond.gen} has the following useful properties
\begin{align}
&\text{Tower Property}:&&
\Eb g(X,Y)
	 =	\Eb \Eb[ g(X,Y) | Y ] , \label{eq:tower} \\
&\text{Taking out what is known}:&&
\Eb[ g(X) h(Y) | Y ]
	 =	 h(Y) \Eb[ g(X) | Y ] , \label{eq:known} \\
&\text{If $X \ind Y$ then}: &&
\Eb[ g(X) | Y ]
	 =	\Eb g(X) . \label{eq:E.ind}
\end{align}


\section{Stochastic Processes}

Throughout this course, we will encounter various processes that evolve randomly in time.  Such processes are known as \textit{Stochastic processes}.  Let us define exactly what a stochastic process is.

\begin{definition}
A \textit{stochastic process} is a collection of random variables $X = (X_t)_{t \in \mathds{T}}$ indexed by some set $\mathds{T}$. 
If $\mathds{T}$ is countable (e.g., $\mathds{T} = \Nb_0 := \{0, 1, 2, \ldots \}$), then the process $X$ is a \textit{discrete time stochastic process}. 
If $\mathds{T}$ is uncountable (e.g., $\mathds{T} = \Rb_+ := [0, \infty)$), then the process $X$ is a \textit{continuous time stochastic process}.
The \textit{state space} $S$ of a stochastic process $X$ is the union of the state spaces of $(X_t)_{t \in \mathds{T}}$.
\end{definition}

\noindent
Note that $X$ is a \textit{process} whereas $X_t$ is the \textit{value} of the process $X$ at time $t$.
We can think of a stochastic process $X : \mathds{T} \times \Om \to \Rb$ in (at least) two ways.  First, for any $t \in \mathds{T}$ we have that $X_t: \Om \to \Rb$ is random variable.  Second, for any $\om \in \Om$, we have that $X(\om) : \mathds{T} \to \Rb$ is a function of time.  Both interpretations can be useful.
Let us take a look at a very simple discrete-time stochastic process.

\begin{example}[Random walk]
Consider a random experiment in which we toss a coin $3$ times.
The sample space $\Om$ of our experiment is
\begin{align}
\Om
	&:=	\{ (HHH), (HHT), (HTH), (THH), (TTH), (THT), (HTT), (TTT) \} ,
\end{align}
Note that the order of the coin tosses matters here; the sequence $(THT)$ is \textit{not} equivalent to $(TTH)$.
A single outcome of $\Om$ will be denoted as $\om =	\om_1 \om_2 \om_3$, where $\om_i$ is the result of the $i$th coin toss.  
For example, one possible element is $\om =	H H T$.
We will assume that each of the coin tosses is independent and that $\Pb(\om_i = H) = p$ and $\Pb(\om_i = T) = 1-p =:q$.
For every $i \in \{ 1, 2, 3 \}$ let us define a random variable $B_i$ as follows
\begin{align}
B_i(\om)
	&=	\left\{ \begin{aligned}
			&+1 &&\text{if $\om_i=H$} , \\
			&-1 &&\text{if $\om_i=T$} .
			\end{aligned} \right.
\end{align} 
Now, let us define a stochastic process $X = (X_i)_{i\in\{0,1,2,3\}}$ as follows
\begin{align}
X_0
	&=	0 , &
X_n(\om)
	&=	\sum_{i=1}^n B_i(\om) , &
n
	&\in \{1, 2, 3 \} .
\end{align}
Clearly, $X$ is a discrete time stochastic process with state space $S = \{-3,-2,-1,0,1,2,3\}$.
Suppose $\om = (HTT)$.
Then we have $B_1 = 1$, $B_2=-1$ and $B_3=-1$.
It follows that $X_1 = 1$, $X_2 = 0$ and $X_3=-1$.
Suppose we fix the time $n=2$.  Then we have
\begin{align}
\Pb(X_2 = 2)
	&=	\Pb( \om_1 = H \cap \om_2 = H ) = p^2 , \\
\Pb(X_2 = 0)
	&=	\Pb( \{\om_1 = H \cap \om_2 = T \} \cup \{ \om_1 = T \cap \om_2 = H \} ) = 2 qp , \\
\Pb(X_2 = -2)
	&=	\Pb( \om_1 = T \cap \om_2 = T ) = q^2 .
\end{align}
If you are able to understand this simple example very well, you will easily be able to understand more complicated stochastic processes as we encounter them.
\end{example}

\noindent
There are a number of properties that a stochastic process $X$ may exhibit.  Two properties that are very important are the \textit{Markov property} and the \textit{Martingale property}, which we now define.

\begin{definition}[Markov]
\label{def:markov}
Let $X = (X_t)_{t \in \mathds{T}}$ be a stochastic process.
We say that \textit{$X$ is a Markov process} (or \textit{$X$ has the Markov property}) if, for any function $g$, we have
\begin{align}
\Eb[ g(X_T) | \Fc_t^X ]
	&=	\Eb[ g(X_T) | X_t ] , &
	&\text{for all}&
0
	&\leq t \leq T < \infty , \label{eq:markov.prop}
\end{align}
where $\Fc_t^X := \{ X_s, 0 \leq s \leq t \}$ is all of the information generated by observing the path of $X$ over the time interval $[0,t]$.
We call $\Fc^X := (\Fc_t^X)_{ t \in \mathds{T} }$ the \textit{history of $X$} or \textit{filtration generated by $X$}.
\footnote{Strictly speaking, $\Fc_t^X$ is the sigma algebra generated by observing the path of $X$ up to time $t$.  But, we need not get into this level of mathematical rigor for this course.}
\end{definition}

\noindent
Suppose $X$ is a discrete-time process.  In this case the filtration generated by $X$ is simply $\Fc_n^X = \{ X_0, X_1, \ldots, X_n \}$ and the Markov property \eqref{eq:markov.prop} becomes
\begin{align}
\Eb[ g(X_N) | X_0, X_1, \ldots, X_n ]
	&=	\Eb[ g(X_N) | X_n ] , &
	&\text{for all}&
0
	&\leq n \leq N < \infty . \label{eq:markov.disc}
\end{align}
Equation \eqref{eq:markov.disc} makes it clear that a Markov process is a process $X$ for which, given the  present $X_n$ the future $g(X_N)$ is independent of the past $\{X_0, X_1, \ldots X_{n-1} \}$.
Note that \textit{this does not mean the future is independent of the past}!    The past obviously affects the present state of a process.  But, for a Markov process, only the present state -- not how we arrive at the present state -- affects the future.  Markov processes are useful because they simplify the computation of conditional expectations greatly.
\\[0.5em]
An alternative way to state the Markov property is that, for any $0 \leq t \leq T < \infty$ and any function $g$, there exists a function $h$ (which depends on $t$, $T$ and $g$) such that
\begin{align}
\Eb[ g(X_T) | \Fc_t^X ]
	&= h(X_t) .
\end{align}

\begin{example}[Random Walk]
\label{ex:rw}
Consider a random experiment in which we toss a coin $N$ times.
The sample space $\Om$ of our experiment is all sequences of heads and tails that are of length $N$.
Thus, a generic outcome $\om = \om_1 \om_2 \ldots \om_N$ where $\om_i \in \{H,T\}$.
As usual, we will assume coin tosses are independent.
For every $i \in \{ 1, 2, \ldots, N \}$ let us define a random variable $B_i$ as follows
\begin{align}
B_i(\om)
	&=	\left\{ \begin{aligned}
			&+1 &&\text{if $\om_i=H$} , \\
			&-1 &&\text{if $\om_i=T$} .
			\end{aligned} \right.
\end{align} 
Now, let us define a stochastic process $X = (X_n)_{n \in \{0,1,2,\ldots,N\}}$ as follows
\begin{align}
X_0(\om)
	&=	0 , &
X_n(\om)
	&=	\sum_{i=1}^n B_i(\om) , &
n
	&\in \{1, 2, \ldots, N \} .
\end{align}
We can clearly see that $X$ is a Markov process because, for any $0 \leq i \leq j \leq N$ we have
\begin{align}
\Eb[ g(X_j) | \Fc_i^X ]
	&=	\Eb[ g(X_j) | X_0, \ldots, X_i ] \\
	&=	\Eb[ g( X_j - X_i + X_i) | X_0, \ldots, X_i ] \\
	&=	\Eb[ g( Y + X_i) | X_i ] , &
Y
	&:=	X_j - X_i = \sum_{k=i+1}^j B_k , \\
	&=	\sum_{k} f_Y(k) g(k + X_i) =: h(X_i), &
f_Y(k)
	&:=	\Pb( Y = k) .
\end{align}
where we have used the fact that $Y$ depends only on $B_{i+1}, \ldots, B_j$, each of which are independent of $X_0, \ldots, X_i$.
\end{example}

\begin{definition}[Martingale]
\label{def:martingale}
Let $X = (X_t)_{t \in \mathds{T}}$ and $Y=(Y_t)_{t \in \mathds{T}}$ be stochastic processes.
We say that \textit{$X$ is a martingale with respect to $\Fc^Y$} (or \textit{$X$ has the martingale property with respect to $\Fc^Y$}) if
\begin{align}
\Eb[ X_T | \Fc_t^Y ]
	&=	X_t , &
	&\text{for all}&
0
	&\leq t \leq T < \infty.
\end{align}
\end{definition}

\noindent
A few notes are in order.  First, a stochastic process $X$ can be a martingale with respect to itself.  In this case, we have
\begin{align}
\Eb[ X_T | \Fc_t^X ]
	&=	X_t , &
	&\text{where}&
0
	&\leq t \leq T < \infty .
\end{align}
Second, observe that the martingale property is defined with respect to a particular probability measure $\Pb$ (because the expectation $\Eb$ is taken with respect to $\Pb$).  It is possible, for example, that $X$ is a martingale with respect to $\Fc^Y$ under $\Pb$ but that $X$ is not a martingale with respect to $\Fc^Y$ under $\Pbt$.  That is, we may simultaneously have
\begin{align}
\Eb[ X_T | \Fc_t^Y ]
	&=	X_t , &
	&\text{and}&
\Ebt[ X_T | \Fc_t^Y ]
	&\neq	X_t ,
\end{align}
Let us look at a simple example.

\begin{example}[Random Walk]
Let us return to the Random walk defined in Example \ref{ex:rw}.
Consider two probability measures $\Pb$ and $\Pbt$ defined as follows
\begin{align}
\Pb(\om_i = H)
	&=	p = 1/2 , &
\Pbt( \om_i = H)
	&=	\pt > 1/2 , &
\om_i
	&\ind \om_j , &
i
	&\neq j .
\end{align}
From the above specification, we have $\Pb( B_i = 1 ) = p = 1/2$ and $\Pbt( B_i = 1 ) = \pt > 1/2$.
The process $X$ is a martingale with respect to $\Fc^X$ under $\Pb$ because, for any $0 \leq i \leq j \leq N$ we have
\begin{align}
\Eb[ X_j | \Fc_i^X ]
	&=	\Eb[ B_j + B_{j-1} + \ldots + B_{i+1} + X_i | \Fc_i^X ] \\
	&=		\Eb[ B_j + B_{j-1} + \ldots + B_{i+1} ] + X_i 
	=		 X_i  ,
\end{align}
where we have used the fact that, for $k>i$, we have $\Eb[ B_k | \Fc_i^X ] = \Eb B_k = 0$ and $\Eb[ X_i | \Fc_i^X ] = X_i$.
On the other hand, $X$ is not a martingale under $\Pbt$ because, for any $0 \leq i < j \leq N$ we have
\begin{align}
\Ebt[ X_j | \Fc_i^X ]
	&=	\Ebt[ B_j + B_{j-1} + \ldots + B_{i+1} + X_i | \Fc_i^X ] \\
	&=		\Ebt[ B_j + B_{j-1} + \ldots + B_{i+1} ] + X_i 
	=		(2 \pt - 1 ) (j - i) + X_i 
	>		X_i .
\end{align}
where we have used the fact that, for $k > i$ we have $\Ebt[ B_k | \Fc_i^X ] = \Ebt B_k = 2\pt-1$ and $\Ebt[ X_i | \Fc_i^X ] = X_i$.
To check your understanding, try to answer the following question: is $X$ a martingale with respect to $\Fc^B$ under $\Pb$ and/or $\Pbt$?
\end{example}

\noindent
\textit{Please do not confuse the Markov property and the Martingale property}; they are entirely separate concepts.
It is possible for a process to be a Markov process and not be a martingale.
It is possible for a process to be a Martingale but not be a Markov process.
It is possible for a process to be both a Markov process and a Martingale.
And it is possible for a process to be neither a Markov process nor a Martingale.

\begin{remark}
At times, we use the simplified notation $\Eb_n$ to indicate an expectation that is conditioned on a history $\Fc_n$.  That is
\begin{align}
\Eb_n Z 
	:=	\Eb[ Z | \Fc_n ]
\end{align}
for some random variable $Z$.
\end{remark}


\section{Exercises}
\label{sec:exer-A}

\begin{exercise}
\label{exer:A-1}
Consider the random variable $X$ defined in Example \ref{ex:die.roll.rv}.  Compute (a) $\Eb X$, (b) $\Eb \ee^X$, (c) $\Ebt X$ and (d) $\Ebt \ee^X$.
\end{exercise}

\begin{exercise}
\label{exer:A-2}
Consider a random experiment in which you toss two distinguishable coins.  The set of possible outcomes is
\begin{align}
\Om
	&=	\{ (HH), (HT), (TH), (TT) \} .
\end{align}
We will denote by $\om = \om_1 \om_2$ an element of $\Om$.  For example, if $\om = (HT)$ then $\om_1 = H$ and $\om_2 = T$.  Now, let us consider two random variables $S_1$ and $S_2$, which are defined as follows
\begin{align}
S_1
	&=	\left\{ \begin{aligned}
			u S_0 &&\text{if $\om_1 = H$} , \\
			d S_0 &&\text{if $\om_1 = T$} , \\
			\end{aligned} \right.  &
S_2
	&=	\left\{ \begin{aligned}
			u S_1 &&\text{if $\om_2 = H$} , \\
			d S_1 &&\text{if $\om_2 = T$}  , \\
			\end{aligned} \right. 
\end{align}
where $0 < d < u < \infty$ and $S_0 > 0$.  Suppose that the two coin tosses are independent so that
\begin{align}
\Pb( \om_1 = H , \om_2 = H )
	&=	\Pb( \om_1 = H ) \cdot \Pb( \om_2 = H ) .
\end{align}
Suppose further that $\Pb(\om_i = H) = p$ for $i = 1,2$ where $p \in (0,1)$.  For convenience, define $q := 1-p$.
(a) What is the joint probability mass function $f_{S_1,S_2}(s_1,s_2)$?
(b) What is the conditional probability mass function $f_{S_1|S_2}(s_1,s_2)$?
(c) What is the conditional probability mass function $f_{S_2|S_1}(s_2,s_1)$?
(d) What is $\Eb (S_2 / S_1)$?
(e) What is $\Eb [ (S_2 / S_1) | S_1 ]$?
\textit{Hint}: before answering part (a), you may want to write out the state spaces of $S_1$ and $S_2$.  Once you have done that, write an expression for $f_{S_1,S_2}(s_1,s_2)$ for each possible combination of $s_1$ and $s_2$.  Note that for some combinations it may be the case that $f_{S_1,S_2}(s_1,s_2) = 0$.
\end{exercise}

\begin{exercise}
\label{exer:A-3}
Expectation $\Eb$ is what is known as a \textit{linear operator}, meaning
\begin{align}
\Eb[ a X + b Y ]
	&=	a \Eb X + b \Eb Y , \label{eq:E.linear}
\end{align}
where $a,b \in \Rb$ are constants.  Prove \eqref{eq:E.linear} when $X$ and $Y$ are (a) discrete random variables, and (b) continuous random variables.
\end{exercise}

\begin{exercise}
\label{exer:A-4}
The \textit{variance} of a random variable $X$, denoted $\Vb X$ is defined as follows
\begin{align}
\Vb X
	&:=	\Eb[ (X - \Eb X)^2 ] .	\label{eq:variance}
\end{align}
The \textit{covariance} of two random variables $X$ and $Y$, denoted $\text{Co}\Vb[X,Y]$, is defined as follows
\begin{align}
\text{Co}\Vb[X,Y]
	&:=	\Eb[ (X - \Eb X)(Y - \Eb Y) ] .	\label{eq:covariance}
\end{align}
Using \eqref{eq:E.linear}, show (a) that
\begin{align}
\Vb X
	&=	\Eb X^2 - ( \Eb X )^2 ,
\end{align}
and (b) that
\begin{align}
\text{Co}\Vb[X,Y]
	&=	\Eb XY - \Eb X \cdot \Eb Y .
\end{align}
Lastly, (c) what is $\text{Co}\Vb[X,Y]$ when $X \ind Y$?
\end{exercise}

\begin{exercise}
\label{exer:A-5}
Prove equations \eqref{eq:tower}, \eqref{eq:known} and \eqref{eq:E.ind} when $X$ and $Y$ are (a) discrete random variables, and (b) continuous random variables.
\end{exercise}

\begin{exercise}
\label{exer:A-6}
Let $X$ and $Y$ be independent uniformly distributed continuous random variables on the intervals $[a,b]$ and $[c,d]$ respectively. Find $\Eb[1/X]$, $\Eb[1/Y]$ and $\Eb[X/Y]$.  You may assume $0 < a < b < \infty$ and $0 < c < d < \infty$.

\end{exercise}




\chapter{What is a financial derivative?}
\begin{definition}
A \textit{derivative} is a financial instrument whose value depends on the value of more basic underlying variables.
\end{definition}

\noindent
Typical \textit{underlyers} are traded assets such as stocks, bonds and currencies.  But underlyers can also be non-traded quantities such as interest rates, exchange rates, average temperatures (yes, weather derivatives really exist!) or a financial indices.
\\[0.5em]
The use of derivatives in finance is so wide-spread that everyone who works in the financial sector must know how they are used and how they are valued.  To give you an idea of just how wide-spread the use of financial derivatives has become, consider that the Chicago Board Options Exchange (CBOE, \url{cboe.com}) trades options, which are types of derivatives, on 100+ stocks.  The options written on certain stocks are even more liquidly traded than the stocks themselves.

\section{Forward Contracts}
The most basic financial derivative is a \textit{forward contract}.
\begin{definition}
A \textit{forward contract} is an agreement between two parties to buy/sell an asset at a certain future date $T$ for a certain price $K$.  The \textit{long side} agrees to buy the asset at time $T$ for price $K$.  The \textit{short side} agrees to sell the asset at time $T$ for price $K$.  The date $T$ is referred to as the \textit{maturity} and the agreed upon price $K$ is referred to as the \textit{delivery price} or \textit{strike}.
The delivery price $K$ is fixed at inception so that the initial value of the forward contract is zero.
\end{definition}
\noindent
Forward contracts can be used to \textit{hedge}, i.e., reduce exposure to the future movements of an underlyer.

\begin{example}[Hedging with long side of a forward contract]
An airline wants to sell tickets for a flight from NYC to London in 3 months.  In order to know if operating the flight will be profitable the airline needs to know how much jet fuel will cost in 3 months.  Obviously, we cannot say today what the price of jet fuel will be in 3 months.  So, the airline is exposed to price fluctuations in the price of jet fuel.  By taking the long side of a forward contract written on jet fuel the airline fixes the price it will pay for jet fuel in 3 months' time.  This allows the airline to figure out today if the flight from NYC to London will be profitable.
\end{example}

\begin{example}[Hedging with short side of a forward contract]
\label{ex:copper_mine_1}
Let $S = (S_t)_{t \geq 0}$ denote the value of one unit of copper and suppose it costs $c$ dollars to excavate and refine a unit of copper.
Let $T$ denote the time it takes to excavate and refine one unit of copper.
If the mine takes the short side of a forward contract on copper, with a maturity date $T$ and delivery price $K$, then the profit/loss the mine incurs when it sells the copper at time $T$ will be $K - c$, which is a constant.
On the other hand, if the mine does not enter into the short side of a forward contract, then the profit/loss incurred by the mine for selling a unit of copper at time $T$ will be $S_T - c$, which is a random number.  If $S_T < c$ the mine will lose money.  If $S_T > c$ the mine will make money. 
Thus, by taking the short side of a forward contract, the copper mine is sacrificing possible future profit in order to reduce risk (i.e., uncertainty).
\end{example}

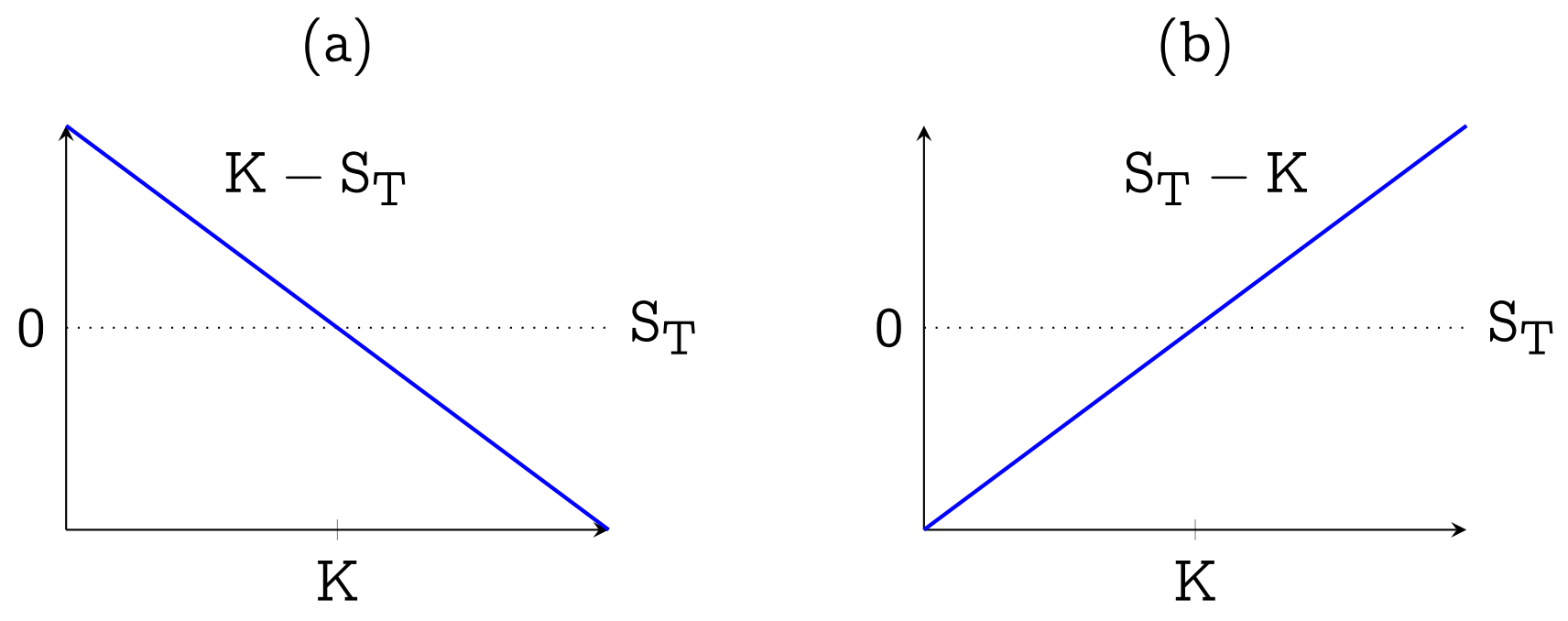
\begin{figure}
    \centering
    \begin{tikzpicture}
\begin{axis}[payoff, title={(a)}, xtick={1}, xticklabels={$K$}]
  \zeroline{$S_T$}
  \addplot coordinates {(0,1) (2,-1)} node[pos=0.25, above right, black] {$K - S_T$};
\end{axis}
\end{tikzpicture}
    \qquad
\begin{tikzpicture}
\begin{axis}[payoff, title={(b)}, xtick={1}, xticklabels={$K$}]
  \zeroline{$S_T$}
  \addplot coordinates {(0,-1) (2,1)} node[pos=0.75, above left, black] {$S_T - K$};
\end{axis}
\end{tikzpicture}
    \caption{(a) Short side payoff. (b) Long side payoff.\label{fig:plot1_3}}
\end{figure}

\noindent
Note that you do \textit{not} need to own the underlying asset $S = (S_t)_{0 \leq t}$ before you enter the short side of a forward contract. 
If you do not own the asset when you enter the short side of a forward contract, you can buy the underlying at market value $S_T$ at the maturity date $T$ and sell it for the strike price $K$.  Thus, the final value of the short side of a forward contract is $K - S_T$.  The final value of the long side of a forward contract at maturity date $T$ is $S_T - K$ because the long-side must buy the asset for $K$ and it can immediately sell the asset for the market price $S_T$.  Figure \ref{fig:plot1_3} plots the final value or \textit{payoff} of the long and short sides of a forward contract, as a function of the underlyer's final value $S_T$.
\\[0.5em]
You can think of entering a forward contract as a bet between two parties.  At maturity, the long side receives $S_T - K$ and the short side pays $S_T - K$ (keep in mind that $S_T - K$ may be negative).  The long and short side agree on the delivery price $K$ when the bet is placed so that the bet is fair in some sense.
\\[0.5em]
As previously mentioned, there is \textit{no cost} to entering a forward contract (i.e., the initial value is zero for both the long and short side by design).  
This raises an important question: \textit{What is the value of the strike price $K$ that makes the initial value of the forward contract zero}?  
We will answer this question shortly.
\\[0.5em]
Note that, even though one enters a forward contract at no cost, as the underlying asset's price $S$ changes over time, the value of the forward contract may become positive or negative.  Another question we will seek to answer in this course is: \textit{what the fair value of a forward contract should be at some time in the future}?  For example, at time $T$ (the maturity date) we know that the value of the long side of the forward contract is $S_T - K$.  But, what is the value of the forward contract at $t \in (0,T)$?

\section{Options}

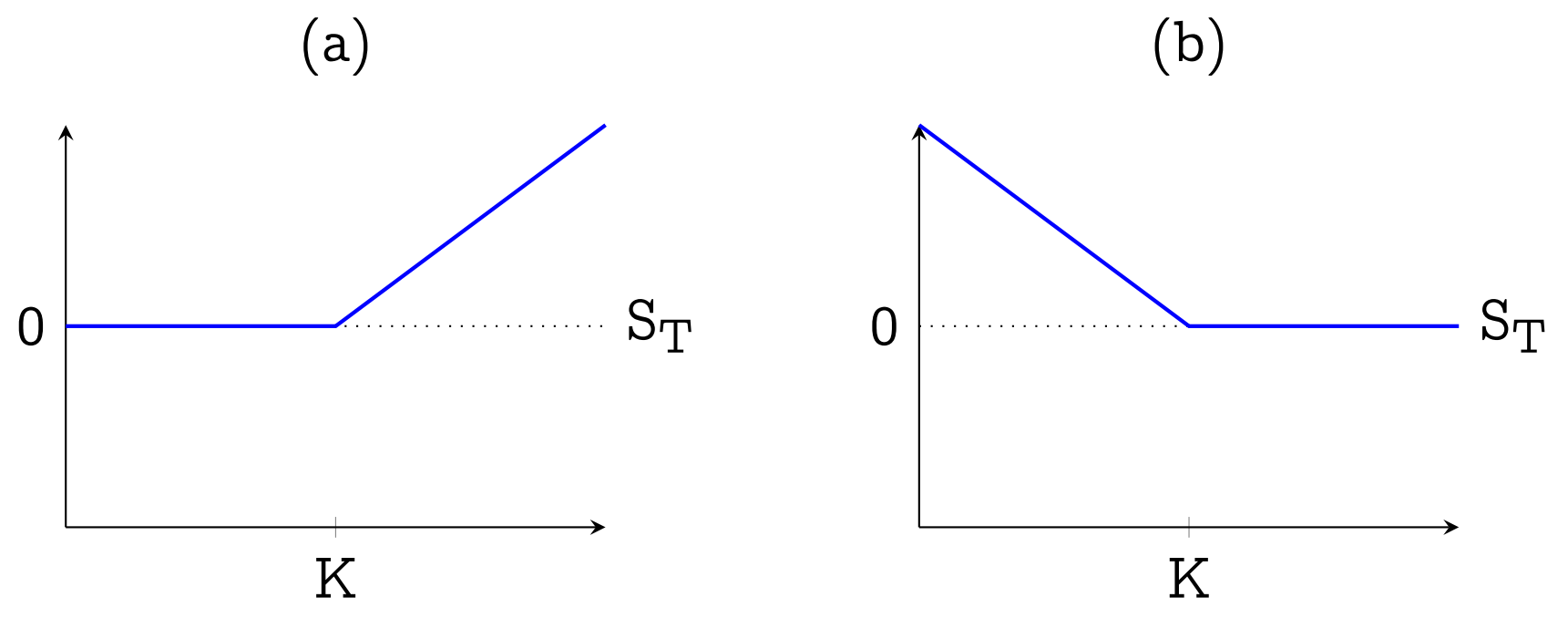
\begin{figure}
    \centering
    \begin{tikzpicture}
\begin{axis}[payoff, title={(a)}, xtick={1}, xticklabels={$K$}]
  \zeroline{$S_T$}
  \addplot coordinates {(0,0) (1,0) (2,1)};
\end{axis}
\end{tikzpicture}
    \qquad
\begin{tikzpicture}
\begin{axis}[payoff, title={(b)}, xtick={1}, xticklabels={$K$}]
  \zeroline{$S_T$}
  \addplot coordinates {(0,1) (1,0) (2,0)};
\end{axis}
\end{tikzpicture}
    \caption{(a) Call payoff. (b) Put payoff.\label{fig:plot1_4}}
\end{figure}

The two most basic types of options are \textit{Calls} and \textit{Puts}.
\begin{itemize}
	\item A \textit{Call option} gives its owner the right, but not the obligation, to \textit{buy} an underlying $S$ for a specified price $K$ at or before a specified date $T$.
	\item A \textit{Put option} gives its owner the right, but not the obligation, to \textit{sell} an underlying $S$ for a specified price $K$ at or before a specified date $T$.
\end{itemize}
We call $K$ the \textit{strike} or \textit{exercise price}.  We call $T$  the \textit{maturity} or \textit{expiration date}.  Calls and Puts can further be classified as either \textit{European} or \textit{American}.
\begin{itemize}
\item \textit{European} options can only be exercised at the maturity date $T$. 
\item \textit{American} options can be exercised at any time $t \leq T$.
\end{itemize}
Although not common, there also exist  \textit{Bermudan} options, which can be exercised only on fixed dates $t_1 < t_2 < t_3 < t_4 < \cdots \leq T$.
The owner of an option is said to be the \textit{buyer} or \textit{on the long side}. The seller of an option is said to be the \textit{writer}, \textit{seller} or \textit{on the short side}.
\\[0.5em]
Unlike a forward contract, which is a binding agreement to buy/sell an underlying $S$ at a specified price $K$, the owner of an option is not required to exercise his right. The owner of an option will only exercise his right if doing so yields a positive payoff.  Consider a European call option.  If at the maturity date $T$, the value of the underlying asset is larger than the strike price (i.e., $S_T > K$), then the owner can (and will) exercise his right to buy the asset for the strike price $K$ and sell the asset at the market price $S_T$.  The payoff in this case is $S_T - K$.  On the other hand, if the value of the underlying asset is less than the strike price (i.e., $S_T < K$), then the owner will not exercise his right to buy the asset because, rather than buy the asset for $K$ he could buy the asset at the market price $S_T$.  Thus, the value of the call option at time $T$ is $(S_T - K)^+$ where $x^+ := \max\{ x, 0\}$.  Similarly, the value of a European Put option at time $T$ is $(K - S_T)^+$.  Take a moment to reason for yourself why this is the case.  In Figure \ref{fig:plot1_4} we plot the final value or \textit{payoff} of a European Call and Put option as a function of the underlyer's final value $S_T$.  
\\[0.5em]
Note that the payoff of both calls and puts are non-negative.  As such, the values of calls and puts must also be non-negative. To own an option, one must \textit{pay} the option seller a \textit{premium} at inception.  
In return for receiving this premium, the seller of an option is obligated to deliver the value of the option to the buyer if and when the option is exercised.
One important question we seek to answer in this course is: \textit{what is the fair price to pay for an option}?

\section{Interest Rates and compounding}
It will be important throughout this course to understand various notions of interest rates.
Let $B = (B_t)_{t \geq 0}$ be the value of a bank account.
Let $r \geq 0$ be the interest rate offered by the bank per unit time.
Below are three different ways that the bank may choose to calculate interest.
\begin{itemize}
	\item \textit{Simple interest}: $B_t = (1 + rt) B_0$
	\item \textit{Compound interest} ($m$ times): $B_t = (1 + rt/m)^m B_0$
	\item \textit{Continuous interest}: $B_t = \lim_{m \rightarrow \infty} (1 + rt/m)^m B_0 = \ee^{rt} B_0$.
\end{itemize}
Unless stated otherwise, \textit{we will always consider continuously compounded interest rates}.  Note that
\begin{align}
(1 + rt) \leq (1 + rt/m)^m \leq \ee^{rt} ,
\end{align}
where the inequalities are strict if $r>0$ and $m > 1$.

\section{Types of traders}
There are three basic types of traders:
\begin{itemize}
	\item \textbf{Hedgers} seek to reduce risk from future movements of underlyings.
	\item \textbf{Speculators} seek to profit by making bets on future movements of underlyings.
	\item \textbf{Arbitrageurs} exploit mispricing in markets in order to make profit with little or no risk.
\end{itemize}
We will give a precise definition of \textit{arbitrage} later in this text.  For now, think of an arbitrage as an investment strategy that requires zero initial investment, never loses money, and yields a strictly positive profit with positive probability at some future date.
\\[0.5em]
Let us see some examples of hedging, speculating and arbitrage.

\begin{example}[Hedging with options]
Consider again the copper mine in Example \ref{ex:copper_mine_1}.  Let $S=(S_t)_{t \geq 0}$ be the price of copper.  Suppose $c$ is the cost to produce one unit of copper.  In order to hedge against copper prices possibly falling, the copper mine might buy a put contract at time $t = 0$ on copper with strike $K$ and maturity $T$ at a price $P_0(T,K)$.  The mine's profit/loss at time $T$ is 
\begin{align}
S_T + (K-S_T)^+ - ( c +  P_0(T,K) )
	&=	\left\{ \begin{aligned}
			& K - ( c + P_0(T,K) ) && \text{if $S_T \leq K$} , \\
			& S_T - ( c +  P_0(T,K) ) && \text{if $K < S_T $} .
			\end{aligned} \right. \\
	&\geq K - ( c +  P_0(T,K) ) .
\end{align}
Thus, the mine limits the downside of its profit/loss to $K - ( c +  P_0(T,K) )$ and leaves the upside unbounded (because there is no limit to how large $S_T$ can be).  Had the mine hedged by taking the short side of a forward contract with delivery price $K$ and maturity $T$, its profit/loss would have been $K - c$, which is a constant larger than $K- ( c +  P_0(T,K) )$ but possibly less than $S_T - ( c +  P_0(T,K) )$.  Whether the mine chooses to hedge with a Put option or a forward contract (or whether it wants to hedge at all) is a matter of the mine's attitude towards risk and view of the future.
\end{example}

\begin{example}[Speculating with options]
Let $S = (S_t)_{t \geq 0}$ be the price of IBM stock.
At time $t=0$, the price of IBM stock is $S_0$. You think the price may go up to $1.2 S_0$ by time $T$. You can speculate on this outcome in the following way:
\begin{itemize}
\item Buy a call option with maturity $T$ and strike $K < 1.2 S_0$ for a price of $C_0(T,K)$.
\item If the price $S_T > K$ then you can buy stock for $K$ by exercising the option and then sell the stock at the market price $S_T$.  In this case, your profit is $S_T - K - C_0(T,K)$
\item If $S_T < K$, you do not exercise your option because, rather than buy the stock for $K$ you could buy it for $S_T$.  In this case, you incur a loss of $C_0(T,K)$.
\end{itemize}
The overall profit/loss of your strategy is the final value of the option minus the price you paid for the call $(S_T - K)^+ - C_0(T,K)$.
\end{example}

\begin{example}[Arbitrage with a forward] \label{ex:forward_arb}
Let $S = (S_t)_{t \geq 0}$ be the price of a stock and let $r>0$ be the (continuously compounded) risk-free rate of interest (i.e., you can lend or borrow at this rate).  Assume the stock pays no dividends.  Suppose that, at time $t \leq T$, you can enter into a forward contract (long or short) with maturity $T$ and delivery price $K \neq S_t \ee^{r(T-t)}$.  We will consider two scenarios: $K > S_t \ee^{r(T-t)}$ and $K < S_t \ee^{r(T-t)}$.
\\[0.5em]
\underline{Case I}:  If $K > S_t \ee^{r(T-t)}$, then you can achieve a risk-free profit by doing the following at time $t$:
\begin{itemize}
\item Enter the short side of the forward contract,
\item Borrow $S_t$ from bank,
\item Buy the stock for $S_t$.
\end{itemize}
The total initial cost of this strategy is zero.
Then at time $T$, you
\begin{itemize}
\item Sell the stock for the delivery price $K > S_t \ee^{r(T-t)}$,
\item Repay the bank what you owe for the loan $S_t \ee^{r(T-t)}$.
\end{itemize}
Your profit is $K - S_t \ee^{r(T-t)} > 0$. Note that, with zero initial investment, you made a guaranteed profit!  This is an arbitrage.
\\[0.5em]
\underline{Case II}:  What would you do if $K < S_t \ee^{r(T-t)}$? At time $t$, you could
\begin{itemize}
\item Enter the long side of a forward contract,
\item Sell the stock for $S_t$ (you do not need to own stock to do this)
\item Put $S_t$ in bank
\end{itemize}
The total initial cost of this strategy is zero.
At time $T$, you close out your position as follows
\begin{itemize}
\item Buy back the stock for the delivery price $K < S_t \ee^{r(T-t)}$,
\item Withdraw the money you invested in the bank, which has grown to $S_t \ee^{r(T-t)}$.
\end{itemize}
In this case, your profit is $S_t \ee^{r(T-t)} - K > 0$. Once again, with zero initial investment, you made a guaranteed profit!  This is an arbitrage.
\end{example}

\noindent
We have derived the following result.

\begin{theorem}
\label{thm:fair-strike-forward}
Suppose the risk-free rate of interest is $r \geq 0$.  
Consider a forward contract with maturity date $T$ written on an asset $S = (S_t)_{t \geq 0}$ that pays no dividends.
Then, at time $t$, the no-arbitrage delivery price $K$ of the forward contract is $K = F_t^T := S_t \ee^{r (T - t)}$.  If $K \neq F_t^T$ there will be an arbitrage opportunity.  We call $F_t^T$ the \emph{fair forward price} or simply \emph{forward price}.
\end{theorem}

\noindent
\textbf{Warning!}  The phrase \textit{forward price} is very misleading.  The forward price $F_t^T$ is \textit{not} the value of a forward contract.  Rather, for a forward contract maturing at time $T$, the forward price $F_t^T$ is the unique value that the delivery price $K$ must be in order to preclude arbitrage.  Really, we should call $F_t^T$ the \textit{fair delivery price} of a forward contract.

\section{Present values of certain cash flows}
We will often talk about \textit{present values} of \textit{certain} or \textit{known} (i.e., non-random) cash flows.  
For example, you may know that your employer pays you on the 1st and the 15th of every month.  What are the present values of these cash flows today?
\\[0.5em]
Suppose you offer to sell me today at time $t$ a cash flow $\text{CF}^T$ to be paid at time $T$ for a price $P < \ee^{-r(T-t)} \text{CF}^T$.  Then, at time $t$, I could borrow $P$ from the bank and purchase the future cash flow.  My initial investment would be zero and, at time $T$, I would have  $\text{CF}^T - P \ee^{r(T-t)} = (\text{CF}^T \ee^{-r(T-t)} - P )\ee^{r(T-t)} >0$.  
This is an arbitrage opportunity.
\\[0.5em]
On the other hand, suppose that I offer to purchase today at time $t$ a cash-flow of $\text{CF}^T$ to be paid at time $T$ for a price $P > \ee^{-r(T-t)} \text{CF}^T$.  Then, at time $t$, you could sell the cash flow to me for $P$ and put your money in the bank.  Your initial investment is zero and, at time $T$, you would have $P \ee^{r(T-t)} - \text{CF}^T = \ee^{r(T-t)} (P - \ee^{-r(T-t)}\text{CF}^T ) > 0$.  Again, this is an arbitrage opportunity.
\\[0.5em]
If we assume there is no arbitrage in the market, then, at time $t \leq T$, the value of the cash flow $\text{CF}^T$ must be $\ee^{-r(T-t)}\textup{CF}^T$.  
We have derived the following result.

\begin{theorem}
\label{thm:present.value}
Assume the risk-free rate of interest is $r \geq 0$.  Denote by $\textup{PV} = (\textup{PV}_t)_{0 \leq t \leq T}$ the no-arbitrage \textbf{present value}  of a certain \textbf{cash flow} $\textup{CF}^T$, to be paid at time $T$.  Then we have
\begin{align}
\textup{PV}_t
	&=	\ee^{-r(T-t)}\textup{CF}^T ,
\end{align}
for any $t \leq T$.
\end{theorem}

\noindent
Keep in mind, Theorem \ref{thm:present.value} is only valid when the cash-flow $\text{CF}^T$ is a known, non-random constant.  You cannot apply Theorem \ref{thm:present.value} to cash-flows that are random.

\begin{definition}
A \textit{zero-coupon bond} pays one unit of currency at time $T$.
\end{definition}

\begin{theorem}
Let $B^T = (B_t^T)_{0 \leq t \leq T}$ denote the value of a zero-coupon bond maturing at time $T$.  Assume the risk free rate of interest is a constant $r>0$.  Then we have $B_t^T = \ee^{-r(T-t)}$.
\end{theorem}

\begin{proof}
The bond pays a certain cash flow of $1$ at time $T$.  The present value of this cash flow $\ee^{-r(T-t)}$ is the bond price.
\end{proof}

\section{The value of a forward contract after inception}
In this section, we will derive the value of the long side of a forward contract.  The value of the short-side of a forward contract is minus the value of the long side.  Throughout this section, we will use the following notation:
\begin{itemize}
	\item The maturity date of a forward contract is $T$.
	\item The underlying asset of the forward contract is $S = (S_t)_{t \geq 0}$.
	\item The risk-free rate of interest will be $r$.
	\item We denote by $F_t^T$ the fair forward price of a forward contract entered at time $t$.  Recall from Example \ref{ex:forward_arb} and Theorem \ref{thm:fair-strike-forward} that $F_t^T = \ee^{r(T-t)} S_t$ assuming the asset $S$ pays no dividends.
	\item We will denote by $f^T = (f_t^T)_{0 \leq t \leq T}$ the value of the long-side of a forward contract that was entered at time $t=0$.  As there is no cost to entering a forward contract, we know that $f_0^T = 0$.  We also know that the final value of the long-side of a forward contract is $f_T^T = S_T - F_0^T$ (because  $F_0^T$ was the fair forward price at time 0 when contract was entered into).  Our goal is to find the value of $f_t^T$ for $t \in (0,T)$.
\end{itemize}
We shall make the following assumptions:
\begin{itemize}
	\item There is no arbitrage in market.
	\item One can lend and borrow at the risk-free rate of interest $r$.
	\item There are no transaction costs.
\end{itemize}

\begin{theorem} 
\label{thm:fairforward}
Under the assumptions listed above, we have
\begin{align}
f_t^T 
	&= \( F_t^T - F_0^T \) \ee^{-r(T-t)}.
\end{align}
\end{theorem}

\begin{proof}
Suppose you enter the long side of a forward contract at time zero.  At time $t$, you enter the short side of a forward contract with maturity $T$.
The value of your portfolio at time $t$ is
\begin{align}
\underbrace{f_t^T}_{\text{value of long side of forward entered at time zero}} + \underbrace{0}_{\text{value of short side of forward entered at time $t$}}
\end{align}
The value of your portfolio at time $T$ is
\begin{align}
\underbrace{\( S_T - F_0^T \) }_{\text{value of long side of forward entered at time zero}} + \underbrace{\( F_t^T - S_T \) }_{\text{value of short side of forward entered at time $t$}}
	&= F_t^T - F_0^T .
\end{align}
Note that the right-hand side above is a deterministic cash flow at time $T$. The present value of this cash flow at time $t$ is
\begin{align}
\( F_t^T - F_0^T \) \ee^{-r(T-t)}
\end{align}
Thus, we conclude that the value at time $t$ of the long side of a forward contract that was entered at time zero is
\begin{align}
f_t^T	
	&=	\( F_t^T - F_0^T \) \ee^{-r(T-t)} ,
\end{align}
as claimed.
\end{proof}

\begin{example}
Suppose $S_0 = 100$ and $r = 0.10$.  Consider a forward contract with a maturity date $T = 1$ year. What is the value at time $t = 6$ months of the long side of a forward contract that was entered at time zero, assuming $S_t = 50$?
From Theorem \ref{thm:fairforward} we have $f_t^T = ( F_t^T - F_0^T ) \ee^{-r(T-t)}$.  And from Theorem \ref{thm:fair-strike-forward} we have $F_t^T = \ee^{r(T-t)} S_t$ and $F_0^T = \ee^{r T} S_0$.  Thus, we have
\begin{align}
f_t^T 
	&= ( \ee^{r(T-t)} S_t -  \ee^{rT} S_0) \ee^{-r(T-t)}
	=  S_t - S_0 \ee^{rt} = 50 - 100 \ee^{0.1 * (1/2)} \approx - 55.13 .
\end{align}
It makes sense that the value is negative.  The long side has agreed to buy the stock at time $T$ for $F_0^T = \ee^{rT} S_0 \approx 110.52$ and yet the value of the stock at time $t$ is only $50$.
\end{example}

\section{Futures Contracts}
\begin{definition}
A \textit{futures contract} written on an underlying $S$ is an agreement between two parties.
The contract specifies a series of dates $t_0, t_1, t_2, \ldots, t_n $ where $t_n = T$.
For $i=1,2,\ldots,n$, the long side of futures contract agrees to receive $\Fh_{t_i}^T - \Fh_{t_{i-1}}^T$ at time $t_i$ while the short side agrees to pay $\Fh_{t_i}^T - \Fh_{t_{i-1}}^T$ on these dates (keep in mind $\Fh_{t_i}^T - \Fh_{t_{i-1}}^T$ could be negative).  The process $\Fh^T = (\Fh_t^T)_{0 \leq t \leq T}$ is called the \textit{futures price}.  The futures price $\Fh_t^T$ is agreed upon at time $t$ so that the cost to enter the futures contract is zero.  The futures price at time $T$ is specified to be $\Fh_T^T = S_T$. 
\end{definition}

\begin{theorem}
Consider a futures contract written on an asset $S = (S_t)_{t \geq 0}$ that pays no dividends.
Assume the risk-free rate of interest is a constant $r$.  Then the futures price is equal to the forward price $\Fh_t^T = F_t^T $ for all $t \in [0,T]$.
\end{theorem}

\begin{proof}
Fix some dates $t_0 = 0 < t_1 < t_2 < \ldots < t_n = T$.  Consider the following strategy: for $i=0,1,\ldots,n-1$, at time $t_i$ go long a total of $\ee^{r(t_{i+1})}$ futures contracts.  Invest/finance all the gains/losses from your strategy in the bank at the risk-free rate of interest.  The total profit/loss from this strategy is
\begin{align}
\sum_{i=0}^{n-1} (\Fh_{t_{i+1}}^T - \Fh_{t_{i}}^T) \ee^{rt_{i+1}}  \ee^{r(T-t_{i+1})}
	&=	\ee^{rT} \sum_{i=0}^{n-1} (\Fh_{t_{i+1}}^T - \Fh_{t_{i}}^T) \\
	&=	\ee^{rT} ( \Fh_T^T - \Fh_0^T ) \\
	&=	\ee^{rT} (S_T - \Fh_0^T) 
\end{align}
where we have used $S_T = \Fh_T^T$.  Now consider another strategy: at time $t=0$ go short a total of $\ee^{rT}$ Forward contracts.  The profit/loss from this strategy is
\begin{align}
\ee^{rT} (F_0^T - S_T) .
\end{align}
If we combine these two strategies into a single portfolio, the final value of our portfolio is
\begin{align}
\ee^{rT} (S_T - \Fh_0^T)  + \ee^{rT} (F_0^T - S_T)  
	&=	\ee^{rT}  ( F_0^T - \Fh_0^T ) .
\end{align}
As there are no costs to enter future and forward contracts, there will be a clear arbitrage opportunity unless the futures price at time zero $\Fh_0^T$ is equal to the forward price at time zero $F_0^T$.
Note that there is nothing special about $t_0=0$.  The same argument works if we start at some $t_0>0$.  
We therefore have that the futures price is equal to the forward price $\Fh_t^T = F_t^T$.
\end{proof}

\begin{example}
Let $r=0.1$.  An investor takes a long position in a futures contract on an underlying with an initial value of $S_0 = 100$.  Suppose the futures contract has settlement dates $t_1 = 0.5$ and $t_2 = T = 1$.  What are the cash flows paid/received by the long side if $S_{t_1} = 80$ and $S_T = 90$?  The investor receives $\Fh_{t_1}^T - \Fh_0^T$ at time $t_1$ and receives $\Fh_T^T - \Fh_{t_1}^T$ at time $T$.  Assuming constant interest rates, the futures price is equal to the forward price.  Thus, we need to compute
\begin{align}
\Fh_0^T 
	&=	F_0^T = S_0 \ee^{r T} \approx 110.52, &
\Fh_{t_1}^T
	&=	F _{t_1}^T = S_{t_1} \ee^{r (T - t_1)} \approx 84.10, &
\Fh_T^T
	&=	F_T^T = S_T = 90 .
\end{align}
The long side receives $\Fh_{t_1}^T - \Fh_0^T = 84.10-110.52 = -26.42$ at time $t_1$ and $\Fh_T^T - \Fh_{t_1}^T = 90 - 84.10=5.90$ at time $T$.
\end{example}

\section{Factors that affect Call and Put Prices}

Throughout this and subsequent sections, we will denote by $C(T,K) = (C_t(T,K))_{0 \leq t \leq T}$ and $P(T,K) = (P_t(T,K))_{0 \leq t \leq T}$, respectively, the prices of a Call and Put options with maturity $T$ and strike $K$, written on an underlying $S = (S_t)_{0 \leq t}$. 
Various factors affect Call and Put prices.  The main factors are
\begin{itemize}
\item The value of the underlying asset $S_t$,
\item The time to maturity $T-t$,
\item The strike price $K$,
\item The \textit{volatility} of the underlying asset $\sigma$
\end{itemize}
We have not yet defined what we mean by \textit{volatility}.  In fact, as we will see in this course, there are many different notions of volatility.  For now, just think of volatility as a rough measure of how large $\Vb( \log S_T | S_t) $ (the conditional variance of $\log S_T$ given $S_t$) is.  A larger volatility $\sig$ implies a larger variance $\Vb ( \log S_T | S_t)$.  
Note that $T-t$ and $K$ are specific to the option contract, while $\sigma$ and $S_t$ are properties of the underlying asset $S$. The interest rate $r$ generally comes from fixed income market (i.e., the bond market).  Let us examine how each of the above factors affect Call and Put prices.
\\[0.5em]
Increasing $\sig^2 (T-t)$ increases the value of both calls and puts.  The reason is that, the larger $\sig^2 (T-t)$ is, the larger the variance $\Vb( \log S_T | S_t)$ becomes.  The larger the variance $\Vb( \log S_T | S_t)$ is, the larger the probability that $S_T$ is \textit{in the money} (i.e., that $S_T > K$ for a Call or that $S_T < K$ for a Put).
\\[0.5em]
Increasing the strike $K$ lowers the value of a Call and raises the value of a Put.  Suppose for example, that $K_2 > K_1$.  Then 
\begin{align}
C_T(T,K_1) = (S_T - K_1)^+ 
	&\geq (S_T - K_2)^+ = C_T(T,K_2) , \\
P_T(T,K_1) = (K_1 - S_T)^+ 
	&\leq (K_2- S_T)^+ = P_T(T,K_2) .
\end{align}
Because $C_T(T,K_1) \geq C_T(T,K_2)$ and $P_T(T,K_1) \leq P_T(T,K_2)$ it follows that $C_t(T,K_1) \geq C_t(T,K_2)$ and $P_t(T,K_1) \leq P_t(T,K_2)$.  If this were not the case, there would be a clear arbitrage opportunity.
\\[0.5em]
Increasing $S_t$ increases the value of a Call and decreases the value of a Put.  The reason is that, the larger $S_t$ is, the higher the probability that a Call finishes in the money (i.e., $S_T > K$) and the lower the probability that a Put finishes in the money (i.e., $S_T < K$).

\section{Bounds for European options}
The core argument that we will use in this and subsequent sections is that, if the value of a portfolio $X^A = (X_t^A)_{t \geq 0}$ is greater or equal to the value of another portfolio $X^B = (X_t^B)_{t \geq 0}$ at time $T$, then the value of $X^A$ must be greater or equal to the value of $X^B$ at any given time $t \leq T$.  More concisely, we have
\begin{align} \label{eq:no_arb_geq}
X_T^A 
	&\geq X_T^B &
	&\Rightarrow&
X_t^A 
	&\geq X_t^B &
\forall t \in [0,T] ,
\end{align}
If the above were not true, there would be an obvious arbitrage opportunity.

\begin{theorem}[Upper bounds for European Calls and Puts]
We have
\begin{align}
C_t(T,K) 
	&\leq S_t , &
P_t(T,K) 
	&\leq K B_t^T .
\end{align}
\end{theorem}

\begin{proof}
We have
\begin{align}
C_T(T,K) = (S_T - K)^+ 
	&\leq S_T &
	&\Rightarrow& 
C_t(T,K) 
	&\leq S_t , \\
P_T(T,K) = (K - S_T)^+ 
	&\leq K B_T^T &
	&\Rightarrow&
P_t(T,K) 
	&\leq K B_t^T .
\end{align}
\end{proof}

\noindent
Establishing Lower bounds for European options requires a little more work (but not much).

\begin{theorem}[Lower bound for European Call and Put Options]
\begin{align}
C_t(T,K) 
	&\geq (S_t - K B_t^T)^+ , \label{eq:call-lower-bound} \\
P_t(T,K) 
	&\geq (K B_t^T - S_t )^+ .	\label{eq:put-lower-bound}
\end{align}
\end{theorem}

\begin{proof}
First, we note that
\begin{align}
C_T(T,K)
	&=	(S_T - K )^+
	\geq S_T - K B_T^T , &
	&\Rightarrow&
C_t(T,K)
	\geq S_t - K B_t^T , \label{eq:call-lb} 
\end{align}
We also know that call prices are non-negative
\begin{align}
C_t(T,K)
	&\geq 0 . \label{eq:call-non-neg}
\end{align}
Combining \eqref{eq:call-lb} and \eqref{eq:call-non-neg} yields the lower bound \eqref{eq:call-lower-bound} for European calls.
\\[0.5em]
Next, we note that
\begin{align}
P_T(T,K)
	&=	(K - S_T )^+
	\geq K B_T^T - S_T , &
	&\Rightarrow&
P_t(T,K)
	\geq K B_t^T - S_t . \label{eq:put-lb}
\end{align}
We also know that put prices are non-negative
\begin{align}
P_t(T,K)
	&\geq 0 .		\label{eq:put-non-neg}
\end{align}
Combining \eqref{eq:put-lb} and\eqref{eq:put-non-neg} gives the lower bound \eqref{eq:put-lower-bound} for European puts.
\end{proof}

\section{Put-Call Parity}
There exists a very important relation between European calls and European puts with the same strike $K$ and maturity $T$.

\begin{theorem}[Put-Call Parity]
The following relation holds between European Call and Put options
\begin{align} \label{eq:put_call_parity}
C_t(T,K) + K B_t^T 
	&= P_t(T,K) + S_t .
\end{align}
\end{theorem}

\begin{proof}
For $t \in [0,T]$, define
\begin{align}
		X_t^A &= C_t(T,K) - P_t(T,K) , &
		X_t^B &= S_t - K B_t^T 
\end{align}
Note that
\begin{align}
		X_T^A &= (S_T - K)^+ - (K - S_T)^+ = S_T - K , &
		X_T^B &= S_T - K B_T^T = S_T - K
\end{align}
Consequently, we have
\begin{align}
X_T^A
	& = X_T^B & 
	&\Rightarrow& 
X_t^A 
	&= X_t^B ,  
\end{align}
and the last equality implies equation \eqref{eq:put_call_parity}.
\end{proof}

\noindent
Note the following special case: if $K = S_t/B_t^T$ then $C_t(T,S_t/B_t^T) = P_t(T,S_t/B_t^T)$.  When $K = S_t/B_t^T$ we say the strike $K$ is \textit{at-the-money}.

\section{American vs European options}
In this section, we denote with a superscript ``${}^A$'' American options and with a superscript ``${}^E$'' European options.
We claim that $C_t^A(T,K) \geq C_t^E(T,K)$ and $P_t^A(T,K) \geq P_t^E(T,K)$. 
The reason is that, the owner of an American option has the same right as the owner of a European option (i.e., the ability to exercise the option at time $T$), but additionally has the right to exercise prior to the maturity date $T$.  The additional optionality makes the American option more valuable than its European counterpart.

\begin{theorem}
An investor should never exercise an American Call (on a non-dividend paying stock) early.  Thus, $C_t^A(T,K) = C_t^E(T,K)$.
\end{theorem}

\begin{proof}
Using the fact that, for a fixed maturity $T$ and strike $K$ an American option is worth at least as much as a European option, and recalling the lower bound for a European Call \eqref{eq:call-lower-bound}, we have for any $t \leq T$ that
\begin{align} \label{eq:am_call_exercise}
C_t^A(T,K) 
	&\geq C_t^E(T,K) &
	&\text{(American call worth more than European call)}\\
	&\geq ( S_t - K B_t^T )^+ &
	&\text{(lower bound for European call \eqref{eq:call-lower-bound})}\\
	&\geq (S_t - K)^+ . &
	&\text{(because $B_t^T \leq 1$ when $r \geq 0$, and $K \geq 0$)}\\
\end{align}
As the value of an American call $C_t^A(T,K)$ is greater than or equal to the payoff from exercising $(S_t-K)^+$, it would be better to sell the American Call rather than exercise it.  We therefore conclude that it is never optimal to exercise an American call option on a non-dividend paying stock prior to the maturity date $T$.
\end{proof}

\noindent
Note that the same argument does \textit{not} apply to an American put.  It may be (and often is) optimal to exercise an American Put option prior to maturity.

\section{Examples of arbitrage}
From the properties established in the previous sections, you should be able to identify some obvious arbitrage opportunities.  In what follows, all options are European-style.

\begin{example}
Suppose $C_t(T,K_1) < C_t(T,K_2)$ with $K_1 < K_2$.  This is an obvious arbitrage.  Buy one share of $C_t(T,K_1)$, sell one share of $C_t(T,K_2)$ and invest $(C_t(T,K_2)-C_t(T,K_1))$ in a bank account.  This strategy requires zero initial investment and at maturity yields a strictly positive value
\begin{align}
(S_T - K_1)^+ - (S_T - K_2)^+ + (C_t(T,K_2) -C_t(T,K_1) ) \ee^{r(T-t)} > 0 .
\end{align}
\end{example}

\begin{example}
Suppose the risk-free rate of interest is zero $r=0$ and $C_t(T,S_t) > P_t(T,S_t)$.  Again, this is an obvious arbitrage.  One can sell one call, buy one put, go long a forward contract, and invest $C_t(T,S_t) - P_t(T,S_t)$ in the bank.  At time $T$ this strategy yields
\begin{align}
&P_T(T,S_t) - C_T(T,S_t) + (S_T - S_t) + (C_t(T,S_t) - P_t(T,S_t)) \\
	&=	(S_t - S_T)^+ - (S_T - S_t)^+ + (S_T - S_t) +  (C_t(T,S_t) - P_t(T,S_t)) \\
	&=	(C_t(T,S_t) - P_t(T,S_t))  > 0 .
\end{align}
\end{example}

\section{Constructing Other Payoffs}
\label{sec:payoff_construction}
In this section, we show that put and call payoffs
\begin{align}
	&\text{Call Payoffs}: &
	\phi_C (S_T;K) &= (S_T - K)^+ , &
	&\text{Put Payoffs}:
	&\phi_P (S_T;K)& = (K - S_T)^+ ,
\end{align}
can be used to construct more general payoffs of the form $\phi(S_T)$, where $\phi$ is an arbitrary function. Let us begin with two very simple examples.
\begin{example}
Defining $\varphi(x) := \varphi_{P}(x;K_1) + \varphi_{C}(x;K_2)$ with $K_2 > K_1$, we obtain the payoff depicted in Figure~\ref{fig:plot1_11}.
\end{example}
\begin{example}
Defining $\varphi(x) := \varphi_{C}(x;K_1) - \varphi_{C}(x;K_2)$ with $K_2 > K_1$, we obtain the payoff depicted in Figure~\ref{fig:plot1_12}.
\end{example}
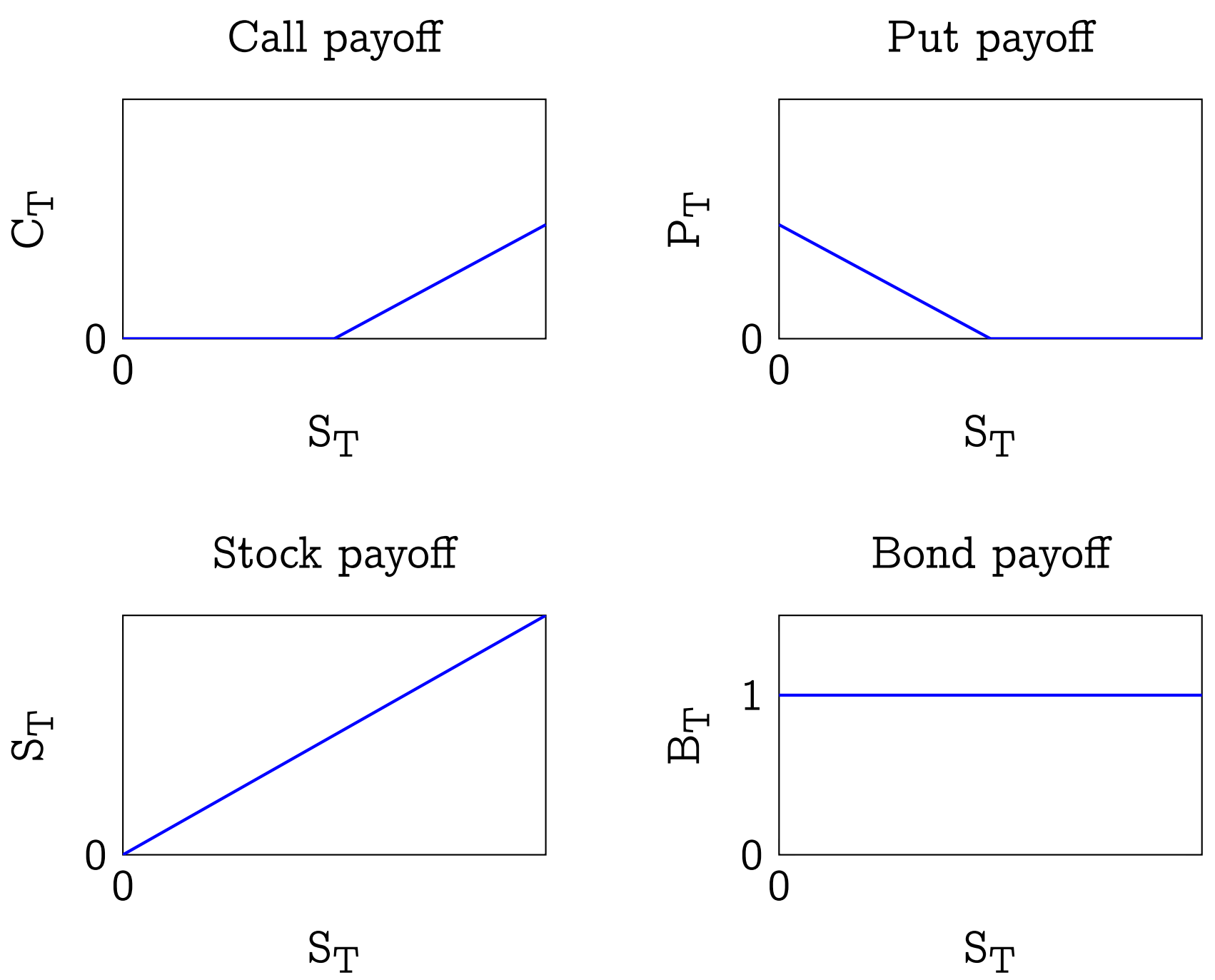
\begin{figure}
    \centering
    \begin{tikzpicture}
\begin{axis}[boxpayoff, title={Call payoff}, ylabel={$C_T$}]
  \addplot coordinates {(0,0) (1,0) (2,0.5)};
\end{axis}
\end{tikzpicture}
    \qquad
\begin{tikzpicture}
\begin{axis}[boxpayoff, title={Put payoff}, ylabel={$P_T$}]
  \addplot coordinates {(0,0.5) (1,0) (2,0)};
\end{axis}
\end{tikzpicture}
    \\[1em]
\begin{tikzpicture}
\begin{axis}[boxpayoff, title={Stock payoff}, ylabel={$S_T$}]
  \addplot coordinates {(0,0) (2,1.05)};
\end{axis}
\end{tikzpicture}
    \qquad
\begin{tikzpicture}
\begin{axis}[boxpayoff, title={Bond payoff}, ylabel={$B_T$}, ymax=1.5, ytick={0,1}]
  \addplot coordinates {(0,1) (2,1)};
\end{axis}
\end{tikzpicture}
    \caption{The payoffs of calls, puts, stocks and bonds.\label{fig:plot1_10}}
\end{figure}
\begin{figure}
    \centering
    \begin{tikzpicture}[baseline]
\begin{axis}[payoff3, anchor=origin, title={(a)}]
  \zeroline{}
  \addplot coordinates {(0,0.8) (0.8,0) (2,0)};
  \node[below] at (axis cs:0.8,0) {$K_1$};
\end{axis}
\end{tikzpicture}\;$+$\;
\begin{tikzpicture}[baseline]
\begin{axis}[payoff3, anchor=origin, title={(b)}]
  \zeroline{}
  \addplot coordinates {(0,0) (1.2,0) (2,0.8)};
  \node[below] at (axis cs:1.2,0) {$K_2$};
\end{axis}
\end{tikzpicture}\;$=$\;
\begin{tikzpicture}[baseline]
\begin{axis}[payoff3, anchor=origin, title={(c)}]
  \zeroline{}
  \addplot coordinates {(0,0.8) (0.8,0) (1.2,0) (2,0.8)};
  \node[below] at (axis cs:0.8,0) {$K_1$};
  \node[below] at (axis cs:1.2,0) {$K_2$};
\end{axis}
\end{tikzpicture}
    \caption{A plot of (a) $\varphi_{P}(\, \cdot \,;K_1)$, (b) $\varphi_{C}(\, \cdot \,;K_2)$, and (c) $\varphi = \varphi_{P}(\, \cdot \,;K_1) + \varphi_{C}(\, \cdot \,;K_2)$.\label{fig:plot1_11}}
\end{figure}
\begin{figure}
    \centering
    \begin{tikzpicture}[baseline]
\begin{axis}[payoff3, anchor=origin, title={(a)}]
  \zeroline{}
  \addplot coordinates {(0,0) (0.8,0) (2,1)};
  \node[below] at (axis cs:0.8,0) {$K_1$};
\end{axis}
\end{tikzpicture}\;$+$\;
\begin{tikzpicture}[baseline]
\begin{axis}[payoff3, anchor=origin, title={(b)}]
  \zeroline{}
  \addplot coordinates {(0,0) (1.2,0) (2,-0.8)};
  \node[above] at (axis cs:1.2,0) {$K_2$};
\end{axis}
\end{tikzpicture}\;$=$\;
\begin{tikzpicture}[baseline]
\begin{axis}[payoff3, anchor=origin, title={(c)}]
  \zeroline{}
  \addplot coordinates {(0,0) (0.8,0) (1.2,0.4) (2,0.4)};
  \node[below] at (axis cs:0.8,0) {$K_1$};
  \node[above] at (axis cs:1.2,0.4) {$K_2$};
\end{axis}
\end{tikzpicture}
    \caption{A plot of (a) $\varphi_{C}(\, \cdot \,;K_1)$, (b) $-\varphi_{C}(\, \cdot \,;K_2)$, and (c) $\varphi = \varphi_{C}(\, \cdot \,;K_1) - \varphi_{C}(\, \cdot \,;K_2)$.\label{fig:plot1_12}}
\end{figure}

\noindent
The following result, from \cite{carrmadan1998}, shows how to construct very general functions $\varphi$ from call and put payoffs.

\begin{theorem}
\label{thm:carr-madan}
For any $\varphi$ that is the difference of convex functions, we have
\begin{align}
\varphi(x)
	&=	\varphi(\lambda) + \varphi'(\lambda) \Big( (x-\lambda)^+ - (\lambda - x)^+ \Big) \\ &\qquad
			+ \int_0^\lambda \varphi''(K) (K-x)^+ \dd K + \int_\lambda^\infty \varphi''(K) (x-K)^+ \dd K . \label{0}
\end{align}
where $\varphi{\prime}$ and $\varphi{\prime\prime}$ denote the first and second derivative of $\varphi$ and $\lam \geq 0$ is an arbitrary constant.
\end{theorem}

\noindent
Before providing a proof of Theorem \ref{thm:carr-madan} we need to introduce 
the \textit{Dirac delta function}, denoted by $\delta$.
There are numerous equivalent definitions of the Dirac delta function.  But, for the purposes of this class, we will define it as follows
\begin{align}
\del(x-z)
	&:= \lim_{\sig \searrow 0} \phi_{z,\sig^2}(x) , &
\phi_{z,\sig^2}(x)
	&:=	\frac{1}{\sqrt{2 \pi \sig^2}} \exp \Big( \frac{-(x-z)^2}{2 \sig^2} \Big) .
\end{align}
Notice that $\phi_{z,\sig^2}(x)$ is the density of a normal random variable with mean $z$ and variance $\sig^2$.  Note also that
\begin{align}
\lim_{\sig \searrow 0} \phi_{z,\sig^2}(x)
	&=	\left\{ \begin{aligned} 
	&	0 ,	&		x \neq z , \\
	&	\infty , & x = z , 
	\end{aligned} \right. &
	&\text{and}&
\int_\Rb \phi_{z,\sig^2}(x) \dd x
	&= 1 .
\end{align}
Thus, $\del$ is an infinitely tall, infinitely narrow function and the area under $\del$ is equal to one.  It follows that $\del$ has the \textit{sifting property}
\begin{align}
\int_a^b \del(x-z) f(x) \dd x
	&=	f(z) , &
a
	&< z < b . \label{eq:sifting}
\end{align}
It may be useful to verify for yourself that \eqref{eq:sifting} holds in a specific example by, e.g., setting $(a,b) = \Rb$ and $f(x) = \ee^{p x}$.  In this case, we have
\begin{align}
\int_\Rb \del(x-z) \ee^{p x} \dd x
	&=	\int_\Rb \lim_{\sigma \searrow 0} \phi_{z,\sig^2}(x) \ee^{p x} \dd x
	=		\lim_{\sigma \searrow 0} \int_\Rb \phi_{z,\sig^2}(x) \ee^{p x} \dd x
	=		\lim_{\sigma \searrow 0} \ee^{p z + p^2 \sig^2 / 2}
	=		\ee^{p z} .
\end{align} 
Taking $(a,b) = (-\infty,x)$ and $f(x) = 1$ in \eqref{eq:sifting}, we have
\begin{align}
\int_{- \infty}^x \del(y-z) \dd y
	&=	\theta(x - z)
	:=	\left\{ \begin{aligned}
			&0 , && x < z , \\
			&1 , && x \geq z ,
			\end{aligned} \right. \label{eq:delta-integral}
\end{align}
where we have introduced the \textit{Heaviside function} $\theta$.  Observe that the integral of the Heaviside function gives a call payoff
\begin{align}
\int_{- \infty}^x \theta(y-z) \dd y
	&=	(x - z)^+ . \label{eq:theta-integral}
\end{align}
Thus, we note from \eqref{eq:delta-integral} that the derivative of a Heaviside function is a delta function and we note from \eqref{eq:theta-integral} that the derivative of a call payoff is a Heaviside function
\begin{align}
\d_x \theta(x-z)
	&=	\del(x-z) , &
\d_x (x - z)^+
	&=	\theta(x - z) .
\end{align}
We are now ready to prove Theorem \ref{thm:carr-madan}.

\begin{proof}[Proof of Theorem \ref{thm:carr-madan}]
Using integration by parts twice, we have
\begin{align}
\int_0^\lambda \varphi''(K) (K-x)^+ \dd K
	&=	(K-x)^+ \varphi'(K) \Big|_0^\lam - \int_0^\lam \varphi'(K) \theta(K-x) \dd K \\
	&=	(K-x)^+ \varphi'(K) \Big|_0^\lam - \varphi(K) \theta(K-x) \Big|_0^\lam + \int_0^\lam \varphi(K) \del(K-x) \dd K \\
	&=	(\lam-x)^+ \varphi'(\lam) - \varphi(\lam) \theta(\lam-x) + \int_0^\lam \varphi(K) \del(K-x) \dd K .  \label{eq:1}
\end{align}
Next, once again using integration by parts, we have
\begin{align}
\int_\lambda^\infty \varphi''(K) (x-K)^+ \dd K
	&=	(x-K)^+ \varphi'(K) \Big|_\lam^\infty + \int_\lam^\infty \varphi'(K) \theta(x-K) \dd K \\
	&=	(x-K)^+ \varphi'(K) \Big|_\lam^\infty + \varphi(K) \theta(x-K) \Big|_\lam^\infty + \int_\lam^\infty \varphi(K) \del(x-K) \dd K \\
	&=	-(x-\lam)^+ \varphi'(\lam) - \varphi(\lam) \theta(x-\lam) + \int_\lam^\infty \varphi(K) \del(x-K) \dd K . \label{eq:2}
\end{align}
Inserting \eqref{eq:1} and \eqref{eq:2} into \eqref{0} and using
\begin{align}
\int_0^\infty \varphi(K) \del(x-K) \dd K
	&=	\varphi(x) ,
\end{align}
 we find that the right-hand side simplifies to $\varphi(x)$, as claimed.
\end{proof}

\noindent
Now, consider a European option whose payoff is $\varphi(S_T)$.  Let $V_t$ be the value of this option at time $t$.  Suppose we can observe call and put prices at every strike $K \in [0,\infty)$.  Denote the time $t$ price of a European Call option with strike $K$ and maturity $T$ by $C_t(T,K)$.  Likewise, denote by $P_t(T,K)$ the price of a European Put option with strike $K$ and maturity $T$.  Finally, let $B_t^T = \ee^{-r(T-t)}$ be the price of a zero-coupon bond that pays $1$ at time $T$.  
We know from the above result that
\begin{align}
\varphi(S_T)
	&=	\varphi(\lambda)  + \varphi'(\lambda) \Big( (S_T-\lambda)^+ - (\lambda - S_T)^+ \Big) \\ &\qquad
			+ \int_0^\lambda \varphi''(K) (K-S_T)^+ \dd K + \int_\lambda^\infty \varphi''(K) (S_T-K)^+ \dd K \\
	&=	\varphi(\lambda) B_T^T + \varphi'(\lambda) \Big( C_T(T,\lam) - P_T(T,\lam) \Big) \\ &\qquad
			+ \int_0^\lambda \varphi''(K) P_T(T,K) \dd K + \int_\lambda^\infty \varphi''(K) C_T(T,K) \dd K
\end{align}
where we have used $(S_T-K)^+ = C_T(T,K)$ and $(K-S_T)^+ = P_T(T,K)$.  Therefore, we must have
\begin{align}
V_t
	&=	B_t^T \varphi(\lambda) + \varphi'(\lambda) \Big( C_t(T,\lam) - P_t(T,\lam) \Big) \\ &\qquad
			+ \int_0^\lambda \varphi''(K) P_t(T,K) \dd K + \int_\lambda^\infty \varphi''(K) C_t(T,K) \dd K . \label{eq:carr-madan}
\end{align}
Note that, everywhere we had a call payoff $C_T(T,K)$ we have replaced it with the corresponding call price $C_t(T,K)$ and likewise for puts and bonds.  What equation \eqref{eq:carr-madan} tells us is that the value $V_t$ of a European derivative with payoff $\varphi(S_T)$ can be obtained by observing call and put prices ($C_t(T,K))_{K \geq 0}$ and $(P_t(T,K))_{K \geq 0}$, respectively).  This is a \textit{model free} result.  We do not need to assume any model for $S$ in order to derive the value $V_t$ of the option.  We only need to observe European call and put prices on the market.

\begin{example}
Assume for simplicity that $r=0$.  In terms of Calls and Puts, what is the time $t=0$ value of a contract that pays $\varphi(S_T) = - 2 \log (S_T/S_0)$ at time $T$?  We have
\begin{align}
\varphi(s) = - 2 \log (s/S_0) , &&
\varphi'(s) = - 2/s , &&
\varphi''(s) =  2/s^2 .
\end{align}
Note that \eqref{eq:carr-madan} holds for any $\lam \geq 0$.  Let's choose $\lam = S_0$.  Note that
\begin{align}
\varphi(S_0) = - 2 \log (S_0 / S_0) = - 2 \log 1 = 0 ,
\end{align}
so the first term in \eqref{eq:carr-madan} is zero.  We also have, by put-call parity that $C_0(T,S_0) - P_0(T,S_0) = 0$ (make sure you understand why), so the second term in \eqref{eq:carr-madan} is zero.  Finally, we are left with
\begin{align}
V_0
	&=	\int_0^{S_0} \frac{2}{K^2} P_0(T,K) \dd K + \int_{S_0}^\infty \frac{2}{K^2} C_0(T,K) \dd K . 
\end{align}
We will come back to this example when we talk about variance swaps.
\end{example}

\begin{example}
Suppose $\varphi(S_T)$ is given by
\begin{align}
\varphi(S_T)
	&=	\left\{ \begin{aligned}
			&0 &&S_T < K_1 , \\
			&S_T - K_1, && K_1 \leq S_T < K_2 , \\
			&K_2 - K_1 , && K_2 \leq S_T .
			\end{aligned} \right.
\end{align}
We can write the time $t$ value of an option with this payoff in terms of calls and puts as follows.  First, we have
\begin{align}
\varphi'(s)
	&=	\theta(s - K_1) - \theta(s - K_2) , &
\varphi''(s)
	&=	\del(s - K_1) - \del(s - K_2) .
\end{align}
Let's choose $\lam < K_1$.  Then, using \eqref{eq:carr-madan} we have
\begin{align}
V_t &= \int_\lambda^\infty \varphi''(K) C_t(T,K) \dd K \\
	&=	\int_\lambda^\infty \del(K - K_1) C_t(T,K) \dd K - \int_\lambda^\infty \del(K - K_2) C_t(T,K) \dd K . \\
	&=	C_t(T,K_1) - C_t(T,K_2) .
\end{align}
\end{example}


\section{Exercises}
\label{sec:exer-1}



\begin{exercise}
\label{exer:1-2}
Assume the risk-free rate of interest is a fixed constant $r>0$.
Suppose that if you hold an asset $S$ then, for $i=1,2, \ldots, n$ you will receive known dividend payments of $D_i$ at time $t_i<T$ (if you short the asset $S$, then you will owe the dividend payments to the person you sold the asset to).
\textbf{(a)} Use no-arbitrage arguments to derive the fair forward price $F_t^T$.  It may help to first find $F_0^T$.
\textbf{(b)} Suppose you enter a forward contract written on $S$ at time zero.  What is the \textit{value} $f_t^T$ of this forward contract at time $t < T$?
\end{exercise}


\begin{exercise}
\label{exer:1-3}
Assume the risk-free rate of interest is a fixed constant $r>0$.
Suppose I offer to pay you a certain cash flow $\text{CF}_1$ at time $t_1$ if you pay me a certain cash flow of $\text{CF}_2$ at time $t_2$.  You think about it for a moment and tell me: ``I'll make that trade if you also pay me $V_0$ today at time zero.'' I think about your offer and say: ``that sounds fair to me.''  
\textbf{(a)} What is $\text{CF}_2$ in terms of $V_0$, $r$, $t_1$, $t_2$ and $\text{CF}_1$?
\textbf{(b)} Suppose $\text{CF}_1>0$ and $\text{CF}_2>0$.  Define $R := \text{CF}_1 / \text{CF}_2$.  Give a condition on $R$ which will guarantee that $V_0>0$.
\textbf{(c)} What is the condition on $R$ that guarantees $V_0>0$ when $\text{CF}_1<0$ and $\text{CF}_2>0$?
\end{exercise}


\begin{exercise}
\label{exer:1-4}
Assume the risk-free rate of interest is a fixed constant $r>0$.
Suppose there is a European contract {that at} time $T$ pays
\begin{align}
\text{Payoff}
	&=	\left\{ \begin{aligned}
			&0 , &&S_T < K_1 , \\
			&S_T - K_1 && K_1 \leq S_T < K_2 , \\
			&K_2 - K_1 && K_2 \leq S_T .
			\end{aligned} \right.
\end{align}
Use the upper and {lower} bounds we derived for European call options to give upper and lower bounds on the price of this contract at time $t < T$.
\end{exercise}

\begin{exercise}
\label{exer:1-5}
Assume the risk-free rate of interest is a fixed constant $r>0$.
We used the argument many times that, if $X_T^A \geq X_T^B$ we must also have $X_t^A \geq X_t^B$ to preclude arbitrage.  Suppose that $X_T^A \geq X_T^B$ and $X_t^A < X_t^B$.  Construct a trading strategy $X$ that satisfies $X_t = 0$ and $X_T \geq 0$.
\end{exercise}


\begin{exercise}
\label{exer:1-6}
Let $H_0$ denote the price at time $t=0$ of a European derivative that pays
\begin{align}
h(S_T) := a \cdot (S_T - K)^+ + b \cdot (K - S_T)^+
\end{align}
at time $T$.  Let $G_0$ denote the price at time $t=0$ of a European derivative that pays
\begin{align}
g(S_T) := c \cdot (S_T - K)^+ + d \cdot (K - S_T)^+
\end{align}
at time $T$.  Here, $(a, b, c, d, K)$ are positive constants such that $ad - bc \neq 0$.\\[0.5em]
\textbf{(a)} Let $C_0(T,K)$ and $P_0(T,K)$, respectively, denote the time $t=0$ price of a European call and put with maturity date $T$ and strike price $K$.  Derive expressions for $C_0(T,K)$ and $P_0(T,K)$ in terms of $H_0$, $G_0$, $a$, $b$, $c$, $d$.\\[0.5em]
\textbf{(b)} Let $B_0^T$ denote the price of a bond at time $t=0$ that pays one unit of currency at time $T$.  Derive an expression for $B_0^T$ in terms of $C_0(T,K)$, $P_0(T,K)$, $K$ and $S_0$.\\[0.5em]
\textit{Hint}: You may find the following helpful.
\begin{align}
A :=
\begin{pmatrix}
a & b \\
c & d
\end{pmatrix}
\quad \Rightarrow \quad
A^{-1} := \frac{1}{ad-bc}
\begin{pmatrix}
d & -b \\
-c & a
\end{pmatrix} .
\end{align}
\end{exercise}


\begin{exercise}
\label{exer:1-7}
Assume that the risk-free rate of interest is a constant $r \geq 0$ and that the stock $S$ pays no dividends. Denote by $B_t^T = \ee^{-r(T-t)}$ the time-$t$ price of a zero-coupon bond that pays one unit of currency at time $T$, and by $C_t(T,K)$ and $P_t(T,K)$ the time-$t$ prices of a European call and a European put, respectively, with maturity date $T$ and strike price $K$. Fix three strikes $0 < K_1 < K_2 < K_3$ and define
\begin{align}
\lam
	&:=	\frac{K_3 - K_2}{K_3 - K_1} \in (0,1) , &
&\text{so that} &
K_2
	&=	\lam K_1 + (1 - \lam) K_3 .
\end{align}
\textbf{(a)} Show that, for every possible value of $S_T$, we have
\begin{align}
\lam (S_T - K_1)^+ + (1 - \lam) (S_T - K_3)^+ - (S_T - K_2)^+
	&\geq	0 .
\end{align}
\textit{Hint}: consider separately the cases $S_T \leq K_1$, $K_1 < S_T \leq K_2$, $K_2 < S_T \leq K_3$ and $S_T > K_3$.\\[0.5em]
\textbf{(b)} State the definition of an arbitrage. Then, assuming the market is free of arbitrage, show that the condition
\begin{align}
C_t(T,K_2)
	&>	\lam \, C_t(T,K_1) + (1 - \lam) \, C_t(T,K_3)
\end{align}
cannot hold, by constructing a portfolio that would be an arbitrage if it did.\\[0.5em]
\textbf{(c)} Show that put prices must also satisfy
\begin{align}
P_t(T,K_2)
	&\leq	\lam \, P_t(T,K_1) + (1 - \lam) \, P_t(T,K_3) .
\end{align}
\textit{Hint}: You do not need to recalculate everything from the beginning. Use directly the conclusion of part (b).
\end{exercise}




\chapter{Binomial model}

Up to now we have not put \textit{any} model on the underlyer $S$, although we did need to make some assumptions on the risk-free rate of interest.  We deduced forward and futures prices, bounds for calls and puts, as well as the Put-Call Parity relation using no arbitrage arguments. 
Model-free results are fantastic when available because they are robust to model misspecification. Unfortunately, we have gone as far as we can with model-free results. In order to price Calls and Puts, we need to assume a model for the dynamics of $S$.  We will begin with a very simple model.  This simple model will help us understand the fundamental concepts of option pricing.  Once we understand this simple model, we will be able to easily understand the more realistic models that will be presented later in this text.

\section{One-Step Binomial Model}
\begin{figure}
  \centering
	\begin{tikzpicture}[sloped]
		\node (a) at ( 0,0) [bag] {$B_0$};
		\node (b) at ( 4,1.5) [bag,text width=10em] {$(1+r)B_0$};
		\node (d) at ( 0,-1.5) [end] {$n=0$};
		\node (e) at ( 4,-1.5) [end] {$n=1$};
		\draw [->] (a) to node [above] {$1$} (b);
		\draw [dotted] (d) to node [below] {} (a);
		\draw [dotted] (e) to node [above] {} (b);
	\end{tikzpicture}
	\caption{The behavior of a bond $B$ in the one-step binomial model.\label{fig:1-step-bond}}
\end{figure}
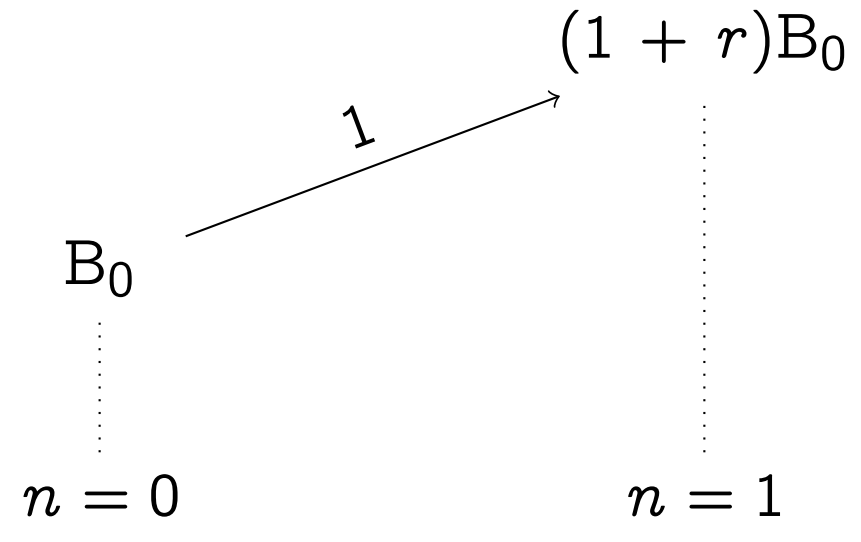

Consider an experiment in which we toss a coin.  Our sample space is $\Om = \{ H,T \}$.  Obviously, there are only two possible outcomes $\om = H$ and $\om = T$.  The probability of these outcomes is $\Pb(H) = p$ and $\Pb(T) = 1-p =: q$.  Now, consider a market on the sample space that includes a stock $S = (S_n)_{n \in \{0,1\}}$ and a bond $B = (B_n)_{n \in \{0,1\}}$.  The value of the stock $S_0$ at time $n=0$ is known.  At time $n=1$, the value of the stock is given by
\begin{align}
S_1
	&=	\left\{ \begin{aligned}
			&u S_0 , & &\text{if $\om = H$} , \\
			&d S_0 , & &\text{if $\om = T$} .
			\end{aligned} \right.
\end{align}
Figure~\ref{fig:1-step-stock} depicts how the stock $S$ behaves in the one-step binomial model.  In case it is not obvious from Figure~\ref{fig:1-step-stock}, the variable $u$ stands for ``up'' and the variable $d$ stands for ``down.''  
The bond has an initial value of $B_0$ and grows to a value of $B_1 = (1+r) B_0$.  
Figure~\ref{fig:1-step-bond} shows how the bond behaves in the one-step binomial model. 
The no-arbitrage assumption implies that 
\begin{align}
0 < d <  1+r  < u . \label{eq:no-arb}
\end{align}
Take a moment to derive why this must be the case!

\begin{figure}
  \centering
	\begin{tikzpicture}[sloped]
		\node (a) at ( 0,0) [bag] {$S_0$};
		\node (b) at ( 4,-1.5) [bag] {$d S_0$};
		\node (c) at ( 4,1.5) [bag] {$u S_0$};
		\node (d) at ( 0,-3) [end] {$n=0$};
		\node (e) at ( 4,-3) [end] {$n=1$};
		\draw [->] (a) to node [below] {$q$} (b);
		\draw [->] (a) to node [above] {$p$} (c);
		\draw [dotted] (d) to node [below] {} (a);
		\draw [dotted] (e) to node [above] {} (b);
	\end{tikzpicture}
	\caption{A one-step binomial tree for a stock $S$.\label{fig:1-step-stock}}
\end{figure}
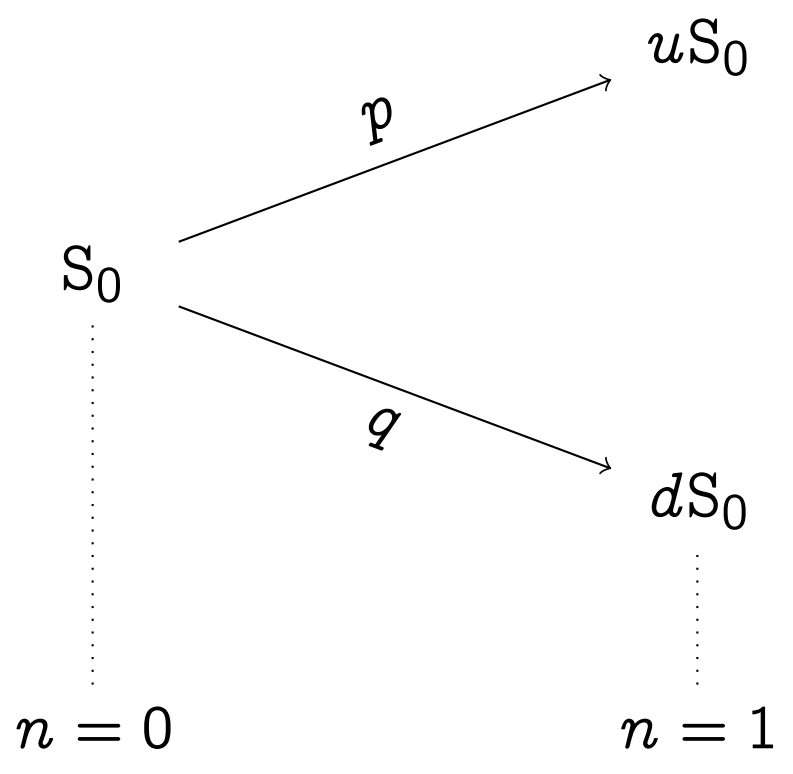

\subsection{Pricing European options by replication}
\label{sec:replication}
We now derive the value of a European option with payoff $\phi(S_1)$.
We will do this by finding a portfolio $X = (X_n)_{n \in \{0,1\}}$ that has the same value at time $n=1$ as the option (i.e., $X_1 = \phi(S_1)$). This approach is called \textit{pricing by replication}.
Let $\Delta_0$ be the number of shares of $S$ in the portfolio at time $n=0$. If the remaining value of the portfolio is invested in bonds, then the portfolio would contain $(X_0 - \Delta_0 S_0)/B_0$ bonds.  We have
\begin{align}
X_0
	&= \underbrace{\Delta_0}_{\text{\# shares}} S_0 + \underbrace{\frac{X_0 - \Delta_0 S_0}{B_0}}_{\text{\# of bonds}} B_0.
\end{align}
Note that both $\Delta_0$ and $(X_0 - \Delta_0 S_0)/B_0$ can be negative.
At time $n=1$ the value of the portfolio is
\begin{align}
X_1 
	&= \Delta_0 S_1 + (X_0 - \Delta_0 S_0) \frac{1}{B_0}B_1  =
		\left\{ \begin{aligned}	
			&\Delta_0 u S_0 + (X_0 - \Delta_0 S_0) (1+r) && \text{if $\om = H$}, \\
			&\Delta_0 d S_0 + (X_0 - \Delta_0 S_0) (1+r) && \text{if $\om = T$}.
		\end{aligned} \right.
\end{align}
Let $V = (V_n)_{n \in \{0,1\}}$ denote the value of the option.  The value of the option at time $n=1$ is
\begin{align}
V_1 = \phi(S_1)
	&=
		\left\{ \begin{aligned}	
			&\phi(u S_0) && \text{if $\om = H$}, \\
			&\phi(d S_0) && \text{if $\om = T$}.
		\end{aligned} \right.
\end{align}
Thus, to create a portfolio for which $X_1 = V_1$ we must have
\begin{align} 
\Delta_0 u S_0 + (X_0 - \Delta_0 S_0) (1+r) 
	&=	\phi(u S_0) , &
\Delta_0 d S_0 + (X_0 - \Delta_0 S_0) (1+r) 
	&=	\phi(d S_0) . \label{eq:option_portfolio_equalities}
\end{align}
Solving for $X_0$ and $\Delta_0$, we obtain
\begin{align}
\Delta_0 
	&= \frac{\phi (u S_0) - \phi (d S_0)}{u S_0 - d S_0}, & 
X_0 
	&= \frac{1}{1+r} \big( \pt \phi (u S_0) + \qt \phi (d S_0) \big), \label{eq:portfolio_0}
\end{align}
where we have introduced $\pt$ and $\qt$, which are given by
\begin{align}
	\pt &= \frac{(1+r) - d}{u - d},  &
	\qt &= 1 - \pt. \label{eq:pt}
\end{align}
Because we have created a portfolio $X$ for which $V_1 = X_1$, we must also have
\begin{align} 
	V_0 = X_0 , \label{eq:option_portfolio_price}
\end{align}
because, if \eqref{eq:option_portfolio_price} did not hold, there would be a clear arbitrage opportunity. Suppose, for example, that $V_0 < X_0$, and try to construct an investment strategy that has zero initial cost and guarantees a profit at time $n=1$.

\begin{remark}
\label{rmk:pq}
Note that the value of the option $V_0$, the initial value of the replication portfolio $X_0$ and the number of shares $\Del_0$ of the stock $S$ do \textit{not} depend on the probabilities $p$ and $q$!  Therefore, we could have changed the probabilities $p$ and $q$ and this would not affect the price of the option.  This is quite an astonishing and unexpected result.  We will use this in the next section.
\end{remark}

\begin{example}
Suppose $S$ evolves as described in the binomial model with $u=1.1$, $d=0.9$ and $S_0=100$ and $r=0$.  What is the price of a Call option with maturity at time $n=1$ with strike $K=100$?  The payoff of the Call is $\phi(S_1) = (S_1 - 100)^+$.  In the event the stock goes up we have $S_1 = u S_0 = 110$ and the call payoff is $\phi(uS_0)=(110 - 100)^+ = 10$.  In the event the stock price goes down, we have $S_1 = d S_0 = 90$ and the call payoff is $\phi(dS_0)=(90 - 100)^+ = 0$.  Thus, we must find $X_0$ and $\Del_0$ so that
\begin{align}
\Delta_0 u S_0 + (X_0 - \Delta_0 S_0) (1+r)
	&= \phi(uS_0) , \\
\Delta_0 110 + (X_0 - \Delta_0 100) 
	&= 10
\end{align}
and
\begin{align}
\Delta_0 d S_0 + (X_0 - \Delta_0 S_0) (1+r)
	&= \phi(dS_0) , \\
\Delta_0 90 + (X_0 - \Delta_0 100)
	&= 0 .
\end{align}
Solving for $X_0$, we find $X_0 = 5$.  We could find the put price in a similar fashion.  But, a faster method to obtain the put price is to use Put-Call parity.  We have $C_0(1,K) - P_0(1,K) = S_0 - K$. So, for $K=S_0=100$ we have $C_0(1,S_0) = P_0(1,S_0) = 5$.
\end{example}

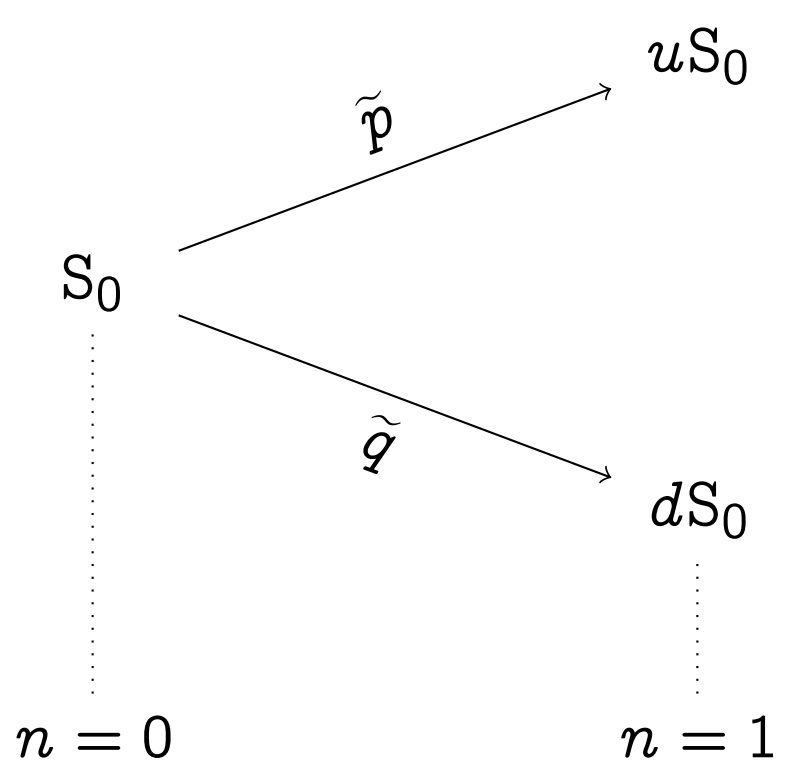
\begin{figure}[t!]
  \centering
	\begin{tikzpicture}[sloped]
		\node (a) at ( 0,0) [bag] {$S_0$};
		\node (b) at ( 4,-1.5) [bag] {$d S_0$};
		\node (c) at ( 4,1.5) [bag] {$u S_0$};
		\node (d) at ( 0,-3) [end] {$n=0$};
		\node (e) at ( 4,-3) [end] {$n=1$};
		\draw [->] (a) to node [below] {$\qt$} (b);
		\draw [->] (a) to node [above] {$\pt$} (c);
		\draw [dotted] (d) to node [below] {} (a);
		\draw [dotted] (e) to node [above] {} (b);
	\end{tikzpicture}
	\caption{A risk-neutral one-step binomial tree for a stock.\label{fig:1-step-stock-RN}}
\end{figure}

\subsection{Risk-Neutral Valuation of European options }
We now provide an alternative method of valuing derivatives known as \textit{risk-neutral valuation}.
Let us consider the binomial model under a new probability measure $\Pbt$.
Let us denote by $\pt$ and $\qt$, respectively, the probabilities of heads and tails
\begin{align}
\Pbt(H)
	&=	\pt , &
\Pbt(T)
	&=	\qt .
\end{align}
Forget that we have previously assigned values to $\pt$ and $\qt$.  Assume for the moment that the values of $\pt$ and $\qt$ are unknown.
The dynamics of $S$ under $\Pbt$ are depicted in Figure \ref{fig:1-step-stock-RN}.
Observe in Figure \ref{fig:1-step-stock-RN} that $\Pbt(S_1 = u S_0) = \pt$ and $\Pbt( S_1 = d S_0) = \qt$ whereas in Figure \ref{fig:1-step-stock} we have $\Pb( S_1 = u S_0) = p$ and $\Pb( S_1 = d S_0) = q$.  In other words, in Figures \ref{fig:1-step-stock} and \ref{fig:1-step-stock-RN} the \textit{same} events can occur (i.e., the stock goes up or the stock goes down), but the \textit{probabilities} of these events have changed.
\\[0.5em]
Now, let us denote by $\Ebt$ an expectation taken under a probability measure $\Pbt$.  
Let us try to find the values of $\pt$ and $\qt$ such that $\Ebt(X_1/B_1) = X_0/B_0$.
Stated another way, let us search for a probability measure $\Pbt$ under which the process $X/B$ is a martingale.
You will see why this is useful in a moment.
For any choice of $\Del_0$, we have
\begin{align}
\Ebt \Big( \frac{X_1}{B_1} \Big)
	&=	\Del_0 \Ebt \Big( \frac{S_1}{B_1} \Big) + \frac{(X_0 - \Del_0 S_0)}{B_0} \frac{B_1}{B_1} \\
	&=	\Del_0 \Big( \pt \frac{u S_0}{B_0 (1+r)} + \qt \frac{dS_0}{B_0 (1+r)} \Big) + \frac{(X_0 - \Del_0 S_0)}{B_0}  .
\end{align}
Thus, in order for $\Ebt(X_1/B_1) = X_0/B_0$, we must have
\begin{align}
\frac{X_0}{B_0}
	&=	\Del_0 \Big( \pt \frac{u S_0}{B_0 (1+r)} + \qt \frac{dS_0}{B_0 (1+r)} \Big) + \frac{(X_0 - \Del_0 S_0)}{B_0} .
\end{align}
Let us solve the above equation for $\pt$ and $\qt$ explicitly.  Multiplying both sides by $B_0$ and subtracting $X_0$ from both sides gives
\begin{align}
0
	&=	\Del_0 \Big( \pt \frac{u S_0}{1+r} + \qt \frac{d S_0}{1+r} \Big) - \Del_0 S_0 .
\end{align}
Dividing both sides by $\Del_0 S_0$ (assuming $\Del_0 \neq 0$) gives
\begin{align}
0
	&=	\frac{\pt u + \qt d}{1+r} - 1
	\qquad \Longleftrightarrow \qquad
\pt u + \qt d
	=	1+r .
\end{align}
Together with the requirement $\pt + \qt = 1$ (i.e., $\qt = 1-\pt$), we find
\begin{align}
\pt
	&=	\frac{(1+r)-d}{u-d} , &
\qt
	&=	1-\pt
	=	\frac{u-(1+r)}{u-d} ,
\end{align}
which are precisely the values stated in \eqref{eq:pt}.
\\[0.5em]
Now, recall from Section \ref{sec:replication} that, by choosing $\Del_0$ and $X_0$ according to \eqref{eq:portfolio_0}, we can set up a replicating portfolio $X$ such that $X_1 = \phi(S_1) = V_1$.  Recall further that neither $X_0$ nor $\Del_0$ depended on the probability measure $\Pb$ (see Remark \ref{rmk:pq}).  Thus, changing from $\Pb$ to $\Pbt$ does not affect the value of an option.  Suppose we did not already know the initial value $X_0$ of the replicating portfolio.  We can use the fact that $X/B$ is a martingale under $\Pbt$ to find $X_0$ rather easily.  We have
\begin{align}
\frac{X_0}{B_0}
	&=	\Ebt \Big( \frac{X_1}{B_1} \Big)
	=		\Ebt \Big( \frac{\phi(S_1)}{B_1} \Big)
	 =	\pt \frac{\phi(u S_0)}{B_0 (1+r)} + \qt \frac{\phi(d S_0)}{B_0 (1+r)} .
\end{align}
Multiplying both sides by $B_0$ we obtain
\begin{align}
X_0 
	&= \frac{1}{(1+r)} \Big( \pt \phi (u S_0) + \qt \phi (d S_0) \Big), 
\end{align}
which agrees with \eqref{eq:portfolio_0}.  Because we know that $X_1 = \phi(S_1) = V_1$, it must be the case that the initial value of the option must be $V_0 = X_0$.  Thus, we have stumbled across a very simple two-step method to price options in the binomial model.
\\[0.5em]
\underline{Step 1}:  Find a probability measure $\Pbt$ (meaning find the values of $\pt$ and $\qt$) so that $X/B$ is a martingale
\begin{align}
\frac{X_0}{B_0}
	&=	\Ebt \( \frac{X_1}{B_1} \) . \label{eq:step1}
\end{align}
\underline{Step 2}: Compute the initial value $V_0$ of an option that pays $\phi(S_1)$ using
\begin{align}
\frac{V_0}{B_0} 
	&= \Ebt \( \frac{V_1}{B_1} \) = \Ebt \( \frac{\phi(S_1)}{B_1} \) . \label{eq:step2}
\end{align}
Computing option prices using the above two-step process is known as \textit{risk-neutral pricing}.
\\[0.5em]
We call $\Pbt$ the \textit{risk neutral probability measure} or \textit{martingale measure}.  We call $B$ the \textit{num\'eraire}.
If we denominate the value of a portfolio $X$ in units of the num\'eraire $B$, then under the risk-neutral measure $\Pbt$, equation \eqref{eq:step1} tells us that, in expectation, the value of $X$ does not change.
\\[0.5em]
In fact, there is nothing special about using the bond $B$ as the num\'eraire.  We can use any strictly positive portfolio as num\'eraire.  For example, if we use $S$ as the num\'eraire, then the above two-step procedure would become
\\[0.5em]
\underline{Step 1}:  Find $\Pbh$ (meaning assign probabilities to events) so that $X/S$ is a martingale
\begin{align}
\frac{X_0}{S_0}
	&=	\Ebh \( \frac{X_1}{S_1} \) . 
\end{align}
\underline{Step 2}: Compute the initial value $V_0$ of an option that pays $\phi(S_1)$ using
\begin{align}
\frac{V_0}{S_0} 
	&= \Ebh \( \frac{V_1}{S_1} \) = \Ebh \( \frac{\phi(S_1)}{S_1} \) . 
\end{align}
Try this for yourself and see that you get the same price $V_0$ of an option when you use $S$ as num\'eraire as when you use $B$ as num\'eraire.

\begin{example}
Let's compute the price of a call option assuming $S_0 = 100$, $K=100$, $u = 1.1$, $d=0.9$ and $r=0$.  First, we must find $\pt$ and $\qt$.  Using \eqref{eq:step1} we have
\begin{align}
\frac{X_0}{B_0}
	&=	\Ebt \[ \frac{X_1}{B_1} \] \\
	&=	\Del_0 \Ebt \[ \frac{S_1}{B_1} \] + \frac{(X_0-\Del_0 S_0)}{B_0}\frac{B_1}{B_1} \\
	&= \Del_0 \Big( \pt \frac{u S_0}{B_0 (1 + r) } + \qt \frac{d S_0}{B_0(1 + r)} \Big) + \frac{(X_0-\Del_0 S_0)}{B_0} \\
	&=	\Del_0  \frac{S_0}{B_0} (1.1 \pt + 0.9 \qt ) + \frac{X_0}{B_0} - \frac{\Del_0 S_0}{B_0}.
\end{align}  
And thus
\begin{align}
1
	&=	(1.1 \pt + 0.9 \qt )  .
\end{align}
Using $\qt = 1-\pt$, we find $\pt = 0.5$ and $\qt = 0.5$.  Next, using \eqref{eq:step2} we have
\begin{align}
\frac{V_0}{B_0}
	&=	\Ebt \[ \frac{V_1}{B_1} \] = \Ebt \[ \frac{(S_1-K)^+}{B_1} \] \\
	&= 	\pt \( \frac{(u S_0 -K)^+}{B_0 (1+r)} \) + \qt \( \frac{(d S_0-K)^+}{B_0 (1+r)} \) \\
	&=	0.5 \( \frac{10}{B_0} \) + 0.5 \( \frac{0}{B_0} \) = \frac{5}{B_0} .
\end{align}
Thus, we have $V_0 = 5$.
\end{example}

\section{$N$-step binomial model}
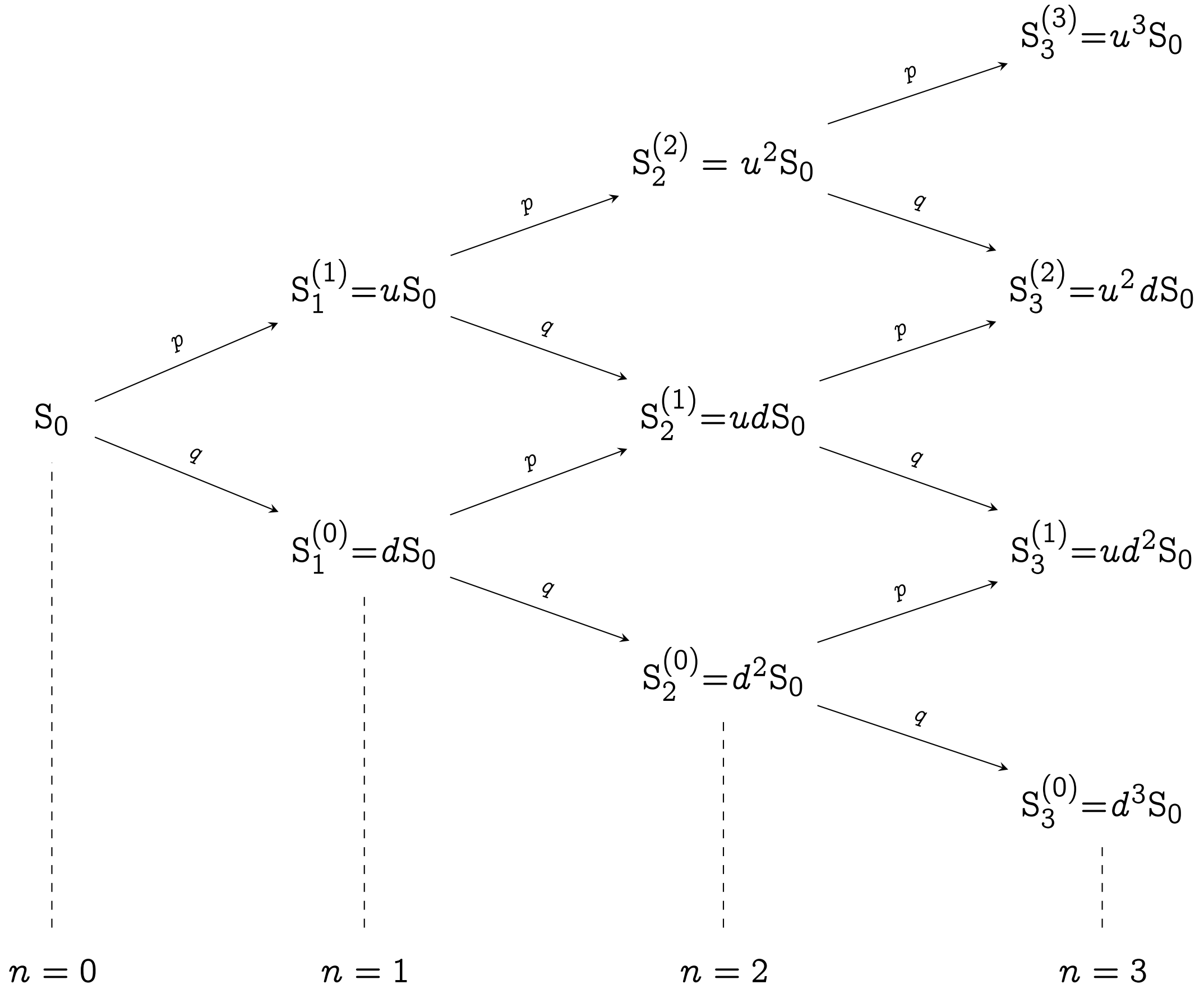
\begin{figure}[ht!]
  \centering
	  \begin{tikzpicture}[>=stealth,sloped]
      \matrix (tree) [%
        matrix of math nodes,
        minimum size=1cm,
        column sep=2cm,
        row sep=0.5cm,
      ]
      {
                  &                             &                                 & |(u3s0)| S_3^{(3)} {=} u^3 S_0  \\
                  &                             & |(u2s0)| S_2^{(2)} = u^2 S_0    &                                 \\
                  & |(us0)| S_1^{(1)} {=} u S_0 &                                 & |(u2ds0)| S_3^{(2)} {=} u^2 d S_0 \\
      |(s0)| S_0  &                             & |(uds0)| S_2^{(1)} {=} u d S_0  &                                 \\
                  & |(ds0)| S_1^{(0)} {=} d S_0 &                                 & |(ud2s0)| S_3^{(1)} {=} u d^2 S_0 \\
                  &                             & |(d2s0)| S_2^{(0)} {=} d^2 S_0  &                                 \\
                  &                             &                                 & |(d3s0)| S_3^{(0)} {=} d^3 S_0  \\
                  &                             &                                 &                                 \\
      |(n0)| n=0  & |(n1)| n=1                  & |(n2)| n=2                      & |(n3)| n=3                      \\
      };
      \tikzstyle{every node}=[midway,above,font=\scriptsize]
      \draw[->] (s0) -- node {$p$} (us0);
      \draw[->] (s0) -- node {$q$} (ds0);
      \draw[->] (us0) -- node {$p$} (u2s0);
      \draw[->] (us0) -- node {$q$} (uds0);
      \draw[->] (ds0) -- node {$p$} (uds0);
      \draw[->] (ds0) -- node {$q$} (d2s0);
      \draw[->] (u2s0) -- node {$p$} (u3s0);
      \draw[->] (u2s0) -- node {$q$} (u2ds0);
      \draw[->] (uds0) -- node {$p$} (u2ds0);
      \draw[->] (uds0) -- node {$q$} (ud2s0);
      \draw[->] (d2s0) -- node {$p$} (ud2s0);
      \draw[->] (d2s0) -- node {$q$} (d3s0);
      \draw[dashed] (n0) -- (s0);
      \draw[dashed] (n1) -- (ds0);
      \draw[dashed] (n2) -- (d2s0);
      \draw[dashed] (n3) -- (d3s0);
  \end{tikzpicture}
	\caption{A 3-step binomial tree for a stock $S$.\label{fig:n-step-S}}
\end{figure}

We are now going to extend the one-period binomial model to an $N$-step binomial model.
Consider an experiment in which we toss a coin $N$ times.  Our sample space is
\begin{align}
\Om
	&=	\{ \text{all sequences of $H$'s and $T$'s of length $N$} \} .
\end{align}
We will denote a sample outcome $\om$ of this experiment as follows
\begin{align}
\om
	&=	\om_1 \om_2 \ldots \om_N , &
\om_i
	&\in \{ H, T \}
\end{align}
For example, if $N = 5$, a possible outcome is $\om = H T T H H$.
We will assume that coin tosses are independent and that $\Pb(H) = p$ and $\Pb(T) = 1-p =:q$.
\\[0.5em]
Now, consider a market on the sample space that includes a stock $S = (S_n)_{n \in \{0,1 \ldots,N\}}$ and a bond $B = (B_n)_{n \in \{0,1,\ldots,N\}}$.  The value of the stock $S_0$ at time $n=0$ is known.  At time $n+1$, the value of the stock is given by
\begin{align}
S_{n+1}
	&=	\left\{ \begin{aligned}
			&u S_n , & &\text{if $\om_{n+1} = H$} , \\
			&d S_n , & &\text{if $\om_{n+1} = T$} .
			\end{aligned} \right.
\end{align}
For example, if $\om = HHT\ldots$ then we would have $S_1 = u S_0$, $S_2 = u^2 S_0$ and $S_3 = u^2 d S_0$.  Figure~\ref{fig:n-step-S} depicts how the stock $S$ behaves in the $3$-step binomial model.  Note that the tree recombines.  Also note in the figure that
\begin{align}
S_n^{(i)}
	&=	u^i d^{n-i} S_0 , &
i
	&\in \{0, 1, \ldots, n \} . \label{eq:Sni}
\end{align} 
Thus subscript $n$ denotes the time and the superscript $(i)$ denotes the number of ``up'' movements in the lattice seen in Figure \ref{fig:n-step-S}.
\\[0.5em]
The bond in the $N$-step binomial model has an initial value of $B_0$.  At time $n+1$, the value of the bond is given by
\begin{align}
B_{n+1}
	&=	(1 + r) B_n .
\end{align}
Thus, we have $B_n = (1+r)^n B_0$.  As in the one-step binomial model, we will assume $0 < d < 1+r < u < \infty$, which guarantees that there is no arbitrage in the market.

\subsection{Pricing  European options by replication}
\label{sec:replication2}
\begin{figure}[ht!]
  \centering
		  \begin{tikzpicture}[>=stealth,sloped]
      \matrix (tree) [%
        matrix of math nodes,
        minimum size=1cm,
        column sep=2cm,
        row sep=0.5cm,
      ]
      {
                  &                             		&                                 		& |(u3s0)| X_3^{(3)} {=} \phi(u^3 S_0)  \\
                  &                             		& |(u2s0)| X_2^{(2)}, \Del_2^{(2)}    &                                 \\
                  & |(us0)| X_1^{(1)}, \Del_1^{(1)} &                                 		& |(u2ds0)| X_3^{(2)} {=} \phi(u^2 d S_0) \\
      |(s0)| X_0  &                            		 & |(uds0)| X_2^{(1)}, \Del_2^{(1)} 	 	&                                 \\
                  & |(ds0)| X_1^{(0)}, \Del_1^{(0)} &                                 		& |(ud2s0)| X_3^{(1)} {=} \phi(u d^2 S_0) \\
                  &                             		& |(d2s0)| X_2^{(0)},  \Del_2^{(0)}  	&                                 \\
                  &                             		&                                 		& |(d3s0)| X_3^{(0)} {=} \phi(d^3 S_0)  \\
                  &                             		&                                 		&                                 \\
      |(n0)| n=0  & |(n1)| n=1                  		& |(n2)| n=2                      		& |(n3)| n=3                      \\
      };
      \tikzstyle{every node}=[midway,above,font=\scriptsize]
      \draw[->] (s0) -- node {$p$} (us0);
      \draw[->] (s0) -- node {$q$} (ds0);
      \draw[->] (us0) -- node {$p$} (u2s0);
      \draw[->] (us0) -- node {$q$} (uds0);
      \draw[->] (ds0) -- node {$p$} (uds0);
      \draw[->] (ds0) -- node {$q$} (d2s0);
      \draw[->] (u2s0) -- node {$p$} (u3s0);
      \draw[->] (u2s0) -- node {$q$} (u2ds0);
      \draw[->] (uds0) -- node {$p$} (u2ds0);
      \draw[->] (uds0) -- node {$q$} (ud2s0);
      \draw[->] (d2s0) -- node {$p$} (ud2s0);
      \draw[->] (d2s0) -- node {$q$} (d3s0);
      \draw[dashed] (n0) -- (s0);
      \draw[dashed] (n1) -- (ds0);
      \draw[dashed] (n2) -- (d2s0);
      \draw[dashed] (n3) -- (d3s0);
  \end{tikzpicture}
	\caption{A $3$-step binomial tree for a portfolio $X$.\label{fig:n-step-X}}
\end{figure}
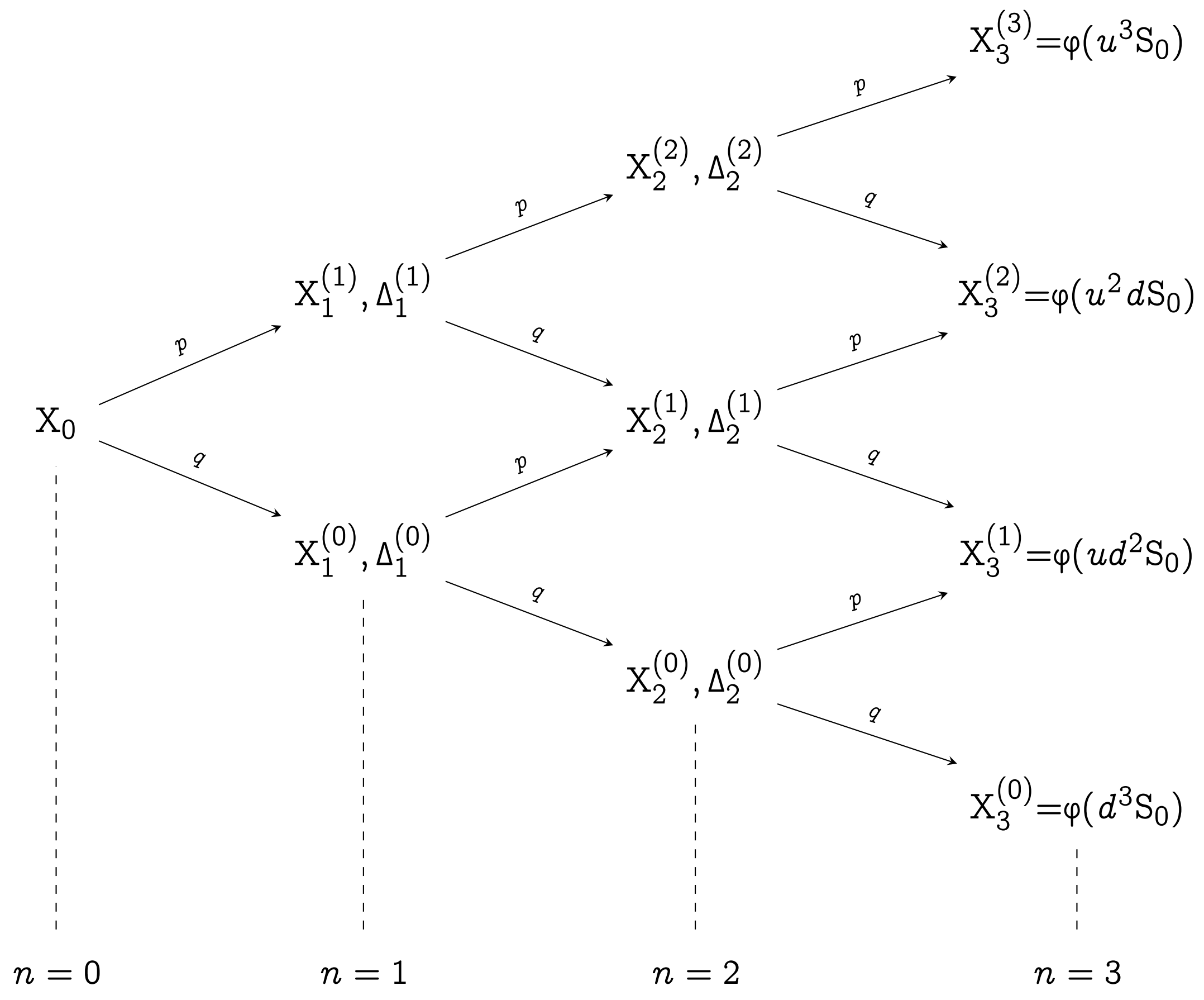

Consider a European option that pays $\phi(S_N)$ at time $N$.
We would like to find a portfolio $X = (X_n)_{n \in \{0,1,\ldots,N\}}$ that replicates the payoff $X_N = \phi(S_N)$.  
Figure \ref{fig:n-step-X} depicts how the portfolio $X$ behaves in the 3-step binomial model.
Assuming replication is possible, then at time $N$ we know that $X_N^{(i)} = \phi(S_N^{(i)})$ for $i=0,1,2, \ldots,N$.  We would like to find $X_n^{(i)}$ and $\Del_n^{(i)}$ for $n=0,1,\ldots, N-1$ and $i=0,1,\ldots,n$.
\\[0.5em]
Suppose the value of the stock and replicating portfolio at time $n$ are $S_n^{(i)}$ and $X_n^{(i)}$, respectively.  If the portfolio contains $\Del_n^{(i)}$ of the stock $S$, then we can write the value $X_n^{(i)}$ as follows
\begin{align}
X_n^{(i)}
	&=	\Del_n^{(i)} S_n^{(i)} + (X_n^{(i)} - \Del_n^{(i)} S_n^{(i)}) \frac{1}{B_n} B_n .
\end{align}
At time $n+1$, that value of the portfolio becomes
\begin{align}
X_{n+1}^{(i+1)} 
	&=	\Del_n^{(i)} S_{n+1}^{(i+1)} + (X_n^{(i)} - \Del_n^{(i)} S_n^{(i)}) (1+r) , && \text{if $\om_{n+1}=H$}, \\
X_{n+1}^{(i)}
	&=	\Del_n^{(i)} S_{n+1}^{(i)} + (X_n^{(i)} - \Del_n^{(i)} S_n^{(i)}) (1+r) , && \text{if $\om_{n+1}=T$}.
\end{align}
Solving the above system of equations for $X_n^{(i)}$ and $\Del_n^{(i)}$, we obtain
\begin{align}
X_n^{(i)}
	&=	\frac{1}{1+r} \Big( \pt X_{n+1}^{(i+1)} + \qt X_{n+1}^{(i)} \Big), &
\Del_n^{(i)}
	&=	\frac{X_{n+1}^{(i+1)} - X_{n+1}^{(i)}}{S_{n+1}^{(i+1)} - S_{n+1}^{(i)} } , \label{eq:portfolio_n}
\end{align}
where $\pt$ and $\qt$ are given by \eqref{eq:pt}.
Using the fact that $X_N^{(i)} = \phi(S_N^{(i)})$, where $S_n^{(i)}$ is given by \eqref{eq:Sni}, we can now use backwards recursion to find $X_n^{(i)}$ and $\Del_n^{(i)}$ at every node.  First we find $X_{N-1}^{(i)}$ and $\Del_{N-1}^{(i)}$ for $i = 0, 1, \ldots, N-1$.  Then we find $X_{N-2}^{(i)}$ and $\Del_{N-2}^{(i)}$ for $i = 0, 1, \ldots, N-2$, etc.  
\\[0.5em]
Because, by construction, the portfolio $X$ replicates the option payoff $X_N = \phi(S_N)$, we conclude that the value $V = (V_n)_{n \in \{0, 1, \ldots, N\}}$ of the derivative asset is given by $V_n = X_n$ for all $n$.  

\begin{remark}
\label{rmk:pq2}
Note once again that, as in the one-step binomial model, in the $N$-step binomial model, the probabilities $\Pb(H)=p$ and $\Pb(T)=q$ do not affect the values of $X_n$, $\Del_n$ or $V_n$!  Thus, we can change the probabilities $p$ and $q$ and this will not affect the values of the option or the replication strategy.  We will use this in the next section.
\end{remark}

\begin{example}
\label{ex:binomial-replication}
Consider a two-step binomial model (i.e., $N=2$) with the following parameters
\begin{align}
S_0
	&=	100, &
u
	&= 	1.1, &
d
	&=	0.9, &
r
	&=	0.00 , &
K
	&=	100, &
\phi(s)
	&=	( s - K )^+ .
\end{align}
Let us find a replicating portfolio $X$ and a hedging strategy $\Del$ that replicates European option that pays $\phi(S_2)$ at time $n=2$.  First we note that
\begin{align}
S_1^{(0)}
	&=	d S_0 = 90 , &
S_1^{(1)}
	&=	u S_0 = 110	, \\
S_2^{(0)}
	&=	d^2 S_0 = 81 , &
S_2^{(1)}
	&=	u d S_0 = 99 , &
S_2^{(2)}
	&=	u^2 S_0 = 121 , &
\end{align}
In order for $X_2 = \phi(S_2)$ we must have
\begin{align}
X_2^{(0)}
	&=	\phi(S_2^{(0)}) = ( 81 - 100 )^+ = 0 , \\
X_2^{(1)}
	&=	\phi(S_2^{(1)}) = ( 99 - 100 )^+ = 0 , \\
X_2^{(2)}
	&=	\phi(S_2^{(2)}) = ( 121 - 100 )^+ = 21 .
\end{align}
Next, using \eqref{eq:portfolio_n} we obtain
\begin{align}
X_1^{(0)}
	&=	\frac{1}{1+r} \Big( \pt X_2^{(1)} + \qt X_2^{(0)} \Big) = 0, &
\Del_1^{(0)}
	&=	\frac{X_2^{(1)}- X_2^{(0)}}{S_2^{(1)} - S_2^{(0)}} = 0 , \\
X_1^{(1)}
	&=	\frac{1}{1+r} \Big( \pt X_2^{(2)} + \qt X_2^{(1)} \Big) = 10.5 , &
\Del_1^{(1)}
	&=	\frac{X_2^{(2)} - X_2^{(1)}}{S_2^{(2)}- S_2^{(1)}} =  \frac{21}{22} , \\
X_0^{(0)}
	&=	\frac{1}{1+r} \Big( \pt X_1^{(1)} + \qt X_1^{(0)} \Big) = 5.25 , &
\Del_0^{(0)}
	&=	\frac{X_1^{(1)} - X_1^{(0)}}{S_1^{(1)}- S_1^{(0)}} = \frac{10.5}{20} ,
\end{align}
where we have used \eqref{eq:pt} to compute $\pt = ( 1 + r - d)/(u-d) = 1/2$ and $\qt = 1- \pt = 1/2$.
Because the portfolio $X$ replicates the payoff of the option $X_2 = \phi(S_2)$, we must have $V_n^{(i)} = X_n^{(i)}$ for all $n$ and $i$, where $V_n^{(i)}$ denotes the value of the option at time $n$ and node $i$.
\end{example}

\subsection{Risk-neutral pricing of  European options}
In this section, we will demonstrate how to value a derivative asset in the $N$-period binomial model using the risk-neutral pricing method.  To begin let us consider the $N$-step binomial model under a probability measure $\Pbt$.  
Note that we are \textit{not} changing how the stock $S$ behaves as a function of $\om$ (i.e., if $\om = HHT\ldots$, then under both $\Pb$ and $\Pbt$ we have $S_1 = uS_0$, $S_2 = u^2 S_0$ and $S_3 = u^2 d S_0$).  We are only changing the \emph{probabilities} of getting a $H$ or $T$ at every step.
Suppose that, under $\Pbt$, the probabilities of heads and tails are given by
\begin{align}
\Pbt(H)
	&=	\pt , &
\Pbt(T)
	&=	\qt ,
\end{align}
where $\pt$ and $\qt$ are, for the moment, unknown.  Let $\Ebt_n$ denote an expectation under $\Pbt$ conditioned on the history of the stock $\Fc_n^S = \{ S_0, S_1, \ldots, S_n \}$.  That is, for any random variable $Z$, we have
\begin{align}
\Ebt_n Z
	&=	\Ebt( Z | \Fc_n^S )
	=		\Ebt( Z | S_0, S_1, \ldots, S_n ) .
\end{align}
Let us search for the values of $\pt$ and $\qt$ such that $\Ebt_n (X_{n+1}/B_{n+1}) = X_n/B_n$.  You will see why this is useful in a bit.  Assuming that $\Del_n$ and $X_n$ can be determined from $\Fc_n^S$ (which is true if one considers only investment strategies that depend on past movements of $S$), we have
\begin{align}
\Ebt_n \( \frac{X_{n+1}}{B_{n+1}} \)
	&=	\Del_n \Ebt_n \( \frac{S_{n+1}}{B_{n+1}} \) + \frac{( X_n - \Del_n S_n )}{B_n} \frac{B_{n+1}}{B_{n+1}} \\
	&=	\Del_n \Big( \pt \frac{u S_n}{B_n (1+r)} + \qt \frac{dS_n}{B_n (1+r)} \Big) + \frac{(X_n - \Del_n S_n)}{B_n} .
\end{align}
Thus, in order for $\Ebt_n (X_{n+1}/B_{n+1}) = X_n/B_n$ we must have
\begin{align}
\frac{X_n}{B_n}
	&=	\Del_n \Big( \pt \frac{u S_n}{B_n (1+r)} + \qt \frac{dS_n}{B_n (1+r)} \Big) + \frac{(X_n - \Del_n S_n)}{B_n} .
\end{align}
Let us solve the above equation for $\pt$ and $\qt$.  Multiplying both sides by $B_n$ and subtracting $X_n$ from both sides gives
\begin{align}
0
	&=	\Del_n \Big( \pt \frac{u S_n}{1+r} + \qt \frac{d S_n}{1+r} \Big) - \Del_n S_n .
\end{align}
Dividing both sides by $\Del_n S_n$ (assuming $\Del_n \neq 0$) gives
\begin{align}
0
	&=	\frac{\pt u + \qt d}{1+r} - 1
	\qquad \Longleftrightarrow \qquad
\pt u + \qt d
	=	1+r .
\end{align}
Together with the requirement $\pt + \qt = 1$, we find that
\begin{align}
\pt
	&=	\frac{(1+r)-d}{u-d} , &
\qt
	&=	1 - \pt
	=	\frac{u - (1+r)}{u-d} ,
\end{align}
which are precisely the values given in \eqref{eq:pt}.
\\[0.5em]
Now, with $\pt$ and $\qt$ given by \eqref{eq:pt}, we have
\begin{align}
\Ebt_n \( \frac{X_{n+m}}{B_{n+m}} \)
	&=	\Ebt_n \Ebt_{n + m -1 } \( \frac{X_{n+m}}{B_{n+m}} \) 
	 =	\Ebt_n \( \frac{X_{n+m-1}}{B_{n+m-1}} \)  
	 =	\ldots
	 =	\Ebt_n \( \frac{X_{n+1}}{B_{n+1}} \) = \frac{X_n}{B_n} . \label{eq:is.martingale}
\end{align}
Thus, for any investment strategy $(X,\Del)$ we have that the process $X/B$ is a martingale under $\Pbt$.
\\[0.5em]
Now, recall from Section \ref{sec:replication2} that, by choosing $\Del_n$ and $X_n$ according to \eqref{eq:portfolio_n}, we can set up a replicating portfolio $X$ such that $X_N = \phi(S_N) = V_N$.  Recall further that neither $X_n$ nor $\Del_n$ depended on the probability measure $\Pb$ (see Remark \ref{rmk:pq2}).  Thus, changing from $\Pb$ to $\Pbt$ does \textit{not} affect the value of an option $V_n$, the value of the replicating portfolio $X_n$ or the trading strategy $\Del_n$.  Suppose we did not already know the value $X_n$ of the replicating portfolio.  We can use the fact that $X/B$ is a martingale under $\Pbt$ to find $X_n$ rather easily.  
We have
\begin{align}
\frac{X_n^{(i)}}{B_n}
	&=	\Ebt_n \Big( \frac{X_{n+1}}{B_{n+1}} \Big)
	 =	\pt \frac{X_{n+1}^{(i+1)}}{B_n (1+r)} + \qt \frac{X_{n+1}^{(i)}}{B_n (1+r)} .
\end{align}
Multiplying both sides by $B_n$ we obtain
\begin{align}
X_n^{(i)}
	&= \frac{1}{(1+r)} \Big( \pt X_{n+1}^{(i+1)}+ \qt X_{n+1}^{(i)} \Big) , 
\end{align}
which agrees with \eqref{eq:portfolio_n}.  We can now use backwards recursion to find $X_n$ at any node.  
Alternatively, we could simply compute
\begin{align}
\frac{X_n}{B_n}
	&=	\Ebt_n \Big( \frac{X_N}{B_N} \Big)
	 =	\Ebt_n \Big( \frac{\phi(S_N)}{B_N} \Big) \\
	&=	\sum_{k=0}^{N-n} \frac{\phi(u^{N-n-k} d^k S_n)}{B_n(1+r)^{N-n}} \Pbt( S_N = u^{N-n-k} d^k S_n | S_n ) , \\
\Pbt( S_N = u^{N-n-k} d^k S_n | S_n )
	&=	\binom{N-n}{k} \pt^{N-n-k} \qt^k .
\end{align}
Though, for large $N-n$ it is typically faster computationally to use backwards recursion.
Because we know that $X_N = \phi(S_N) = V_N$, it must be the case that $V_n = X_n$ for all $n$.  Thus, we have stumbled across a very simple two-step method to price options in the binomial model.
\\[0.5em]
\underline{Step 1}:  Find a probability measure $\Pbt$ (meaning find the values of $\pt$ and $\qt$) so that $X/B$ is a martingale
\begin{align}
\frac{X_n}{B_n}
	&=	\Ebt_n \( \frac{X_{n+1}}{B_{n+1}} \) . \label{eq:step1a}
\end{align}
Note that, from \eqref{eq:is.martingale}, equation \eqref{eq:step1a} is sufficient to show that $X/B$ is a martingale under $\Pbt$.
\\[0.5em]
\underline{Step 2}: Compute the value $V_n$ of an option that pays $\phi(S_N)$ using
\begin{align}
\frac{V_n}{B_n} 
	&= \Ebt_n \( \frac{V_{n+1}}{B_{n+1}} \) , &
n
	&=	0, 1, \ldots, N-1 , &
V_N
	&=	 \phi(S_N) . \label{eq:step2a}
\end{align}
Computing option prices using the above two-step process is known as \textit{risk-neutral pricing}.
\\[0.5em]
As in the one-step binomial model, we call $\Pbt$ the \textit{risk neutral probability measure} or \textit{martingale measure}.  We call $B$ the \textit{num\'eraire}.
If we denominate the value of a portfolio $X$ in units of the num\'eraire $B$, then under the risk-neutral measure $\Pbt$, equation \eqref{eq:step1a} tells us that, in expectation, the value of $X$ does not change.
\\[0.5em]
Also, as in the one-period case, there is nothing special about using the bond $B$ as the num\'eraire.  We can use any strictly positive portfolio as num\'eraire.  For example, if we use $S$ as the num\'eraire, then the above two-step procedure would become
\\[0.5em]
\underline{Step 1}:  Find $\Pbh$ (meaning assign probabilities to events) so that $X/S$ is a martingale
\begin{align}
\frac{X_n}{S_n}
	&=	\Ebh_n \( \frac{X_{n+1}}{S_{n+1}} \) . 
\end{align}
\underline{Step 2}: Compute the value $V_n$ of an option that pays $\phi(S_N)$ using
\begin{align}
\frac{V_n}{S_n} 
	&= \Ebh_n \( \frac{V_{n+1}}{S_{n+1}} \) , &
n
	&=	0, 1, \ldots, N-1 , &
V_N
	&=	 \phi(S_N) . \label{eq:step2a-S}
\end{align}

\begin{example}
As in Example \ref{ex:binomial-replication}, consider the two-step binomial model $N=2$.  Assume $r=0$, $S_0 = 100$, $u=1.1$ and $d=0.9$.  What is the price of a call option with strike $K=100$?  At time $n = 2$ the stock $S$ can be one of three values $u^2 S_0 = 121$, $u d S_0 = 99 $, and $d^2 S_0 =81$.  The payoff $\phi(S_{2}) = (S_{2}-K)^+$ in these three states is $(u^2 S_0 - K)^+ = 21$, $(ud S_0 - K)^+=0$ and $(d^2 S_0 - K)^+=0$.  With $u$ and $d$ as given, we have $\pt = \qt = 0.5$.  Thus, defining
\begin{align}
V_{1}^{(1)} 
	&:=	\Ebt [ \phi(S_{2}) | S_{1} = u S_0 ]
	=		\pt \, \cdot \,  21 + \qt \, \cdot \,  0 = 10.5 , \\
V_{1}^{(0)} 
	&:=	\Ebt [ \phi(S_{2 }) | S_{1} = d S_0 ]
	=		\pt \, \cdot \,  0 + \qt \, \cdot \,  0 = 0 ,
\end{align}
the value of the option $V_0^{(0)}$ is given by
\begin{align}
V_0^{(0)}
	&=	\Ebt [ V_{1} ]
	=		\pt V_{1}^{(1)}  + \qt V_{1}^{(0)} 
	=		\pt \, \cdot \,  10.5 + \qt \, \cdot \,  0 
	=		5.25 .
\end{align}
Note the value of $V_0^{(0)}$ obtained above is the same as the value we obtained for the replicating portfolio $X_0^{(0)}$ in Example \ref{ex:binomial-replication}.
\end{example}

\subsection{Risk-neutral pricing of American Options}
\label{sec:binomial-american}
Recall that an American option gives its holder the right, but not the obligation, to exercise the option at any time before or at maturity.  
It will be optimal for the owner of an American option to exercise his right only if the value of immediately exercising the option is greater than the value of holding the option.   From \eqref{eq:step2a}, we have that
\begin{align}
\{ \text{value of holding the option at time $n$} \}
	&=	B_n \Ebt_n \( \frac{V_{n+1}}{B_{n+1}} \)  . \label{eq:value.hold}
\end{align}
On the other hand, we also have that
\begin{align}
\{\text{value of exercising the option at time $n$} \}
	&=	\phi(S_n) . \label{eq:value.exericise}
\end{align}
The value of the American option at time $n$ is the maximum of \eqref{eq:value.hold} and \eqref{eq:value.exericise}.  Thus, we have
\begin{align}
V_n
	&=	\text{max} \left\{ \phi(S_n) , B_n \Ebt_n \( \frac{V_{n+1}}{B_{n+1}} \) \right\} , &
n
	&=	0, 1, \ldots , N-1 , &
V_N
	&=	\phi(S_N) . \label{eq:value.american}
\end{align}
Using \eqref{eq:value.american}, we can value an American option using backwards recursion.  The following example illustrates this procedure.

\begin{example}
\label{ex:american}
Consider an American Put option in the two-step binomial model.  We fix the following parameters
\begin{align}
u &= 1.2, &
d &= 0.8, &
1+r	&= \ee^{0.05} \approx 1.0513, &
S_0
	&=	50 , &
K 
	&= 52 .
\end{align}
With $u$, $d$ and $r$ as given, we find
\begin{align}
\pt
	&=	\frac{(1+r) - d}{u - d} \approx  0.6282 , &
\qt
	&=	1- \pt \approx 0.3718.
\end{align}
Next, let us find the value of $S$ at each of the nodes.  Using the notation of Figure \ref{fig:n-step-S}, we have
\begin{align}
S_0	&=	50, &
S_1^{(1)} &= u S_0 = 60, &
S_1^{(0)} &= d S_0 = 40, \\
S_2^{(2)} &= u^2 S_0 = 72, &
S_2^{(1)} &= u d S_0 = 48, &
S_2^{(0)} &= d^2 S_0 = 32, 
\end{align}
Setting $\phi(s) = (K - s)^+$, the value of exercising the option at each node is
\begin{align}
\phi(S_0)	&=	2 , &
\phi(S_1^{(1)}) &= \phi( u S_0) = 0, &
\phi(S_1^{(0)}) &= \phi(d S_0) = 12, \\
\phi(S_2^{(2)}) &= \phi(u^2 S_0) = 0, &
\phi(S_2^{(1)}) &= \phi(u d S_0) = 4, &
\phi(S_2^{(0)}) &= \phi(d^2 S_0) = 20, 
\end{align}
Now, let us figure out the value of the option at time $n=1$.  At the top node we have
\begin{align}
V_{1}^{(1)}
	&=	\max \{ \phi( u S_0) , \frac{1}{1+r} \Ebt[ \phi(S_{2}) | S_{1} = u S_0 = 60 ] \} \\
	&=	\max \{ \phi( u S_0) , \frac{1}{1+r} \( \pt \phi(S_{2}^{(2)}) + \qt \phi(S_{2}^{(1)})\) \} \\
	&=	\max \{ 0 , 1.4147 \} 
	=		 1.4147 .
\end{align}
So, if $S_{1} = u S_0$, then it is better to hold the option to maturity (this should be obvious, since the payoff from exercising is zero).
At the bottom node we have
\begin{align}
V_{1}^{(0)}
	&=	\max \{ \phi( d S_0) , \frac{1}{1+r} \Ebt[ \phi(S_{2}) | S_{1} = d S_0 = 40 ] \} \\
	&=	\max \{ \phi( d S_0) , \frac{1}{1+r} \( \pt \phi(S_{2}^{(1)}) + \qt \phi(S_{2}^{(0)})\) \} \\
	&=	\max \{ 12 ,  9.4636 \} 
	=		 12 .
\end{align}
As the exercise value is greater than the value of holding the option, if $S_{1} = d S_0$, then it is better to exercise the option.  Finally, let us compute the value of the option at time zero.  We have
\begin{align}
V_0
	&=	\max \{ \phi( S_0) , \frac{1}{1+r} \Ebt[ V_{1} | S_0 = 50 ] \} \\
	&=	\max \{ \phi( S_0) , \frac{1}{1+r} \( \pt V_{1}^{(1)} + \qt V_{1}^{(0)}\) \} \\
	&=	\max \{ 2 ,  5.0894 \} 
	=		5.0894 .
\end{align}
As the value of holding the option is greater than the value of exercising the option, it is best to hold the option at time $n=0$.
\end{example}

\noindent
We have shown how to value an American option using risk-neutral valuation.  A natural question to ask is: how do we replicate the option?  Denoting by $X$ the value of the replicating portfolio and by $V$ the value of the option, we would like for $X_n = V_n$ for every $n \leq \nu$, where $\nu$ is the optimal exercise date.  To accomplish this, just as in the European case, we try to set up a replicating portfolio.
Suppose the value of the stock and replicating portfolio at time $n$ are $S_n^{(i)}$ and $X_n^{(i)} = V_n^{(i)}$, respectively.  If the portfolio contains $\Del_n^{(i)}$ shares of the stock $S$, then we can write the value $X_n^{(i)}$ as follows
\begin{align}
X_n^{(i)}
	&=	\Del_n^{(i)} S_n^{(i)} + (V_n^{(i)} - \Del_n^{(i)} S_n^{(i)}) \frac{1}{B_n} B_n .
\end{align}
At time $n+1$, that value of the portfolio becomes
\begin{align}
X_{n + 1}
	&= 	\left\{ \begin{aligned}
			&\Del_n^{(i)} S_{n+1}^{(i+1)} + (V_n^{(i)} - \Del_n^{(i)} S_n^{(i)}) (1+r), && \text{if $\om_{n+1}=H$},\\
			&\Del_n^{(i)} S_{n+1}^{(i)} + (V_n^{(i)} - \Del_n^{(i)} S_n^{(i)}) (1+r), && \text{if $\om_{n+1}=T$}.
			\end{aligned} \right. 
\end{align}
We would like to have
\begin{align}
X_{n + 1}
	&= 	\left\{ \begin{aligned}
			&V_{n+1}^{(i+1)}, && \text{if $\om_{n+1}=H$},\\
			&V_{n+1}^{(i)}, && \text{if $\om_{n+1}=T$}.
			\end{aligned} \right. 
\end{align}
In order for this to occur, the following two equations must be satisfied
\begin{align}
V_{n+1}^{(i+1)} 
	&=	\Del_n^{(i)} S_{n+1}^{(i+1)} + (V_n^{(i)} - \Del_n^{(i)} S_n^{(i)}) (1+r) , \label{eq:american-1} \\
V_{n+1}^{(i)}
	&=	\Del_n^{(i)} S_{n+1}^{(i)} + (V_n^{(i)} - \Del_n^{(i)} S_n^{(i)}) (1+r) . \label{eq:american-2}
\end{align}
Solving either one of these equations for $\Del_n^{(i)}$, we obtain
\begin{align}
\Del_n^{(i)}
	&=	\frac{V_{n+1}^{(i+1)} - V_{n+1}^{(i)}}{S_{n+1}^{(i+1)} - S_{n+1}^{(i)} } , 
\end{align}
where $V_n^{(i)}$ denotes the value of the American option at time $n$ assuming $S_n = S_n^{(i)}$.
\\[0.5em]
Now, what happens if we arrive at the optimal exercise date $\nu$ and the holder of the option does not exercise?  In this case, we can actually hedge in such a way that $X_{\nu+1} \geq V_{\nu+1}$.  To see this, let us try to choose $X_\nu$ and $\Del_\nu$ so that $X_{\nu+1} = V_{\nu+1}$.  The equations we must solve are
\begin{align}
V_{\nu+1}^{(i+1)} 
	&=	\Del_\nu^{(i)} S_{\nu+1}^{(i+1)} + (X_\nu^{(i)} - \Del_\nu^{(i)} S_\nu^{(i)}) (1+r) , \label{eq:american-3} \\
V_{\nu+1}^{(i)}
	&=	\Del_\nu^{(i)} S_{\nu+1}^{(i)} + (X_\nu^{(i)} - \Del_\nu^{(i)} S_\nu^{(i)}) (1+r) . \label{eq:american-4}
\end{align}
Suppose we use equations \eqref{eq:american-3}-\eqref{eq:american-4} to find $X_\nu^{(i)}$ and $\Del_\nu^{(i)}$, what we would find is that $X_\nu^{(i)} \leq V_{\nu}^{(i)}$.  What this means is that, if we have been hedging properly up to time $\nu$, the value of the hedging portfolio $V_\nu$ is \textit{more} than what is needed $X_\nu$ to replicate the value of the option $V_{\nu+1}$ at time $\nu+1$.  So, the bank that sold the option can now make money by doing the following:
(i) buy $\Del_\nu^{(i)}$ shares of $S$ and (ii) put the rest of the money in the bank account.

\begin{example}
Let us compute the $\Del_0$ of the put option discussed in Example \ref{ex:american} above.  Noting that it is \textit{not} optimal to exercise at time $n=0$, we have
\begin{align}
\Delta_{0} 
	&= \frac{V_{1}^{(1)} - V_{1}^{(0)}}{S_{1}^{(1)} - S_{1}^{(0)}} 
	 = \frac{V_{1}^{(1)} - V_{1}^{(0)}}{u S_0 - d S_0} 
	=	\frac{ 1.4147 - 12}{60 - 40}
	=		-0.5292 .
\end{align}
Notice that $\Del_0$ is negative.  
Now, what happens at time $n=1$ at node $0$?  We know that it \textit{is} optimal to exercise because $V_1^{(0)} = \phi(S_1^{(0)}) = 12$.  If we use equations \eqref{eq:american-3}-\eqref{eq:american-4} to find $X_1^{(0)}$ and $\Del_1^{(0)}$, we obtain
\begin{align}
X_1^{(0)}
	&=	\frac{1}{1+r} \( \pt \phi(S_{2}^{(1)}) + \qt \phi(S_{2}^{(0)})\) =  9.4636  , \\
\Del_1^{(0)}
	&=	\frac{V_{2}^{(1)} - V_{2}^{(0)}}{S_{2}^{(1)} - S_{2}^{(0)}} = \frac{4 - 20}{48 - 32} = - 1 .
\end{align}
Thus, at time $n=1$ and node $0$, we know that a portfolio with a value of 9.4636 at time that holds $\Del_1^{(0)} = -1$ shares of $S$ will replicate the value of the option at time $n=2$ (both at node $0$ and at node $1$).  Therefore, if the holder of the option does not exercise at time $n=1$ and node $0$, the seller of the option can stick the extra cash $V_1^{(0)} - X_1^{(0)} = 12 - 9.4636 = 2.5364$ into the bank account and make a profit $2.5364 (1+r) \approx 2.6665$.
\end{example}

\subsection{Calibration to data}
Consider a derivative written on $S = (S_t)_{0 \leq t \leq T}$ that matures at time $T$.
Suppose we wish to use the $N$-step binomial model to describe the value of $S$ and $B$ at times
\begin{align}
	&\{ t_0, t_1, \ldots , t_N \} , &
t_n
	&:=	n \del t , &
\del t
	&:=	T/N ,
\end{align}
We have
\begin{align}
S_{t_{n+1}}
	&=	\left\{ \begin{aligned}
			& u S_{t_n} & \om_{n+1} &= H , \\
			& d S_{t_n} & \om_{n+1} &= T ,
			\end{aligned} \right.  &
B_{t_{n+1}}
	&=	B_{t_n} \ee^{r \del t} .
\end{align}
Here, $r$ is the continuously compounded rate of interest per unit time.
Suppose, from time series data, we have an estimate for expected rate of return of a stock $\mu$ and the expected variance $\sig^2$ per unit time $\del t$.  Then, under the real world measure $\Pb$ (because we observe $S$ under $\Pb$ -- \textit{not} under $\Pbt$) we have
\begin{align}
\Eb \[ \left. \frac{S_{t_n + \delta t}}{S_{t_n}} \right| S_{t_n}\] 
	&\approx \frac{1}{N} \sum_{n=0}^{N-1} \frac{S_{t_{n+1}}}{S_{t_n}}
	=:	\ee^{\mu \delta t} , \\
\Vb \[ \left. \frac{S_{t_n + \delta t}}{S_{t_n}} \right| S_{t_n} \] 
	&\approx \frac{1}{N-1} \sum_{n=0}^{N-1} \Big( \frac{S_{t_{n+1}}}{S_{t_n}} - \ee^{\mu \del t} \Big)^2
	=:	\sigma^2 \delta t.
\end{align}
We  want to calibrate a binomial tree (i.e., choose parameters $u$, $d$, $p$ and $q$) so that it has the sample mean and sample variance given above.  That is, we would like
\begin{align}
\ee^{\mu \delta t}
	&=	\Eb \[ \left. \frac{S_{{t_n} + \delta t}}{S_{t_n}} \right| S_{t_n}\] 
	 = 	p u + q d , \label{eq:mean-match} \\
\sigma^2 \del t
	&=	\Vb \[ \left. \frac{S_{{t_n} + \delta t}}{S_{t_n}} \right| S_{t_n} \] 
	 = (p u^2 + q d^2) - (p u + d q)^2 
	=	p q ( u - d )^2 . \label{eq:xi}
\end{align}
where we have used $p+q=1$.  From \eqref{eq:mean-match} and $p+q=1$ we find that $p$ and $q$ must satisfy
\begin{align} 
p 
	&= \frac{\ee^{\mu \delta t} - d}{u - d} , &
q
	&=	\frac{u - \ee^{\mu \delta t}}{u - d}.	\label{eq:calibrated_p}
\end{align}
Having obtained $p$ and $q$, we turn our attention to $u$ and $d$.
Inserting \eqref{eq:calibrated_p} into \eqref{eq:xi}, we obtain
\begin{align}
\sig^2 \del t
	&= ( \ee^{\mu \delta t} - d )( u - \ee^{\mu \delta t} ) . \label{eq:xi-2}
\end{align}
We would like to find $u$ and $d$ so that \eqref{eq:xi-2} is satisfied.  
As we have only one equation to be satisfied and two unknowns, we have a bit of flexibility in how we choose $u$ and $d$.
We guess that
\begin{align}
u &= \ee^{\sigma \sqrt{\delta t}} , &
d &= \ee^{-\sigma \sqrt{\delta t}}. \label{eq:ud}
\end{align}
Let us suppose that  $\delta t \ll 1$.  This is reasonable because, for any $T$ we can make $\del t$ as small as we like by choosing $N$ sufficiently large.  Expanding the exponential up to $\Oc(\del t)$, we have
\begin{align} 
\ee^{\mu \delta t} &= 1 + \mu \delta t + \Oc(\delta t^2), \label{eq:taylor1} \\ 
	u =  \ee^{\sig \sqrt{\del t}} &= 1 + \sigma \sqrt{\delta t} + \frac{1}{2} \sigma^2 \delta t + \Oc(\delta t^{3/2}), \label{eq:taylor2} \\
	d =  \ee^{- \sig \sqrt{\del t}} &= 1 - \sigma \sqrt{\delta t} + \frac{1}{2} \sigma^2 \delta t + \Oc(\delta t^{3/2}). \label{eq:taylor3}
\end{align}
Inserting \eqref{eq:taylor1}, \eqref{eq:taylor2} and \eqref{eq:taylor3} into the right-hand side of \eqref{eq:xi-2} yields
\begin{align}
( \ee^{\mu \delta t} - d )( u - \ee^{\mu \delta t} )
	&=	\sig^2 \del t + \Oc( \del t^2) .
\end{align}
Thus, by choosing $p$ and $q$ according to \eqref{eq:calibrated_p} and by choosing $u$ and $d$ according to \eqref{eq:ud}, we have by construction that
\begin{align}
\Eb \[ \left. \frac{S_{{t_n} + \delta t}}{S_{t_n}} \right| S_{t_n}\] 
	&= \ee^{\mu \delta t}, &
\Vb \[ \left. \frac{S_{{t_n} + \delta t}}{S_{t_n}} \right| S_{t_n} \] 
	&= \sigma^2 \delta t + \Oc(\delta t^{2}).
\end{align}
In other words, the $N$-step binomial tree has, to a high degree of accuracy, the correct statistical properties of $S$ under $\Pb$.
\\[0.5em]
We now consider the $N$-step binomial model under $\Pbt$ with $u$ and $d$ as chosen in \eqref{eq:ud}.  Recall that $\pt$ and $\qt$ are given by \eqref{eq:pt} assuming $B_{n+1} = (1+r) B_n$.  But, we have specified above that $B_{t_{n+1}} = \ee^{r \del t} B_{t_n}$.  So, in the expression for $\pt$ we must replace $(1+r)$ with $\ee^{r \del t}$, which leads to
\begin{align}
\pt
	&=	\frac{ \ee^{r \del t} - d}{u - d} .
\end{align}
Recall that the no-arbitrage condition in the binomial model is $0 < d < 1+r < u < \infty$, which in the present setting becomes
\begin{align}
0 < \ee^{- \sig \sqrt{\del t}} < \ee^{r \del t} < \ee^{\sig \sqrt{\del t}} < \infty .
\end{align}
No matter what the values of $r$ and $\sig$ are, the above inequalities will be satisfied for $\del t < (\sig / r)^2$.  Thus, one should always choose $N$ sufficiently large so that $\del t = T/N$ satisfies this constraint.


\section{Exercises}
\label{sec:exer-2}

\begin{figure}[htb!]
  \centering
	\begin{tikzpicture}[sloped]
		\node (a) at ( 0,0) [] {$S_0$};
		\node (b) at ( 4,-1.5) [] {$d S_0$};
		\node (c) at ( 4,1.5) [] {$u S_0$};
		\node (d) at ( 0,-3) [] {$n=0$};
		\node (e) at ( 4,-3) [] {$n=1$};
		\draw [->] (a) to node [below] {$q$} (b);
		\draw [->] (a) to node [above] {$p$} (c);
		\draw [dotted] (d) to node [below] {} (a);
		\draw [dotted] (e) to node [above] {} (b);
	\end{tikzpicture}
	\begin{tikzpicture}[sloped]
		\node (a) at ( 0,0) [] {$B_0$};
		\node (b) at ( 4,1.5) [] {$B_0 (1+r)$};
		\node (d) at ( 0,-1.5) [] {$n=0$};
		\node (e) at ( 4,-1.5) [] {$n=1$};
		\draw [->] (a) to node [above] {$1$} (b);
		\draw [dotted] (d) to node [below] {} (a);
		\draw [dotted] (e) to node [above] {} (b);
	\end{tikzpicture}
\caption{One-period binomial model}
\label{fig:1-step}
\end{figure}
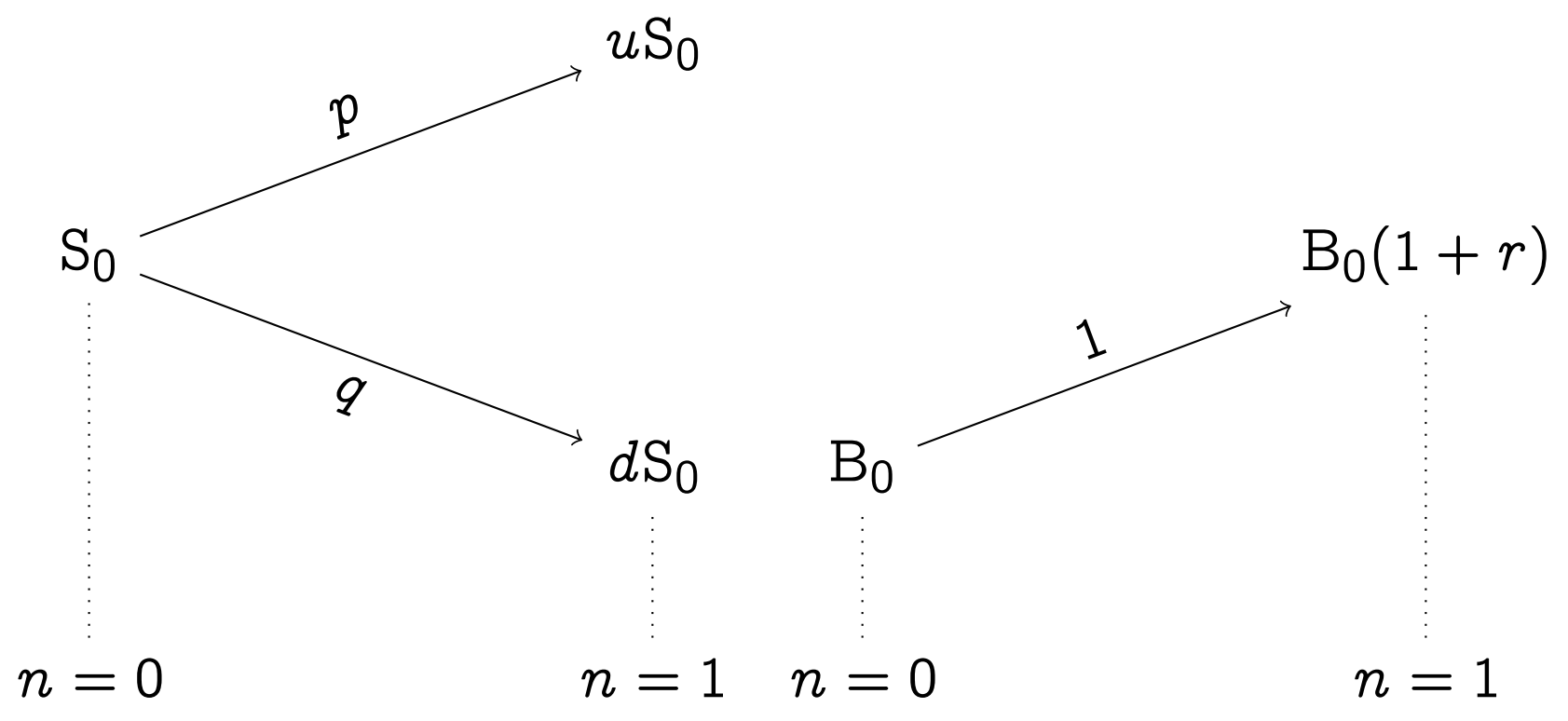

\begin{exercise}
\label{exer:2-1}
Consider the one-step binomial model, depicted in Figure \ref{fig:1-step}.  Show that if
\begin{align}
0 < d < u < 1+r< \infty , &&
\text{or}&&
0 < 1+r < d < u < \infty ,&&
\end{align} 
then you can construct a portfolio $X$ that satisfies $X_0 = 0$ and $X_1 > 0$.
\end{exercise}


\begin{exercise}
\label{exer:2-2}
Consider the one-step binomial model, depicted in Figure \ref{fig:1-step}.  Suppose that, if you own a share of $S$, you will receive a dividend payment of $a S_0$ at time $n=1$ (and if you short the stock, you owe this dividend payment to the owner).  Assume
\begin{align}
0 < d + a < 1 + r < u + a < \infty .
\end{align} 
\textbf{(a)} Construct a portfolio $X$ (meaning find the initial value $X_0$ and the number of shares $\Del_0$ invested in $S$ at time zero) so that $X_1 = \phi(S_1)$, where $\phi$ is an arbitrary payoff function. 
\textbf{(b)} Find a probability measure $\Pbt$ (meaning find $\pt$ and $\qt$) such that $\Ebt (X_1 / B_1) = X_0 / B_0$.
\textbf{(c)} Using your answer to part (b), compute $B_0 \Ebt ( \phi(S_1) / B_1 )$.  How does this compare to the value of $X_0$ you found in part (a)?
\textbf{(d)} Is it true that $\Ebt (S_1/B_1) = S_0/B_0$?  Try to explain why or why not.
\end{exercise}


\begin{exercise}
\label{exer:2-3}
Consider the one-step binomial model, depicted in Figure \ref{fig:1-step}.  Assume no dividends are paid (i.e., $a=0$). 
\textbf{(a)}  In class, we introduced a two-step procedure to find the time-zero value $V_0$ of an option that pays $\phi(S_1)$ at time $n=1$.  Step 1: find $\Pbt$ (meaning find $\pt$ and $\qt$) such that $\Ebt (X_1 / B_1) = X_0 / B_0$ for all portfolios $X$.  Step 2: Using your answer from Step 1, compute $V_0$ using $V_0 / B_0 = \Ebt ( V_1 / B_1 )$ where $V_1 = \phi(S_1)$. Use this method to compute $V_0$.
\textbf{(b)}  We are now going to introduce a different (but similar) two-step procedure to find the time-zero value $V_0$ of an option that pays $\phi(S_1)$ at time $n=1$.  Step 1: find $\Pbh$ (meaning find $\ph$ and $\qh$) such that $\Ebh (X_1 / S_1) = X_0 / S_0$ for all portfolios $X$.  Step 2: Using your answer from Step 1, compute $V_0$ using $V_0 / S_0 = \Ebh ( V_1 / S_1 )$ where $V_1 = \phi(S_1)$. Use this method to compute $V_0$.  The expressions you obtain for $V_0$ in parts (a) and (b) should be the same.
\end{exercise}

\begin{exercise}
\label{exer:2-4}
Consider the multi-period binomial model for a stock $S$ and bond $B$.  At each step we have
\begin{align}
S_{(n+1)\del t}
	&=	\left\{ \begin{aligned}
			&u S_{n \del t} &&\text{with probability $p$}\\
			&d S_{n \del t} &&\text{with probability $q$}
			\end{aligned} \right. , &
B_{(n+1)\del t}
	&=	\ee^{r \del t} B_{n \del t} .
\end{align}
We will set
\begin{align}
\del t
	&=	T/N, &
u
	&=	\ee^{\sig \sqrt{\del t}} , &
d
	&=	\ee^{-\sig \sqrt{\del t}} .
\end{align}
We call $\sig$ the \textit{volatility} of $S$.  Write a program to price and hedge a European put.  Your inputs should be the initial stock price $S_0$, the maturity date $T$, the strike $K$, the risk free rate of interest $r$, the volatility $\sig$, and the number of steps in your binomial tree $N$.  Your outputs should be the value of the option (or replicating portfolio) at time zero $X_0$, and the number of shares of $S$ you should have in the replicating portfolio at time zero $\Del_0$. 
\\[0.5em]
Your program should read the input from a text file called \textbf{input.txt}. The input file \textbf{input.txt} will consist of a single line listing $S_0$, $T$, $K$, $r$, $\sig$, $N$ in that order, separated by commas. Your program should write the solution to a text file called \textbf{output.txt}. Your output file should list $X_0$ and $\Del_0$ on a single line, separated by a comma.  You may use any programming language. 
\end{exercise}


\begin{exercise}
\label{exer:2-5}
Modify your program from Exercise \ref{exer:2-4} to price and hedge an American put.  Your inputs should be the initial stock price $S_0$, the maturity date $T$, the strike $K$, the risk free rate of interest $r$, the volatility $\sig$, and the number of steps in your binomial tree $N$.  Your outputs should be the value of the option (or replicating portfolio) at time zero $X_0$, and the number of shares of $S$ you should have in the replicating portfolio at time zero $\Del_0$.
\\[0.5em]
Your program should read the input from a text file called \textbf{input.txt}. The input file \textbf{input.txt} will consist of a single line listing $S_0$, $T$, $K$, $r$, $\sig$, $N$ in that order, separated by commas. Your program should write the solution to a text file called \textbf{output.txt}. Your output file should list $X_0$ and $\Del_0$ on a single line, separated by a comma.    You may use any programming language. 

\end{exercise}


\begin{exercise}
\label{exer:2-6}
Assume $\phi$ is a convex function and $\theta$ is a concave function.  Recall Jensen's Inequality: for any random variable $Z$ and any probability measure,
\begin{align}
\Eb[\phi(Z)] \geq \phi(\Eb[Z]) , \qquad \Eb[\theta(Z)] \leq \theta(\Eb[Z]) .
\end{align}
Consider the one-step binomial model, depicted in Figure \ref{fig:1-step}.  Assume $r = 0$ and $0 < d < 1 < u < \infty$.\\[0.5em]
\textbf{(a)} What is the value at time $n=0$ of an American option that pays $\phi(S_n)$ if it is exercised at time $n$?\\[0.5em]
\textbf{(b)} What is the value at time $n=0$ of an American option that pays $\theta(S_n)$ if it is exercised at time $n$?
\end{exercise}


\begin{exercise}
\label{exer:2-7}
Consider a foreign exchange market consisting of an American bond $A$ (in units of Dollars), a British bond $B$ (in units of Pounds), a stock $S$ (in units of Pounds), and an exchange rate $Z$ (in units of Dollars per Pound), with one-period dynamics
\begin{align}
A_0 \to A_1 = (1+a) A_0 , \quad
B_0 \to B_1 = (1+b) B_0 , \quad
Z_0 \to Z_1 = (1+z) Z_0 , \quad
S_0 \to
\left\{
\begin{aligned}
& u S_0 && \text{with prob.\ } p , \\
& d S_0 && \text{with prob.\ } q .
\end{aligned}
\right.
\end{align}
Suppose you are an American investor with initial wealth $X_0$ Dollars.  At time $n=0$ you purchase $\Del_0$ shares of $S$ and $\Gam_0$ British bonds, financing the position with American bonds.\\[0.5em]
\textbf{(a)} Derive an expression for $X_1$ in Dollars in terms of $(S_0, S_1, A_0, A_1, B_0, B_1, Z_0, Z_1, \Del_0, \Gam_0)$.\\[0.5em]
\textbf{(b)} Assume there is no arbitrage.  Find an expression for $z$ in terms of $(a,b)$.\\[0.5em]
\textbf{(c)} Find the risk-neutral probabilities $\pt := \Pbt(S_1 = u S_0)$ and $\qt := \Pbt(S_1 = d S_0)$ such that $X_0/A_0 = \Ebt[X_1/A_1]$.\\[0.5em]
\textbf{(d)} Let $V_0$ denote the time $n=0$ value in Dollars of an option that pays $\phi(S_1)$ Pounds at time $n=1$.  What is $V_0$?  Please leave your answer in terms of $\pt$ and $\qt$.
\end{exercise}


\begin{exercise}
\label{exer:2-8}
Consider the one-step binomial model, depicted in Figure \ref{fig:1-step}.  Assume $0 < d < (1+r) < u < \infty$ and that there is no arbitrage.  A leveraged portfolio with initial value $L_0 > 0$ and leverage ratio $\beta$ is managed as follows: at time $n=0$ the portfolio manager invests $\beta L_0$ dollars in the stock $S$.  To finance this position, the manager lends or borrows from the bank ($\beta$ could be positive or negative).\\[0.5em]
\textbf{(a)} Give an expression for the value of $L_1$ in terms of $\beta$, $L_0$, $S_0$, $B_0$, $S_1$ and $B_1$.\\[0.5em]
\textbf{(b)} Consider an option that pays $\phi(L_1/L_0)$ at time $n=1$.  Find a portfolio that replicates this option.  In other words, find the initial value $X_0$ and the number of shares $\Del_0$ invested in the stock $S$ so that $X_1 = \phi(L_1/L_0)$.
\end{exercise}


\begin{exercise}
\label{exer:2-9}
Consider a market with a stochastic interest rate.  A stock goes up from $S_0$ at time $n=0$ to $u S_0$ at time $n=1$ with probability $p$, or goes down to $d S_0$ with probability $q=1-p$.  If the stock goes up, then the bond grows from $B_0$ at time $n=0$ to $B_0(1+r_u)$ at time $n=1$.  If the stock goes down, then the bond grows from $B_0$ to $B_0(1+r_d)$.  Assume
\begin{align}
0 < d < 1+r_d < 1+r_u < u \qquad \text{and} \qquad d(1+r_u) < u(1+r_d) .
\end{align}
\textbf{(a)} Write a system of equations that you would solve in order to find risk-neutral probabilities $\pt$ and $\qt$ such that $X_0/B_0 = \Ebt[X_1/B_1]$ for all portfolios $X$.  You do \emph{not} need to find $\pt$ and $\qt$.\\[0.5em]
\textbf{(b)} Find the value of an option with payoff $\phi(S_1)$ at time $n=1$.  Please write your answer in terms of $\pt$ and $\qt$.
\end{exercise}


\begin{exercise}
\label{exer:2-10}
Consider a stock $S$ and bond $B$ in the binomial model with $0 < d < (1+r) < u$, and a \emph{perpetual contract} between a long side and a short side.  At time $n=0$, the long side pays the short side an initial amount $V_0$.  Both sides have the right to terminate the contract at any time.  If the contract is terminated at time $n$, the short side pays the long side $\phi(S_n)$.  If the contract is not terminated at time $n$, the long side must pay the short side a cash flow $F_n$ at time $n+1$.  Assume no arbitrage.\\[0.5em]
\textbf{(a)} What is $V_0$?\\[0.5em]
\textbf{(b)} Suppose the short side wishes to maintain a portfolio whose value $X$ satisfies $X_n = \phi(S_n)$ for all $n$.  If $\Del_n$ is the number of shares of $S$ in the portfolio at time $n$, what should $\Del_n$ be?\\[0.5em]
\textbf{(c)} What is $F_n$?  You may leave $\Del_n$ in your expression for $F_n$.
\end{exercise}


\begin{exercise}
\label{exer:2-11}
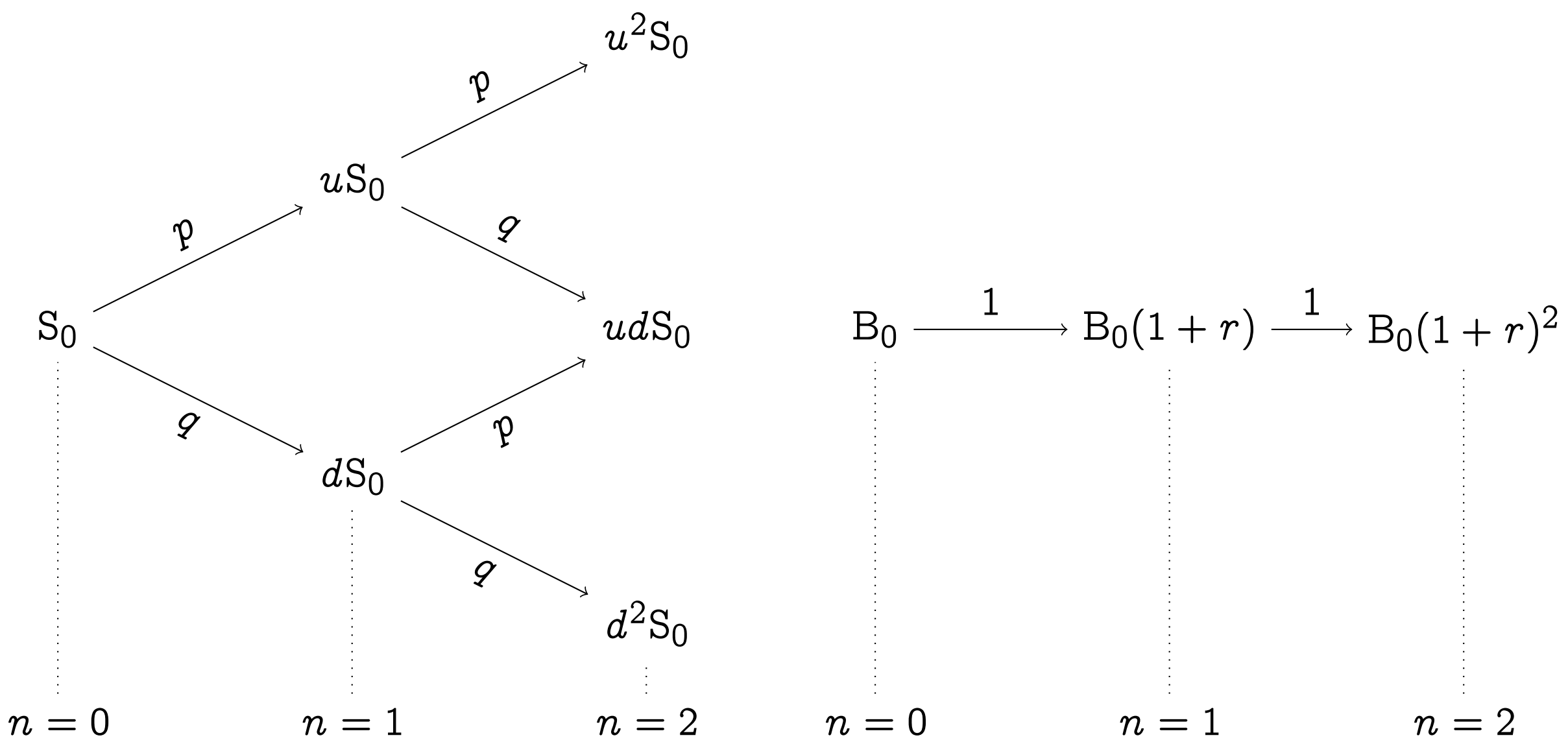
\begin{figure}[htb!]
  \centering
	\begin{tikzpicture}[sloped]
		\node (a) at ( 0,0) [] {$S_0$};
		\node (b) at ( 3,1.5) [] {$u S_0$};
		\node (c) at ( 3,-1.5) [] {$d S_0$};
		\node (d) at ( 6,3) [] {$u^2 S_0$};
		\node (e) at ( 6,0) [] {$u d S_0$};
		\node (f) at ( 6,-3) [] {$d^2 S_0$};
		\node (g) at ( 0,-4) [] {$n=0$};
		\node (h) at ( 3,-4) [] {$n=1$};
		\node (i) at ( 6,-4) [] {$n=2$};
		\draw [->] (a) to node [above] {$p$} (b);
		\draw [->] (a) to node [below] {$q$} (c);
		\draw [->] (b) to node [above] {$p$} (d);
		\draw [->] (b) to node [above] {$q$} (e);
		\draw [->] (c) to node [below] {$p$} (e);
		\draw [->] (c) to node [below] {$q$} (f);
		\draw [dotted] (g) to node [below] {} (a);
		\draw [dotted] (h) to node [above] {} (c);
		\draw [dotted] (i) to node [above] {} (f);
	\end{tikzpicture}
	\qquad
	\begin{tikzpicture}[sloped]
		\node (a) at ( 0,0) [] {$B_0$};
		\node (b) at ( 3,0) [] {$B_0(1+r)$};
		\node (c) at ( 6,0) [] {$B_0(1+r)^2$};
		\node (d) at ( 0,-4) [] {$n=0$};
		\node (e) at ( 3,-4) [] {$n=1$};
		\node (f) at ( 6,-4) [] {$n=2$};
		\draw [->] (a) to node [above] {$1$} (b);
		\draw [->] (b) to node [above] {$1$} (c);
		\draw [dotted] (d) to node [below] {} (a);
		\draw [dotted] (e) to node [above] {} (b);
		\draw [dotted] (f) to node [above] {} (c);
	\end{tikzpicture}
\caption{Two-period binomial model under the physical probability measure $\Pb$ (Exercise \ref{exer:2-11}).}
\label{fig:exer-2-step}
\end{figure}
Consider the two-period binomial model depicted in Figure \ref{fig:exer-2-step}. We toss a coin twice, $\om = \om_1 \om_2$ with $\om_i \in \{H,T\}$, and assume the coin tosses are independent with $\Pb(\om_i = H) = p \in (0,1)$ and $q = 1-p$. The dynamics of the stock $S$ and the bond $B$ are
\begin{align}
S_{n+1}
	&=	\begin{cases}
			u S_n , & \om_{n+1} = H , \\
			d S_n , & \om_{n+1} = T ,
			\end{cases} &
B_{n+1}
	&=	(1+r) B_n , &
n
	&=	0,1,
\end{align}
with $S_0 = 100$, $u = 2$, $d = \tfrac{1}{2}$ and $r = \tfrac{1}{4}$. An \textit{(arithmetic) Asian call} pays $(A-K)^+$ at time $n=2$, where
\begin{align}
A
	&:=	\tfrac{1}{2} ( S_1 + S_2 ) , &
K
	&=	100 .
\end{align}
Let $V_n$ denote the value of the Asian call at time $n$.\\[0.5em]
\textbf{(a)} Derive a probability measure $\Pbt$ (i.e., find $\pt := \Pbt(\om_i = H)$ and $\qt := \Pbt(\om_i = T)$) under which $X/B = (X_n/B_n)_{n=0,1,2}$ is a martingale for all portfolios $X$ of the stock $S$ and the bond $B$.\\[0.5em]
\textbf{(b)} Compute $V_1$ at each node at time $n=1$ and compute $V_0$. Explain why, in contrast with a European option that pays $\varphi(S_2)$, the value of the Asian call at time $n=2$ is not determined by the value of $S_2$ alone.\\[0.5em]
\textbf{(c)} Find a portfolio $X$ that replicates the Asian call. That is, find the number of shares $\Del_0$ of $S$ held at time $n=0$ and the number of shares $\Del_1$ held at each node at time $n=1$.
\end{exercise}




\chapter{Fundamental Theorems of Asset Pricing}

In this chapter, we will discuss the 1st and 2nd Fundamental Theorems of asset pricing.

\section{1st Fundamental Theorem of Asset Pricing}
Although we have often referred to arbitrage and the lack of it in this text, we have yet to give a mathematical definition of exactly what an arbitrage is.  Below, we provide a precise mathematical definition of what we mean by an \textit{arbitrage}.
\begin{definition}[Arbitrage] \label{def:arb}
	An \textit{arbitrage} is a portfolio $X$ satisfying the following three points:
	\begin{enumerate}
		\item $X_0 = 0$ (starting from nothing),
		\item $X_T \geq 0$ (probability of losing money is zero),
		\item $\Pb (X_T > 0) > 0$ (probability of making something is greater than zero),
	\end{enumerate}
for some $T > 0$.
\end{definition}
\noindent
Note that if $X$ is an arbitrage under $\Pb$ then it is also an arbitrage under any equivalent measure $\Pbt$ because equivalent measures agree on events that occur with probability one.
\\[0.5em]
Suppose that the no-arbitrage condition in the binomial model $d < 1+r < u$ is not satisfied.  Then we can show that there is an arbitrage. For example, assume $0 < 1+r  < d  < u$. Let us construct a portfolio $X$ that satisfies the three points in Definition~\ref{def:arb} above.  At time zero, we buy $\Del_0 > 0$ shares of $S$ and sell $\Del_0 S_0 / B_0$ bonds. The value of this portfolio at time zero is the
\begin{align}
	X_0 
		&= \Del_0 S_0 - \frac{\Del_0 S_0}{B_0} B_0  = 0 .
\end{align}
At time $n=1$, the value of the portfolio is
\begin{align}
	X_1 
	&= 			\Del_0 S_1 - \frac{\Del_0 S_0}{B_0} B_1 
	\geq		\Del_0 d S_0 - \Del_0 S_0 (1+r)
	=				\Del_0 S_0 ( d - (1+r) ) > 0 .
\end{align}
As, by assumption, $1+r < d$, we see that $\Pb (X_1 \geq 0) = 1$ and $\Pb (X_1 > 0) > 0$. As the three points in Definition~\ref{def:arb} are fulfilled, we see that $X$ is an arbitrage. The case $0 < d  < u < 1+r$ is treated similarly.
\\[0.5em]
We also showed that, in the binomial model, we could price assets by changing to a probability measure $\Pbt$ under which $X/B$ is a martingale for all portfolios $X$.  In searching for $\Pbt$, we found that
\begin{align}
\Ebt \( \frac{X_1}{B_1} \) 
	&= \frac{X_0}{B_0} &
	&\Leftrightarrow&
\pt 
	&= \frac{(1+r) - d}{u - d} . \label{reminder:pt}
\end{align}
Note now that, if $0 < 1+r < d < u$ and therefore there is an arbitrage, we have from \eqref{reminder:pt} that $\pt < 0$.
As probabilities cannot be negative, we conclude that the martingale measure $\Pbt$ does not exist!
In fact, there is a connection between the existence of $\Pbt$ and the absence of arbitrage.

\begin{theorem}[1st fundamental theorem of asset pricing]
\label{thm:1st}
A market defined under a probability measure $\Pb$ is free of arbitrage if and only if there exists a measure $\Pbt$, equivalent to $\Pb$, under which all traded portfolios, denominated in units of a common \textit{num\'eraire} $N$, are martingales.
\end{theorem}

\noindent
Note that Theorem \ref{thm:1st} applies to \emph{all} market models -- not only the binomial model.
\\[0.5em]
Theorem \ref{thm:1st} gives us a very simple way to check if a market is free of arbitrage.
\\[0.5em]
\textbf{Step 1}:  Choose a num\'eraire asset $N$.  (Typically, we choose the num\'eraire asset $N$ to be the bond $B$, but we do not have to make this choice).
\\[0.5em]
\textbf{Step 2}:  Search for a probability measure $\Pbt$ under which, for \textit{all} portfolios $X$, we have the martingale property for $X/N$
\begin{align}
\frac{X_t}{N_t}
	&=	\Ebt \[ \frac{X_T}{N_T} \Big| \Fc_t \] . \label{eq:martingale.again}
\end{align}
Here, the filtration $\Fc_t$ represents all of the information in the market that is available at time $t$.  For example, in the binomial model, $\Fc_t = \Fc_t^S$.  But, in a more complicated market $\Fc_t$ could include the history of various stocks, economic indicators, interest rates, etc..
\\[0.5em]
Again, by ``search for a probability measure $\Pbt$'' we mean ``assign probabilities to events.''  For example, in the binomial model, in searching for $\Pbt$ we may assign the probabilities $\pt$ and $\qt$ in Figure \ref{fig:1-step-stock-RN}.  Note, we are \textit{not} allowed to change $u$, $d$, $r$, etc., as these are not probabilities of events.  If we cannot find a probability measure $\Pbt$ under which \eqref{eq:martingale.again} holds, then the market contains an arbitrage.
\\[0.5em]
Theorem \ref{thm:1st} also gives us a way to price derivative assets in a way that guarantees we do not introduce arbitrage into the market.  Suppose $V_t$ is the value of a derivative asset with payoff $V_T = \phi(A_T)$.  If we assign a value $V_t$ using
\begin{align}
\frac{V_t}{N_t}
	&=	\Ebt \[ \frac{V_T}{N_T} \Big| \Fc_t \]
	=		\Ebt \[ \frac{\phi(A_T)}{N_T} \Big| \Fc_t \] \label{eq:pricing.formula}
\end{align}
then we are guaranteed to not introduce arbitrage into the market.

\begin{example}[Binomial model with dividends]
Consider, the one-period binomial model.  Suppose that, at time $n=1$, the owner of one share of $S$ receives a dividend payment of $a S_0$.  If we assume the market contains only a stock and a bond $B$, then the value of any portfolio $X$ at time $n=0$ is given by
\begin{align}
X_0
	&=	\Del_0 S_0 + (X_0 - \Del_0 S_0)\frac{1}{B_0} B_0 ,
\end{align}
and at time $n=1$ the value of $X$ becomes
\begin{align}
X_1
	&=	\Del_0 ( S_1 + a S_0) + (X_0 - \Del_0 S_0)\frac{1}{B_0} B_1
\end{align}
Let us choose the bond $B$ as num\'eraire (as we typically do) and search for a probability measure $\Pbt$ under which $X/B$ is a martingale.  We would like
\begin{align}
\frac{X_0}{B_0}
	&=	\Ebt \frac{X_1}{B_1} \\
	&=	\Del_0 \( \Ebt \frac{S_1}{B_1} + a \frac{S_0}{B_1} \) + (X_0 - \Del_0 S_0)\frac{1}{B_0} \frac{B_1}{B_1} \\
	&=	\Del_0 \( \pt \frac{u S_0}{B_0(1+r)} + \qt \frac{d S_0}{B_0(1+r)} + a \frac{S_0}{B_0(1+r)} \) + (X_0 - \Del_0 S_0)\frac{1}{B_0} .
\end{align}
For the above equation to hold, we must have
\begin{align}
\pt u + \qt d + a
	&=	1+r
\end{align}
Solving the above equation for $\pt$ and using $\pt + \qt = 1$ we obtain
\begin{align}
\pt
	&=	\frac{(1 + r - a) - d}{u-d} , &
\qt
	&=	1 - \pt . \label{eq:pt.div}
\end{align}
What are the no-arbitrage conditions?  In order for $\Pbt$ to exist, we must have $\pt, \qt \in (0,1)$.  Thus the no arbitrage condition is
\begin{align}
d
	& < 1 + r - a < u .
\end{align}
Now, assume the no-arbitrage conditions are satisfied.  Suppose we wish to price a derivative asset that pays $V_1 = \phi(S_1)$ at time $n=1$.  In order for the market to be free of arbitrage, the initial value $V_0$ of this asset must satisfy
\begin{align}
\frac{V_0}{B_0}
	&=	\Ebt \frac{V_1}{B_1}
	=		\Ebt \frac{\phi(S_1)}{B_1}
	=		\pt \frac{\phi(u S_0)}{B_0(1+r)} + \qt \frac{\phi(d S_0)}{B_0(1+r)} ,
\end{align}
where $\pt$ and $\qt$ are given by \eqref{eq:pt.div}.
\end{example}

\begin{example}[Forward contracts revisited]
Let us find the fair forward price and the fair value of a forward contract after inception using our pricing formula \eqref{eq:pricing.formula}.
First, we choose as num\'eraire, a bond, whose value is given by
\begin{align}
B_t
	&=	B_0 \ee^{rt}
\end{align}
Let $V_t$ be the value of a forward contract with payoff $S_T - K$.  We know that, in order for there to be no arbitrage in the market, there must be some probability measure $\Pbt$, under which $(V_t/B_t)$ is a martingale.  Thus, for all $t \leq T$ we have
\begin{align}
\frac{V_t}{B_t}
	&=	\Ebt \[ \frac{V_T}{B_T} \Big| \Fc_t \]  
	 =	\Ebt \[ \frac{S_T-K}{B_T} \Big| \Fc_t \]  \\
	&=	\Ebt \[ \frac{S_T}{B_T} \Big| \Fc_t \] - \frac{K}{B_T} 
	 =	\frac{S_t}{B_t} - \frac{K}{B_T} , \\
V_t
	&=	S_t - \frac{K B_t}{B_T}  
	 =	S_t - K \ee^{-r(T-t)} .
\end{align}
So, we have found the fair value of a forward contract with strike $K$ at time $t$.  Recall that the fair forward price $F_t^T$ (not to be confused with the value of a forward contract) is the strike $K$ that makes the value of the forward contract equal to zero.  Setting $V_t = 0$ in the last equation above, we have
\begin{align}
0
	&=	S_t - K \ee^{-r(T-t)}  &
	&\Rightarrow&
K
	&=	S_t \ee^{r(T-t)} = F_t^T .
\end{align}
In particular, at time zero, we have $K = \ee^{rT}S_0$.  Note that we did not need a model for $S$ in order to determine the value of a forward contract or the fair forward price.
\end{example}




\section{2nd Fundamental Theorem of Asset Pricing}
In the one-period binomial model, the risk-neutral measure $\Pbt$ was unique; there was one and \textit{only} one way to assign risk-neutral probabilities
\begin{align}
\pt &= \frac{(1+r) - d}{u - d} , &
\qt &= 1-\pt .
\end{align}
However, for market models in general, when we search for a risk-neutral measure $\Pbt$, we may find that it is not unique.  Let us see a simple example.

\subsection{The trinomial model as an example}
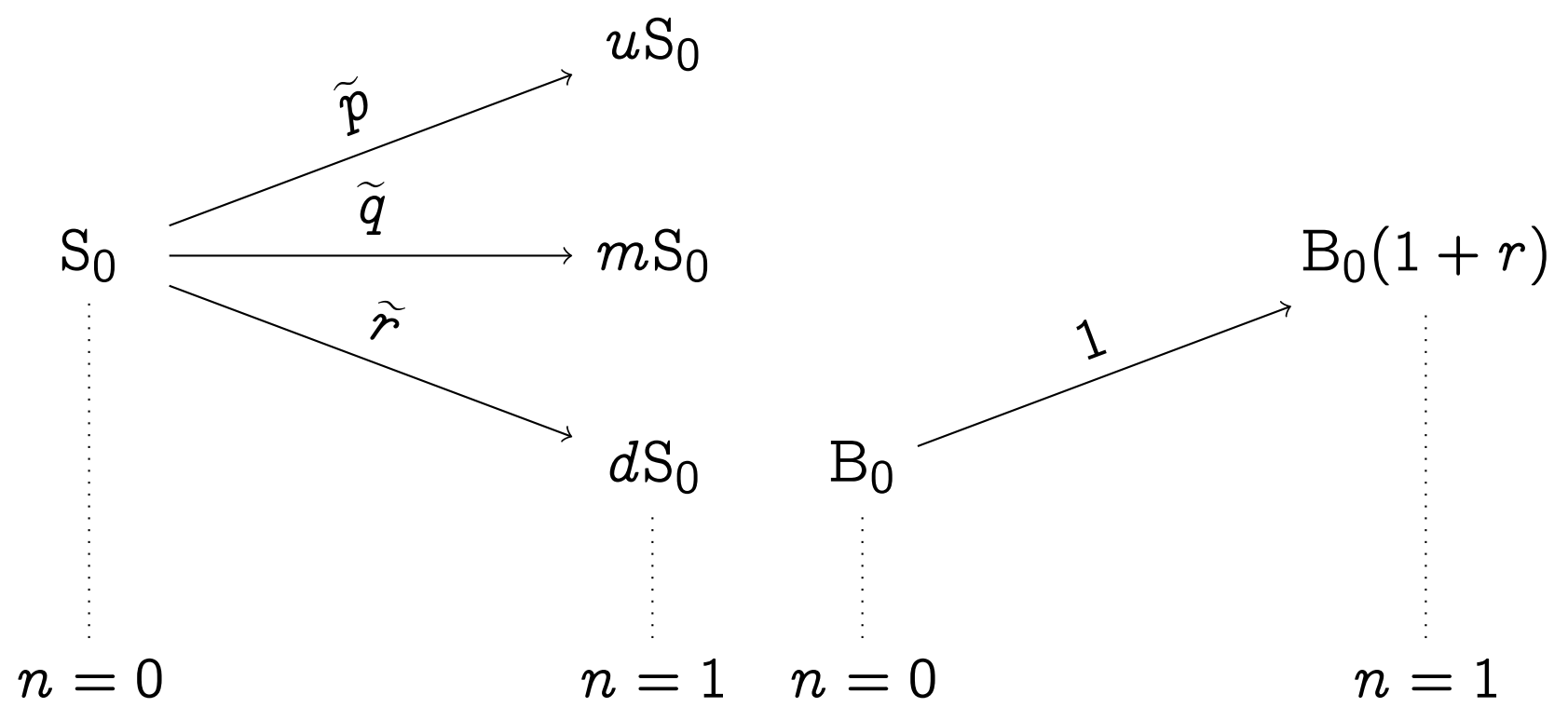
\begin{figure}[ht!]
  \centering
	\begin{tikzpicture}[sloped]
		\node (a) at ( 0,0) [bag] {$S_0$};
		\node (b) at ( 4,-1.5) [bag] {$d S_0$};
		\node (c) at ( 4,0) [bag] {$m S_0$};
		\node (d) at ( 4,1.5) [bag] {$u S_0$};
		\node (e) at ( 0,-3) [end] {$n=0$};
		\node (f) at ( 4,-3) [end] {$n=1$};
		\draw [->] (a) to node [above] {$\rt$} (b);
		\draw [->] (a) to node [above] {$\qt$} (c);
		\draw [->] (a) to node [above] {$\pt$} (d);
		\draw [dotted] (e) to node [below] {} (a);
		\draw [dotted] (f) to node [above] {} (b);
	\end{tikzpicture}
		\begin{tikzpicture}[sloped]
		\node (a) at ( 0,0) [] {$B_0$};
		\node (b) at ( 4,1.5) [] {$B_0 (1+r)$};
		\node (d) at ( 0,-1.5) [] {$n=0$};
		\node (e) at ( 4,-1.5) [] {$n=1$};
		\draw [->] (a) to node [above] {$1$} (b);
		\draw [dotted] (d) to node [below] {} (a);
		\draw [dotted] (e) to node [above] {} (b);
	\end{tikzpicture}
	\caption{A one-step trinomial tree for a stock $S$ and bond $B$ under $\Pbt$.
	\label{fig:1-step-stock-tri}}
\end{figure}

In the trinomial model, the stock can take one of three different paths at each node. This is illustrated in Figure~\ref{fig:1-step-stock-tri}.
With reference to the numbers in the Figure \ref{fig:1-step-stock-tri}, we make the following assumptions
\begin{align}
		0 &< d <1+r< u, &
		0 &< d < m < u.
\end{align}
We will now try to find a risk-neutral measure $\Pbt$ (i.e., we search for probabilities $\pt, \qt, \rt$ such that $X/B$ is a martingale under $\Pbt$). We would like
\begin{align}
\frac{X_0}{B_0}
	&=	\Ebt \[ \frac{X_1}{B_1} \] = \Del_0 \frac{(\pt u + \qt m + \rt d) S_0}{B_0 (1+r)} + \frac{X_0}{B_0} - \frac{\Del_0 S_0}{B_0} ,
\end{align}
where we have used $X_1 = \Del_0 S_1 + (X_0 - \Del_0 S_0)(1+r)$.
Thus, we must have
\begin{align}
		\pt u + \qt m + \rt d &= 1+r , &
		\pt + \qt + \rt &= 1 .
\end{align}
We have two equations and three unknowns. In general, we cannot solve this system uniquely for $\pt, \qt, \rt$.  However, if we fix $\rt$, then we can solve for $\pt (\rt)$ and $\qt (\rt)$ as functions of $\rt$.  Doing this, we obtain
\begin{align}
\pt (\rt) &= \frac{(1+r) - m - \rt d +  m \rt}{u - m} , &
\qt (\rt) &= 1 - \pt (\rt) - \rt . \label{eq:pt.rt}
\end{align}
Let us denote by $\Pbt (\rt)$ the risk-neutral measure in which $\pt(\rt)$ and $\qt(\rt)$ are given by \eqref{eq:pt.rt}.
As $\rt$ can take any value as long as $\rt$, $\pt(\rt)$ and $\qt(\rt)$ are all in $(0,1)$ (so that $\Pbt(\rt) \sim \Pb$) 
there are \textit{infinitely many} probability measures $\Pbt (\rt)$!
\\[0.5em]
Now, what happens if we try to replicate a European option with payoff $h(S_1)$ in the trinomial model? At time $n=0$ the value of a portfolio $X$ is given by
\begin{align}
X_0 
	&= \Del_0 S_0 +\frac{X_0 - \Delta_0 S_0}{B_0} B_0 ,
\end{align}
and at time $n=1$ the value becomes
\begin{align}
X_1 &= \Delta_0 S_1 + \frac{X_0 - \Delta_0 S_0}{B_0} B_1 \\
		&=	\left\{
		\begin{aligned}
			&\Delta_0 u S_0 + (X_0 - \Delta_0 S_0) (1+r) && \text{if $S_1 = u S_0$} ,\\
			&\Delta_0 m S_0 + (X_0 - \Delta_0 S_0) (1+r) && \text{if $S_1 = m S_0$} , \\
			&\Delta_0 d S_0 + (X_0 - \Delta_0 S_0) (1+r) && \text{if $S_1 = d S_0$} .
		\end{aligned} \right.
\end{align}
At time $n=1$ the value of the derivative asset is
\begin{align}
h(S_1) 		
	&=	\left\{
		\begin{aligned}
			&h(u S_0) && \text{if $S_1 = u S_0$} ,\\
			&h(m S_0) && \text{if $S_1 = m S_0$} , \\
			&h(d S_0) && \text{if $S_1 = d S_0$} .
		\end{aligned} \right.
\end{align}
In order for $X_1 = h(S_1)$ we must have
\begin{align}
h(u S_0)
	&=	\Delta_0 u S_0 + (X_0 - \Delta_0 S_0) (1+r)  , \\
h(m S_0)
	&=	\Delta_0 m S_0 + (X_0 - \Delta_0 S_0) (1+r) , \\
h(d S_0)
	&=	\Delta_0 d S_0 + (X_0 - \Delta_0 S_0) (1+r)  .
\end{align}
We have three equations and two unknowns ($\Del_0$ and $X_0$).
The above system of equations is over-determined.
Unless we are very lucky, there will not be any combination of $X_0$ and $\Del_0$ for which the above three equations will be satisfied.
Thus, we will not be able find a portfolio $X$ that replicates the derivative payoff $h(S_1)$. 
\\[0.5em]
There is a connection between the uniqueness of risk-neutral measures $\Pbt$ and the ability to perfectly hedge.  Before describing this connection, let us first introduce the notion of a complete market.

\begin{definition}
 A market that is free of arbitrage is said to be \textit{complete} if every derivative asset can be perfectly replicated.  If a market is not complete, we say it is \textit{incomplete}.
\end{definition}

\noindent
The binomial model is an example of a complete market.  The trinomial model is an example of an incomplete market.

\begin{theorem}[Second fundamental Theorem of Asset Pricing]
\label{thm:second}
 A market, defined under a probability measure $\Pb$, which is free of arbitrage, is complete if and only if there exists a \underline{unique} equivalent probability measure $\Pbt$ under which all portfolios, denominated in units of a common num\'eraire, are martingales.
\end{theorem}

\noindent
Theorem \ref{thm:second} applies to \textit{all} market models -- not just the binomial and trinomial models.
\\[0.5em]
Theorem \ref{thm:second} gives us an easy way to check if a market is complete.  We start with a market, defined under some probability measure $\Pb$.  We then choose a num\'eraire asset $N$ (typically, we choose the num\'eraire to be the bond $N=B$).  We now search for a probability measure $\Pbt$, equivalent to $\Pb$, under which \textit{all} portfolios $X$ are martingales
\begin{align}
\frac{X_t}{N_t}
	&=	\Ebt \[ \frac{X_T}{N_T} | \Fc_t \] .  \label{eq:martingale.again2}
\end{align}
If there is a unique way to choose $\Pbt$ (as there is in the binomial model), then the market is complete and every derivative asset can be hedged perfectly.  If there are multiple ways to choose $\Pbt$ such that \eqref{eq:martingale.again2} is satisfied (as in the trinomial model), then the market is incomplete, and derivative assets cannot be hedged perfectly.  This brings up the following question:
\\[0.5em]
\textbf{Question}: How do we price derivative assets in incomplete markets?
\\[0.5em]
In fact, there is no unique answer to this question.  Below, using the trinomial model as an example, we will discuss four possible answers.

\subsection{Answer 1: the Market chooses a pricing measure}
Consider the trinomial model depicted in Figure \ref{fig:1-step-stock-tri}.
Suppose we observe from market quotes the price $V_0$ of a derivative asset with payoff $h(S_1)$. 
The market $(S,B)$ is \textit{incomplete} because we cannot replicate the  payoff $h(S_1)$ by trading only the stock $S$ and the bond $B$. 
However, the market $(S, B, V)$ is \textit{complete} because we can replicate any payoff $g(S_1)$ by trading the stock $S$, the bond $B$ and the option $V$ with payoff $h(S_1)$. To see this, observe that a hedging portfolio at time $n=0$ is given by:
\begin{align}
X_0 
	&= \Delta_0 S_0 + \Gamma_0 V_0 + \frac{X_0 - \Delta_0 S_0 - \Gamma_0 V_0}{B_0} B_0.
\end{align}
At time $n=1$, the portfolio's value becomes
\begin{align}
X_1 
	&= \Delta_0 S_1 + \Gamma_0 V_1 + \frac{X_0 - \Delta_0 S_0 - \Gamma_0 V_0}{B_0} B_1 \\
	&=	\Delta_0 S_1 + \Gamma_0 h(S_1) + \(X_0 - \Delta_0 S_0 - \Gamma_0 V_0\) (1+r) .
\end{align}
In order for $X_1 = g(S_1)$ the following three equations must be satisfied
\begin{align}
g(u S_0)
	&=	\Delta_0 u S_0 + \Gamma_0 h(u S_0) + \(X_0 - \Delta_0 S_0 - \Gamma_0 V_0\) (1+r) , \\
g(m S_0)
	&=	\Delta_0 m S_0 + \Gamma_0 h(m S_0) + \(X_0 - \Delta_0 S_0 - \Gamma_0 V_0\) (1+r) , \\
g(d S_0)
	&=	\Delta_0 d S_0 + \Gamma_0 h(d S_0) + \(X_0 - \Delta_0 S_0 - \Gamma_0 V_0\) (1+r) ,
\end{align}
With three equations, we can now solve for the three unknowns in our hedging portfolio: $\Delta_0$, $X_0$ and $\Gamma_0$.
The price $P_0$ of the option with payoff $g(S_1)$ is the initial value of the replicating portfolio $X_0$.
\\[0.5em]
Another way to think about this is as follows: when we see that a derivative asset with payoff $h(S_1)$ has a price of $V_0$, we can actually observe the particular probability measure under which the market is pricing assets.  To see this, note that, if the market is free of arbitrage, then we know that
\begin{align}
\frac{V_0}{B_0} 
	&= \Ebt^{(\rt)} \[ \frac{V_1}{B_1} \] = \Ebt^{(\rt)} \[ \frac{h(S_1)}{B_1} \],
\end{align}
for some choice of $\rt$.  Solving for $V_0$, we obtain
\begin{align}
V_0 
	&= \frac{1}{1+r} \Big( \pt (\rt) h(u S_0) + \qt (\rt) h(m S_0) + \rt h(d S_0) \Big) , \label{eq:V=rt}
\end{align}
where $\pt(\rt)$ and $\qt(\rt)$ are given by \eqref{eq:pt.rt}.
Using \eqref{eq:V=rt} we can solve for $\rt$ in terms of $V_0$. This defines uniquely the pricing measure $\Pbt^{(\rt)}$.  In a sense, the market has \textit{chosen} the pricing measure.
\\[0.5em]
Suppose now, that a new derivative with payoff $g(S_1)$ is introduced to the market. If there is no arbitrage in the market, then the initial price $P_0$ of this asset \textit{must} be given by
\begin{align}
\frac{P_0}{B_0} 
	&= \Ebt^{(\rt)} \[ \frac{P_1}{B_1} \] = \Ebt^{(\rt)} \[ \frac{g(S_1)}{B_1} \] .
\end{align}
Solving for $P_0$, we obtain
\begin{align}
P_0 
	&= \frac{1}{1+r}  \Big( \pt (\rt) g(u S_0) + \qt (\rt) g(m S_0) + \rt g(d S_0) \Big) , \label{eq:P0-trinomial}
\end{align}
As we already know the value of $\rt$ in terms of $V_0$, we have the value of $P_0$ explicitly.

\begin{example}
\label{ex:range}
Consider the trinomial model with the following parameters
\begin{align}
u
	&=	1.3 , &
m
	&=	1.0 , &
d
	&=	0.7 , &
r
	&=	0 , &
S_0
	&=	100 , &
B_0
	&=	1 .
\end{align}
Suppose we observe from market quotes that the price of a call option with strike $K_1 = 100$ is $V_0 = 10$.  Let us find the price $P_0$ of a put option with strike $K_2 = 90$.  To do this, we try to construct a portfolio $X$ consisting of $(S,B,V)$ such that $X_1 = P_1 = (K_2 - S_1)^+$.  At time zero we have
\begin{align}
X_0
	&=	\Del_0 S_0 + \Gam_0 V_0 + (X_0 - \Del_0 S_0 - \Gam_0 V_0) B_0 .
\end{align}
At time $n=1$ the value of this portfolio is
\begin{align}
X_1
	&=	\Del_0 S_1 + \Gam_0 V_1 + (X_0 - \Del_0 S_0 - \Gam_0 V_0) B_1 \\
	&=	\Del_0 S_1 + \Gam_0 (S_1 - K_1)^+ + (X_0 - \Del_0 S_0 - \Gam_0 V_0) .
\end{align}
Setting $X_1 = (K_2 - S_1)^+$ gives us three equations with three unknowns ($X_0$, $\Del_0$, $\Gam_0$).  We have
\begin{align}
\Del_0 u S_0 + \Gam_0 (u S_0 - K_1)^+ + (X_0 - \Del_0 S_0 - \Gam_0 V_0)
	&=	(K_2 - u S_0)^+ , \\
\Del_0 m S_0 + \Gam_0 (m S_0 - K_1)^+ + (X_0 - \Del_0 S_0 - \Gam_0 V_0)
	&=	(K_2 - m S_0 )^+ , \\
\Del_0 d S_0 + \Gam_0 (d S_0 - K_1)^+ + (X_0 - \Del_0 S_0 - \Gam_0 V_0)
	&=	(K_2 - d S_0)^+ .
\end{align}
Solving this system of equations, we obtain
\begin{align}
X_0
	&=	20/3 , &
\Del_0
	&=	- 2/3 , &
\Gam_0
	&=	2/3 .
\end{align}
As the portfolio replicates the payoff of the Put, we conclude that $P_0 = X_0 = 20/3$.
\\[0.5em]
As an alternative method of finding $P_0$, we could have used risk-neutral pricing.  We know that
\begin{align}
V_0
	&=	\Ebt^{(\rt)}[ V_1 ] = \Ebt^{(\rt)}[ (S_1 - K_1)^+ ]\\
	&=	\pt(\rt) ( u S_0 - K_1 )^+ + \qt(\rt) ( m S_0 - K_1 )^+ + \rt ( d S_0 - K_1 )^+ .  \label{eq:eqn4rt}
\end{align}
Solving \eqref{eq:eqn4rt} for $\rt$ we find
\begin{align}
\rt
	&=	1/3 , &
	&\text{and thus, from \eqref{eq:pt.rt}, we have} &
\pt
	&=	1/3 , &
\qt
	&=	1/3 .
\end{align}
To find the price $P_0$ of the put option with strike $K_2$ we compute
\begin{align}
P_0
	&=	\Ebt^{(\rt)}[ P_1 ] = \Ebt^{(\rt)}[ (K_2 - S_1)^+ ]\\
	&=	\pt(\rt) ( K_2 - u S_0 )^+ + \qt(\rt) ( K_2 - m S_0 )^+ + \rt ( K_2 - d S_0 )^+ \\
	&=	20/3 .
\end{align}
Note that we obtained the same value for $P_0$ as we obtained using the replication method.
\end{example}

\subsection{Answer 2: Super-hedging}

\begin{definition}
Consider a derivative asset that pays $h(S_T)$ at time $T$.
A portfolio strategy $X$ is a \textit{super-hedge} for $h(S_T)$ if $X_T \geq h(S_T)$.
\end{definition}

\noindent
If the bank that sells a derivative with payoff $h(S_T)$ wishes to guarantee that it does not lose money, then it should charge at least the smallest initial value $X_0$ of super-hedging portfolio for the derivative.  That is, the bank should charge at least the \textit{minimum super-hedging price} $\overline{V}_0$, which is given mathematically by
\begin{align}
\overline{V}_0
	&:=	\inf \{ X_0 : X_T \geq h(S_T) \} . \label{def:superhedge}
\end{align}
How does one find $\overline{V}_0$?  The following theorem addresses this question.

\begin{theorem}[Cost of super-hedging] \label{thm:superhedge}
Consider a market defined under a probability measure $\Pb$.  
The minimum super-hedging price $\overline{V}_0$ of a derivative that pays $h(S_T)$ at time $T$ is given by
\begin{align}
\overline{V}_0
	&=	\sup_{\Pbt \sim \Pb} B_0 \Ebt \Big( \frac{h(S_T)}{B_T} \Big) , \label{eq:superhedge}
\end{align}
where the supremum is taken over all risk-neutral measures $\Pbt$ that are equivalent to $\Pb$, with the bond $B$ as num\'eraire.
\end{theorem}

\noindent
We will not prove Theorem \ref{thm:superhedge}.  Rather, we will simply comment on the intuition behind the theorem.  We know from the first fundamental Theorem of asset pricing that, if $\Pbt$ is a risk-neutral measure equivalent to $\Pb$, then if we define $V_0$ via
\begin{align}
V_0
	&=	B_0 \Ebt \frac{h(S_T)}{B_T} ,
\end{align}
then $V_0$ is a no-arbitrage price for a derivative that pays $h(S_T)$.  The range of no-arbitrage prices for the derivative that pays $h(S_T)$ is then given by
\begin{align}
\{ \text{no-arbitrage range} \}
	&=	\bigg( \inf_{\Pbt \sim \Pb} B_0 \Ebt \frac{h(S_T)}{B_T} , \sup_{\Pbt \sim \Pb} B_0 \Ebt \frac{h(S_T)}{B_T} \bigg) ,
\end{align}
where the $\inf$ and $\sup$ are taken over all risk-neutral measures $\Pbt$ equivalent to $\Pb$.
If a bank sells a derivative for any amount $X_0 > \sup_{\Pbt \sim \Pb} B_0 \Ebt[h(S_T)/B_T]$ then there is an arbitrage in the market.  In particular, there is a trading strategy that satisfies $\Pb( X_T - h(S_T) > 0) > 0$ and $\Pb( X_T - h(S_T) \geq 0) = 1$.
Thus, we conclude the super-hedging price \eqref{def:superhedge} must be given by the upper end of the no-arbitrage range, which is what Theorem \ref{thm:superhedge} states.
\\[0.5em]
In the context of the trinomial model depicted in Figure \ref{fig:1-step-stock-tri}, the minimum super-hedging price $\overline{V}_0$ of a derivative that pays $h(S_1)$ at time $n=1$ is given by
\begin{align}
\overline{V}_0
		&= \sup_{\Pbt (\rt) \sim \Pb} \left\{ B_0 \Ebt^{(\rt)} \[ \frac{h(S_1)}{B_1} \] \right\} \\
		&= \sup_{\rt}  \left\{ \frac{1}{1+r} \Big( \pt (\rt) h(u S_0) +  \qt (\rt) h (m S_0) + \rt h(d S_0)\Big)	\right\}.	 \label{eq:superhedge-trinomial}
\end{align}
where $\pt(\rt)$ and $\qt(\rt)$ are given by \eqref{eq:pt.rt}.  Note that the supremum is likely to occur at the end-point of the valid interval for $\rt$, which is why we take a $\sup$ rather than a $\max$.

\begin{example}
As in Example \ref{ex:range}, consider the trinomial model with the following parameters
\begin{align}
u
	&=	1.3 , &
m
	&=	1.0 , &
d
	&=	0.7 , &
r
	&=	0 , &
S_0
	&=	100 , &
B_0
	&=	1 .
\end{align}
What are the possible no-arbitrage values of a call option with strike $K_1 = 100$?
As $r = 0$ we have that $B_1 = B_0 = 1$.  As such, all traded assets are martingales under some risk-neutral measure $\Pbt(\rt)$.  Thus, we have
\begin{align}
V_0(\rt)
	&=	\Ebt^{(\rt)}[ V_1 ] = \Ebt^{(\rt)}[ (S_1 - K_1)^+ ]\\
	&=	\pt(\rt) ( u S_0 - K_1 )^+ + \qt(\rt) ( m S_0 - K_1 )^+ + \rt ( d S_0 - K_1 )^+ \\
	&=	\pt(\rt) ( u S_0 - K_1 )^+ ,
\end{align}
where $\pt(\rt)$ and $\qt(\rt)$ are given by \eqref{eq:pt.rt}.  
Note that we have indicated the dependence of $V_0$ on $\rt$.
With $u$, $m$ and $d$ as given, we have
\begin{align}
\pt(\rt)
	&=	\rt, &
\qt(\rt)
	&=	1 - 2 \rt .
\end{align}
In order for $\Pbt^{(\rt)}$ to be an probability measure equivalent to $\Pb$, we need that $\pt,\rt,\qt \in (0,1)$.  Thus, we must have
\begin{align}
\rt 
	&\in (0,1/2) .
\end{align}
If $\rt$ falls outside of this range then $\qt$ would not be in $(0,1)$.  Thus, the no-arbitrage range of prices is
\begin{align}
V_0 \in \Big( \inf_{\rt \in (0,1/2)} \pt(\rt) ( u S_0 - K_1 )^+, \sup_{\rt \in (0,1/2)} \pt(\rt) ( u S_0 - K_1 )^+ \Big)
	&=	( 0 , 15 )
\end{align}
The minimal cost to super-hedge the call option is just the upper end of the range
\begin{align}
\overline{V}_0
	=		\sup_{\Pbt (\rt)}  \left\{ \pt (\rt) (u S_0 - K_1)^+	\right\} 
	&=  \sup_{\rt \in (0,1/2)}  \left\{ \rt \, \cdot \, 30	\right\}
	=		15 .
\end{align}
Thus, if a bank sells the claim for $\overline{V}_0 = 15$ it is guaranteed to not lose money (with a proper hedging strategy).
\end{example}

\subsection{Answer 3: Indifference pricing}

\begin{figure}
    \centering
    \begin{tikzpicture}
\begin{axis}[width=8cm, height=7cm, axis lines=left, xmin=0, xmax=3, ymin=0, ymax=1.05,
             xtick=\empty, ytick=\empty, xlabel={$x$}, ylabel={$U(x)$}]
  \addplot[blue, thick, domain=0:3, samples=100] {1-exp(-1.2*x)};
\end{axis}
\end{tikzpicture}
    \caption{Example of a utility function $U$.\label{fig:plot2_1}}
\end{figure}
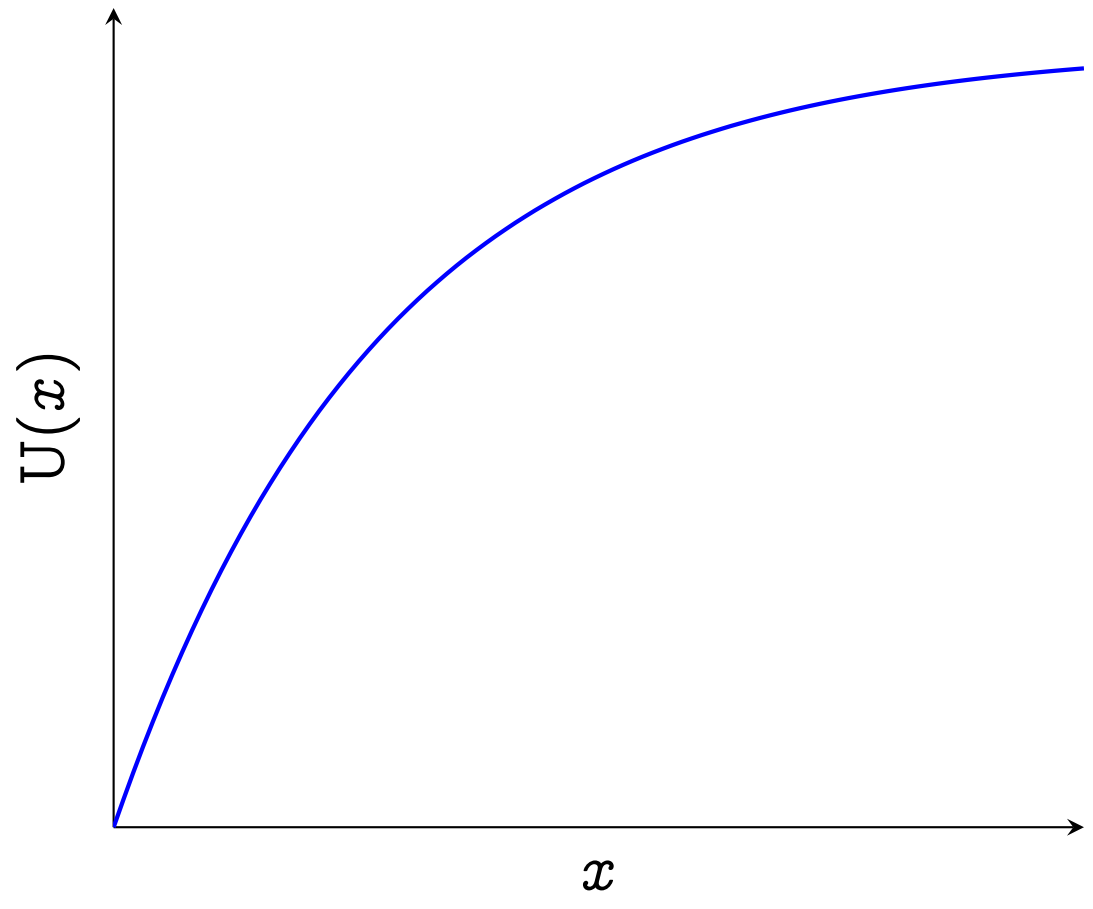

According to standard economic theory, every individual has a \textit{utility function} $U$ that, roughly speaking, maps one's wealth to one's happiness.
All utility functions are increasing, because more money makes an individual happier (again, according to economic theory).
Additionally, all utility functions are concave, because the marginal happiness one gains from making an additional unit of currency goes down the more money one already has.  For example, increasing one's wealth from \$10 to \$20 makes one much happier than increasing one's wealth from \$10000 to \$10010.  
Figure \ref{fig:plot2_1} plots a generic utility function.
An example of a utility function is $U(x) = -\ee^{-\gamma x}$.
\\[0.5em]
The idea behind \textit{indifference pricing} is that an investor will only buy an option if it increases his expected future utility.
 Let $x$ be the initial wealth of an investor and consider a claim with payoff $h(S_T)$.  We define an investor's \textit{value function} by
\begin{align}
v(x,k) 
	&:= \sup_{\Del} \Eb \[ U(X_T^\Delta + k h(S_T)) | X_0^\Del = x \], \label{eq:value-function}
\end{align}
where the $\sup$ is taken over all investment strategies $\Del$.  For example, in a one-period model, if an investor purchases $\Del_0$ shares of an asset $S$ at time $n=0$ and puts the rest of his money in a bond, then the final value of this portfolio will be
\begin{align}
X_1^\Del
	&=	\Del_0 S_1 + (X_0^\Del - \Del_0 S_0 ) \frac{B_1}{B_0} .
\end{align} 
Thus, an investment strategy corresponds merely to choosing a value for $\Del_0$.
In an $N$-period model, we have
\begin{align}
X_{n+1}^\Del
	&=	\Del_n S_{n+1} + (X_n^\Del - \Del_n S_n ) \frac{B_{n+1}}{B_n} .
\end{align}
In this case, an investment strategy $\Del$ would involve choosing $\Del_n$ at time $n$ for $n=0, 1, \ldots, N-1$.
In a continuous time model, an investment strategy $\Del$ would involve choosing $\Del_t$ for all $t \in [0,T]$ where $T$ is some target date in the future.
\\[0.5em]
Note that the expectation in \eqref{eq:value-function} is taken with respect to the real-world probability measure $\Pb$ -- \textit{not} a risk-neutral measure.

\begin{definition}[Indifference price]
The \textit{indifference price} for $k$ claims with payoff $h(S_T)$ is the unique solution $p$ of the following equation
\begin{align}
v(x,0)
		&= v(x-k p, k) .  \label{eq:indiff}
\end{align}
where the function $v$ is defined in \eqref{eq:value-function}.
\end{definition}

\noindent
The left-hand side of \eqref{eq:indiff} is the maximum expected utility that an investor with initial wealth $x$ can achieve by dynamically trading only in a stock $S$ and a bond $B$.  The right-hand side of \eqref{eq:indiff} is the maximum expected utility that an investor with initial wealth $x$ can achieve by purchasing $k$ claims with payoff $h(S_T)$ for a unit price $p$ at time zero (leaving him with $x- k p$), and dynamically trading the stock $S$ and the bond $B$.
Thus, equation \eqref{eq:indiff} gives the price $p$ of the claim that pays $h(S_T)$ such that the investor with initial wealth $x$ can achieve the same expected utility either with or without buying the claim.  In other words, the above equation gives the price $p$ that makes the investor \textit{indifferent} to the existence of the claim.  Let us investigate how to solve \eqref{eq:indiff} through an example. 

\begin{example}
Consider the trinomial model with the following parameters
\begin{align}
u
	&=	1.3, &
m
	&=	1.1, &
d
	&=	0.9 , \\
\Pb( S_1 = u S_0 )
	&=	1/3, &
\Pb( S_1 = m S_0 )
	&=	1/3, &
\Pb( S_1 = d S_0 )
	&=	1/3 , \\
S_0
	&=	100 , &
B_0
	&=	100 , &
B_1
	&=	110 .
\end{align}
Assume the utility function of an investor is given by $U(x) = - \exp( - x/10 )$.
Let us find the indifference price for one Call option with payoff $h(S_1) = (S_1 - K)^+$ where $K = 110$ for an investor with initial wealth $x$.  First we find the value function of the investor assuming he cannot invest in the call option.   We have
\begin{align}
v(x,0)
	&=	\sup_{\Del_0} \Eb [ U(X_1^\Del) | X_0^\Del = x ], &
X_1^\Del
	&=	\Del_0 S_1 + (x - \Del_0 S_0) \frac{B_1}{B_0} .
\end{align}
Computing the expectation  and simplifying, we obtain
\begin{align}
v(x,0)
	&=	\sup_{\Del_0} \frac{-\ee^{-1.1x/10}}{3} \( \ee^{ -20 \Del_0 / 10 } + 1 + \ee^{ 20 \Del_0 / 10 } \) \\
	&=	\sup_{\Del_0} \frac{-\ee^{-1.1x/10}}{3} \( 2 \cosh ( 20 \Del_0 / 10 ) + 1  \) .
\end{align}
To find the optimal $\Del_0$, denoted $\Del_0^*$, we differentiate the argument of the $\sup$ and set it equal to zero.  We have
\begin{align}
0
	&=	\frac{\d}{\d \Del_0} \frac{-\ee^{-1.1x/10}}{3} \( 2 \cosh ( 20 \Del_0 / 10 ) + 1  \) \Big|_{\Del_0=\Del_0^*} \\
	&=	\frac{-\ee^{-1.1x/10}}{3} 2 \sinh ( 20 \Del_0^* / 10 )  &
	&\Rightarrow&
\Del_0^*
	&=	0 .
\end{align}
Thus, we have
\begin{align}
v(x,0)
	&=	\frac{-\ee^{-1.1x/10}}{3} \( 2 \cosh ( 20 \Del_0 / 10 ) + 1  \) \Big|_{\Del_0=\Del_0^*}
	=		-\ee^{-1.1x/10} .
\end{align}
Next, we compute the value function of the investor assuming he purchases one call option for price $p$.  We have
\begin{align}
v(x-p,1)
	&=	\sup_{\Del_0} \Eb [ U(X_1^\Del + h(S_1)) | X_0^\Del = x - p ] , &
X_1^\Del
	&=	\Del_0 S_1 + (x - p - \Del_0 S_0) \frac{B_1}{B_0} .
\end{align}
Computing the expectation and simplifying, we obtain
\begin{align}
v(x-p,1)
	&=	\sup_{\Del_0}  \frac{-\ee^{-1.1(x-p)/10}}{3} \( \ee^{ 20 \Del_0 / 10 } + 1 + \ee^{ -20 (\Del_0 + 1) / 10 } \) .
\end{align}
To find the optimal $\Del_0$, once again denoted $\Del_0^*$, we differentiate the argument of the $\sup$ and set it equal to zero.  We have
\begin{align}
0
	&=	\frac{\d}{\d \Del_0} \frac{-e^{-1.1(x-p)/10}}{3} \( \ee^{ 20 \Del_0 / 10 } + 1 + \ee^{ - 20 (\Del_0 + 1) / 10 } \) \Big|_{\Del_0=\Del_0^*} \\
	&=	\frac{-\ee^{-1.1(x-p)/10}}{3} \frac{20}{10}\( \ee^{ 20 \Del_0^* / 10 } - \ee^{ - 20 (\Del_0^* + 1) / 10 } \) &
	&\Rightarrow&
\Del_0^*
	&=	-1/2 .
\end{align}
Thus, we have
\begin{align}
v(x-p,1)
	&=	\frac{-\ee^{-1.1(x-p)/10}}{3} \( \ee^{ 20 \Del_0 / 10 } + 1 + \ee^{ - 20 (\Del_0 + 1) / 10 } \) \Big|_{\Del_0 = \Del_0^*}
	=		\frac{-\ee^{-1.1(x-p)/10}}{3} ( 1 + 2 / \ee ) .
\end{align}
Finally, to find the indifference price we set $v(x,0) = v(x-p,1)$ and solve for $p$.  We have
\begin{align}
-\ee^{-1.1x/10}
	&=	\frac{-\ee^{-1.1(x-p)/10}}{3} ( 1 + 2 / \ee ) &
	&\Rightarrow&
p
	&=	(10/1.1) \log \( \tfrac{3 }{1 + 2/\ee} \) \approx 4.97 .
\end{align}
This example was taken directly from: \url{https://en.wikipedia.org/wiki/Indifference_price#Example}.
\end{example}

\noindent
A few final remarks about indifference prices.  First, the indifference price for any claim is guaranteed to exist.  In particular, if one is in a complete market, the indifference price is equal to the unique no-arbitrage price.  Second, indifference pricing is a \textit{nonlinear} pricing mechanism.  In other words, if $p(k)$ is the price per claim for $k$ claims that each pay $h(S_T)$ at time $T$, we have that $p(k) \neq p(k')$ when $k \neq k'$.  In other words, the indifference price for $k$ claims is not equal (in general) to $k$ times the price of one claim.  Lastly, indifference prices do not introduce arbitrage into the market.

\subsection{Answer 4: Mean-variance hedging}
The last method (among those we will discuss) for pricing derivatives in incomplete markets is \textit{mean-variance hedging}.  The bank charges the initial value $x^*$ of the portfolio $X^{\Del^*}$ that minimizes the expected squared hedging error (under the real-world measure $\Pb$).  That is, the bank charges $x^*$ for the derivative that pays $h(S_T)$ where
\begin{align}
(x^*,\Del^*)
	&=	\operatorname*{arg min}_{x,\Del} \Eb \Big[ \big( X_T^\Del - h(S_T) \big)^2 \Big| X_0^\Del = x \Big] .
\end{align}
Note that the minimum is taken over all values of initial wealth $x$ and all investment strategies $\Del$.  Because $\Eb [ Z^2 ] = (\Eb Z)^2 + \Vb Z$, the criterion penalizes both a systematic shortfall (or surplus) and the variance of the hedging error.
\\[0.5em]
In a one-period model, such as the trinomial model, the investment strategy $\Del$ corresponds to choosing only $\Del_0$.  Thus, we can find $x^*$ and $\Del^*$ rather easily.  Define
\begin{align}
J(x,\Del_0)
	&=	\Eb \Big[ \big( X_1^\Del - h(S_1) \big)^2 \Big| X_0^\Del = x \Big] , \label{eq:J-one-period} \\
X_1^\Del
	&=	\Del_0 S_1 + (x - \Del_0 S_0)\frac{B_1}{B_0} .
\end{align}
To find the optimal values of $x$ and $\Del_0$, simply solve the following equations for $(x^*,\Del_0^*)$
\begin{align}
\frac{\d J(x,\Del_0)}{\d x} \Big|_{(x,\Del_0)=(x^*,\Del_0^*)}
	&=	0 , &
\frac{\d J(x,\Del_0)}{\d \Del_0} \Big|_{(x,\Del_0)=(x^*,\Del_0^*)}
	&=	0 .
\end{align}
As $J$ is quadratic in $(x,\Del_0)$, these are two linear equations.  The down-side of this approach is that the resulting derivative price $x^*$ may be arbitrageable.


\section{Exercises}
\label{sec:exer-3}


\begin{exercise}
\label{exer:3-2}
Consider the following model for a stock market with a stock $S$ and a bond $B$.  At time zero, the stock price is worth $S_0$ and the bond is worth $B_0$.  At time $T$, under the real world probability measure $\Pb$ we have
\begin{align}
\Pb( S_T = u S_0 )
	&=	p , &
\Pb( S_T = d S_0 )
	&=	q , &
\Pb( B_T = R B_0 )
	&=	p , &
\Pb( B_T = r B_0 )
	&=	q .
\end{align}
Here, we assume that the stock and bond price move \textit{together}, meaning that 
\begin{align}
S_T = u S_0  \quad \Leftrightarrow \quad B_T = R B_0 , 
\end{align}
\textbf{(a)}
Find a probability measure $\Pbt$ (meaning assign probabilities to $\pt$ and $\qt$) such that $X/B$ is a martingale for all portfolios $X$.
\\[0.5em]
\textbf{(b)}
Is the market $(S,B)$ complete?  Explain your answer.
\\[0.5em]
For the rest of this exercise, suppose that $u = 12/10$, $R = 11/10$, $r=1$, $d = 8/10$, $S_0 = 100$ and $B_0 = 1$.
\\[0.5em]
\textbf{(c)}
What are upper and lower bounds for the price of a call option with payoff $(S_T - 100)^+$? If the upper and lower bounds are the same, explain why this is the case.
\\[0.5em]
\textbf{(d)}
State the definition of an \textit{arbitrage}.  Show that if the price $V_0$ falls above the upper bound for the call price (which you obtain in part (c)), then there is an arbitrage.  State exactly what the arbitrage strategy is (i.e., state what should you be buying and selling at time zero and show that this strategy gives you a guaranteed profit.)?
\end{exercise}


\begin{exercise}
\label{exer:3-3}
In this exercise, we consider a market with two assets, a money-market account $M$ and a defaultable bond $B$.  We toss a coin $N$ times, i.e., $\om = \om_1 \om_2 \ldots \om_N$ where $\om_i \in \{H,T\}$ with $\Pb(\om_i=H)=p \in (0,1)$.  Assume coin tosses are independent.  The dynamics of $M$ and $B$ are as follows
\begin{align}
B_{n+1}
	&=	\left\{ \begin{aligned}
			&(1+R) B_n && \om_{n+1} = H , \\
			&0					&& \om_{n+1} = T ,
			\end{aligned} \right. &
M_{n+1}
	&=	(1+r) M_n .
\end{align}
(a) Derive conditions under which this market has no arbitrage.  You may assume that $r > -1$ and $R > -1$.  \\[0.5em]
(b) Assuming the market has no arbitrage, is the market complete or incomplete?  Why?\\[0.5em]
(c) Let $V$ represent the value of a derivative that pays $\phi(B_N)$ at time $N$.  What is $V_n$ where $n \leq N$?  You may want to consider two cases: $B_n > 0$ and $B_n = 0$.\\[0.5em]
(d) Let $X$ be a self-financing portfolio with dynamics as follows
\begin{align}
X_{n+1}
	&=	\Del_n B_{n+1} + (X_n - \Del_n B_n) \frac{1}{M_n}M_{n+1}.
\end{align}
What must $X_0$ and $\Del_n$ be in order for $X_N = \phi(B_N)$?
\end{exercise}

\begin{exercise}
\label{exer:3-4}
We will examine a weather derivative market.
Suppose the temperature $T=(T_n)_{n \in \{0,1\}}$ is modeled as follows: at time $n=0$ the temperature is $T_0$.  At time $n=1$ the temperature satisfies $\Pb(T_1 = u T_0) = p$ and $\Pb(T_1 = d T_0 ) = q = 1-p$, where $\Pb$ is the real-world probability measure.  You obviously cannot trade temperature directly, as temperature is not an asset that you can buy or sell.  However, you can buy a weather derivative that pays $g(T_1)$ at time $n=1$.  You can also lend or borrow from a bank account $B = (B_n)_{n \in \{0,1\}}$ where $B_1 = (1+r) B_0$.  Assume there is no arbitrage in the market.  
 \\[0.5em]
(a) Without any additional information other than what is given above, derive upper and lower bounds for the value $V_0$ of the weather derivative.  You may assume that $g$ is a positive and increasing function.
 \\[0.5em]
(b) Suppose, now, that you can observe $V_0$.  Derive the price $P_0$ of a derivative that pays $h(T_1)$ at time $n=1$.
\end{exercise}


\begin{exercise}
\label{exer:3-5}
Assume $0 < d < m < u < \infty$, $0 < d < 1+r < u < \infty$, and that there is no arbitrage.  Consider the one-period trinomial model with states $S_1 \in \{u S_0, m S_0, d S_0\}$ under a risk-neutral measure, and suppose the market is also selling an option with payoff $h(S_1)$ for a price $V_0$.  Assume you can trade the stock $S$, the bond $B$, and the option with payoff $h(S_1)$.\\[0.5em]
\textbf{(a)} Write a system of equations that you would solve in order to find a portfolio $X$ such that $X_1 = g(S_1)$, where $g$ is some given function.  Clearly state which variables you would solve for.  You do not need to solve the system of equations.\\[0.5em]
\textbf{(b)} Suppose $h(S_1) = \alpha S_1$ where $\alpha$ is a real constant.  Is it possible to determine $V_0$ uniquely?  If so, what is $V_0$?\\[0.5em]
\textbf{(c)} In general, can the system of equations you derived in part (a) be solved when $h(S_1) = \alpha S_1$?  Explain your answer.
\end{exercise}


\begin{exercise}
\label{exer:3-6}
Consider a one-period market with two stocks $S$ and $Z$ and a bank account $B$, whose values at time $n=0$ are $S_0$, $Z_0$ and $B_0$, respectively.  At time $n=1$ we have four possible outcomes for $S$ and $Z$, whose (strictly positive) probabilities under the real-world probability measure $\Pb$ are
\begin{align}
\Pb( S_1 = u S_0 , Z_1 = a Z_0 ) &= p_{ua} , &
\Pb( S_1 = u S_0 , Z_1 = b Z_0 ) &= p_{ub} , \\
\Pb( S_1 = d S_0 , Z_1 = a Z_0 ) &= p_{da} , &
\Pb( S_1 = d S_0 , Z_1 = b Z_0 ) &= p_{db} ,
\end{align}
where $p_{ua} + p_{ub} + p_{da} + p_{db} = 1$.  The bank account at time $n=1$ grows to $B_1 = B_0(1+r)$.  Assume throughout that the market does not admit arbitrage.\\[0.5em]
\textbf{(a)} Let $X$ denote the value of a portfolio that, at time $n=0$, holds $\Del_0$ shares of $S$ and $\Gam_0$ shares of $Z$.  Write an expression for $X_1$ assuming the initial value of the portfolio is $X_0$.\\[0.5em]
\textbf{(b)} Consider a financial derivative that pays $g(S_1,Z_1)$ at time $n=1$.  Write a system of equations that would need to be satisfied by $(X_0,\Del_0,\Gam_0)$ in order for $X_1 = g(S_1,Z_1)$ for any choice of $g$.  You do not need to solve this system.\\[0.5em]
\textbf{(c)} Let $\Pbt$ denote a probability measure equivalent to $\Pb$ with probabilities $\pt_{ua}$, $\pt_{ub}$, $\pt_{da}$, $\pt_{db}$.  Write a system of equations that would need to be satisfied by $(\pt_{ua}, \pt_{ub}, \pt_{da}, \pt_{db})$ in order for $X_0/B_0 = \Ebt(X_1/B_1)$.  You do not need to solve this system.\\[0.5em]
\textbf{(d)} Is the market $(S,Z,B)$ complete?  State why or why not.\\[0.5em]
\textbf{(e)} Suppose $g(S_1,Z_1) = \alpha S_1 + \beta Z_1 + \gam$, where $(\alpha,\beta,\gam)$ are constants.  Is there a unique price for this financial derivative?  If so, what is it?\\[0.5em]
\textbf{(f)} Now, let $V$ denote the value of a financial derivative that pays $h(S_1,Z_1)$ at time $n=1$, and suppose the market is selling this derivative for a price $V_0$.  Let $Y$ denote the value of a portfolio that at time $n=0$ holds $\Del_0$ shares of $S$, $\Gam_0$ shares of $Z$ and $\Phi_0$ shares of $V$.  Write an expression for $Y_1$ assuming the initial value of the portfolio is $Y_0$.\\[0.5em]
\textbf{(g)} Consider a financial derivative that pays $f(S_1,Z_1)$ at time $n=1$.  Write a system of equations that would need to be satisfied by $(Y_0,\Del_0,\Gam_0,\Phi_0)$ in order for $Y_1 = f(S_1,Z_1)$ for any choice of $f$.  You do not need to solve this system.\\[0.5em]
\textbf{(h)} Write a system of equations that would need to be satisfied by $(\pt_{ua}, \pt_{ub}, \pt_{da}, \pt_{db})$ in order for $Y_0/B_0 = \Ebt(Y_1/B_1)$.  You do not need to solve this system.\\[0.5em]
\textbf{(i)} Assume that $h(S_1,Z_1) \neq \alpha S_1 + \beta Z_1 + \gam$ for any $(\alpha,\beta,\gam)$.  Is the market $(S,Z,V,B)$ complete?  State why or why not.
\end{exercise}

\begin{exercise}
\label{exer:3-7}
\begin{figure}[htb!]
  \centering
	\begin{tikzpicture}[sloped]
		\node (a) at ( 0,0) [] {$S_0$};
		\node (b) at ( 4,1.5) [] {$u S_0$};
		\node (c) at ( 4,0) [] {$m S_0$};
		\node (d) at ( 4,-1.5) [] {$d S_0$};
		\node (e) at ( 0,-3) [] {$n=0$};
		\node (f) at ( 4,-3) [] {$n=1$};
		\draw [->] (a) to node [above] {$p_u$} (b);
		\draw [->] (a) to node [above] {$p_m$} (c);
		\draw [->] (a) to node [below] {$p_d$} (d);
		\draw [dotted] (e) to node [below] {} (a);
		\draw [dotted] (f) to node [above] {} (d);
	\end{tikzpicture}
	\qquad
	\begin{tikzpicture}[sloped]
		\node (a) at ( 0,0) [] {$B_0$};
		\node (b) at ( 4,0) [] {$B_0$};
		\node (d) at ( 0,-3) [] {$n=0$};
		\node (e) at ( 4,-3) [] {$n=1$};
		\draw [->] (a) to node [above] {$1$} (b);
		\draw [dotted] (d) to node [below] {} (a);
		\draw [dotted] (e) to node [above] {} (b);
	\end{tikzpicture}
\caption{One-period trinomial model under the physical probability measure $\Pb$ (Exercise \ref{exer:3-7}).}
\label{fig:exer-trinomial}
\end{figure}
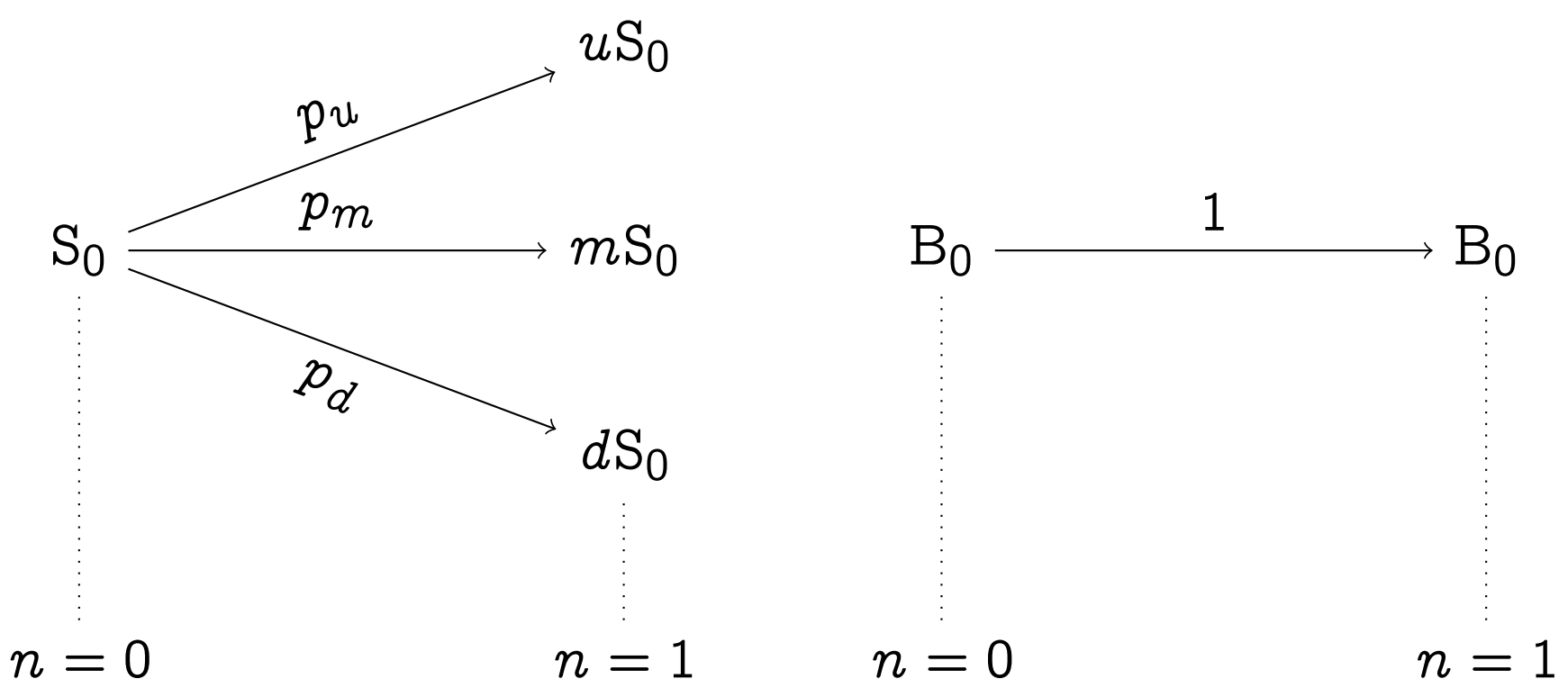
Consider the one-period trinomial model depicted in Figure \ref{fig:exer-trinomial}, with $S_0 = 100$, $u = 1.2$, $m = 1$, $d = 0.7$, $r = 0$ and $B_0 = 1$. Under the physical measure, $\Pb(S_1 = u S_0) = p_u$, $\Pb(S_1 = m S_0) = p_m$ and $\Pb(S_1 = d S_0) = p_d$, where $p_u$, $p_m$ and $p_d$ are strictly positive and $p_u + p_m + p_d = 1$. For a probability measure $\Pbt$, we write
\begin{align}
\pt
	&:=	\Pbt(S_1 = u S_0) , &
\qt
	&:=	\Pbt(S_1 = m S_0) , &
\rt
	&:=	\Pbt(S_1 = d S_0) .
\end{align}
\textbf{(a)(i)} Find all probability measures $\Pbt$ \textit{equivalent} to $\Pb$ under which $X/B$ is a martingale for all portfolios $X$ of the stock $S$ and the bond $B$. Express $\pt$ and $\qt$ as functions of $\rt$.\\[0.5em]
\textbf{(a)(ii)} Is the market $(S,B)$ free of arbitrage? Is it complete? Justify your answers.\\[0.5em]
\textbf{(b)} Consider a European call and a European put, both with strike $K = 100$ and maturity $n=1$.\\[0.5em]
\textbf{(b)(i)} Find the range of no-arbitrage prices of the call and the range of no-arbitrage prices of the put.\\[0.5em]
\textbf{(b)(ii)} Explain why the two ranges must coincide. \textit{Hint}: How can you compare calls and puts?\\[0.5em]
\textbf{(b)(iii)} What is the minimal super-hedging price of the call?\\[0.5em]
\textbf{(c)} Suppose now that the market sells the call with strike $K = 100$ for a price $C_0 = 6$.\\[0.5em]
\textbf{(c)(i)} Find the unique no-arbitrage price of a derivative that pays $|S_1 - 100|$ at time $n=1$.\\[0.5em]
\textbf{(c)(ii)} Find a portfolio that holds $\Del_0$ shares of $S$ and $\Gam_0$ calls with strike $100$, invests the rest in the bond, and replicates the payoff $|S_1 - 100|$. Verify that its initial value equals the price found in (c)(i).\\[0.5em]
\textbf{(c)(iii)} Suppose that, instead of the call, the market quotes the price $V_0$ of a derivative that pays $h(S_1) = 2 S_1 + 10$ at time $n=1$. Could you use this quote to determine uniquely the price of the derivative that pays $|S_1 - 100|$? Explain.
\end{exercise}




\chapter{Brownian motion, It\^o processes and SDEs}

\section{Brownian motion}
Louis Bachelier is credited as being the first person to study the process that we call \textit{Brownian motion}.  He used the process as a model for stock price movements in his thesis \textit{The Theory of Speculation}, which was published in 1900.  Brownian motion is also referred to as a \textit{Wiener process} in honor of Norbert Wiener, who proved a number of important mathematical properties of Brownian motion.  

\begin{definition}[Brownian Motion]
\label{def:bm}
A \textit{Brownian motion} or \textit{Wiener Process}, defined under a probability measure $\Pb$, is a process $W = (W_t)_{t \geq 0}$ that satisfies the following
\begin{enumerate}
\item The process starts from zero: $W_0 = 0$.
\item Non-overlapping increments are independent: $W_T - W_t$ $\ind$ $W_u - W_s$ for all $0 \leq s \leq u \leq t \leq T$. 
\item Increments are stationary and normally distributed: $W_T - W_t \sim W_{T-t} \sim N(0, T-t)$.
\item The process $W$ has continuous sample paths.
\end{enumerate}
\end{definition}

\noindent
Recall that
\begin{align}
Z 
	&\sim N(\mu, \sigma^2)&
	&\Rightarrow&
\Pb(Z \in \dd z ) 
	&= \frac{1}{\sqrt{2 \pi \sigma^2}} \exp\[-\frac{(z-\mu)^2}{2 \sigma^2}\] \dd z ,
\end{align}
Before discussing some important properties of Brownian motion, we need to introduce some important concepts.

\begin{definition}
We say that a random process $\Del = (\Del_t)_{t \geq 0}$ is \textit{adapted} to a filtration $\Fc=(\Fc_t)_{t \geq 0}$, if, for all $t \geq 0$, the information in $\Fc_t$ is sufficient to determine the value of $\Del_t$.  In this case, we write $\Del_t \in \Fc_t$ and say $\Del_t$ is \textit{$\Fc_t$-measurable}.
\end{definition}

\noindent
Some examples should help clarify the concept of adaptedness.

\begin{example}
Suppose $\Del_t = f(X_t)$ for some function $f$. Then we clearly have $\Del_t \in \Fc_t^X$ because, by observing $\{X_s, 0 \leq s \leq t\}$ we can determine $\Del_t = f(X_t)$ uniquely.
\end{example}

\begin{example}
Suppose $\Del_t = \max_{s \in [0,t]} Y_s$.  Then we clearly have $\Del_t \in \Fc_t^Y$ because, by observing $\{Y_s, 0 \leq s \leq t\}$ we can determine $\Del_t = \max_{s \in [0,t]} Y_s$.
\end{example}

\begin{definition}[Filtration for $W$]
We say that $\Fc = (\Fc_t)_{t \geq 0}$ is a \textit{filtration for $W$} if (i) $W$ is adapted to $\Fc$ and (ii) for all $0 \leq s \leq t \leq T$ we have that $W_T - W_t \ind \Fc_s$.
\end{definition}

\noindent
Roughly speaking, $\Fc = (\Fc_t)_{t \geq 0}$ is a filtration for $W$ if the information in $\Fc_t$ is sufficient to determine $\{ W_s, 0 \leq s \leq t\}$ and if the information in $\Fc_t$ gives us no information about $W_T - W_t$ for any $T \geq t$.  Again, some examples will help clarify things a bit.

\begin{example}
The \textit{natural filtration} $\Fc^W = (\Fc_t^W)_{t \geq 0}$ of a Brownian motion $W$, is a filtration for $W$ because (i) the information in $\Fc_t^W = \{ W_s, 0 \leq s \leq t\}$ is sufficient to determine $W_t$ (i.e., $W$ is adapted to $\Fc^W$) and (ii) the increment $W_T - W_t$ is independent of $\Fc_t^W = \{ W_s, 0 \leq s \leq t \}$.
\end{example}

\begin{example}
Suppose $\Fc^\pm = (\Fc_t^\pm)_{t \geq 0}$ is defined by $\Fc_t^\pm = \{ W_s, 0 \leq s \leq t \pm \del \}$ where $\del >0$ is a constant
(if $t \in [0,\del]$ then assume $\Fc_t^-$ has no information).
Then $\Fc_t^+$ is \textit{not} a filtration for $W$ because $W_{t+\del} - W_t$ is \textit{not} independent of $\Fc_t^+ = \{ W_s , 0 \leq s \leq t+\del\}$.  Likewise, $\Fc_t^-$ is \textit{not} a filtration for $W$ because the information in $\Fc_t^- = \{ W_s, 0 \leq s \leq t-\del \}$ is not sufficient to determine $W_t$ (thus, $W$ is not adapted to $\Fc^-$).
\end{example}

\begin{assumption}
Throughout this chapter, $\Fc = (\Fc_t)_{t \geq 0}$ will always represent a filtration for $W$.
\end{assumption}

\noindent
With the important technicalities out of the way, let us discuss two important properties of Brownian motion.
\begin{enumerate}
\item  Brownian motion is a Markov process.  That is, for any function $g$ and for any $0 \leq t \leq T < \infty$ we have $\Eb [ g(W_T) | \Fc_t] = \Eb [ g(W_T) | W_t]$.
\item Brownian motion is a martingale under $\Pb$ with respect to $\Fc = (\Fc_t)_{t \geq 0}$.  That is, for any $0 \leq t \leq T < \infty$ we have $\Eb [W_T | \Fc_t]  = W_t$.
\end{enumerate}

\noindent
We can easily derive the martingale property as follows
\begin{align}
\Eb [W_T | \Fc_t ]
		&= \Eb[W_T | W_t] 
		 = \Eb[(W_T - W_t) + W_t | W_t] 
		 = \Eb[W_T - W_t] + W_t 
		 = W_t .
\end{align}
The third equality follows from the fact that $W_T - W_t \ind W_t$.  The fourth equality follows from the fact that $W_T - W_t \sim N(0,T-t)$ and thus $\Eb[W_T-W_t]=0$.

\begin{example}
Let $X = (X_t)_{t \geq 0}$ be defined by $X_t:= W_t^2 - t$.  Let us show that $X$ is a martingale under $\Pb$ with respect to $\Fc$.  We have
\begin{align}
\Eb [X_T | \Fc_t] 
		&= \Eb [W_T^2 - T | W_t] \\
		&= \Eb [ (W_T - W_t)^2 + 2 W_T W_t - W_t^2 - T | W_t] \\
		&= \Eb [ (W_T - W_t)^2 | W_t] + 2 W_t \Eb[W_T | W_t]  - W_t^2 - T \\
		&= T - t + 2 W_t W_t - W_t^2 - T \\
		&= W_t^2 - t = X_t .
\end{align}
Make sure you understand how each equality follows from the properties of Brownian motion!
\end{example}

\begin{example}[Generalized Brownian motion]
The \textit{Generalized Brownian motion} is a process $X = (X_t)_{t \geq 0}$ that is given by
\begin{align}
X_t 
	&:= \mu t + \sigma W_t,
\end{align}
for some constants $\mu \in \Rb$ and $\sig > 0$.  Note that $X_t$ is normally distributed because $W_t$ is normally distributed and an affine transformation of a normally distributed random variable results in a normally distributed random variable. 
We can compute the mean and variance of $X_t$ directly.  We have
\begin{align}
\Eb[X_t] 
	&= \mu t + \sigma \Eb[W_t] = \mu t  , \\
\Vb[X_t] 
	&= \Eb [(X_t - \Eb[X_t])^2]
	 = \sigma^2 \Eb [W_t^2] = \sigma^2 t  .	
\end{align}
Thus, we have $X_t \sim N(\mu t, \sig^2 t)$.
\end{example}

\section{It\^o integrals and It\^o processes}
Brownian motion can be used as a building block to create more general processes known as \textit{It\^o integrals}.  
Presumably, you are familiar with a Riemann integral
\begin{align}
\int_0^T f(t) \dd t
	&:= \lim_{\del t \to 0} \sum_{j=0}^{n-1} f(t_j^*) ( t_{j+1} - t_j ) , &
t_j^*
	&\in [t_j, t_{j+1}] , \\
\del t
	&:=	\max_j | t_{j+1} - t_j | , &
0
	&= t_0 < t_1 < \ldots < t_n = T . \label{eq:delta.t}
\end{align}
where the particular choice of $t_j^*$ does not matter.  You may be somewhat less familiar with the Riemann-Stieltjes integral
\begin{align}
\int_0^T f(t) \dd g(t)
	&:= \lim_{\del t \to 0} \sum_{j=0}^{n-1} f(t_j^*) ( g(t_{j+1}) - g(t_j) ) , &
t_j^*
	&\in [t_j, t_{j+1}] .
\end{align}
where, as long as $g$ is piecewise differentiable, the particular choice of $t_j^*$ does not matter.
Similar to the Riemann-Stieltjes integral, we can define an \textit{It\^o integral}.  

\begin{definition}[It\^o Integral]
Suppose that $\sig = (\sig_t)_{t \geq 0}$ is adapted to a filtration $\Fc$ of a Brownian motion $W$.  We define
\begin{align}
\int_0^T \sig_t \dd W_t
	&:= \lim_{\del t \to 0} \sum_{j=0}^{n-1} \sig_{t_j} ( W_{t_{j+1}} - W_{t_j} )  , \label{eq:ito.def}
\end{align}
where $\del t$ is as described in \eqref{eq:delta.t}.
\end{definition}

\noindent
Believe it or not, it is actually important that we have written $\sig_{t_j}$ in \eqref{eq:ito.def} and \textit{not} $\sig_{t_j^*}$ for some $t_j^* \in (t_j,t_{j+1}]$.
The reason, without getting too much in the weeds, is that paths of Brownian motion $W$ are not differentiable (\textit{anywhere}!).
\\[0.5em]
It is not terribly important that you remember the definition of an It\^o integral.  But, it is important that you remember a few properties.  In what follows, let $J = (J_t)_{t \geq 0}$ be the following It\^o integral
\begin{align}
J_t := \int_0^t \sig_s \dd W_s ,  
\end{align}
and let $\Fc = (\Fc_t)_{t \geq 0}$ be a filtration for $W$.  Then the following holds:
\begin{enumerate}
\item The process $J$ is adapted to $\Fc$.  That is, $J_t \in \Fc_t$ for all $t \geq 0$.
\item The process $J$ is a martingale with respect to $\Fc$ (i.e., $\Eb[ J_T | \Fc_t ] = J_t$ for any $0 \leq t \leq T < \infty$).
\item The \textit{It\^o isometry} holds: $\Eb J_T^2 = \Eb \int_0^T \sig_t^2 \dd t$.
\item If the process $\sig = (\sig_t)_{t \geq 0}$ is a deterministic (i.e., not random) function $\sig_t = \sig(t)$, then the It\^o integral is normally distributed
$ J_T \sim N( 0 , \Sigma^2(T) )$ where $\Sigma^2(T) = \int_0^T \sig^2(t) \dd t$.
\end{enumerate}

\noindent
Having defined It\^o integrals and discussed some of their properties, we are now in a position to describe what we mean by an \textit{It\^o process}.  

\begin{definition}[It\^o Process]
An \textit{It\^o process} is a process of the form
\begin{align}
X_T
	&=	X_0 + \int_0^T \mu_t \dd t + \int_0^T \sig_t \dd W_t , \label{eq:ito-integral}
\end{align}
where $\mu = (\mu_t)_{t \geq 0}$ and $\sig = (\sig_t)_{t \geq 0}$ are adapted to a filtration $\Fc$ for a Brownian motion $W$.
\end{definition}

\noindent
Very often, we will write the process $X$ in differential form:
\begin{align}
\dd X_t 
	&= \mu_t \dd t + \sigma_t \dd W_t . \label{eq:ito-differential}
\end{align}
Equation \eqref{eq:ito-differential} gives us an intuitive idea of how $X$ behaves in a small increment $\dd t$ of time.
Roughly speaking, we have
\begin{align}
X_{t + \dd t} - X_t 
	&\approx \mu_t \dd t + \sig_t \underbrace{(W_{t + \dd t} - W_t)}_{\sim N (0, \dd t)}.
\end{align}
But, keep in mind that, when we write ``$\dd X_t = \ldots$'' as we have in \eqref{eq:ito-differential}, what we really mean is that $X_T$ is given by \eqref{eq:ito-integral}.
\\[0.5em]
It will be important in subsequent chapters for us to be able to identify when an It\^o process is a martingale.  We know that
\begin{align}
\dd X_t
	&=	\sig_t \dd W_t &
	&\Rightarrow&
	&\text{$X$ is a martingale}, \label{is-a-martingale}
\end{align}
because It\^o integrals are martingales.  Equation \eqref{is-a-martingale} gives us a \textit{sufficient condition} for any It\^o process $X$ to be a martingale.  Although we will not prove it, it turns out that \eqref{is-a-martingale} is also a \textit{necessary condition} for $X$ to be a martingale.  In other words, we have
\begin{align}
\dd X_t
	&=	\mu_t \dd t + \sig_t \dd W_t , &
\mu
	&\neq 0 &
	&\Rightarrow&
	&\text{$X$ is not a martingale}. \label{is-not-a-martingale}
\end{align}
For an It\^o process $X$ to be a martingale, the ``$\dd t$''-term \textit{must} be zero.  To see why this is the case, note that
\begin{align}
\Eb( X_{t + \dd t} | \Fc_t )
	&\approx	X_t + \mu_t \dd t + \sig_t \underbrace{ \Eb( W_{t + \dd t} - W_t | \Fc_t ) }_{=0} 
	=	X_t + \mu_t \dd t .
\end{align}
If $\mu_t \neq 0$ then $\Eb( X_{t + \dd t} | \Fc_t ) \neq X_t $ and thus $X$ is not a martingale.

\section{It\^o's Lemma}
Suppose $X=(X_t)_{t \geq 0}$ is an It\^o process that, in differential form, is given by \eqref{eq:ito-differential}.
We can define another process $f(X)=(f(X_t))_{t \geq 0}$.  
We expect that $\dd f(X_t)$ is something like
\begin{align}
\dd f(X_t) 
	&= g_t \dd t + h_t \dd W_t .
\end{align}
for some processes $g = (g_t)_{t \geq 0}$ and $h = (h_t)_{t \geq 0}$.  \textit{It\^o's Lemma}, presented below, tells us how to find the processes $g$ and $h$.
\\[0.5em]
Consider first a deterministic (i.e., not random) process that satisfies the following Ordinary Differential Equation (ODE)
\begin{align}
\frac{\dd x(t)}{\dd t}
	&=	\mu(t) , &
	&\text{or, in differential notation,}&
\dd x(t)
	&=	\mu(t) \dd t .
\end{align}
Then, we have by the
chain rule that
\begin{align}
\frac{\dd f(x(t))}{\dd t}
	&=	f'(x(t)) \frac{\dd x(t)}{\dd t} .
\end{align}
Multiplying through by $\dd t$ we obtain
\begin{align}
\dd f(x(t))  
	&=	f'(x(t)) \dd x(t)  
	 =	f'(x(t)) \mu(t) \dd t .
\end{align}
The above computation \textit{suggests} that, when $X$ is given by \eqref{eq:ito-differential}, we have
\begin{align}
\dd f(X_t) 
	&\stackrel{?}{=}	f'(X_t) \dd X_t 
	\stackrel{?}{=}	f'(X_t) \Big( \mu_t \dd t  + \sig_t \dd W_t \Big) . 
	& \label{eq:integral}
\end{align}
Unfortunately, this is not quite correct.  It turns out that we are missing a term of order $\dd t$ in \eqref{eq:integral}.  To understand why this is the case, consider what happens to the process $f(X)$ in an infinitesimally small increment of time.  We have
\begin{align}
\dd f(X_t)
	&= f(X_t + \dd X_t) - f(X_t) \\
	&= f(X_t) + f'(X_t) \dd X_t + \tfrac{1}{2} f''(X_t) (\dd X_t)^2 + \ldots - f(X_t) \\
	&= f'(X_t) ( \mu_t \dd t + \sig_t \dd W_t ) + \tfrac{1}{2} f''(X_t) ( \mu_t \dd t + \sig_t \dd W_t )^2 + \ldots \\
	&= f'(X_t) ( \mu_t \dd t + \sig_t \dd W_t ) + \tfrac{1}{2} f''(X_t) ( \mu_t^2 (\dd t)^2 + 2 \mu_t \sig_t \dd t \dd W_t + \sig_t^2 (\dd W_t)^2 ) + \ldots. \label{eq:df}
\end{align}
Now, recall that
\begin{align}
\Eb  (\dd W_t )^2
	&=	\Eb (W_{t + \dd t} - W_t )^2
	 =	\dd t .
\end{align}
This implies that $\dd W_t$ is $\Oc(\sqrt{\dd t})$ and thus 
\begin{align}
&\text{$(\dd W_t)^2$ is $\Oc(\dd t)$}, &
&\text{$\dd t\dd W_t$ is $\Oc(\dd t^{3/2})$}, &
&\text{$(\dd t)^2$ is $\Oc(\dd t^2)$}.
\end{align}
Because the $f'(X_t)$-term in \eqref{eq:df} has a $\Oc(\dd t)$ term, we need to keep the $\dd W_t^2$ term, which is $\Oc(\dd t)$.  
On the other hand, we can ignore the $\dd t \dd W_t$ and $\dd t^2$ terms, which are of $\Oc(\dd t^{3/2})$ and $\Oc(\dd t^2)$, respectively.
Thus, we have from \eqref{eq:df} that
\begin{align}
\dd f(X_t)
	&= f'(X_t) ( \mu_t \dd t + \sig_t \dd W_t ) + \tfrac{1}{2} f''(X_t) \sig_t^2 (\dd W_t)^2 . \label{eq:Ito-form-0}
\end{align}
For reasons that we will not go into here,
\footnote{In fact, we are using $(\dd W_t)^2$ as a proxy for an infinitesimal increment of the \textit{quadratic variation} of a Brownian motion, denoted $\dd[W,W]_t$.  Rigorously, we have $\dd [W,W]_t = \dd t$.}
it turns out that, not only do we have $\Eb (\dd W_t )^2 = \dd t$, we also have for an infinitesimally small $\dd t$ that
\begin{align}
(\dd W_t)^2
	&= \dd t. \label{eq:dW2=dt}
\end{align}
Inserting \eqref{eq:dW2=dt} into \eqref{eq:Ito-form-0}, we obtain It\^o's Lemma.

\begin{theorem}[It\^{o}'s Lemma, 1 dimension]
Let $\mu_t, \sigma_t$ be stochastic processes adapted to a filtration $\Fc$ for a Brownian motion $W$. Suppose $X$ is the following It\^o process given by \eqref{eq:ito-differential}.  Then we have
\begin{align}
\dd f(X_t) 
	&= \big( \mu_t f'(X_t) + \frac{1}{2} \sigma_t^2 f''(X_t) \big) \dd t + f'(X_t) \sigma_t \dd W_t , \label{eq:ito-1d}
\end{align}
for any function $f$ that is twice differentiable.
\end{theorem}

\noindent
An easy way to remember It\^o's Lemma is to use the following two-step procedure.
\\[0.5em]	
\underline{Step 1}: Define $\dd f(X_t) := f(X_t + \dd X_t) - f(X_t)$ and expand $f(X_t + \dd X_t)$ as a Taylor series about $X_t$ to second order:
\begin{align}
\dd f(X_t) = f'(X_t) \dd X_t + \frac{1}{2} f''(X_t) (\dd X_t)^2 . \label{eq:taylor}
\end{align}
\underline{Step 2}: Insert the expression \eqref{eq:ito-differential} for $\dd X_t$ and use the rules 
\begin{align}
(\dd t)^2 &= 0, &
\dd t \dd W_t &= 0, &
(\dd W_t)^2 &= \dd t. \label{eq:rules}
\end{align}

\begin{example}
Suppose that $X$ is given by \eqref{eq:ito-differential}.  What is $\dd f(X_t)$ when $f(x) = x^p$?  We have $f'(x) = p x^{p-1}$ and $f''(x) = p (p-1) x^{p-2}$.  Thus, from \eqref{eq:taylor}, we obtain
\begin{align}
\dd (X_t^p) 
	&= p X_t^{p-1} \dd X_t + \frac{1}{2} p (p-1) X_t^{p-2} (\dd X_t)^2 .
\end{align}
Now, we insert the expression \eqref{eq:ito-differential} for $\dd X_t$ and use the rules in \eqref{eq:rules}.  We have
\begin{align}
\dd (X_t^p) 
	&= p X_t^{p-1} ( \mu_t \dd t + \sig_t \dd W_t ) + \frac{1}{2} p (p-1) X_t^{p-2} ( \mu_t \dd t + \sig_t \dd W_t )^2 \\
	&= p X_t^{p-1} ( \mu_t \dd t + \sig_t \dd W_t ) + \frac{1}{2} p (p-1) X_t^{p-2} ( \mu_t^2 (\dd t)^2 + 2 \mu_t \sig_t \dd t \dd W_t + \sig_t^2 (\dd W_t)^2 ) \\
	&= p X_t^{p-1} ( \mu_t \dd t + \sig_t \dd W_t ) + \frac{1}{2} p (p-1) X_t^{p-2} \sig_t^2 \dd t \\
	&= \Big( p X_t^{p-1} \mu_t + \frac{1}{2} p (p-1) X_t^{p-2} \sig_t^2 \Big) \dd t + p X_t^{p-1} \sig_t \dd W_t .
\end{align}
Alternatively, we could have derived this expression using \eqref{eq:ito-1d}.  Whether you choose to use \eqref{eq:ito-1d} or the two-step method described above is up to you.
\\[0.5em]
Now, let us define $Y := X^p$.  Then, assuming the dynamics of $X$ are such that $X_t > 0$ for all $t \geq 0$, we have $X = Y^{1/p}$.  Inserting this expression for $X$ into the above expression, we obtain an It\^o process for $Y$
\begin{align}
\dd Y_t 
	&= \Big( p Y_t^{(p-1)/p} \mu_t + \frac{1}{2} p (p-1) Y_t^{(p-2)/p} \sig_t^2 \Big) \dd t + p Y_t^{(p-1)/p} \sig_t \dd W_t .
\end{align}
\end{example}

\section{Extension to multiple dimensions}

It\^o's Lemma generalizes in a straightforward manner to multiple dimensions.  We provide the main result without proof below.

\begin{theorem}[$d$-dimensional It\^o Formula]
Consider a $d$-dimensional process 
\begin{align}
X 
	= (X_t^{(1)}, X_t^{(2)}, \ldots, X_t^{(d)})_{t \geq 0}, 
\end{align}
whose $i$th component is given by
\begin{align}
\dd X_t^{(i)}
	&=	\mu_t^{(i)} \dd t + \sum_{j=1}^n \sig_t^{(i,j)} \dd W_t^{(j)} , &
i
	&=	1, 2, \ldots, d , \label{eq:ito-differential-d}
\end{align}
where $W = (W_t^{(1)}, W_t^{(2)}, \ldots, W_t^{(n)})_{t \geq 0}$ is a $n$-dimensional Brownian motion with independent components (i.e., $W^{(j)}$ is a one-dimensional Brownian motion for all $j$ and $W^{(j)} \ind W^{(k)}$ if $j \neq k$).  Suppose $f: \Rb^d \to \Rb$.  Then we have
\begin{align}
\dd f(X_t)
	&=	\Big( \sum_{i=1}^d \mu_t^{(i)} \d_{x_i} f(X_t)  + \frac{1}{2} \sum_{i=1}^d \sum_{j=1}^d (\sig_t \sig_t^T)^{(i,j)} \d_{x_i} \d_{x_j} f(X_t) \Big) \dd t \\ & \quad
			+ \sum_{i=1}^d \sum_{j=1}^n \sig_t^{(i,j)} \d_{x_i} f(X_t) \dd W_t^{(j)} . \label{eq:ito-d}
\end{align}
where we have used the notation
\begin{align}
\d_{x_i} 
	&:= \frac{\d}{\d x_i} , &
(\sig_t \sig_t^T)^{(i,j)}
	&=	\sum_{k=1}^n \sig_t^{(i,k)} \sig_t^{(j,k)} .
\end{align}
\end{theorem}

\noindent
As in the one-dimensional case, rather than remember the multi-dimensional It\^o formula \eqref{eq:ito-d} we can simply use the following two-step procedure
\\[0.5em]
\underline{Step 1}: Define $\dd f(X_t) := f(X_t + \dd X_t) - f(X_t)$ and expand $f(X_t + \dd X_t)$ as a Taylor series about $X_t$ to second order:
\begin{align}
\dd f(X_t) = \sum_{i=1}^d \d_{x_i}f(X_t) \dd X_t^{(i)} + \frac{1}{2} \sum_{i=1}^d \sum_{j=1}^d \d_{x_i} \d_{x_j} f(X_t) \dd X_t^{(i)} \dd X_t^{(j)} . \label{eq:taylor-d}
\end{align}
\underline{Step 2}: Insert the expression \eqref{eq:ito-differential-d} for $\dd X_t^{(i)}$ and use the rules 
\begin{align}
(\dd t)^2 &= 0, &
\dd t \dd W_t^{(i)} &= 0, &
\dd W_t^{(i)} \dd W_t^{(j)} &= \del_{i,j} \dd t , \label{eq:rules-d}
\end{align}
where $\del_{i,j}$ is a Kronecker delta (i.e., $\del_{i,j} = 0$ if $i \neq j$ and $\del_{i,i} = 1$).

\begin{example}[Product Rule]
Suppose that $X$ and $Y$ are It\^o processes of the form
\begin{align}
\dd X_t
	&=	\mu_t \dd t + \sig_t \dd W_t^{(1)} , &
\dd Y_t
	&=	\alpha_t \dd t + \beta_t \dd W_t^{(1)} + \gam_t \dd W_t^{(2)} , \label{eq:dXdY}
\end{align}
where $W^{(1)}$ and $W^{(2)}$ are independent Brownian motions.
What is $\dd (X_t Y_t)$?  Setting $f(x,y) = xy$, we have
\begin{align}
{\d_x} (xy) &= y, &
{\d_y} (xy)&= x, &
{\d_x^2} (xy) &= 0 , &
{\d_y^2} (xy) &= 0, & 
{\d_x \d_y}(xy) &= 1 .
\end{align}
Thus, from \eqref{eq:taylor-d} we have
\begin{align}
\dd (X_t Y_t) 
	&=	Y_t \dd X_t + X_t \dd Y_t + \dd X_t \dd Y_t . \label{eq:product-rule}
\end{align}
Equation \eqref{eq:product-rule} is known as the \textit{product rule} for It\^o processes.
Next, using the expressions for $\dd X_t$ and $\dd Y_t$ given in \eqref{eq:dXdY} and the rules provided in \eqref{eq:rules-d}, we obtain
\begin{align}
\dd (X_t Y_t) 
	&=	Y_t (\mu_t \dd t + \sig_t \dd W_t^{(1)}) + X_t (\alpha_t \dd t + \beta_t \dd W_t^{(1)} + \gam_t \dd W_t^{(2)}) \\ & \quad
			+ (\mu_t \dd t + \sig_t \dd W_t^{(1)}) (\alpha_t dt + \beta_t \dd W_t^{(1)} + \gam_t \dd W_t^{(2)}) \\
	&=	Y_t (\mu_t \dd t + \sig_t \dd W_t^{(1)}) + X_t (\alpha_t \dd t + \beta_t \dd W_t^{(1)} + \gam_t \dd W_t^{(2)}) + \sig_t \beta_t \dd t .
\end{align}
Of course, we could have obtained the above expression by using \eqref{eq:ito-d} directly.
\end{example}

\begin{example}
Suppose $X$ is given by \eqref{eq:ito-differential}.  
Let us compute $\dd f(t,X_t)$.  Note that $f$ is a function of two variables.  We Taylor expand $\dd f(t,X_t) := f(t + \dd t, X_t + \dd X_t) - f(t,X_t)$ to second order about $(t,X_t)$ to obtain
\begin{align}
\dd f(t,X_t)
	&=	\frac{\d f}{\d x} \dd X_t + \frac{\d f}{\d t} \dd t 
			+ \frac{1}{2} \Big( \frac{\d^2 f}{\d x^2} (\dd X_t)^2 + 2 \frac{\d^2 f}{\d x \d t} \dd X_t \dd t + \frac{\d^2 f}{\d t^2} \dd t^2 \Big) .
\end{align}
Now, we insert the expression for $\dd X_t$ and use the rules $(\dd t)^2 = 0$, $\dd W_t \dd t = 0$ and $(\dd W_t)^2 = \dd t$ and obtain
\begin{align}
\dd f(t,X_t)
	&=	\[ \frac{\d f}{\d t}(t,X_t) + \mu_t \frac{\d f}{\d x}(t,X_t) + \frac{1}{2} \sigma_t^2 \frac{\d^2 f}{\d x^2}(t,X_t) \] \dd t
			+ \sig_t \frac{\d f}{\d x}(t,X_t)  \dd W_t.
\end{align}
\end{example}

\section{Stochastic Differential Equations}

Consider an It\^o process $X = (X_t)_{t \geq 0}$ whose coefficients $\mu$ and $\sig$ are deterministic functions of $t$ and $X_t$.  Then the dynamics of $X$ are of the form
\begin{align}
\dd X_t
	&=	\mu(t,X_t) \dd t + \sig(t,X_t) \dd W_t . \label{eq:SDE}
\end{align}
In the simplest case, $X$ lives in $\Rb$ and $W = (W_t)_{t \geq 0}$ is a one-dimensional Brownian motion.  But, more generally, we could have $X \in \Rb^d$ and $W \in \Rb^n$, in which case we would have $\mu : \Rb_+ \times \Rb^d \to \Rb^d$ and $\sig : \Rb_+ \times \Rb^d \to \Rb^{d \times n}$.  The $i$th component of $X$ would then be given by
\begin{align}
\dd X_t^{(i)}
	&=	\mu^{(i)}(t,X_t)  \dd t + \sum_{j=1}^n \sig^{(i,j)}(t,X_t)  \dd W_t^{(j)} , &
i
	&=	1, 2, \ldots, d . \label{eq:SDE-d}
\end{align} 
We call an equation of the form \eqref{eq:SDE} a \textit{Stochastic Differential Equation} (SDE).  
The (strong) solution of an SDE is a functional of the path of a Brownian motion $W$.  That is $X_T = F[\{W_t,0\leq t \leq T\}]$.

\begin{assumption}
Throughout this text, we will assume every SDE has a unique strong solution.
\end{assumption}

\noindent
We will not prove the following very important theorem.

\begin{theorem}
\label{thm:sde-markov}
If an SDE of the form \eqref{eq:SDE} has a unique solution $X$, then $X$ is a Markov process.
\end{theorem}

\noindent
What Theorem \ref{thm:sde-markov} tells us is that, for any function $g$, and for any $T > t$, there exists a function $u$ such that
\begin{align}
\Eb[ g(X_T) | \Fc_t ]
	&=	\Eb[ g(X_T) | X_t ] = u(t,X_t) . \label{eq:Markov-SDE}
\end{align}
Let us see some examples of SDEs.

\begin{example}[Geometric Brownian Motion]
\label{ex:gbm}
Suppose $S = (S_t)_{t \geq 0}$ is the solution of the following SDE
\begin{align}
\dd S_t
	&=	\mu S_t \dd t + \sig S_t \dd W_t .
\end{align}
For reasons that will become apparent later, let us compute $\dd (\log S_t)$.  We have
\begin{align}
f(s) 
	&=	\log s , &
f'(s)
	&=	\frac{1}{s} , &
f''(s)
	&=	\frac{-1}{s^2} .
\end{align}
Thus, from It\^o's Lemma, we have
\begin{align}
\dd \log S_t
	&=	\frac{1}{S_t} \dd S_t + \frac{1}{2} \frac{-1}{S_t^2} \sig^2 S_t^2 \dd t \\
	&=	\Big( \mu - \frac{1}{2} \sig^2 \Big) \dd t + \sig \dd W_t .
\end{align}
Integrating from $0$ to $T$ we obtain
\begin{align}
\log S_T - \log S_0
	&=	\log(S_T/S_0)
	=		\Big( \mu - \frac{1}{2} \sig^2 \Big) T + \sig W_T  .
\end{align}
Taking the exponential of both sides, and multiplying both sides by $S_0$, we obtain
\begin{align}
S_T
	&=	S_0 \exp \( \Big( \mu - \frac{1}{2} \sig^2 \Big) T + \sig W_T \) .
\end{align}
We now have an explicit expression for $S_T$ in terms of $T$ and $W_T$.
As we shall see, people often model the price of a stock as a geometric Brownian motion.
\end{example}

\begin{example}[Ornstein-Uhlenbeck Process]
\label{ex:OU}
An Ornstein--Uhlenbeck (OU) process is the solution $X$ of an SDE of the following form
\begin{align}
\dd X_t 
	&= \kappa (\theta - X_t) \dd t + \sigma \dd W_t.
\end{align}
Let us define $Y_t := X_t - \theta$. Then we have $\dd X_t = \dd Y_t$ and thus
\begin{align}
\dd Y_t 
	&=	\kappa (\theta - X_t) \dd t + \sigma \dd W_t
	= 	-\kappa Y_t \dd t + \sigma  \dd W_t.
\end{align}
Now define $Z_t := \ee^{\kappa t} Y_t$. Noting that $\dd (\ee^{\kappa t}) = \kappa \ee^{\kappa t} \dd t$, we have
\begin{align}
\dd Z_t 
	&=	\kappa \ee^{\kappa t} Y_t \dd t + \ee^{\kappa t} \dd Y_t 
	= \ee^{\kappa t} \sigma \, dW_t.
\end{align}
Integrating from $0$ to $T$, we obtain
\begin{align}
Z_T - Z_0 
		&= \int_0^T \ee^{\kappa t} \sigma \dd W_t .
\end{align}
Next, using $Z_t := \ee^{\kappa t} Y_t$, we have
\begin{align}
\ee^{\kappa T} Y_T - Y_0
	&=  \int_0^T \ee^{\kappa t} \sigma \dd W_t .
\end{align}
Using $Y_t := X_t - \theta$, we obtain
\begin{align}
\ee^{\kappa T} (X_T - \theta) - (X_0 - \theta)
	&=  \int_0^T \ee^{\kappa t} \sigma \dd W_t .
\end{align}
Finally, solving for $X_T$, we find
\begin{align}
X_T 
	&=	\theta + \ee^{-\kappa T} (X_0 - \theta) + \int_0^T \ee^{-\kappa(T-t)} \sigma \dd W_t.
\end{align}
We now have an explicit expression for $X_T$ in terms of $T$ and $(W_s)_{0 \leq s \leq T}$.  Note that $X_T$ is a normally distributed random variable because the It\^o integral has a deterministic integrand.  Let us find the mean and variance of $X_T$.  We have
\begin{align}
\Eb X_T = \theta + \ee^{-\kappa T} (X_0 - \theta),
\end{align}
where we have used the fact that the expected value of an It\^o integral is zero.
Next, we compute the variance of $X_T$.  We have
\begin{align}
\Vb X_T
	&= \Eb[(X_T - \Eb[X_T])^2]  
	 = \Eb \[ \( \int_0^T \ee^{-\kappa(T-t)} \sigma \dd W_t \)^2 \]  
	 = \int_0^T \ee^{-2\kappa(T-t)} \sigma^2 \dd t  
	 = \frac{\sigma^2}{2\kappa} (1 - \ee^{-2\kappa T}) ,
\end{align}
where the third equality follows from the It\^o isometry.
\end{example}

\section{Simulating an SDE}
Many SDEs do not have explicit solutions.  In such cases, it is often useful to simulate approximate sample paths of the solution.
Consider the following SDE
\begin{align}
\dd X_t 
	&= \mu(t,X_t) \dd t + \sigma(t,X_t) \dd W_t .
\end{align}
In a small interval of time $\dd t$, the process $X$ behaves approximately as follows
\begin{align} 
X_{t + \delta t} - X_t
	&\approx \mu (t, X_t) \delta t + \sigma(t,X_t) (W_{t+ \delta t} - W_t) ,
\end{align} 
where $W_{t+ \delta t} - W_t \sim N(0, \del t).$  The above equation provides a mechanism for simulating (at least approximately) sample paths of $X$.
\\[0.5em]
\underline{Step 1}: Divide the time interval of interest $[0,T]$ into intervals of size $\del t := T/N$, where $N$ is a fixed positive integer.
\\[0.5em]
\underline{Step 2}: For $n = 0, 1, \ldots, N-1$, set
\begin{align}
\Xh_{(n+1) \delta t} = \Xh_{n\del t} + \mu (n \del t, \Xh_{n \del t}) \delta t + \sigma( n \del t,\Xh_{n \del t}) \sqrt{\del t} Z_{n+1} ,
\end{align}
where the $(Z_n)_{n \in \{1,2,\ldots,N\}}$ are iid standard normal random variables $Z_n \sim N(0,1)$.
\\[0.5em]
The resulting discrete time process $\Xh = (\Xh_{n\del t})_{n \in \{0,1,\ldots,N\}}$ has approximately the same distribution as the solution $X$ of the SDE
\begin{align}
(\Xh_0,\Xh_{\del t}, \ldots, \Xh_{N \del t})
	&\stackrel{\Dc}{\approx} (X_0,X_{\del t}, \ldots, X_{N \del t}) .
\end{align}
\begin{example}
See the solution to Exercise \ref{exer:5-1}, parts (c), (d) and (f), for Python code that implements this scheme.
\end{example}

\noindent
In some situations, it may be better to make a variable transformation before running the simulation.  For example, consider a Geometric Brownian motion
\begin{align}
\dd S_t 
	&= \mu S_t \dd t + \sigma S_t \dd W_t .
\end{align}
If we simulate this process directly, we are making a small error at every step because, for every $n$ we are making the approximation
\begin{align}
\int_{n \del t}^{(n+1) \del t} \mu S_u \dd u + \int_{n \del t}^{(n+1) \del t} \sigma S_u \dd W_u
	&\approx \mu S_{n \del t} \del t +  \sigma S_{n \del t} (W_{(n+1)\del t} - W_{n \del t}) .
\end{align}
Now, suppose we define a new process $X$ as follows $X_t = \log S_t$.  Then the dynamics of $X$ are
\begin{align}
\dd X_t 
	&= (\mu - \frac{1}{2} \sigma^2) \dd t + \sigma \dd W_t.
\end{align}
Now, if we simulate $X$ we make no error because, for every $n$ we have exactly
\begin{align}
\int_{n \del t}^{(n+1) \del t} (\mu - \tfrac{1}{2} \sig^2) \dd u + \int_{n \del t}^{(n+1) \del t} \sigma \dd W_u
	&= (\mu - \tfrac{1}{2} \sig^2) \del t + \sigma (W_{(n+1)\del t} - W_{n \del t}) .
\end{align}
Thus, in this case, it is better to simulate sample paths of $X$ rather than sample paths of $S$.   Once we have generated a sample path of $X$, we can obtain a sample path of $S$ by setting $S_t = \ee^{X_t}$.


\section{Exercises}
\label{sec:exer-4}

\begin{exercise}
\label{exer:4-1}
Consider a model for a stock and a money market account
\begin{align}
\dd S_t
	&=	\mu S_t \dd t + \sig S_t \dd W_t , \\
\dd M_t
	&=	r M_t \dd t .
\end{align}
Define $X_t = S_t/M_t$.  Compute $\dd X_t$.  You should be able to write your answer as an SDE involving only $X_t$ (neither $S$ nor $M$ should appear in your answer).
\end{exercise}


\begin{exercise}
\label{exer:4-2}
Consider a model for two stocks
\begin{align}
\dd S_t
	&=	\mu S_t \dd t + \sig S_t \dd W_t , \\
\dd Y_t
	&=	\alpha Y_t \dd t + \beta Y_t ( \rho \dd W_t + \sqrt{1-\rho^2} \dd B_t ) ,
\end{align}
where $\rho \in (-1,1)$ and $W$ and $B$ are independent Brownian motions.
Define $Z_t = S_t/Y_t$.  Compute $\dd Z_t$.  You should be able to write your answer as an SDE involving only $Z_t$ (neither $S$ nor $Y$ should appear in your answer).
\end{exercise}


\begin{exercise}
\label{exer:4-3}
Consider a model for a stock $S$.
\begin{align}
\dd S_t
	&=	\mu S_t \dd t + \sig S_t \dd W_t , 
\end{align}
Define $P_t = S_t^\beta$.  Compute $\dd P_t$.  You should be able to write your answer as an SDE involving only $P_t$ (i.e., $S$ should not appear in your answer).  Find all values of $\beta$ for which the process $P$ is a martingale.
\end{exercise}


\begin{exercise}
\label{exer:4-4}
Consider the following two-dimensional SDE
\begin{align}
\dd X_t
	&=	- \tfrac{1}{2} X_t \dd t - Y_t \dd W_t , &
\dd Y_t
	&=	- \tfrac{1}{2} Y_t \dd t + X_t \dd W_t  . \label{fivepointfoursde}
\end{align}
\textbf{(a)} Compute $\dd (X_t^2 + Y_t^2)$.\\
\textbf{(b)} Show that $X_t = \cos W_t$ and $Y_t = \sin W_t$ is a solution of the above SDE (i.e., compute $\dd X_t$ and $\dd Y_t$ and show that you obtain the SDE above).\\
\textbf{(c)} Using your result from part (b), explain your answer to part (a).
\end{exercise}


\begin{exercise}
\label{exer:4-5}
Suppose $X$ satisfies
\begin{align}
\dd X_t
	&=	\mu(X_t) \dd t + \sig(X_t) \dd W_t .
\end{align}
Define $M_t = \ee^{- \lam t} \phi(X_t)$.  Derive an ordinary differential equation that $\phi$ must satisfy in order for $M$ to be a martingale.
\end{exercise}


\begin{exercise}
\label{exer:4-6}
Suppose $X$ satisfies
\begin{align}
\dd X_t
	&=	\mu \dd t + \sig \dd W_t .
\end{align}
Define $Y$, the exponential moving average of $X$, by $Y_t = \lam \int_{-\infty }^t \ee^{-\lam(t-s)} X_s \dd s$, where $\lam > 0$.  Next, define $Z_t = X_t - Y_t$.  Derive an SDE for $Z$.
\end{exercise}


\begin{exercise}
\label{exer:4-7}
Suppose $Z$ is given by
\begin{align}
Z_t
	&=	\exp \Big( - \tfrac{1}{2}\int_0^t \Del_s^2 \dd s + \int_0^t \Del_s \dd W_s \Big) .
\end{align}
where $\Del$ is bounded and adapted to the filtration generated by observing the Brownian motion $W$.  What is $\Eb Z_t$?
\end{exercise}




\chapter{Black-Scholes model}

Consider a financial market with a stock $S = (S_t)_{0 \leq t \leq T}$ and a bond $B = (B_t)_{0 \leq t \leq T}$, whose dynamics under the real-world probability measure $\Pb$ are given by
\begin{align}
\dd S_t 
	&= \mu S_t \dd t + \sigma S_t \dd W_t , &
\dd B_t 
	&= r B_t \dd t ,  \label{eq:BS-model}
\end{align}
where $W = (W_t)_{0 \leq t \leq T}$ is a Brownian motion.  Throughout this chapter, $\Fc = (\Fc_t)_{0 \leq t \leq T}$ will always be a filtration for the Brownian motion $W$.  Expression \eqref{eq:BS-model} is the well-known Black-Scholes model.  In this chapter, we derive the price of a European option in the Black-Scholes setting.  Just as in the Binomial model, we will present two methods (i) pricing by replication, and (ii) risk-neutral pricing.

\section{Pricing by replication}
Consider a European option that pays $\phi(S_T)$ at time $T$.  We will try to construct a portfolio $X = (X_t)_{0 \leq t \leq T}$ that replicates the option payoff: $X_T = \phi(S_T)$.  Let us denote by $\Del = (\Del_t)_{ 0 \leq t \leq T}$ the number of shares of $S$ we hold in our portfolio.  If changes in the value of $X$ are due only to the changes in the value of $S$ and $B$ (i.e., the portfolio is self-financing), then the dynamics of $X$ must be as follows
\begin{align}
\dd X_t 
	&=	\Del_t \dd S_t + (X_t - \Del_t S_t) \frac{1}{B_t} \dd B_t \\
	&=	\Del_t ( \mu S_t \dd t + \sig S_t \dd W_t ) + (X_t - \Del_t S_t) r \dd t \\
	&=	\Del_t S_t (\mu - r) \dd t + \Del_t S_t \sig \dd W_t + r X_t \dd t , \label{eq:BS-dX}
\end{align}
where, in the second equality, we have used \eqref{eq:BS-model}.  It will simplify the computations that follow if we derive the dynamics of $(X/B)$.  Using the product rule, we compute
\begin{align}
\dd \Big( \frac{X_t}{B_t} \Big)
	&=	\frac{1}{B_t} \dd X_t + X_t \dd \big( \frac{1}{B_t} \Big) + \dd X_t \dd \big( \frac{1}{B_t} \Big) \\
	&=	\frac{1}{B_t} \dd X_t + \frac{-X_t}{B_t^2} \dd B_t \\
	&=	\frac{\Del_t S_t}{B_t} (\mu - r) \dd t +  \frac{\Del_t S_t}{B_t}  \sig \dd W_t . \label{eq:XoverB}
\end{align}
where  we have used \eqref{eq:BS-model} and \eqref{eq:BS-dX}.
\\[0.5em]
Let $V = (V_t)_{0 \leq t \leq T}$ denote the value of the option with payoff $\phi(S_T)$.
The option value $V$ will fluctuate as $S$ moves randomly. Thus, it is reasonable to expect that the option value $V$ is a function $v$ of time and the stock price: $V_t = v(t, S_t)$.  Obviously, $V$ should depend on other parameters as well (e.g., $r$ and $\sig$).  But these other parameters are constant whereas $t$ and $S$ evolve in time. Using the multidimensional version of It\^o's Lemma, we compute the dynamics of $V$ as follows
\begin{align}
\dd V_t 
	&= \dd v(t,S_t) \\
	&=	\frac{\d v}{\d t}(t,S_t) \dd t + \frac{\d v}{\d s}(t,S_t) \dd S_t + \frac{1}{2} \frac{\d^2 v}{\d s^2}(t,S_t) ( \dd S_t )^2 \\
	&= \[ \frac{\d v}{\d t}(t,S_t) + \mu S_t \frac{\d v}{\d s}(t,S_t) + \frac{1}{2} \sigma^2 S_t^2 \frac{\d^2 v}{\d s^2}(t,S_t) \] \dd t 
			+ \frac{\d v}{\d s} (t,S_t) \sigma S_t \dd W_t , \label{eq:BS-dV}
\end{align}
where we have used \eqref{eq:BS-model}.  Once again, it will simplify the computations that follow if we derive the dynamics of $(V/B)$.  Using the product rule, we obtain
\begin{align}
\dd \Big( \frac{V_t}{B_t} \Big)
	&=	\Big( \frac{1}{B_t} \Big) \dd V_t + V_t  \dd \Big( \frac{1}{B_t} \Big) + \dd V_t \dd \Big( \frac{1}{B_t} \Big) \\
	&= 	\frac{1}{B_t} \dd V_t + \frac{-V_t}{B_t^2} \dd B_t \\
	&=	\frac{1}{B_t} \[ \frac{\d v}{\d t}(t,S_t) + \mu S_t \frac{\d v}{\d s}(t,S_t) + \frac{1}{2} \sigma^2 S_t^2 \frac{\d^2 v}{\d s^2}(t,S_t) - r v(t,S_t) \] \dd t 
			\\ &\quad
			+ \frac{1}{B_t}  \frac{\d v}{\d s} (t,S_t) \sigma S_t \dd W_t . \label{eq:VoverB}
\end{align}
where we have used \eqref{eq:BS-model} and \eqref{eq:BS-dV}, as well as  $V_t = v(t,S_t)$.
\\[0.5em]
Recall, we are seeking a portfolio $X$ that replicates the option value $V$.  Suppose that
\begin{align}
X_0
	&=	V_0 &
&\text{and}&
\dd \Big( \frac{X_t}{B_t} \Big)
	&=	\dd \Big( \frac{V_t}{B_t} \Big) . \label{eq:XequalV}
\end{align}
Then we would have $X_t =	V_t$ for all $t \in [0,T]$ because
\begin{align}
\frac{X_t}{B_t}
	&=	\frac{X_0}{B_0} + \int_0^t \dd \Big( \frac{X_s}{B_s} \Big)
	 =	\frac{V_0}{B_0} + \int_0^t \dd \Big( \frac{V_s}{B_s} \Big) 
	 =  \frac{V_t}{B_t} .
\end{align}
In order for \eqref{eq:XequalV} to hold, we must have
\begin{align}
0
	&=	\dd \Big( \frac{V_t}{B_t} \Big) - \dd \Big( \frac{X_t}{B_t} \Big) \\
	&=	\frac{1}{B_t} \[ \frac{\d v}{\d t}(t,S_t) + \mu S_t \frac{\d v}{\d s}(t,S_t) + \frac{1}{2} \sigma^2 S_t^2 \frac{\d^2 v}{\d s^2}(t,S_t) - r v(t,S_t)
			- (\mu - r ) \Del_t S_t \] \dd t 
			\\ &\quad
			+ \frac{1}{B_t}  \Big( \frac{\d v}{\d s} (t,S_t) - \Del_t \Big) \sigma S_t \dd W_t . \label{eq:diff}
\end{align}
The $\dd W_t$-term in \eqref{eq:diff} will equal zero if and only if we set
\begin{align}
\Del_t
	&=	\frac{\d v}{\d s} (t,S_t) . \label{eq:Delta}
\end{align}
Inserting expression \eqref{eq:Delta} for $\Del_t$ into \eqref{eq:diff}, we obtain
\begin{align}
0
	&=	\frac{1}{B_t} \[ \frac{\d v}{\d t}(t,S_t) + r S_t \frac{\d v}{\d s}(t,S_t) + \frac{1}{2} \sigma^2 S_t^2 \frac{\d^2 v}{\d s^2}(t,S_t) - r v(t,S_t) \] \dd t . 
			\label{eq:dt-term}
\end{align}
The $\dd t$-term will equal zero if and only if the function $v$ satisfies 
\begin{align}
0
	&=	\frac{\d v}{\d t} + r s \frac{\d v}{\d s} + \frac{1}{2} \sigma^2 s^2 \frac{\d^2 v}{\d s^2} - r v . \label{eq:BS-PDE}
\end{align}
Equation \eqref{eq:BS-PDE} is the famous \textit{Black-Scholes Partial Differential Equation} (PDE).  In order to find a solution $v$ of \eqref{eq:BS-PDE} we need a \textit{terminal condition} (TC) for the function $v$.  This can be obtained as follows
\begin{align}
v(T,S_T)
	&=	V_T = \phi(S_T) , &
	&\Rightarrow&
v(T,s)
	&=	\phi(s) . \label{eq:BS-TC}
\end{align}
The terminal condition \eqref{eq:BS-TC} together with the PDE \eqref{eq:BS-PDE} are sufficient to determine the function $v$ uniquely.  We will solve the PDE \eqref{eq:BS-PDE} in a subsequent section.  For now, let us review what we have done.
\begin{enumerate}
\item We have supposed that the dynamics of a stock $S$ and a bond $B$ are given by \eqref{eq:BS-model}.
\item We have considered a European option that pays $\phi(S_T)$ at time $T$.  The value $V = (V_t)_{0 \leq t \leq T}$ of this option is given by $V_t = v(t,S_t)$ where the function $v$ satisfies PDE \eqref{eq:BS-PDE} and terminal condition \eqref{eq:BS-TC}.
\item If the initial value of a portfolio $X$ is given by $X_0 = V_0 = v(0,S_0)$ and if the number of shares  $\Del$ of $S$ one holds in this portfolio is given by \eqref{eq:Delta}, then the portfolio $X$ replicates the value $V$ of the option.  In particular, we have $X_T = V_T = \phi(S_T)$. 
\end{enumerate}

\section{Risk-Neutral Pricing}

The 1st fundamental theorem of asset pricing (Theorem \ref{thm:1st}) states that a market defined under a probability measure $\Pb$ is free of arbitrage if and only if there exists a probability measure $\Pbt$, equivalent to $\Pb$, under which all portfolios $X$, denominated in units of a common num\'eraire $N$ are martingales.  
In the Black-Scholes setting \eqref{eq:BS-model} this leads to a very simple procedure for finding the value $V = (V_t)_{0 \leq t \leq T}$ of an option that pays $\phi(S_T)$ at time $T$.
\begin{enumerate}
\item Choose a num\'eraire.  We will choose the bond $B=(B_t)_{0 \leq t \leq T}$, as this choice leads to the easiest computations.
\item Search for a probability measure $\Pbt$, under which $(X/B)$ is a martingale for all portfolios $X$.  That is, find $\Pbt$ such that
\begin{align}
\frac{X_t}{B_t}
	&=	\Ebt \Big( \frac{X_T}{B_T} \Big| \Fc_t \Big) , 
\end{align} 
where $X$ is given by \eqref{eq:BS-dX}.
\item The value $V$ of the option can then be computed using
\begin{align}
\frac{V_t}{B_t}
	&=	\Ebt \Big( \frac{V_T}{B_T} \Big| \Fc_t \Big) 
	 = 	\Ebt \Big( \frac{\phi(S_T)}{B_T} \Big| \Fc_t \Big) . \label{eq:V-price}
\end{align}
\end{enumerate}

\noindent
In the binomial model, ``search for a probability measure $\Pbt$'' meant that we should look for probabilities $\pt$ and $\qt$ that the stock would go up and down.  
But, what does it mean to ``search for a probability measure $\Pbt$'' when the market is described by SDEs \eqref{eq:BS-model}?  The following Theorem provides an answer to this question.

\begin{theorem}[Girsanov's Theorem]
\label{thm:girsanov}
Suppose $W = (W_t^{(1)},\ldots,W_t^{(d)})_{0 \leq t \leq T}$ is a $d$-dimensional Brownian motion with independent components under a probability measure $\Pb$.  Define a process $\Wt^{(\gam)} = (\Wt_t^{(\gam,1)},\ldots,\Wt_t^{(\gam,d)})_{0 \leq t \leq T}$ by
\begin{align}
\Wt_t^{(\gam)}
	&=	\int_0^t \gam_s \dd s + W_t , \label{eq:W-tilde}
\end{align}
where $\gam = (\gam_t^{(1)},\ldots,\gam_t^{(d)})_{0 \leq t \leq T}$ is adapted to a filtration $\Fc$ for $W$.
Then there exists a probability measure $\Pbt^{(\gam)}$, equivalent to $\Pb$, under which $\Wt^{(\gam)}$ is a $d$-dimensional Brownian motion with independent components.
\end{theorem}

\noindent
Let us see how to use Theorem \ref{thm:girsanov} to write the dynamics of $(X/B)$ under a risk-neutral measure $\Pbt^{(\gam)}$.  Using equation \eqref{eq:XoverB}, the dynamics of $X/B$ are given by
\begin{align}
\dd \Big( \frac{X_t}{B_t} \Big)
	&=	\frac{\Del_t S_t}{B_t} (\mu - r) \dd t +  \frac{\Del_t S_t}{B_t}  \sig \dd W_t \\
	&=	\frac{\Del_t S_t}{B_t} (\mu - r) \dd t +  \frac{\Del_t S_t}{B_t}  \sig ( \dd \Wt^{(\gam)}_t - \gam_t \dd t) \\
	&=	\frac{\Del_t S_t}{B_t} (\mu - r - \sig \gam_t) \dd t +  \frac{\Del_t S_t}{B_t}  \sig \dd \Wt^{(\gam)}_t ,
\end{align}
where, in the second equality, we have used \eqref{eq:W-tilde}.  Now, we know from Theorem \ref{thm:girsanov} that, for any choice of $\gam$, there exists a probability $\Pbt^{(\gam)}$ under which $\Wt^{(\gam)}$ is a Brownian motion.  If we choose $\gam$ such that the $\dd t$-term equals zero, then $X/B$ will be a martingale under $\Pbt^{(\gam)}$.  Thus, we choose
\begin{align}
\gam_t
	&=	\frac{\mu - r}{\sig} . \label{eq:gamma}
\end{align}
Now, what are the dynamics of $S$ under $\Pbt^{(\gam)}$?  We have
\begin{align}
\dd S_t
	&=	\mu S_t \dd t + \sig S_t \dd W_t \\
	&=	\mu S_t \dd t + \sig S_t ( \dd \Wt^{(\gam)}_t - \gam_t \dd t) \\
	&=	( \mu - \sig \gam_t ) S_t \dd t + \sig S_t \dd \Wt^{(\gam)}_t \\
	&=	r S_t \dd t + \sig S_t \dd \Wt^{(\gam)}_t ,
\end{align}
where, in the last line, we have used \eqref{eq:gamma}.  Under $\Pbt^{(\gam)}$, the stock price $S$ is a geometric Brownian motion with drift $r$ and volatility $\sig$.  We found an explicit expression for $S$ in Example \ref{ex:gbm}.  We have
\begin{align}
S_T
	&=	S_t \exp \Big[ (r - \tfrac{1}{2} \sig^2)(T - t) + \sig (\Wt^{(\gam)}_T - \Wt^{(\gam)}_t) \Big] .
\end{align}
We can now compute the value $V$ of the option.  From \eqref{eq:V-price}, we have
\begin{align}
V_t
	&=	B_t \Ebt^{(\gam)} \Big( \frac{\phi(S_T)}{B_T} \Big| \Fc_t \Big) \\
	&=	\ee^{-r(T-t)} \Ebt^{(\gam)} \Big( \phi \Big( S_t \exp \Big[ (r - \tfrac{1}{2} \sig^2)(T - t) + \sig (\Wt^{(\gam)}_T - \Wt^{(\gam)}_t) \Big] \Big) \Big| S_t \Big) \\
	&=	\int_\Rb \frac{\ee^{-r(T-t)} }{\sqrt{2 \pi \sig^2(T-t)}}  \exp \Big( \frac{- y^2}{2 \sig^2 (T-t)} \Big)
			\phi \Big( S_t \exp \Big[ (r - \tfrac{1}{2} \sig^2)(T - t) + y \Big] \Big) \dd y =: v(t,S_t) , \label{eq:BS-V}
\end{align}
where, in the second equality, we have used the fact that $S$ is a Markov process, and in the third equality, we have used the fact that $\sig ( \Wt^{(\gam)}_T - \Wt^{(\gam)}_t ) \sim N(0,\sig^2(T-t))$ under $\Pbt^{(\gam)}$.  For all but a few particular choices of $\phi$, the integral in \eqref{eq:BS-V} must be computed numerically.  The function $v$ defined in \eqref{eq:BS-V} is, in fact, the unique solution of the Black-Scholes PDE \eqref{eq:BS-PDE} and TC \eqref{eq:BS-TC}.  In the next section, we will show how to find this function using PDE methods.

\begin{remark}
\label{rmk:drift}
Observe that \textit{the drift $\mu$ of $S$ does not appear} in the price \eqref{eq:BS-V} of the option nor does it appear in the Black-Scholes PDE \eqref{eq:BS-PDE}.  In fact, although we have assumed that the drift of $S$ was a constant $\mu$ in \eqref{eq:BS-model}, we could have assumed that the drift of $S$ was a stochastic process $\mu = (\mu_t)_{0 \leq t \leq T}$, and this would not have had any effect at all on the option price.  The fact that $\mu$ plays no role in the pricing of options is analogous to the real-world probabilities $p$ and $q$ playing no role in the binomial setting.
\end{remark}

\section{Solving Black--Scholes PDE}
Recall that the Black--Scholes PDE and TC are
\begin{align}
0
	&=	\(\d_t - r + r s \d_s + \frac{1}{2} \sigma^2 s^2 \d_s^2 \) v , &
v(T,s) 
	&= \phi(s) , \label{eq:BS-PDE-2}
\end{align}
where, we have re-introduced the notation $\d_s^n := \frac{\d^n}{\d s^n}$ and likewise for other partial derivatives.
We will solve this PDE in a series of steps.
\\[0.5em]
\underline{Step 1: Change of Variables}.
Suppose that
\begin{align}
v(t,s) &= \ee^{-r(T-t)} u(\tau(t), x(t,s)) , \label{eq:u.def}
\end{align}
where the functions $\tau$ and $x$ are defined as follows
\begin{align}
\tau(t) 
	&:= T-t , &
x(t,s) 
	&:= \frac{1}{\sig} \[ \log s + \( r - \frac{1}{2} \sigma^2 \)(T-t) \] .
\end{align}
Let us see what the function $u$ must satisfy.  Using the chain rule, we find
\begin{align}
\frac{\d v}{\d t} 
	&= \ee^{-r(T-t)}\[r u + \d_\tau u \cdot \d_t \tau + \d_x u \cdot \d_t x\] , \\
\frac{\d v}{\d s} 
	&= \ee^{-r(T-t)}\d_x u \cdot \d_s x , \\
\frac{\d^2 v}{\d s^2} 
	&=	\ee^{-r(T-t)} \[ \d^2_x u \cdot (\d_s x)^2 + \d_x u \cdot \d_s^2 x \] ,
\end{align}
where
\begin{align}
\d_t \tau(t)
	&=	-1 , &
\d_t x(t,s)
	&=	\frac{-1}{\sig} \( r - \frac{1}{2} \sigma^2 \) , &
\d_s x(t,s)
	&=	\frac{1}{\sig s} , &
\d_s^2 x(t,s)
	&=	\frac{- 1}{\sig s^2} .
\end{align}
Plugging the above expressions in to the Black-Scholes PDE \eqref{eq:BS-PDE-2} yields
\begin{align}
0	
	&=	(-\d_\tau + \frac{1}{2} \d_x^2) u , &
u(0,\sig^{-1} \log s)
	&=	\phi(s) &
	&\Rightarrow&
u(0,x) = \phi(\ee^{\sig x}) . \label{eq:u.pde}
\end{align}
We have obtained a PDE and initial condition (IC) for the function $u$.
\\[0.5em]
\underline{Step 2: Fundamental solution}.
We introduce a function $\Gamma$, which is given by
\begin{align}
	\Gamma(\tau,x,y) := \frac{1}{\sqrt{2 \pi \tau}} \ee^{\frac{-(x-y)^2}{2\tau}}.
\end{align}
One can check by direct computation that $\Gam$ satisfies 
\begin{align}
0
	&=	(-\d_\tau + \frac{1}{2}\d_x^2) \Gamma , &
\Gam(0,x,y)
	&=	\del(x-y) .
\end{align}
Thus, $\Gam$ satisfies the PDE in \eqref{eq:u.pde}, but not the IC.
We call $\Gamma$ the \textit{fundamental solution} of $(-\d_\tau + \tfrac{1}{2} \d_x^2 )$.
\\[0.5em]
\underline{Step 3: Solution of the PDE}.
We claim that the solution $u$ of \eqref{eq:u.pde} is given by
\begin{align}
	u(\tau,x) = \int_\Rb \dd y \, \Gamma(\tau,x,y) \phi(e^{\sig y}) .
\end{align}
To see this, note that
\begin{align}
	u(0,x) = \int_\Rb \dd y \underbrace{\Gamma(0,x,y)}_{\delta(x-y)} \phi(e^{\sig y}) = \phi(\ee^{\sig x}) .
\end{align}
Thus, $u$ satisfies the IC $u(0,x) = \phi(\ee^{\sig x})$.  Next, note that
\begin{align}
(-\d_\tau + \frac{1}{2} \d_x^2) u (\tau, x) 
		&= \int_\Rb \underbrace{(-\d_\tau + \frac{1}{2}\d_x^2) \Gamma(\tau,x,y)}_{=0} \phi(e^{\sig y}) \dd y = 0 .
\end{align}
Hence, the function $u$ also satisfies the PDE in \eqref{eq:u.pde}.
\\[0.5em]
\underline{Step 4: Undo the change of variables}.
Finally, we need to write the function $v$ in terms of $t$ and $s$.  Using \eqref{eq:u.def}, we have
\begin{align}
v(t,s)
	&=	\ee^{-r(T-t)} u(\tau(t),x(t,s)) \\
	&=	\ee^{-r(T-t)} \int_\Rb \Gamma(\tau(t),x(t,s),y) \phi(e^{\sig y}) \dd y \\
	&=	\int_\Rb \frac{\ee^{-r(T-t)} }{\sqrt{2\pi(T-t)}} \exp \( \frac{-(\log s + (r-\sig^2/2)(T-t) - \sig y )^2}{2 \sig^2 (T-t)}\)\phi(\ee^{\sig y}) \dd y \\
	&=	\int_\Rb \frac{\ee^{-r(T-t)} }{z \sqrt{2\pi \sig^2 (T-t)}} \exp \( \frac{-(\log s + (r-\sig^2/2)(T-t) - \log z )^2}{2 \sig^2 (T-t)}\)
			\phi(z) \dd z \label{eq:v.pde}
\end{align}
One can show that the expression \eqref{eq:v.pde} agrees with the expression given for $v$ in \eqref{eq:BS-V}.

\section{Black-Scholes formulas for European Calls and Puts}
For call and put options we have $\phi(s) = (s-K)^+$ and $\phi(s) = (K-s)^+$.  In these cases the above integral \eqref{eq:v.pde} can be computed explicitly, leading to the following expressions for call and put prices
\begin{align}
C^\BS(t,S_t;\sig,T,K) 
	&= S_t \Phi(d_+) - K \Phi(d_-) \ee^{-r(T-t)}, \label{eq:BS-Call} \\
P^\BS(t,S_t;\sig,T,K) 
	&= K \ee^{-r(T-t)} \Phi(-d_-) - S_t \Phi(-d_+) , \label{eq:BS-Put}
\end{align}
where $\Phi$ and $d_\pm = d_\pm(t,S_t;\sig,T,K)$ are given by
\begin{align}
d_{\pm} (t,S_t;\sig,T,K)
	&:= \frac{1}{\sigma \sqrt{T-t}} \[ \log \( \frac{S_t}{K} \) + \( r \pm \frac{1}{2} \sigma^2 \)(T-t) \], &
\Phi(x) 
	&:= \int_{-\infty}^x \frac{\ee^{\frac{-z^2}{2}}}{\sqrt{2\pi}} \dd z . \label{eq:d-Phi}
\end{align}
Note that the function $\Phi$ is the \textit{cumulative distribution function of a standard normal random variable}.
When it causes no confusion, we may omit the arguments $(\sig,T,K)$ from $C^\BS$ and $P^\BS$.
Note that
\begin{align}
\Del_t^\text{BS-Call} = \frac{\d C^\BS}{\d s} (t,S_t;\sig,T,K) 
	&= \Phi(d_+(t,S_t;\sig,T,K)) \in (0,1), \label{eq:BS-Delta-Call} \\
\Del_t^\text{BS-Put} = \frac{\d P^\BS}{\d s} (t,S_t;\sig,T,K) 
	&= \Phi(d_+(t,S_t;\sig,T,K)) - 1 \in (-1,0) . \label{eq:BS-Delta-Put}
\end{align}
These are the deltas of calls and puts, respectively, (i.e. the number of shares $\Delta$ of the stock $S$ one should hold in a portfolio $X$ in order to replicate a Call or Put).  Thus, to replicate a Call, one always holds a positive number of shares of $S$ and to replicate a Put, one always holds a negative number of shares of $S$.  Take a moment to think about why this is the case intuitively (\textit{hint}: think about what happens to the value of a Call or Put when the underlying stock $S$ goes up).

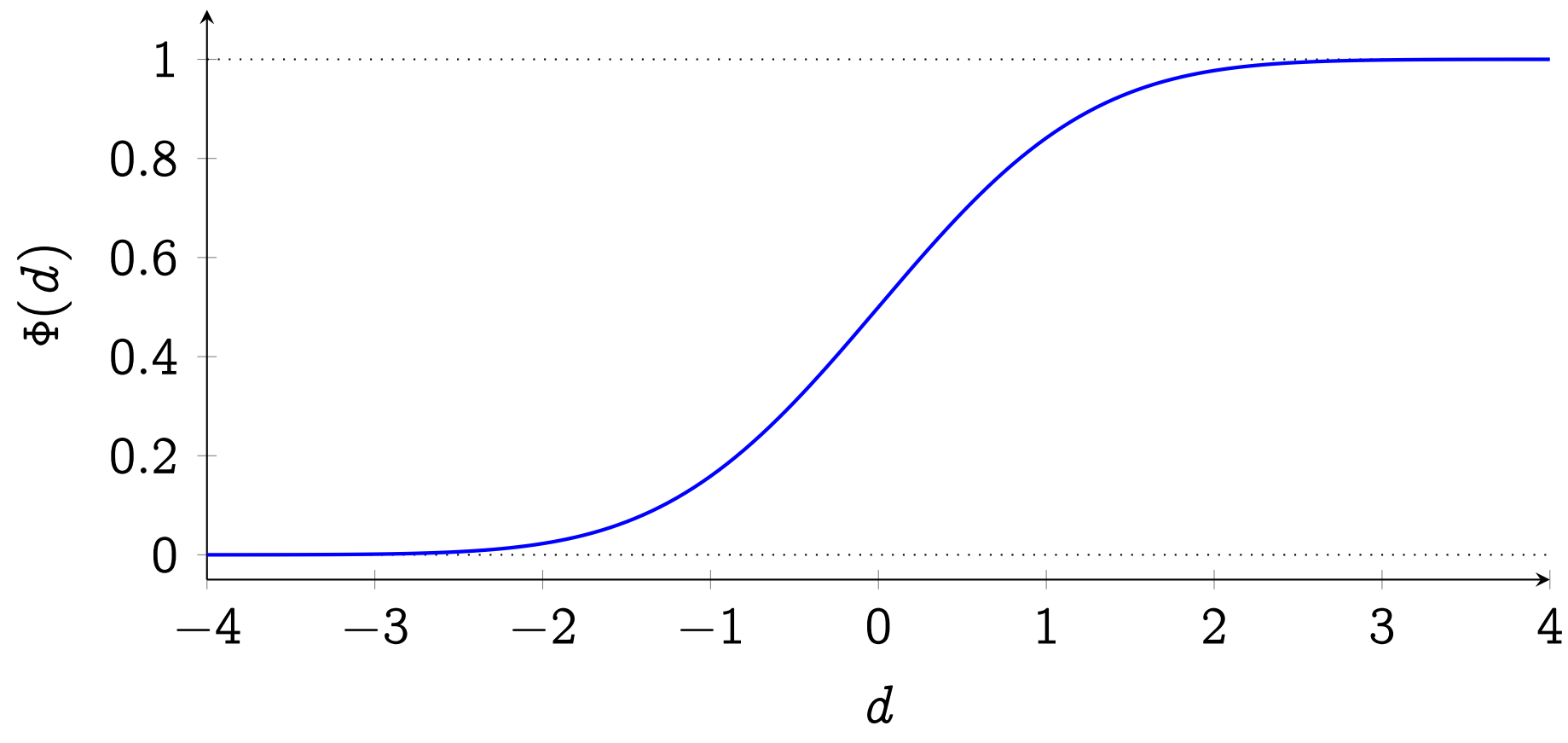
\begin{figure}[hb!]
    \centering
    \begin{tikzpicture}
\begin{axis}[width=12cm, height=6cm, axis lines=left, xmin=-4, xmax=4, ymin=-0.05, ymax=1.1,
             xlabel={$d$}, ylabel={$\Phi(d)$}, ytick={0,0.2,...,1}]
  \draw[zeroline] (axis cs:-4,0) -- (axis cs:4,0);
  \draw[zeroline] (axis cs:-4,1) -- (axis cs:4,1);
  \addplot[blue, thick, domain=-4:4, samples=161] {normcdf(x)};
\end{axis}
\end{tikzpicture}
    \caption{The cumulative distribution function $\Phi$ of a standard normal random variable. \label{fig:plot2_6}}
\end{figure}

\section{Implied Volatility}

\begin{figure}[hb!]
\centering
\begin{tikzpicture}
\begin{axis}[width=12cm, height=6cm, axis lines=left, xmin=-1.6, xmax=1.6, ymin=0, ymax=1.1,
             xtick={-0.6,0,0.6}, xticklabels={$\leftarrow$ ITM, ATM, OTM $\rightarrow$},
             ytick=\empty, xlabel={Option moneyness}, ylabel={Implied volatility}]
  \addplot[blue, thick, domain=-1.3:1.3, samples=100] {0.15+0.55*x^2};
\end{axis}
\end{tikzpicture}
\caption{This is an illustration of the so called volatility smile. The implied volatility tends to be lower for at-the-money (ATM) options, and then increase the further into the money (ITM) or out of the money (OTM) the option gets. \label{fig:plot2_5}}
\end{figure}
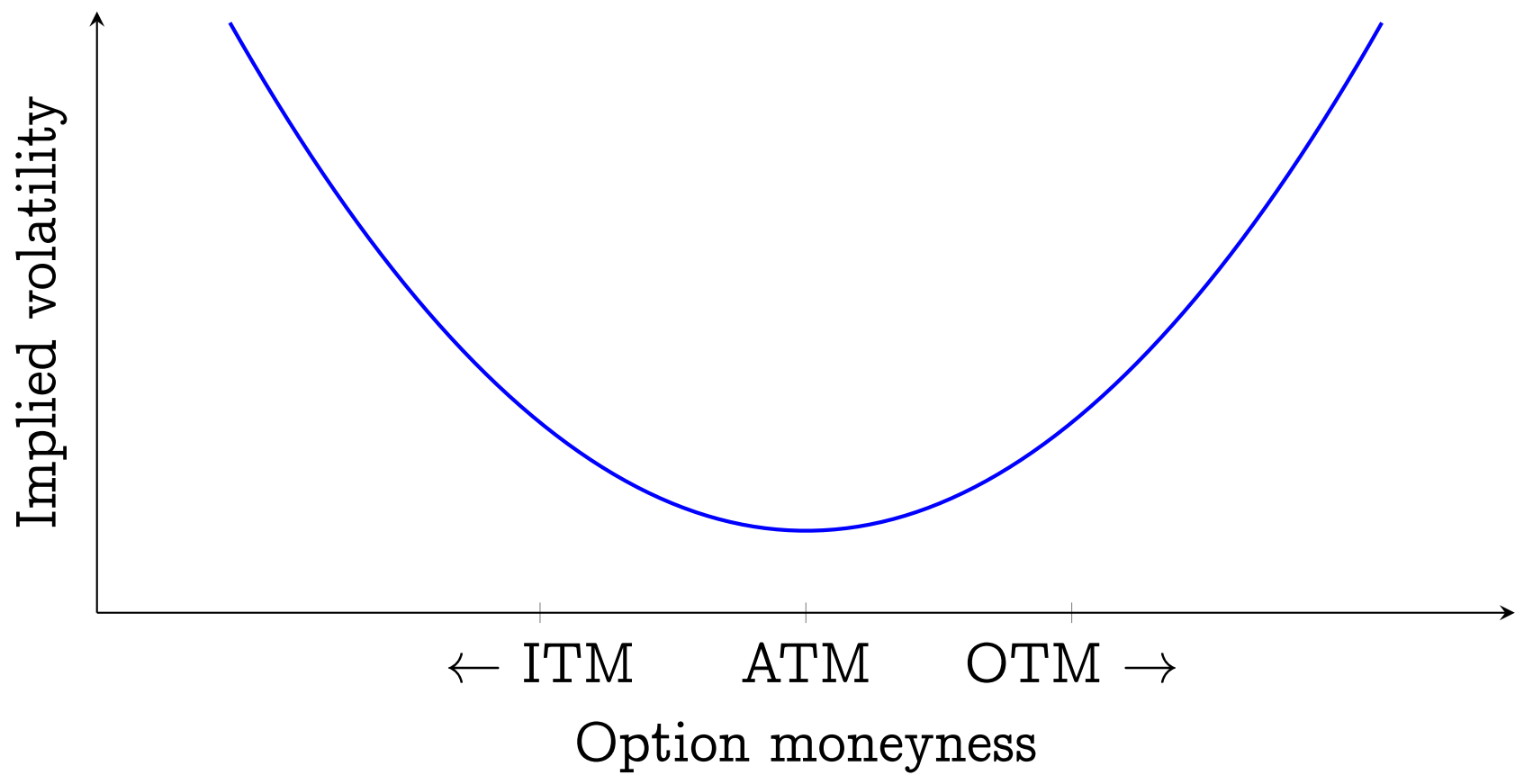

\noindent
Note from \eqref{eq:BS-Call} and \eqref{eq:BS-Put}, that the Black-Scholes price of a Call or Put depends on five \textit{observable} variables $(t,S_t,T,K,r)$ and one \textit{unobservable} variable $\sig$.  This permits us to make the following definition.

\begin{definition}[Implied volatility]
Let $C_t(T,K)$ be the price of a European Call option, written on an underlyer $S$ with strike $K$ and maturity $T$, as quoted on the market.  We define the \textit{Implied volatility} $\sigma(T,K)$ of this option as the unique positive solution of
\begin{align}
		C^\BS(t,S_t;\sig_t(T,K),T, K) = C_t(T,K) , \label{eq:iv.def}
\end{align}
where $C^\BS$ is given by \eqref{eq:BS-Call}.  
\end{definition}
\noindent
The implied volatility of a Put with price $P_t(T,K)$ is defined analogously, i.e., as the unique positive solution $\sig(T,K)$ of
\begin{align}
		P^\BS(t,S_t;\sig_t(T,K),T, K) = P_t(T,K) .
\end{align}
It turns out that a Put and Call with the same strike and maturity must have the same implied volatility.  To see why this is the case, suppose that we have $C_t(T,K) = C^\BS(t,S_t;\sig_c,T,K)$ and $P_t(T,K) = P^\BS(t,S_t;\sig_p,T,K)$.  Then using Put-Call parity we have
\begin{align}
S_t - B_t^T K
	&=	C_t(T,K) - P_t(T,K) \\
	&=	C^\BS(t,S_t;\sig_c,T,K) - P^\BS(t,S_t;\sig_p,T,K) \\
	&=	C^\BS(t,S_t;\sig_c,T,K) - P^\BS(t,S_t;\sig_c,T,K) + P^\BS(t,S_t;\sig_c,T,K) - P^\BS(t,S_t;\sig_p,T,K) \\
	&=	S_t - B_t^T K + P^\BS(t,S_t;\sig_c,T,K) - P^\BS(t,S_t;\sig_p,T,K) .
\end{align}
Thus, we have  
\begin{align}
P^\BS(t,S_t;\sig_c,T,K) 
	&= P^\BS(t,S_t;\sig_p,T,K) ,
\end{align}
which is true if and only if $\sig_c = \sig_p$.
\\[0.5em]
We can think of the implied volatility $\sig_t(T,K)$ as the volatility that one must insert into the Black--Scholes Call formula $C^\BS(t,S_t;\sig_t(T,K),T, K)$ in order for it to equal the market price $C_t(T,K)$ of a Call.  
There is no analytical solution $\sig_t(T,K)$ of \eqref{eq:iv.def}.  Thus, one must find $\sig_t(T,K)$ numerically.  
\\[0.5em]
It is important to recognize that the Black-Scholes Call and Put prices are derived assuming a very particular \textit{model} \eqref{eq:BS-model} for $S$, namely, that $S$ is a geometric Brownian motion. \textit{If} this model captured the true dynamics of $S$, the implied volatility $\sig_t(T,K)$ for all $(T,K)$ would equal the volatility $\sig$ of $S$ in \eqref{eq:BS-model}.  However, when we use market options data to find the implied volatilities $\sig_t(T,K)$ for different maturities $T$ and strikes $K$, we observe that implied volatility $\sig_t(T,K)$ is \textit{not} constant.  For a fixed $T$, what we typically observe is that the implied volatility $\sig_t(T,K)$ is a convex function of $K$,  as illustrated in Figure \ref{fig:plot2_5}.
\\[0.5em]
Thus, we have strong evidence that the Black-Scholes model does \textit{not} correctly capture the dynamics of $S$.  Nevertheless, the Black-Scholes model is a good first approximation for the dynamics of $S$.  And, having a solid understanding of how to price and replicate options in a Black-Scholes setting helps us to understand how to price and replicate options in more complicated settings.  Moreover, option prices are typically quoted in units of Black-Scholes implied volatility rather than in units of dollars and cents.
\\[0.5em]
The implied volatility surface $(T,K) \rightarrow \sig_t(T,K)$ is not completely arbitrary. 
For example, using the fact that Call prices are strictly decreasing in $K$ and Put prices are strictly increasing in $K$, we can find bounds on the slope of the implied volatility.  We have
\begin{align}
0 \geq \frac{\d C_t(T,K)}{\d K} 
	&= \underbrace{\frac{\d C^\BS}{\d K}}_{<0} + \underbrace{ \frac{\d C^\BS}{\d \sigma} }_{>0} \frac{\d \sigma}{\d K}   &
	&\Rightarrow &
\frac{\d \sigma}{\d K} 
	&\leq \frac{-\d C^\BS / \d K}{\d C^\BS / \d \sigma} , \label{eq:4}
\end{align}
where we have used the short-hand $C^\BS = C^\BS(t,S_t;\sig,T,K)$.
We can derive a lower bound for $\d \sig(T,K)/\d K$ using Puts.  We have
\begin{align} 
0 \leq \frac{\d P_t(T,K)}{\d K} 
	&= \underbrace{\frac{\d P^\BS}{\d K}}_{>0} + \underbrace{\frac{\d P^\BS}{\d \sigma}}_{>0} \frac{\d \sigma}{\d K} , &
	&\Rightarrow &
\frac{\d \sigma}{\d K} 
	&\geq \frac{-\d P^\BS / \d K}{\d P^\BS / \d \sigma} , \label{eq:5}
\end{align}
where we have used the short-hand $P^\BS = P^\BS(t,S_t;\sig,T,K)$.
From \eqref{eq:4}, \eqref{eq:5} and the Black-Scholes formulas for Calls \eqref{eq:BS-Call} and Puts \eqref{eq:BS-Put}, we obtain
\begin{align} 
\frac{-\sqrt{2 \pi}}{S_t \sqrt{T-t}} (1 - \Phi(d_-)) \ee^{\frac{1}{2}d_+^2} 
	&\leq \frac{\d \sigma}{\d K} 
	 \leq \frac{\sqrt{2 \pi}}{S_t \sqrt{T-t}} \Phi(d_-) \ee^{\frac{1}{2} d_+^2}.
\end{align}
where we have taken $r=0$ for simplicity.  Keep in mind that $d_{\pm}$ depends on $(t,S_t,\sig,T,K)$, as indicated in \eqref{eq:d-Phi}.  Thus, the bounds on the slope of implied volatility depend on the level of implied volatility.


\section{Exercises}
\label{sec:exer-5}

\begin{exercise}
\label{exer:5-1}
Suppose that the stock price $S$ and the bond price $B$ under the physical measure $\Pb$ are given by
\begin{align}
\dd S_t
	&=	\mu S_t \dd t + \sig S_t \dd W_t , &
\dd B_t
	&=	r B_t \dd t .
\end{align}
where $W$ is a Brownian motion under $\Pb$.\\
\textbf{(a)} Consider a claim that pays $\phi(S_T) = S_T^p$ at time $T$.  Derive an explicit expression for $V_t = v(t,S_t)$, the value of the claim at time $t < T$.\\
\textbf{(b)} Let $X$ be a portfolio that replicates the option payoff
\begin{align}
\dd X_t
	&=	\Del_t \dd S_t + (X_t - \Del_t S_t) \frac{1}{B_t} \dd B_t .
\end{align}
Derive an explicit expression for $\Del_t$ as a function of $(t,S_t)$.  Also, derive an expression for $X_0$ as a function of $S_0$.\\
\textbf{(c)} Write a Monte Carlo algorithm to plot sample paths of $B$ and $S$.  
Your inputs should be $(S_0, B_0, \mu, \sig, r, T, N)$ where $N$ is the number of time steps in your Monte Carlo simulation (i.e., your time step should be $\del t = T/N$).  The output of your program should be a plot of paths of $S$ and $B$.  \textit{Hint}: you may wish to simulate $Z := \log S$ and $Y = \log B$ rather than $S$ and $B$ and then set $S_t = \ee^{Z_t}$ and $B_t = \ee^{Y_t}$.\\
\textbf{(d)} Modify your Monte Carlo algorithm to plot sample paths of $V$ and $X$.  Your inputs should be $(S_0, B_0, \mu, \sig, r, T, p, N)$.
The output of your program should be a plot of paths of $V$ and $X$.  \textit{Hint}: if the paths of $V$ and $X$ are not almost identical, then there is an error in your program.\\
\textbf{(e)} Explain why the paths of $V$ and $X$ do not overlap exactly.\\
\textbf{(f)} Using your program from part (d), run $M$ sample paths and plot a histogram of $X_T - V_T$.\\
\textbf{Instructions:} Parts (c), (d), and (f) ask you to perform simulations and plot the results. For these parts, use the following parameters
\begin{align}
S_0
	&=	1 , &
B_0
	&=	1 , &
\mu
	&=	0.1 , &
\sig
	&=	0.2 , &
r
	&=	0.03 , &
T
	&=	1 , &
p
	&=	-2 , &
N
	&=	100 , &
M
	&=	1000 .
\end{align}
Your programs should read the input from a text file called \textbf{input.txt}, consisting of a single line listing $S_0, B_0, \mu, \sig, r, T, p, N, M$ in that order, separated by commas.  Some parts do not use all of the inputs (e.g., only part (f) requires $M$).
\end{exercise}


\begin{exercise}
\label{exer:5-2}
Consider the Black-Scholes model for a stock $S$ and a bond $B$
\begin{align}
\dd S_t
	&=	\mu S_t \dd t + \sig S_t \dd W_t , &
\dd B_t
	&=	r B_t \dd t .
\end{align}
Suppose that, if you own one share of $S$ then you receive a cash flow of $q S_t \dd t$ during the time interval $\dd t$.\\
\textbf{(a)} Write the dynamics of a portfolio $X$ of an investor who holds $\Del_t$ shares of $S$ at time $t$.\\
\textbf{(b)} Consider a claim that pays $\phi(S_T)$ at time $T$.  Let $V_t = v(t,S_t)$ be the price of this claim at time $t$.  Using a replication argument, derive a PDE for the function $v(t,s)$.  Also state the terminal condition for $v$ at time $T$.\\
\textbf{(c)} When dividends are involved, the no-arbitrage condition is that $(X/B)$ is a martingale for \textit{all} portfolios $X$ under the risk-neutral measure $\Pbt^{(\gam)}$.  If we have
\begin{align}
\dd S_t
	&=	\widetilde{\mu} S_t \dd t + \sig S_t \dd \Wt^{(\gam)}_t,
\end{align}
under $\Pbt^{(\gam)}$, then what should $\widetilde{\mu}$ be so that $(X/B)$ is a martingale?  Again, do not forget the $S$ pays dividends.  
\end{exercise}


\begin{exercise}
\label{exer:5-3}
\textit{Challenge Question}:
Consider a global market consisting of a domestic bond $B$ (in Dollars) and foreign bond $C$ (in Euros) and an exchange rate $Q$ (in Dollars per Euro), whose dynamics under the physical measure $\Pb$ are given by
\begin{align}
\dd B_t
	&=	r B_t \dd t, \\
\dd C_t
	&=	R C_t \dd t , \\
\dd Q_t
	&=	\mu Q_t \dd t + \sig Q_t \dd W_t .
\end{align}
Derive an expression for the value (in Dollars) of an option that pays $\phi(Q_T C_T)$ in Dollars at time $T$.  Derive a PDE for the option value.
\end{exercise}

\begin{exercise}
\label{exer:5-4}
Derive the Black-Scholes price of a Call option assuming that the underlying stock $S$ pays a dividend of $q S_t \dd t$ during the interval $[t, t+ \dd t)$.  \textit{Hint}: If a trader holds a single share of the stock $S$ at all times, then the change in the value of his portfolio is $\dd X_t = \dd S_t + q S_t \dd t$.
\end{exercise}

\begin{exercise}
\label{exer:5-5}
Many cryptocurrency exchanges offer derivatives known as \textit{perpetual options} or just \textit{perps}, for short.  Consider a perp written on an underlyer $S$.  The owner of a perp can exercise his option at any time $t \geq 0$ and receive $\phi(S_t)$, where the function $\phi$ is specified in the contract.  In exchange for this right, the owner of a perp must pay an initial amount $V_0$ to the exchange as well as a cash-flow $C_t$ per unit time called the \textit{funding rate}.  The exchange will use the initial payment $V_0$ as well as the cash-flow $C_t$ and dynamic trading in order to create a portfolio $X$ that replicates the perp payoff.  The goal of this exercise is to determine the dynamics of the replicating portfolio $X$ as well as the funding rate $C_t$.  Throughout this exercise, we will assume that the dynamics of the money-market account $M$ and the underlying asset $S$ (typically Bitcoin or Ethereum) are as follows
\begin{align}
\dd M_t
	&=	r_t M_t \dd t , &
\dd S_t
	&=	\mu_t S_t \dd t + \sig_t S_t \dd W_t .
\end{align}
(a) Let $\Del_t$ denote the number of shares of $S$ held in the replicating portfolio $X$ at time $t$.  Write the dynamics of the replicating portfolio $X$ (i.e., $\dd X_t = \ldots$).  Note that your answer should include the funding rate received by the exchange from the owner of the perp.\\
(b) Let $V = (V_t)_{t \geq 0}$ be the value of the perp.  Write the dynamics of $V$ (i.e., $\dd V_t = \ldots$).\\
(c) We would like the replicating portfolio $X$ to satisfy $X_t = V_t$ for all $t$.  Using your answers from parts (a) and (b) find $\Del_t$, the funding rate $C_t$ as well as the initial value of the portfolio $X_0$.\\
CHALLENGE: Suppose the perp pays $\phi(S_t,Y_t)$ where the dynamics of $Y$ are
\begin{align}
\dd Y_t
	&=	b_t Y_t \dd t + a_t Y_t ( \rho \dd W_t + \rhob \dd B_t ) ,
\end{align}
where $W$ and $B$ are independent Brownian motions and $\rhob := \sqrt{1-\rho^2}$.  What is the funding rate $C_t$ for this perp?  And, if we denote by $\Del_t$ and $\Gam_t$ the shares owned of $S$ and $Y$ respectively at time $t$, then what should these quantities be in order for the portfolio $X$ to replicate the value of the perp?
\end{exercise}

\begin{exercise}
\label{exer:5-6}
Consider the Black-Scholes model
\begin{align}
\dd S_t
	&=	r S_t \dd t + \sig S_t \dd \Wt_t , &
\dd B_t
	&=	r B_t \dd t ,
\end{align}
where $\Wt$ is a Brownian motion under the risk-neutral probability measure $\Pbt$.  Let $V$ be the value of a financial derivative that pays $S_T^p$ at time $T$.
\\
(a) We know that $V_t/B_t = \Ebt[ V_T / B_T | \Fc_t]$ where $V_T = \phi(S_T)$.  We also know, from the Markov property of $S$ that $V_t = v(t,S_t)$ for some function $v$.  Find $v(t,S_t)$.
\\
(b)  Under $\Pbt$ we have that $X/B$ is a martingale for all portfolios $X$.  Consider now, a measure $\Pbh$ under which $X/V$ is a martingale for all portfolios $X$.  What are the dynamics of $S$ under $\Pbh$?  That is, find $\dd S_t = (\ldots ) \dd t + (\ldots) \dd \Wh_t$ where $\Wh$ is a Brownian motion under $\Pbh$.
\\
(c)  Let $U$ be the value of a derivative that pays $S_T^q$ at time $T$.  Compute $U_t=u(t,S_t)$ using $U_t / V_t = \Ebh[ U_T / V_T | \Fc_t ]$.  
\\
(d)  Using your answers from parts (a) and (c), verify that, if you set $p = q$, then we have $U_t = V_t$.
\end{exercise}


\begin{exercise}
\label{exer:5-7}
Consider the Black-Scholes model for a stock $S$ and a bond $B$, written under the real-world measure $\Pb$:
\begin{align}
\dd S_t &= \mu S_t \dd t + \sig S_t \dd W_t , \qquad \dd B_t = r B_t \dd t ,
\end{align}
where $W$ is a Brownian motion under $\Pb$.  Let $u(t,S_t)$ be the value at time $t$ of a European contract with payoff $\phi(S_T) = (\log S_T)^2$.\\[0.5em]
\textbf{(a)} What should the dynamics of $\log S$ be under the risk-neutral pricing measure $\Pbt$ (using the bond $B$ as numeraire)?  That is, find $\dd (\log S_t) = (\ldots)\dd t + (\ldots) \dd \Wt_t$.  Explain how you obtain your answer.\\[0.5em]
\textbf{(b)} Derive an explicit expression for $u(t,S_t)$.\\[0.5em]
\textbf{(c)} What should $X_0$ and $\Del_t$ be in order for $X_T = \phi(S_T) = (\log S_T)^2$?
\end{exercise}


\begin{exercise}
\label{exer:5-8}
Suppose the dynamics of a stock $S$ and a bond $B$ under the physical probability measure $\Pb$ are given by
\begin{align}
\dd S_t &= \mu(S_t) \dd t + \sig(S_t) \dd W_t , \qquad \dd B_t = r B_t \dd t .
\end{align}
You wish to create a portfolio $X$ that replicates the value of a call option with maturity $T_1$ and strike $K_1$.  However, you are not allowed to trade the asset $S$.  Instead, you are allowed to trade only the bond $B$ and a put option with maturity $T_2$ and strike $K_2$, where $T_2 > T_1$.  Denote by $C(t,S_t)$ and $P(t,S_t)$ the time $t$ price of the call and put option, respectively.\\[0.5em]
\textbf{(a)} Compute the dynamics of $C(t,S_t)$ under $\Pb$.  That is, give an expression for $\dd C(t,S_t) = (\ldots)\dd t + (\ldots)\dd W_t$.  You may leave your answer in terms of partial derivatives of $C(t,S_t)$.\\[0.5em]
\textbf{(b)} Suppose you have a portfolio $X$ with initial value $X_0 = C(0,S_0)$.  At time $t$ you hold $\Del_t$ put options and invest the rest of your money in a bond.  Write the dynamics of $X_t$ under $\Pb$.  You may leave your answer in terms of partial derivatives of $P(t,S_t)$.\\[0.5em]
\textbf{(c)} What should $\Del_t$ be if you want to ensure $X_{T_1} = (S_{T_1} - K_1)^+$ at time $T_1$?  You may leave your answer in terms of partial derivatives of $C(t,S_t)$ and $P(t,S_t)$.
\end{exercise}


\begin{exercise}
\label{exer:5-9}
Consider a foreign exchange market with two assets: an American money-market account $M$ (in units of US dollars) and a British money-market account $B$ (in units of pounds).  Denote by $Q$ the exchange rate between dollars and pounds (in units of \textdollar$/$\textsterling).  The dynamics under the real-world probability measure $\Pb$ are
\begin{align}
\dd M_t &= r M_t \dd t , &
\dd B_t &= R B_t \dd t , &
\dd Q_t &= \mu Q_t \dd t + \sig Q_t \dd W_t ,
\end{align}
where $W$ is a Brownian motion under $\Pb$.\\[0.5em]
\textbf{(a)} The value in US dollars of the British money-market account is $Q_t B_t$.  Let $X_t$ denote the value in dollars of a portfolio that holds $\Del_t$ shares of $Q_t B_t$.  What are the dynamics of $X$?\\[0.5em]
\textbf{(b)} If we choose $M$ as our numeraire, then under a risk-neutral measure $\Pbt$ the process $X/M$ must be a martingale.  What are the dynamics of $Q$ under $\Pbt$?\\[0.5em]
\textbf{(c)} Consider an option that pays $\log(Q_T B_T)$ at time $T$.  What is the value $V_t$ of this option at time $t$?
\end{exercise}




\chapter{Greeks}
The \textit{Greeks} are (in addition to being an ancient civilization and the inventors of democracy), the sensitivities of portfolios or options with respect to objects that affect the portfolios' or options' values such as, e.g., the value of the underlying $S$, the time $t$, or the volatility $\sig$.  If this seems a bit abstract at the moment, things will become more clear as we go through some examples.
\\[0.5em]
Throughout this chapter, we will consider a stock (or some other underlyer) $S = (S_t)_{t \geq 0}$ and a bank account $B = (B_t)_{t \geq 0}$ whose dynamics are as in the Black-Scholes setting.  Under the physical probability measure $\Pb$ we have
\begin{align}
\dd S_t
	&=	\mu S_t \dd t + \sig S_t \dd W_t , &
\dd B_t
	&=	r B_t \dd t .
\end{align}
The value of a bank account $B$ and zero-coupon bond $B^T = (B_t^T)_{t \geq 0}$ are explicit functions of $t$
\begin{align}
B_t \equiv B(t)
	&=	B_0 \ee^{r t} , &
B_t^T \equiv B(t;T)
	&=	\ee^{-r (T-t)} .
\end{align}
The value $V = (V_t)_{t \geq 0}$ of any derivative written on $S$ in this setting is a function of time, the stock's price and the volatility $\sig$: $V_t = v(t,S_t;\sig)$.  Calls and Puts are additionally parametrized by their strikes and maturity dates.  Thus, we have
\begin{align}
C_t(T,K)
	&=	C^\BS(t,S_t;\sig,T,K) , &
P_t(T,K)
	&=	P^\BS(t,S_t;\sig,T,K) .
\end{align}
where $C^\BS$ and $P^\BS$ are the Black-Scholes prices of calls and puts, respectively.  
We may sometimes limit the dependence on $\sig$, $T$ and $K$ when it causes no confusion.
Although we will limit ourselves in this chapter to the Black-Scholes setting, the definitions we give below for the Greeks apply to all market models.

\section{Delta}
Consider a portfolio $X = (X_t)_{t \geq 0}$ of traded assets.  For example, a portfolio $X$ could consist of $\alpha$ shares of a stock $S$, $\beta$ Call options with maturity $T$ and strike $K$, and $\gamma$ put options with the same strike and maturity.  Because the call and put values can be expressed as functions of $t$, $S_t$ and $\sig$, the portfolio value can be expressed as a function of these variables as well: $X_t = X(t,S_t;\sig)$.  Thus, we have
\begin{align}
X(t,S_t;\sig)
	&=	\alpha S_t + \beta C^\BS(t,S_t;\sig,T,K) + \gamma P^\BS(t,S_t;\sig,T,K) . \label{eq:X-example}
\end{align}
We define the \textit{Delta} $\Delta = (\Del_t)_{t \geq 0}$ of a portfolio to be the derivative of $X$ with respect to the stock price $S$.  
Just as $X_t$ can be expressed as a function of $t$, $S_t$ and $\sig$, so can the delta: $\Del_t = \Del(t,S_t;\sig)$.  That is
\begin{align}
\Delta(t,S_t;\sig)
	&:=	\frac{\d X}{\d s}(t,S_t;\sig).
\end{align}
For example, if $X$ is given by \eqref{eq:X-example}, then we have
\begin{align}
\Delta(t,S_t;\sig)
	&=	\frac{\d}{\d s} X(t,S_t;\sig) \\
	&=	\frac{\d}{\d s} \( \alpha S_t + \beta C^\BS(t,S_t;\sig,T,K) + \gamma P^\BS(t,S_t;\sig,T,K) \) \\
	&=	\alpha + \beta \frac{\d}{\d s}C^\BS(t,S_t;\sig,T,K) + \gamma \frac{\d}{\d s} P^\BS(t,S_t;\sig,T,K) .
\end{align}
As previously mentioned, for a portfolio consisting of a single Call option, we have
\begin{align}
\Del^\text{BS-Call}(t,S_t;\sig) = \frac{\d C^\text{BS}}{\d s}(t,S_t;\sig,T,K) = \Phi(d_+(t,S_t;\sig,T,K)) .
\end{align}
Why is it important to know the $\Del$ of a call option?  Recall that, in the Black-Scholes setting, $\Del^\BS$ is the number of shares of $S$ one should hold in a self-financing portfolio $X$ in order to replicate the call payoff.  In fact, this is true for any option in the Black-Scholes setting.

\section{Gamma}
We define the \textit{Gamma} $\Gamma = (\Gam_t)_{t \geq 0}$ of a portfolio to be the second derivative of $X$ with respect to the stock price $S$.  Alternatively, we can define it as the first derivative of the $\Delta$ with respect to the stock price $S$.  As with $X$ and $\Del$, we can express $\Gamma$ as a function of $t$, $S_t$ and $\sig$.  We have
\begin{align}
\Gamma(t,S_t;\sig)
	&:=	\frac{\d^2 X}{\d s^2}(t,S_t;\sig)
	=		\frac{\d \Del}{\d s}(t,S_t;\sig) .
\end{align}
The Gamma of a single Call option 
is given by
\begin{align}
\Gamma^\text{BS-Call} (t,S_t;\sig)
	&= \frac{\d^2 C^\text{BS-Call}}{\d s^2} (t,S_t;\sig,T,K) = \frac{\phi(d_+(t,S_t;\sig,T,K))}{S_t \sigma \sqrt{T-t}},  &
\phi(d) 
	&= \frac{\ee^{-d^2/2}}{\sqrt{2 \pi}} ,
\end{align}
Why is it important to know the $\Gamma$ of a call option?
In order to replicate a call option payoff, one has to adjust the number of shares $\Delta$ one has in a portfolio as $S$ moves randomly.  The $\Gamma$ tells us how much $\Delta$ changes for a given change in $S$.  If the $\Gamma$ is large, then a small move in $S$ will result in a large change in $\Del$.  When would this happen?
Well, note from \eqref{eq:d-Phi} that, if the option is at-the-money, i.e., $S_t = K$, then
\begin{align}
d_+(t,K;\sig,T,K)
	&= \( \frac{r}{\sig} + \frac{\sig}{2} \) \sqrt{T-t} \to 0 , &
	&\text{and thus}&
\phi(d_+(t,K;\sig,T,K))
	&\to 1/\sqrt{2 \pi} , &
	&\text{as $t \to T$},
\end{align}
which implies that
\begin{align}
\Gamma^\text{BS-Call}(t,K;\sig,T,K)
	&= \frac{\phi(d_+(t,K;\sig,T,K))}{K \sigma \sqrt{T-t}}
	\to \infty , &
	&\text{as $t \to T$}.
\end{align}
This presents a \textit{huge} problem when trying to replicate a Call option. If $S_t \approx K$, then as $t \to T$ a small change in $S$ results in a large change in  $\Delta^\text{BS-Call}(t,S_t;\sig,T,K)$.  On the other hand, if $S_t$ stays away from $K$, then $|d_+| \to \infty$ as $t \to T$, so $\Gamma^\text{BS-Call}(t,S_t;\sig,T,K) \to 0$, and thus, replication becomes nearly static.

\section{Theta}
Another important Greek is the \textit{Theta} $\Theta = (\Theta_t)_{t \geq 0}$, which we define as the derivative of $X$ with respect to time $t$.  As with all other Greeks, it can be expressed as a function of $t$, $S_t$ and $\sig$.  We have
\begin{align}
\Theta(t,S_t;\sig)
	:&= \frac{\d X}{\d t}(t,S_t;\sig) .
\end{align}
In the Black-Scholes setting, the $\Theta$ of a single call option is given by
\begin{align}
\Theta^\text{BS-Call}(t,S_t;\sig,T,K)
	&= \frac{\d C^\BS}{\d t}(t,S_t;\sig,T,K) \\
	&= - S_t \phi(d_+(t,S_t;\sig,T,K)) \frac{\sigma}{2 \sqrt{T-t}} - r K \ee^{-r (T-t)} \Phi(d_-(t,S_t;\sig,T,K)) .
\end{align}
Note that the $\Theta$ of a Call on a non-dividend-paying stock is negative.  This is \textit{not} true for all options: for example, a deep in-the-money European Put with $r > 0$ has a positive $\Theta$.
\\[0.5em]
Here, we should mention that some people define the $\Theta$ of a single option to be derivative of the option value with respect to the time to maturity $\tau := T-t$.  This definition can be problematic because it cannot be extended to a portfolio of options with differing maturities, which is why we define the $\Theta$ in this text to be the derivative of the portfolio value with respect to time $t$.

\section{It\^o's Lemma in terms of the Greeks}
Note that we can write It\^o's Lemma in terms of the Greeks
\begin{align}
\dd X(t,S_t;\sig)
	&=	\frac{\d X}{\d t} (t,S_t;\sig)\dd t + \frac{\d X}{\d s}(t,S_t;\sig) \dd S_t 
			+ \frac{1}{2} \frac{\d^2 X}{\d s^2}(t,S_t;\sig) ( \dd S_t )^2 \\
	&=	\Theta(t,S_t;\sig) \dd t + \Del(t,S_t;\sig) \dd S_t + \frac{1}{2} \Gam(t,S_t;\sig) ( \dd S_t )^2 \\
	&=	\Theta(t,S_t;\sig) \dd t + \Del(t,S_t;\sig) \Big( \mu S_t \dd t 
			+ \sig S_t \dd W_t \Big) + \frac{1}{2} \Gam(t,S_t;\sig) \sig^2 S_t^2 \dd t \\
	&=	\Big( \Theta(t,S_t;\sig) + \mu S_t \Del(t,S_t;\sig) + \frac{1}{2} \sig^2 S_t^2 \Gam(t,S_t;\sig) \Big) \dd t 
			+ \sig S_t \Del(t,S_t;\sig)  \dd W_t . 
\end{align}

\section{Greek-neutral Hedging}
A portfolio $X$ is said to be \textit{Greek-neutral} if a given Greek of a portfolio is zero.  For example
\begin{align}
\text{$\Delta$-neutral}:&&
\Delta(t,S_t;\sig)
	&=	\frac{\d X}{\d s}(t,S_t;\sig) = 0 , \\
\text{$\Gam$-neutral}:&&
\Gamma(t,S_t;\sig)
	&=	\frac{\d^2 X}{\d s^2}(t,S_t;\sig) = 0 .
\end{align}
Let us see why it can be advantageous to have a portfolio that is neutral to certain Greeks.
\\[0.5em]
Suppose a bank has sold a Call option $C$ and owns $\alpha$ shares of $S$.  Then, at time $t$, the value of the bank's portfolio is
\begin{align}
X(t,S_t;\sig)
	&=	- C^\BS(t,S_t;\sig) + \alpha S_t ,
\end{align}
where, to ease notation, we have omitted the dependence of the Call price on the maturity $T$ and strike $K$.
What would happen to the value of this portfolio if the stock price were to jump from $S_t$ to $S_t + \del$ instantaneously?  
In the Black-Scholes model, stock prices cannot jump.  However, in reality, the prices of stocks can and do jump.  So, it is worth considering what happens in this case.
The change in the value of the portfolio would be
\begin{align}
X(t,S_t+\del;\sig) - X(t,S_t;\sig)
	&=	- \( C^\BS(t,S_t + \del;\sig) - C^\BS(t,S_t;\sig) \) + \alpha \( (S_t + \del) - S_t  \)\\
	&=	- \( \frac{\d C^\BS}{\d s}(t,S_t;\sig) \del + \frac{1}{2} \frac{\d^2 C^\BS}{\d s^2} (t,S_t;\sig) \del ^2 + \Oc(\del^3) \) + \alpha \del ,
\end{align}
where, in the second equality, we have expanded $C^\BS(t,S_t + \del;\sig)$ as a Taylor series about $S_t$.
Suppose now, that the bank had chosen $\alpha$ so that the portfolio $X$ was $\Delta$-neutral
\begin{align}
0
	&=	\Delta(t,S_t;\sig)
	=		\frac{\d X}{\d s}(t,S_t;\sig)
	=		- \frac{\d C^\BS}{\d s}(t,S_t;\sig) + \alpha , &
&\Rightarrow&
\alpha
	&=	\frac{\d C^\BS}{\d s}(t,S_t;\sig) .
\end{align}
Note that $\alpha$ is equal to the number of shares of $S$ that the bank would hold in a self-financing portfolio if it were trying to replicate a Call payoff.
In the above scenario, the change in the value of the portfolio $X$ would have been
\begin{align}
X(t,S_t+\del;\sig) - X(t,S_t;\sig)
	&=	- \frac{1}{2} \frac{\d^2 C^\BS}{\d s^2}(t,S_t;\sig) \del^2 + \Oc(\del^3) ,
\end{align}
because the order $\Oc(\del)$ terms have canceled each other out.
Thus, a small change in the stock price $\del$ only affects the portfolio value at order $\Oc(\del^2)$.  Remember for $\del<1$ we have $\del^2 < \del$.  Unfortunately, for the bank, call prices are convex as a function of $S$ (this is true in general for calls and puts, not just for Calls in the Black-Scholes framework).  Thus, for a sufficiently small $\del$ we have
\begin{align}
\frac{\d^2 C^\BS}{\d s^2}(t,S_t;\sig)
	&> 0 , &
	&\Rightarrow&
X(t,S_t+\del;\sig) - X(t,S_t;\sig)
	&= - \frac{1}{2} \frac{\d^2 C^\BS}{\d s^2}(t,S_t;\sig) \del^2 + \Oc(\del^3) < 0 .
\end{align}
So, if the stock jumps from $S_t$ to $S_t + \del $ the bank loses money.  If the bank had constructed a portfolio that was both $\Delta$-neutral \textit{and} $\Gamma$-neutral, the gains or losses of the portfolio would have been pushed to $\Oc(\del^3)$.  Note that, to construct a portfolio that is both $\Del$-neutral and $\Gam$-neutral, the bank would need to add a third asset to the portfolio.

\begin{example}
Suppose a bank sells a call option with maturity date $T$ and strike $K$.  In order to replicate the payoff of the option by trading the stock $S$ and bank account $B$, the bank must maintain a portfolio whose total value is zero and whose delta is zero.  We have
\begin{align}
0
	&=	X(t,S_t,\sig) 
	=	- C^\text{BS}(t,S_t;\sig,T,K) + \alpha S_t + \beta B_t , \\
0
	&=	\frac{\d X}{\d s}(t,S_t,\sig) 
	=	- \frac{\d C^\text{BS}}{\d s}(t,S_t;\sig,T,K) + \alpha . 
\end{align}
Solving for $\alpha$ and $\beta$, we find
\begin{align}
\alpha
	&=	\frac{\d C^\BS}{\d s}(t,S_t;\sig,T,K) = \Del^\text{BS-Call}(t,S_t;\sig,T,K) , \\
\beta
	&=	( C^\text{BS}(t,S_t;\sig,T,K) - \alpha S_t ) / B_t .
\end{align}
While this hedging strategy theoretically will replicate the call option, the problem is that the Gamma of this portfolio
\begin{align}
\frac{\d^2 X}{\d s^2}(t,S_t;\sig)
	&=	- \frac{\d^2 C^\BS}{\d s^2}(t,S_t;\sig,T,K) = - \Gam^\text{BS-Call}(t,S_t;\sig,T,K)
\end{align}
 can be very high, especially when $S_t \approx K$, and this makes $\Delta$-hedging difficult because the number of shares of $S$ that the bank must hold will fluctuate greatly.  As an alternative to simply trading the stock $S$ and the bank account $B$, the bank could, for example, use some of the funds from the sale of a call option to buy $\gam$ call options with a maturity date $\Tb > T$ and the same strike $K$.  The bank must still maintain a portfolio whose value and whose $\Delta$ is zero
\begin{align}
0
	&=	X(t,S_t,\sig) 
	=	- C^\text{BS}(t,S_t;\sig,T,K) + \alpha S_t + \beta B_t + \gam C^\text{BS}(t,S_t;\sig,\Tb,K), \\
0
	&=	\frac{\d X}{\d s}(t,S_t,\sig) 
	=	- \frac{\d C^\text{BS}}{\d s}(t,S_t;\sig,T,K) + \alpha + \gam \frac{\d C^\text{BS}}{\d s}(t,S_t;\sig,\Tb,K). 
\end{align}
Solving for $\alpha$ and $\beta$, we find
\begin{align}
\alpha
	&=	\frac{\d C^\BS}{\d s}(t,S_t;\sig,T,K) - \gam \frac{\d C^\BS}{\d s}(t,S_t;\sig,\Tb,K) \\
	&= 	\Del^\text{BS-Call}(t,S_t;\sig,T,K) - \gam \Del^\text{BS-Call}(t,S_t;\sig,\Tb,K) , \\
\beta
	&=	( C^\text{BS}(t,S_t;\sig,T,K) - \gam C^\text{BS}(t,S_t;\sig,\Tb,K)  - \alpha S_t ) / B_t .
\end{align}
The gamma of the portfolio, is given by
\begin{align}
\frac{\d^2 X}{\d s^2}(t,S_t;\sig)
	&=	- \frac{\d^2 C^\BS}{\d s^2}(t,S_t;\sig,T,K) + \gam \frac{\d^2 C^\BS}{\d s^2}(t,S_t;\sig,\Tb,K) \\
	&=	- \Gam^\text{BS-Call}(t,S_t;\sig,T,K) + \gam \Gam^\text{BS-Call}(t,S_t;\sig,\Tb,K) .
\end{align}
Now, the portfolio can be made Gamma-neutral by setting
\begin{align}
\gam
	&= \frac{ \Gam^\text{BS-Call}(t,S_t;\sig,T,K) }{ \Gam^\text{BS-Call}(t,S_t;\sig,\Tb,K) } ,
\end{align}
which makes replication of the call with maturity $T$ much easier.
\end{example}

\section{Relationship between $\Delta$, $\Theta$ and $\Gamma$}
Let $X$ be a portfolio of derivatives written on $S$.  The value of the portfolio at time $t$ is
\begin{align}
X(t,S_t;\sig) 
		&= \sum_i \alpha_i v_i (t, S_t;\sig),
\end{align}
where $v_i(t,S_t;\sig)$ is the price of a financial derivative with payoff $\phi_i(S_{T_i})$ at time $T_i$ and $\alpha_i$ is the number of shares of the $i$th derivative in the portfolio. Each $v_i$ must satisfy the Black--Scholes PDE:
\begin{align} \label{eq:7}
0
		&= \(\d_t + r s \d_s + \frac{1}{2} \sigma^2 s^2 \d_s^2 - r\) v_i(t,s;\sig) .
\end{align}
Note that the price of a bond $B_t^T \equiv B(t;T) = \ee^{-r(T-t)}$ also satisfies this PDE, as can easily be checked by direct computation
\begin{align}
0
	&=	\(\d_t + r s \d_s + \frac{1}{2} \sigma^2 s^2 \d_s^2 - r\) \ee^{-r(T-t)} .
\end{align}
Let us define the Black-Scholes operator
\begin{align}
\Ac^\BS 
&:= \( \d_t + r s \d_s + \frac{1}{2} \sigma^2 s^2 \d_s^2 - r \) .
\end{align}
By linearity we have
\begin{align}
\Ac^\BS X(t,S_t;\sig) 
	&= \sum_i \alpha_i \Ac^\BS v_i (t, S_t;\sig) = 0 .
\end{align}
Thus, we see that $X$ satisfies the Black-Scholes PDE:
\begin{align}
0
	&=	\( \d_t + r S_t \d_s + \frac{1}{2} \sigma^2 S_t^2 \d_s^2 - r \) X(t,S_t;\sig) = 0 .
\end{align}
Writing the partial derivatives in terms of the Greeks, we have
\begin{align}
0
		&=	\Theta(t,S_t;\sig) + r S_t \Delta(t,S_t;\sig) + \frac{1}{2} \sigma^2 S_t^2 \Gam(t,S_t;\sig) - r X(t,S_t;\sig) .
\end{align}
If the portfolio is $\Delta$-neutral (which can be achieved by choosing the $\alpha_i$'s so that $\sum_i \alpha_i \d_s v_i(t,S_t) = 0$), then
\begin{align}
0
		&=	\Theta(t,S_t;\sig) + \frac{1}{2} \sigma^2 S_t^2 \Gam(t,S_t;\sig) - r X(t,S_t;\sig)  .
\end{align}
Although we have derived the above in the Black-Scholes setting, the result is roughly true in any diffusion model for $S$.

\section{Vega}
We define the \textit{Vega} $\Vc=(\Vc_t)_{t \geq 0}$ of a portfolio to be the derivative of the portfolio value with respect to the volatility:
\begin{align}
\Vc(t,S_t;\sig)
	&:=	\frac{\d X}{\d \sigma}(t,S_t;\sig) .
\end{align}
Note that ``Vega'' is not actually a Greek letter.  But it sounds Greek!  
Because the volatility is not observed directly, a trader may wish to know how the value of his portfolio changes with $\sig$.
In the Black--Scholes setting, the $\mathscr{V}$ of a single Call option satisfies 
\begin{align}
\Vc^\text{BS-Call}(t,S_t;\sig,T,K)
	&=	\frac{\d C^\text{BS}}{\d \sigma}(t,S_t;\sig,T,K)  
	 =  (T-t) \sigma S_t^2 \frac{\d^2 C^\BS}{\d s^2}(t,S_t;\sig,T,K) \\
	&= (T-t) \sigma S_t^2 \Gamma^\text{BS-Call}(t,S_t;\sig,T,K) .
\end{align}
Thus, there is a relationship between the $\Vc^\text{BS-Call}$ and the $\Gamma^\text{BS-Call}$.  In fact, the above relation holds for \textit{all} European options in the Black-Scholes setting -- not just Calls.


\section{Exercises}
\label{sec:exer-6}

\begin{exercise}
\label{exer:6-1}
This exercise is meant to give you an intuitive feel for the Greeks.
Assume for simplicity that the risk-free rate of interest is zero $r=0$.
Suppose $S$ is modeled as in the Black-Scholes model $\dd S_t = \mu S_t \dd t + \sig S_t \dd W_t$.
The value of a call option at time $t$ given $S_t = s$ is given by
\begin{align}
C^\BS(t,s;\sig,T,K)
	&=	s \Phi ( d_+ ) - K \Phi( d_- ) , &
d_\pm
	&=	\frac{1}{\sig \sqrt{T-t}} \( \log \frac{s}{K} \pm \frac{1}{2} \sig^2(T-t) \) .
\end{align}
For all of the plots in this problem, assume $\sig = 0.2$, $K = 1$, $T=1$.  For computational work, leave $\sig$, $K$, and $T$ as unknown constants.\\
\textbf{(a)} On a single graph, plot $C^\BS(t,s;T,K)$ as a function of $s$ for $t = 0$, $t = T/2$ and $t = 3 T / 4$.  Also plot the call payoff $(s-K)^+$ as a function of $s$ on the same graph.\\
\textbf{(b)} Derive an expression for the $\Del^\BS(t,s;\sig,T,K)$ by direct computation (do \textit{not} just copy formulas from the notes).  On a single graph, plot $\Del^\BS(t,s;\sig,T,K)$ as a function of $s$ for $t = 0$, $t = T/2$ and $t = 3 T / 4$.\\
\textbf{(c)} Derive an expression for the $\Gam^\BS(t,s;\sig,T,K)$ by direct computation (do \textit{not} just copy formulas from the notes).  On a single graph, plot $\Gam^\BS(t,s;\sig,T,K)$ as a function of $s$ for $t = 0$, $t = T/2$ and $t = 3 T / 4$.\\
\textbf{(d)} Derive an expression for the Vega $\Vc^\BS(t,s;\sig,T,K)$ by direct computation (do \textit{not} just copy formulas from the notes).  Show that $\Vc^\BS(t,s;\sig,T,K) = (T-t) \sig s^2 \Gam^\BS(t,s;\sig,T,K)$.
\end{exercise}


\begin{exercise}
\label{exer:6-2}
Assume for simplicity that the risk-free rate of interest is zero $r=0$.
Suppose that the price of call options are given by $C(t,S_t;T,K)$ for some function $C$.  
Suppose further that you can observe the implied volatility smile $\sig(T,K)$ at every $T$ and $K$.  In particular, assume that $\sig(T,K) = I(T-t,\log(K/S_t))$ for some function $I(\tau,x)$.  Let $C^\BS(t,S_t;\sig,T,K)$ denote the Black-Scholes price of a call option at time $t$ with maturity $T$ and strike $K$, assuming a fixed volatility $\sig$.  And let $\Del^\BS(t,S_t;\sig,T,K)$, $\Theta^\BS(t,S_t;\sig,T,K)$ and $\Vc^\BS(t,S_t;\sig,T,K)$ denote the Black-Scholes Delta, Theta and Vega respectively.\\
\textbf{(a)} Derive an expression for the $\Del$ of a call option.  Your answer should involve Black-Scholes Greeks and derivatives of $I$.\\
\textbf{(b)} Derive an expression for the $\Theta$ of a call option.  Your answer should involve Black-Scholes Greeks and derivatives of $I$.
\end{exercise}


\begin{exercise}
\label{exer:6-3}
Assume that $\dd S_t = \mu S_t \dd t + \sig S_t \dd W_t$ and $\dd B_t = r B_t \dd t$.
Suppose you have a portfolio $X$ that consists of $\alpha$ shares of a stock $S$, $\beta$ call options $C^\BS$, $\gam$ put options $P^\BS$ and $\del$ bonds $B$.  Do \textit{not} assume $r = 0$.  Note that the call and put options are written on $S$.  \\
\textbf{(a)}  Suppose the Call and the Put option have the same strike and maturity.  Is it possible to have a portfolio $X$ that is $\Gam$-neutral but is not $\Vc$-neutral?\\
\textbf{(b)}  Consider a portfolio $Y$ that has the following Greeks: $\Del^Y$, $\Gam^Y$, $\Vc^Y$ and $\Theta^Y$.  Find a system of equations for $\alpha$, $\beta$, $\gam$ and $\del$ so that the portfolio $(X+Y)$ is $\Del$-neutral, $\Gam$-neutral, $\Theta$-neutral and $\Vc$-neutral.  Note that the call and put may have different strikes and maturities.  BONUS (not required):  Find $\alpha$, $\beta$, $\gam$ and $\del$.  Your answer should involve partial derivatives of $P$ and $C$ with respect to $s$, $\sig$ and $t$.
\end{exercise}


\begin{exercise}
\label{exer:6-4}
Consider the Black-Scholes model
\begin{align}
\dd S_t &= r S_t \dd t + \sig S_t \dd \Wt_t , \qquad \dd B_t = r B_t \dd t ,
\end{align}
with risk-free rate zero (i.e., $r = 0$).  Denote by $V_t^{(p)}$ the value at time $t$ of a power option that pays $S_T^p$ at time $T$, and by $\Del_t^{(p)}$ its Delta.\\[0.5em]
\textbf{(a)} Compute $V_t^{(p)} = v^{(p)}(t,S_t)$ explicitly.\\[0.5em]
\textbf{(b)} Compute $\Del_t^{(p)}$ explicitly.\\[0.5em]
\textbf{(c)} Suppose a portfolio consists of one share of $V_t^{(p)}$, $\alpha$ shares of $V_t^{(q)}$ and $\beta$ shares of $V_t^{(k)}$, where $p, q, k \in \Rb$.  Write a system of equations you would solve in order to find $\alpha$ and $\beta$ so that the portfolio has zero value and is Delta-neutral.  Please give your answer in matrix form
\begin{align}
\begin{pmatrix} a_{11} & a_{12} \\ a_{21} & a_{22} \end{pmatrix}
\begin{pmatrix} \alpha \\ \beta \end{pmatrix}
=
\begin{pmatrix} b_{11} \\ b_{21} \end{pmatrix} .
\end{align}
\end{exercise}




\chapter{Stochastic volatility}
If we assume that stock prices are strictly positive processes and have continuous sample paths, then the most general model for $S$ is
\begin{align}
\dd S_t
	&=	\mu_t S_t \dd t + \sig_t S_t \dd W_t , \label{eq:S-general}
\end{align}
where $W=(W_t)_{t \geq 0}$ is a Brownian motion under the real-world probability measure $\Pb$ and $\mu=(\mu_t)_{t \geq 0}$ and $\sig=(\sig_t)_{t \geq 0}$ are stochastic processes adapted to a filtration $\Fc$ of $W$.  In the Black-Scholes framework, the volatility process $\sig$ was constant.  In this chapter, we will explore some of the consequences of allowing for the possibility that $\sig$ is stochastic.

\section{$\Delta$-hedging when volatility is stochastic}
Suppose that, under $\Pb$, a stock $S$ has the following dynamics given by \eqref{eq:S-general}.
Suppose, further, that the market prices a call option \textit{as if} volatility were a constant
\begin{align}
\dd S_t
	&=	\mu_t S_t \dd t + \sig S_t \dd W_t . &
	&(\text{assumption of Market participants}) \label{eq:market-assumption}
\end{align}
Note that \eqref{eq:market-assumption} implies that the price of a Call option is the Black-Scholes price $C_t(T,K) = C^\BS(t,S_t;\sig,T,K)$.
Henceforth, we will omit the $(T,K)$-dependence, as it will not play a role in the computations below.
\\[0.5em]
If a banker sells a call option $C^{BS}(t,S_t; \sigma)$ and $\Delta$-hedges, what will the profit or loss (P\&L) of his portfolio be at maturity? For simplicity, let us assume that $r=0$.  The dynamics of the banker's portfolio $X$ are
\begin{align}
\dd X_t 
	&= \Delta_t \dd S_t + (X_t - \Delta_t S_t)\frac{1}{B_t} \dd B_t \\
	&=	\frac{\d C^{\text{BS}}}{\d s} (t, S_t;\sigma) \dd S_t 
\end{align}
where, in the second equality, we have used $\Del_t = \d_s C^\BS(t,S_t;\sig)$ and $\dd B_t = 0$.
The dynamics of the Call option are
\begin{align}
\dd C^\BS(t, S_t; \sigma) 
	&= 	\frac{\d C^\BS}{\d t} (t, S_t; \sigma) \dd t + \frac{\d C^\BS}{\d s} (t, S_t; \sigma) \dd S_t 
			+ \frac{1}{2} \frac{\d^2 C^\BS}{\d s^2} (t, S_t; \sigma) ( \dd S_t )^2 \\
	&= 	\frac{\d C^\BS}{\d t} (t, S_t; \sigma) \dd t + \frac{\d C^\BS}{\d s} (t, S_t; \sigma) \dd S_t 
			+ \frac{1}{2} \frac{\d^2 C^\BS}{\d s^2} (t, S_t; \sigma) \sigma^2_t S_t^2  \dd t .
\end{align}
Combining the above equations, we obtain
\begin{align}
\dd X_t -\dd C^\BS(t,S_t;\sig) 
	&= 	- \[  \frac{\d C^\BS}{\d t} (t, S_t; \sigma) + \frac{1}{2} \frac{\d^2 C^\BS}{\d s^2} (t, S_t; \sigma) \sigma^2_t S_t^2 \] \dd t \\
	&= 	- \[  \frac{\d C^\BS}{\d t} (t, S_t; \sigma) + \frac{1}{2} \frac{\d^2 C^\BS}{\d s^2} (t, S_t; \sigma) \sigma^2 S_t^2 \] \dd t \\ &\quad
			+ \frac{1}{2} (\sigma^2 - \sigma^2_t) S_t^2 \frac{\d^2 C^\BS}{\d s^2} (t, S_t; \sigma) \dd t \\
	&=	\frac{1}{2} (\sigma^2 - \sigma^2_t) S_t^2  \frac{\d^2 C^\BS}{\d s^2} (t, S_t; \sigma) \dd t ,
\end{align}
where, in the 2nd equality, the term in square brackets is equal to zero $\[ \ldots \] = 0$ because $C^\BS$ satisfies the Black-Scholes PDE.  Finally, we obtain
\begin{align}
\text{P\&L}_T 
	&= X_T - C^\BS(T,S_T;\sig) \\
	&= X_0 - C^\BS(0,S_0;\sig) + \int_0^T \big( \dd X_t - \dd C(t,S_t;\sig) \big) \\
	&= \int_0^T \(\sigma^2 - \sigma_t^2\) \frac{1}{2} S_t^2 \frac{\d^2 C^{\text{BS}}}{\d s^2} (t, S_t; \sigma) \dd t ,
\end{align}
where we have used the fact that $X_0 = C^\BS(0,S_0;\sig)$.  Thus the final profit or loss depends on the fixed volatility $\sig$ that the market assumes and the path that the volatility process $(\sig_t)_{0 \leq t \leq T}$ actually takes.  Note that $\d_s^2 C^\BS > 0$.  Thus, for a banker that sells an option and $\Delta$-hedges, he would prefer that the market prices options with a large $\sig$.

\section{Variance Swaps}
As we showed in the previous section, $\Delta$-hedging in an environment where volatility is stochastic leaves one exposed to fluctuations in volatility.  The desire to reduce volatility risk has led to the development of various volatility derivatives.  
The simplest of these derivatives is the \textit{variance swap}. 
Let us denote by $\text{VS} = (\text{VS}_t)_{0 \leq t \leq T}$ the value of the long side of a variance swap.   
The payoff of the long side of a variance swap $\text{VS}_T$ is
\begin{align}
\text{VS}_T
	&= \sum_{i=0}^{n-1} \( \log S_{t_{i+1}} - \log S_{t_i} \)^2 - K , \label{eq:VS-payoff}
\end{align} 
where $0 = t_0 < t_1 < t_2 < \ldots < t_n = T$ are some pre-specified dates, and $K$ is a constant, which is determined at inception so that the initial value of the variance swap is zero.
Typically $\del t := t_{i+1} - t_i$ is one trading day.
The sum in \eqref{eq:VS-payoff} is referred to as the \textit{floating leg} and the constant $K$ is referred to as the \textit{fixed leg}.
Note that as $\del t \rightarrow 0$ the floating leg converges to an integral
\begin{align} 
\lim_{\del t \to 0}\sum_{i=0}^{n-1} \( \log S_{t_{i+1}} - \log S_{t_i} \)^2 
	&= \int_0^T (\dd \log S_t)^2 . \label{eq:VS-limit}
\end{align}
Assuming $S$ has dynamics of the form \eqref{eq:S-general}, we have from It\^o's Lemma that
\begin{align}
\dd \log S_t 
	&= \( \mu_t - \frac{1}{2} \sigma_t^2 \) \dd t + \sigma_t \dd W_t &
	&\Rightarrow&
(\dd \log S_t)^2 
	&= \sigma_t^2 \dd t. \label{eq:d-log-s}
\end{align}
Inserting \eqref{eq:d-log-s} into \eqref{eq:VS-limit} yields
\begin{align}
\lim_{\del t \to 0} \text{VS}_T
	&=	\int_0^T \sigma_t^2 \dd t - K. \label{eq:approx-VS-payoff} 
\end{align}
Henceforth, we shall assume that the payoff of the long side of a variance swap is given by the right-hand side of \eqref{eq:approx-VS-payoff}.
The fair strike $K$ is given by setting the initial value of the variance swap to zero and solving for $K$.  We have
\begin{align}
0 = \text{VS}_0
	&=	\frac{1}{B_T} \Ebt \[ \int_0^T \sigma_t^2 \dd t - K \] &
	&\Rightarrow &
K 
	&= \Ebt \[ \int_0^T \sigma_t^2 \dd t \] , \label{eq:fair-strike-VS}
\end{align}
where $\Ebt$ denotes an expectation taken with respect to the market's chosen risk-neutral (i.e., pricing) measure $\Pbt$.
\\[0.5em]
Let us see how we can replicate the floating leg $\int_0^T \sigma_t^2 \dd t$ of a variance swap.
For simplicity, we will once again assume that $r = 0$.  Under the pricing measure $\Pbt$, the stock $S$ has dynamics of the form
\begin{align}
\dd S_t 
	&= \sigma_t S_t \dd \Wt_t ,
\end{align}
where $\Wt = (\Wt_t)_{0 \leq t \leq T}$ is a Brownian motion under $\Pbt$.  Using It\^o's Lemma, we have
\begin{align}
\dd \log S_t 
	&= -\frac{1}{2} \sigma_t^2 \dd t + \sigma_t \dd \Wt_t &
	&\Rightarrow& 
\sigma_t^2 \dd t
	&= 2 \sigma_t \dd \Wt_t - 2 \dd \log S_t
	 = \frac{2}{S_t} \dd S_t - 2 \dd \log S_t .
\end{align}
Integrating from $0$ to $T$, we obtain
\begin{align} \label{eq:8}
\int_0^T \sigma_t^2 \dd t 
	&= \int_0^T \frac{2}{S_t} \dd S_t - 2 \log \( \frac{S_T}{S_0 }\).
\end{align}
Thus, to replicate the floating leg of a variance swap, one can do the following
\begin{enumerate}
\item At time $t=0$, buy a European \textit{log contract} with payoff
\begin{align}	
\phi(S_T) 
	&= - 2 \log \( \frac{S_T}{S_0} \) ,
\end{align}
\item	For all $0 \leq t \leq T$ hold $\Del_t = \frac{2}{S_t}$ shares of $S$.
\item Lend and borrow from the bank as needed to finance the position.  
\end{enumerate}
At maturity $T$, the above portfolio will have a value equal to the floating leg of the variance swap $\int_0^T \sigma_t^2 \dd t$.
Note that, because $\Pbt \sim \Pb$, the replication strategy above holds under $\Pb$ as well.  
\\[0.5em]
Now, let us determine the fair strike $K$ of a VS.  From \eqref{eq:fair-strike-VS} and \eqref{eq:8}, we have
\begin{align}
K
	&=	\Ebt \[ \int_0^T \sigma_t^2 \dd t \]  
	 = \Ebt \[ \int_0^T \frac{2}{S_t}\dd S_t \] - 2 \Ebt \[ \log \(\frac{S_T}{S_0}\) \] \\
	&= \Ebt \[ \int_0^T \frac{2}{S_t} \sigma_t S_t \dd \Wt_t \] - 2 \Ebt \[ \log \(\frac{S_T}{S_0}\) \]  
	 =  - 2 \Ebt \[ \log \(\frac{S_T}{S_0}\) \] .
\end{align}
The above result says that the fair strike $K$ of a variance swap is equal to the price $- 2 \Ebt \[ \log \(S_T/S_0\) \]$ of a (path-independent) European log contract payoff.   In reality, one cannot buy log contracts in the market.  However, one can create a synthetic $\log$ contract from European Calls and Puts.
\\[0.5em]
Recall from Section~\ref{sec:payoff_construction} that, for any $\kappa > 0$ and any reasonable function $\phi$ we have
\begin{align}
\phi(S_T)
	&=	\phi(\kappa) + \phi'(\kappa) \Big( (S_T-\kappa)^+ - (\kappa - S_T)^+ \Big) \\ &\qquad
			+ \int_0^\kappa \phi''(K) (K-S_T)^+ \dd K + \int_\kappa^\infty \phi''(K) (S_T-K)^+ \dd K .
\end{align}
By taking the risk-neutral expectation of both sides of the above equation, we obtain
\begin{align}
\Ebt \[ \phi(S_T) \]
	&=	\phi(\kappa) + \phi'(\kappa) \Big( C_0(T,\kappa) - P_0(T,\kappa) \Big) \\ &\qquad
			+ \int_0^\kappa \phi''(K) P_0(T,K) \dd K + \int_\kappa^\infty \phi''(K) C_0(T,K) \dd K, 
\end{align}
where we have used $\Ebt[(K-S_T)^+] = P_0(T,K)$ and $\Ebt[(S_T-K)^+] = C_0(T,K)$.
Setting $\phi(S_T) = -2 \log( S_T/S_0)$ and choosing $\kappa = S_0$ yields
\begin{align}
\Ebt \[ - 2 \log \frac{S_T}{S_0} \]
	&=	-2 \log  \frac{S_0}{S_0} + \frac{-2}{S_0} \Big( C_0(T,S_0) - P_0(T,S_0) \Big) \\ &\qquad
			+ \int_0^{S_0} \frac{2}{K^2} P_0(T,K) \dd K + \int_{S_0}^\infty \frac{2}{K^2} C_0(T,K) \dd K \\
	&=	\int_0^{S_0} \frac{2}{K^2} P_0(T,K) \dd K + \int_{S_0}^\infty \frac{2}{K^2} C_0(T,K) \dd K , \label{eq:9}
\end{align}
where we have used
\begin{align}
\log (S_0/S_0) 
	&= 0 &
	&\text{and}&
C_0(T,S_0) - P_0(T,S_0) 
	&= \Ebt (S_T - S_0) = 0 .
\end{align}
As $K = \Ebt [ \int_0^T \sig^2_t \dd t] = \Ebt[ -2 \log (S_T/S_0) ]$, we see from \eqref{eq:9} that the market's forward looking expectation of realized volatility over the interval $[0,T]$ can be deduced by observing European Call and Put prices that mature at time $T$.

\section{VIX}
You may have heard of the CBOE's \textit{volatility index} or VIX.  The VIX is a discretized approximation of the integral given in \eqref{eq:9}, normalized by the length $T$ of the averaging window (30 days for the VIX).  Roughly, assuming $r = 0$, we have
\begin{align}
\text{VIX}_t^2 
	&= \frac{1}{T} \Big( \sum_{K_i < S_t} \frac{2}{K_i^2} P_t (t+T,K_i) \Delta K_i + \sum_{K_i > S_t} \frac{2}{K_i^2} C_t (t+T, K_i) \Delta K_i \Big) , &
\Delta K_i
	&=	K_{i+1} - K_i .
\end{align}
Thus, the $\text{VIX}_t^2$ is an approximate measure of the market's forward-looking risk-neutral expectation of the average squared volatility over $[t,t+T]$, that is $\frac{1}{T}\Ebt [ \int_t^{t+T} \sig^2_s \dd s | \Fc_t]$.


\section{Exercises}
\label{sec:exer-7}

\noindent
\textbf{Instructions:} Exercise \ref{exer:7-1} asks you to write code that produces a plot. Your program should read the input from a text file called \textbf{input.txt}, consisting of a single line listing $S_0, K,T,\sig_1, \sig_2, \mu, N,M$ in that order, separated by commas.

\begin{exercise}
\label{exer:7-1}
Assume the risk free rate of interest is zero $r=0$.
Suppose that the dynamics of $S$ are given by $\dd S_t = \mu S_t \dd t + \sig_1 S_t \dd W_t$.  Suppose the market prices options assuming that the dynamics of $S$ are given by $\dd S_t = \mu S_t \dd t + \sig_2 S_t \dd W_t$ where $\sig_2 > \sig_1$.  Assume you \textit{sell} a call option for the Black-Scholes price $C^\BS(t,S_t;T,K,\sig_2)$ and you $\Del$-hedge by holding $\Del^\BS(t,S_t;T,K,\sig_2)$ shares of $S$ at all times $t \in [0,T]$.  Thus, the dynamics of your replicating portfolio $X$ are
\begin{align}
\dd X_t
	&=	\Del^\BS(t,S_t;T,K,\sig_2) \dd S_t + (X_t - \Del^\BS(t,S_t;T,K,\sig_2) S_t) \frac{1}{B_t} \dd B_t , &
X_0
	&=	C^\BS(0,S_0;T,K,\sig_2) .
\end{align}
The value of the option, according to the market is $C^\BS(t,S_t;T,K,\sig_2)$ for all $t \in [0,T]$.  \\
Write a Monte Carlo algorithm that plots the value of your hedging portfolio $X_t$ and the value of the option $C^\BS(t,S_t;T,K,\sig_2)$ as a function of time.  Plot a histogram of $(X_T - (S_T - K)^+)$ for $M$ sample paths.  
Remember to simulate the paths of $S$ using the true dynamics (i.e., with $\sig_1$).
Use the following parameters:
\begin{align}
S_0 & = K = T = 1, &
\sig_1
	&=	0.2 , &
\sig_2
	&=	0.3 , &
\mu
	&=	0.1 , &
N
	&=	1000, &
M
	&=	1000 ,
\end{align}
where $N$ is the number of time-steps ($\del t = T/N$) and $M$ is the number of sample paths.
\end{exercise}


\begin{exercise}
\label{exer:7-2}
Let the risk-free rate of interest be zero $r=0$ so that the value of a bond is a constant $B_t = 1$.
Assume that, under the risk-neutral pricing measure $\Pbt$, the dynamics of a stock $S$ are given by
\begin{align}
\dd S_t
	&=	\sig_t S_t \dd \Wt_t ,
\end{align}
where $(\sig_t)_{t \geq 0}$ is a stochastic process.  Consider a contract that pays $\int_0^T \sig_t^2 \dd t$ at time $T$ (i.e., the floating leg of a variance swap).  In class, we showed that
\begin{align}
\int_0^T \sig_t^2 \dd t
	&=	\int_0^T \frac{2}{S_t} \dd S_t - 2 \log \( \frac{S_T}{S_0} \) . \label{eq:RV}
\end{align}
And thus, to replicate the payoff $\int_0^T \sig_t^2 \dd t$ one should buy a European contract that pays $-2 \log ( S_T / S_0 )$, hold $\Del_t = 2/S_t$ shares of $S_t$ at all times $t$, and lend or borrow from the bank as needed.  What is the number of bonds $\Gam_t$ one should hold at time $t$?  Your answer should involve a stochastic integral with respect to $S$.
\end{exercise}




\chapter{Dupire's Local Volatility}
It would be preferable to have some model for $S$ that results in option prices that agree with market quotes.  As previously mentioned, the Black-Scholes model does not accomplish this task.  In this chapter, we will introduce a model for $S$ that induces option prices that are consistent with market quotes.
\\[0.5em]
Consider a model in which the volatility of a stock is a deterministic function of time and the asset's price $\sig_t = f(t,S_t)$.  If we assume the risk-free rate of interest is zero ($r=0$), then the dynamics of $S$ under the risk-neutral probability measure $\Pbt$ are of the form
\begin{align}
\dd S_t 
	&= f(t,S_t) S_t \dd \Wt_t , \label{eq:lv-model}
\end{align}
where $\Wt$ is a $\Pbt$ Brownian motion.  The price of a call option in this setting is a function $C$ of time, the stock price, the maturity and strike $C(t,S_t;T,K)$.
Assume that, at time $t$, we can observe the price of a Call option $C_t(T,K)$ for every $T \geq t$ and $K \geq 0$.  We would like to find a function $f$ such that the model-induced option prices $C(t,S_t;T,K)$ match the observed market prices $C_t(T,K)$.  That is, we would like
\begin{align}
C(t,S_t;T,K)
	&=	C_t(T,K) , &
	&\forall \, T \geq t, \, K \geq 0 .
\end{align}
Let us consider a European option with payoff $\phi(S_T)$ and let us denote by $V = (V_t)_{t \geq 0}$ the value of this option.
We know from the first fundamental Theorem of asset pricing that $V/B$ is a martingale under $\Pbt$ with respect to a filtration $\Fc$ for the Brownian motion $\Wt$.
As we have assumed $r=0$, this implies the bond price $B$ is a constant, and thus $V$ is a martingale.  Explicitly, we have
\begin{align}
V_t
	&=	\Ebt [ V_T | \Fc_t ] =  \Ebt [ \phi(S_T) | \Fc_t ] .
\end{align}
Now, the stock price $S$, as the solution of SDE \eqref{eq:lv-model}, is a Markov process.  It follows that there exists a function $v$ such that $V_t = v(t,S_t)$.  Taking the differential of $v(t,S_t)$ we find that 
\begin{align}
\dd v(t,S_t) 
	&= \frac{\d v}{\d t} (t,S_t) \dd t + \frac{\d v}{\d s} (t,S_t) \dd S_t + \frac{1}{2} \frac{\d^2 v}{\d s^2}(t,S_t) ( \dd S_t )^2 \\
	&= \( \d_t + \frac{1}{2} f^2(t,S_t) S_t^2 \d_s^2 \) v(t,S_t) \dd t + f(t,S_t) S_t \d_s v(t,S_t) \dd \Wt_t .
\end{align}
Because $(v(t,S_t))_{0 \leq t \leq T}$ is a martingale, the $\dd t$-term above must equal zero. 
Thus, we obtain a pricing PDE for the function $v$
\begin{align}
0
	&=	\( \d_t + \frac{1}{2} f^2 (t,s) s^2 \d_s^2 \) v , &
v(T,s) 
	&= 	\phi(s) , \label{eq:local-vol-pde}
\end{align}
where the terminal condition $v(T,s) = \phi(s)$ comes from the fact that $V_T = v(T,S_T) = \Ebt[ \phi(S_T) | \Fc_T^S] = \phi(S_T)$.
When $\phi(s) = (s - K)^+$ we have  
\begin{align}
0
	&=	\( \d_t + \frac{1}{2} f^2 (t,s) s^2 \d_s^2 \) C(t,s;T,K) , &
C(T,s;T,K) 
	&= (s-K)^+.
\end{align}
Differentiating the above equations twice with respect to $K$ yields
\begin{align} \label{eq:localvol1}
0
	&=	\( \d_t + \frac{1}{2} f^2 (t,s) s^2 \d_s^2 \) \d_K^2 C(t,s;T,K) , &
\d_K^2 C(T,s;T,K) 
	&= \del (s-K) .
\end{align}
where we have used that $\d_K^2(s-K)^+ = \delta(s-K)$.
Now, note that 
\begin{align}
C(t,s;T,K)
	&= \Ebt [(S_T - K)^+ | S_t = s ]
	=		\int_0^\infty \pt (t,s;T,y) (y - K)^+ \dd y ,
\end{align}
where $\pt(t,s;T,y) \dd y := \Pbt (S_T \in \dd y | S_t = s )$ is the (unknown) transition density of $S$.
Differentiating the above equation twice with respect to $K$, we obtain
\begin{align} 
\d^2_K C(t,s;T,K) 
	&= \int_0^\infty \pt(t,s;T,y) \d_K^2 (y-K)^+ \dd y\\
	&= \int_0^\infty \pt(t,s;T,y) \delta(y-K) \dd y\\
	&= \pt (t,s;T,K) . \label{eq:localvol2}
\end{align}
Combining \eqref{eq:localvol1} and \eqref{eq:localvol2}, we have
\begin{align}
0
	&=	\(\d_t + \frac{1}{2} s^2 f^2(t,s) \d_{s}^2\) \pt (t,s;T,y) , &
\pt(T,s;T,y) 
	&= 	\delta(s-y) . \label{eq:KBE}
\end{align}
Equation \eqref{eq:KBE} is known as the \textit{Kolmogorov Backward equation} (KBE).
Note that the KBE is a PDE and terminal condition for $\pt$ in the \textit{backward variables} $(t,s)$.  
The forward variables $(T,y)$ are merely parameters.
For reasons that will become apparent later, we will need to know what $\pt$ satisfies in \textit{forward variables} $(T,y)$.  
To this end, let us fix a $\overline{T} > t$ and let us consider a general function $u$ of time and the stock price.  Using the short-hand notation
$\Ebt_{t,s}\, \cdot \, := \Ebt[ \, \cdot \, | S_t = s ]$ we have
\begin{align}
\Ebt_{t,s} u(\overline{T},S_{\overline{T}}) 
	&= u(t,s) + \Ebt_{t,s} \int_t^{\overline{T}} \dd u(T,S_{T}) \\
	&= u(t,s) + \Ebt_{t,s} \int_t^{\overline{T}} \(\d_{T} + \frac{1}{2}f^2(T,S_{T})S^2_{T}  \d_{s}^2\) u(T,S_{T}) \dd T  
		 + \Ebt_{t,s} \int_t^{\overline{T}} (\ldots) \dd \Wt_T \\
	&= u(t,s) + \int_0^\infty \int_t^{\overline{T}} \left\{ \pt(t,s,T,y)\[ \d_{T} + \frac{1}{2} f^2(T,y) y^2 \d_y^2 \] u(T,y) \right\} \dd T \dd y ,
\end{align}
where the precise form of $( \ldots )$ is not important, as the expected value of an It\^o integral is zero.
Integrating the first term in the integral by parts once with respect to $T$, and integrating the second term in the integral by parts twice with respect to $y$, we obtain
\begin{align}
\Ebt_{t,s} u(\overline{T},S_{\overline{T}}) 
	&= u(t,s) + \int_0^\infty \int_t^{\overline{T}} u(T,y) \[ -\d_{T} + \frac{1}{2} \d^2_y f^2 (T,y) y^2  \] \pt (t,s,T,y) \dd T \dd y \\ & \quad
		+ \int_0^\infty u(T,y) \pt (t,s,T,y) \dd y \Big|_{T=t}^{T=\overline{T}} \\
	&= u(t,s) + \int_0^\infty \int_t^{\overline{T}} u(T,y) \[ -\d_{T} + \frac{1}{2} \d^2_y f^2 (T,y) y^2 \] \pt (t,s,T,y) \dd T \dd y \\ & \quad
		+ \Ebt_{t,s} u(\overline{T},S_{\overline{T}})  - u(t,s) ,
\end{align}
where we have used $\lim_{y \to 0} \pt(t,s;T,y) = 0$ and $\lim_{y \to \infty} \pt(t,s;T,y) = 0$.
Canceling the terms $\Ebt_{t,s} u(\overline{T},S_{\overline{T}}) $ and $u(t,s)$, which both appear twice above with opposite signs, we obtain
\begin{align}
0 = \int_0^\infty \int_t^{\overline{T}} u(T,y) \[ -\d_{T} + \frac{1}{2} \d^2_y f^2 (T,y) y^2 \] \pt (t,s,T,y) \dd T \dd y .
\end{align}
The above equation must hold for \textit{all} functions $u$.  The only way for this to happen is if $\pt(t,s;T,y)$ satisfies
\begin{align} \label{eq:KFE}
 0 
	&= -\d_{T} \pt (t,s,T,y) + \frac{1}{2} \d_y^2 \Big( f^2(T,y) y^2  \pt(t,s,T,y) \Big), &
\pt (t,s;t,y) 
	&= \delta(s-y) .
\end{align}
The above equation is known as the \textit{Kolmogorov Forward Equation} (KFE).
The KFE is a PDE and initial condition in forward variables $(T,y)$.
The backward variables $(t,s)$ serve merely as parameters.
\\[0.5em]
Now we return to the call option.  If we differentiate the value of a Call option with respect to the maturity date $T$, we obtain
\begin{align}
\d_T C(t,s;T,K) 
	&= \int_0^\infty \d_T \pt(t,s,T,y) (y-K)^+ \dd y  \\
	&= \int_0^\infty (y-K)^+ \d_y^2 \( \frac{1}{2}f^2(T,y) y^2 \pt (t,s;T,y) \) \dd y &
	&\text{(by \eqref{eq:KFE})} \\
	&= \int_0^\infty \frac{1}{2} f^2(T,y) y^2 \pt (t,s;T,y) \d_y^2 (y-K)^+ \dd y &
	&\text{(integration by parts)}\\
	&= \int_0^\infty \frac{1}{2} f^2(T,y) y^2 \pt (t,s;T,y) \del(y-K) \dd y  \\
	&= \frac{1}{2} f^2(T,K) K^2 \pt (t,s;T,K)  \\
	&= \frac{1}{2} f^2(T,K) K^2 \d_K^2 C(t,s,T,K) . &
	&\text{(by \eqref{eq:localvol2})}
\end{align}
Solving the above equation for $f^2 (T,K)$ yields
\begin{align}
	f^2(T,K)= \frac{\d_T C(t,s,T,K)}{\tfrac{1}{2} K^2 \d_K^2 C(t,s,T,K)}. \label{eq:dupire}
\end{align}
Equation \eqref{eq:dupire} is \textit{Dupire's Formula}. It also holds if we replace Call prices with Put prices because $\d_K^2 (K-S_T)^+ = \delta(K-S_T)$.  The formula tells us how to construct the function $f(t,s)$ from (observed Market) Call and Put prices so that the model \eqref{eq:lv-model} produces option prices that are consistent with market quotes.

\section{Practical considerations}
Implementation of Dupire's Formula requires one to differentiate Call and/or Put prices with respect to $T$ and $K$. This, in turn, requires a smooth surface of call prices $(T,K) \mapsto C(T,K)$. In practice, we only have access to Market quotes at discrete maturities and strikes: $(T_i, K_j) \mapsto C(T_i, K_j)$. We will discuss two possible ways to handle this.
\\[0.5em]
\textbf{Finite Difference Approximations}.
For simplicity, let us assume that Call prices are quoted at uniformly spaced intervals in $T$ and $K$.  We will denote the sizes of the maturity intervals as $\Del T$ and the sizes of the strike intervals as $\Del K$.
In order to obtain an approximation for $\d_T C(T,K)$ we expand $C(T + \Del T, K)$ as a Taylor series about $(T,K)$.  We have 
\begin{align}
C(T \pm \Delta T, K) 
	&=	C(T,K) \pm \d_T C(T,K) \Delta T + \frac{1}{2!} \d_T^2 C(T,K) \Delta T^2 + \Oc(\Del T^3) .
\end{align}
From the above equation, we obtain
\begin{align}
\d_T C(T,K)
	&=	\frac{C(T+\Delta T, K) - C(T - \Del T,K)}{2 \Delta T} + \Oc( \Del T^2 ) . \label{eq:central.diff-1}
\end{align}
Equation \eqref{eq:central.diff-1} is known as the \textit{central difference approximation} of $\d_T C(T,K)$.
To obtain an approximation for $\d_K^2 C(T,K)$  we again use a Taylor series.  We have
\begin{align}
C(T,K \pm \Delta K) 
	&= C(T,K) \pm \d_K C(T,K) \Delta K + \frac{1}{2!} \d_K^2 C(T,K) \Delta K^2 \pm \frac{1}{3!} \d_K^3 C(T,K) \Delta K^3 + \Oc(\Delta K^4)
\end{align}
From the above equation, we obtain
\begin{align}
\d_K^2 C(T,K)  
	&= \frac{C(T,K+\Delta K) - 2 C(T,K) + C(T, K-\Delta K)}{\Delta K^2} + \Oc(\Del K^2) . \label{eq:central.diff-2}
\end{align}
Equation \eqref{eq:central.diff-2} is known as the \textit{central difference approximation} of $\d_K^2 C(T,K)$.
\\[0.5em]
\textbf{Smooth Interpolation}.
Rather than use finite difference methods, one might instead choose to fit the volatility surface with a parametric curve in $K$ for every fixed maturity $T_i$.  Consider the \textit{Stochastic Volatility Inspired} (SVI) parametrization of implied volatility
\begin{align}
\sigma^\text{SVI} (T,k) 
	:= \sqrt{\[ a(T) + b(T) \( \rho(T) (k - x - m(T)) + \sqrt{(k-x-m(T))^2 + \xi^2(T)} \) \]},
\end{align}
where $k := \log K$, $x = \log S_0$ and  $a(T)$, $b(T)$, $\rho(T)$, $m(T)$ and $\xi(T)$ are parameters to be obtained via calibration.
Now, consider the following procedure
\begin{enumerate}
\item For each maturity $T_i$, find the parameters $a(T_i)$, $b(T_i)$, $\rho(T_i)$, $m(T_i)$ and $\xi(T_i)$ that give the best fit (in the least squares sense) to the market's implied volatility smile.  This yields a call price
\begin{align}
C(T_i,K)
	&=	C^\text{BS}(0,S_0;\sigma^\text{SVI}(T_i,\log K),T_i,K)
\end{align}
for each maturity $T_i$ and every $K \geq 0$.
\item Use central difference \eqref{eq:central.diff-1} to obtain an approximation of $\d_T C(T_i,K)$.
\item To obtain an approximation of $\d_K^2 C(T_i,K)$ simply differentiate call prices analytically
\begin{align}
\d_K^2 C(T_i,K) 
	&= \d_K^2 C^\text{BS}(0,S_0;\sigma^\text{SVI}(T_i,\log K),T_i,K).
\end{align}
\end{enumerate}

\section{Advantages and Disadvantages of local volatility}
One \textit{advantage} of the local volatility framework (aside from the fact that it induces option prices that fit market quotes) is that the local volatility model is a \textit{complete market model} (i.e., derivatives can be perfectly replicated).  Assuming that $r=0$ for simplicity the dynamics of a portfolio $X$ and an option price $v(t,S_t)$ are given by
\begin{align}
\dd X_t 
	&= \Delta_t \dd S_t + (X_t - \Delta_t S_t) \frac{1}{B_t}  \dd B_t
	=	 \Delta_t \dd S_t , \\
\dd v(t,S_t) 
	&= \( \d_t + \frac{1}{2}f^2(t,S_t)  S_t^2  \d_s^2 \) v(t,S_t) \dd t  + \d_s v(t,S_t) \dd S_t 
	 =	\d_s v(t,S_t) \dd S_t ,
\end{align}
where we have used $\dd B_t = 0$ and the fact that the $\dd t$-term is zero by \eqref{eq:local-vol-pde}.
Thus, if we choose 
\begin{align}
\Del_t
	&=	\d_s v(t,S_t) , &
	&\text{then we have}&
\dd X_t
	&=	\dd v(t,S_t) .
\end{align}
If we further have $X_0 = v(0,S_0)$ then we will have $X_t = v(t,S_t)$ for all $t \in [0,T]$ where $T$ is the maturity date of the option.
\\[0.5em]
The main \textit{disadvantage} of the local volatility framework is that 
the volatility function $f(t,s)$ must be recalibrated daily! That is, if we find a local volatility function $f(t,s)=f_1(t,s)$ on day 1, then we very often find a different local volatility function $f(t,s) = f_2(t,s)$ on day two, and still a different volatility function $f(t,s) = f_3(t,s)$ on day three and so on.  This, of course, brings into question the use of the local volatility framework.


\section{Exercises}
\label{sec:exer-8}

\begin{exercise}
\label{exer:8-1}
Assume for simplicity that the risk-free rate of interest is zero $r = 0$.  Suppose we model the stock price $S$ under the risk-neutral pricing measure as
\begin{align}
\dd S_t
	&=	f(t,S_t) S_t \dd \Wt_t .
\end{align}
In order for the dynamics of $S$ to be consistent with option prices on the market, Dupire's formula must be satisfied
\begin{align}
f^2(T,K)
	&=	\frac{\d_T C(t,s,T,K)}{\tfrac{1}{2} K^2 \d_K^2 C(t,s,T,K)} ,
\end{align}
where $C(t,s,T,K)$ denotes the value of a call option with maturity $T$ and strike $K$ given $S_t = s$.  Suppose that, for every $T$ and $K$, the implied volatility of an option is a function of $T$ only: $\sig(T)$ (i.e., there is no $K$-dependence). \\
\textbf{(a)} Derive an expression for $f(T,K)$.  In particular, show that $f(T,K)$ has no dependence on $K$.\\
\textbf{(b)} Show that
\begin{align}
\sig^2(T)
	&=	\frac{1}{T-t} \int_t^T f^2(u,K) \dd u .
\end{align}
\end{exercise}




\chapter{One-factor Stochastic Volatility Models}

In a standard (one-factor) \textit{stochastic volatility} model the dynamics of a stock price $S$ under the physical measure $\Pb$ are of the form
\begin{align}
\dd S_t 
	&= \mu(Y_t) S_t \dd t + \sigma(Y_t) S_t \dd W_t^S , &
\dd Y_t 
	&= \alpha (Y_t) \dd t + \beta(Y_t) \dd W_t^Y , \label{eq:vol-process} \\
\dd W_t^Y 
	&=	\rho \dd W_t^S + \rhob \dd W_t , &
\rhob
	&=	\sqrt{1-\rho^2} , \label{eq:WY}
\end{align}
where $W^S = (W_t^S)_{t \geq 0}$ and $W = (W_t)_{t \geq 0}$ are independent, one-dimensional Brownian motions.
One can check that $W^Y = (W_t^Y)_{t \geq 0}$, defined by \eqref{eq:WY}, is a Brownian motion (i.e., it satisfies the properties given in the definition of a Brownian motion).  Furthermore, we have $\dd W_t^S \dd W_t^Y = \rho \dd t$.  Thus, we say that $W^S$ and $W^Y$ are \textit{correlated} and we call $\rho \in [-1,1]$ the \textit{correlation coefficient}.
We shall refer to $Y$ as the \textit{driver of volatility} or \textit{volatility driving process} because $\sig(Y)$ is the volatility of $S$.
In most (but not all) stochastic volatility models, the $Y$ process is \textit{mean-reverting} meaning that $Y$ fluctuates around a long-run mean (like the OU process) that we discussed in Example \ref{ex:OU}.  
As in the Black-Scholes setting, the drift $\mu$ plays no role in the pricing of options.
Let us see a few examples of one-factor stochastic volatility models.

\begin{example}[Heston]
\label{ex:heston}
In the stochastic volatility model of \cite{heston1993}, the coefficient functions $\alpha$, $\beta$ and $\sig$ in \eqref{eq:vol-process} are given by
\begin{align}
\alpha(y) 
	&= \kappa (\theta - y), &
\beta(y) 
	&= \delta \sqrt{y}, &
\sigma(y) 
	&= \sqrt{y}.
\end{align}
The $Y$ process in the Heston model is commonly known as a \textit{Cox-Ingersoll-Ross} (CIR) process.
The {Heston model} is well-known in industry and in academia due to the fact that there exists a closed-form formula for European option prices.
\end{example}
	
\begin{example}[Exponential OU]
In the Exponential Ornstein-Uhlenbeck (ExpOU) stochastic volatility model, the coefficient functions $\alpha$, $\beta$ and $\sig$ in \eqref{eq:vol-process} are given by 
\begin{align}
\alpha (y) 
	&= m (\theta - y) , &
\beta(y) 
	&= \delta , &
\sigma(y) 
	&= \exp(y) .
\end{align}
European option prices in this setting can be computed in closed-form only when $S$ and $Y$ are uncorrelated: $\rho = 0$.
\end{example}
	
\begin{example}[3/2's]
In the 3/2's stochastic volatility model, the coefficient functions $\alpha$, $\beta$ and $\sig$ in \eqref{eq:vol-process} are given by
\begin{align}
\alpha (y) 
	&= y \kappa (\theta -y) , &
\beta(y) 
	&= \delta y \sqrt{y} , &
\sigma(y) 
	&= \sqrt{y} .
\end{align}
European option prices in this setting can be computed in closed-form involving special functions.
\end{example}

\section{Pricing by replication}
We would like to derive a pricing formula and replication strategy for European options in the stochastic volatility setting \eqref{eq:vol-process}. 
In the Black-Scholes and local volatility settings, we were able to replicate a derivative with a self-financing portfolio that contained only two assets -- the underlying stock $S$ and a bond $B$.  Perfect replication of the derivative asset was possible because the derivative price depended on a single Brownian motion $W$, which we could hedge by trading the stock.  
In the stochastic volatility setting \eqref{eq:vol-process}, the derivative asset's price depends on two sources of randomness $W^S$  and  $W^Y$. This means that we can no longer replicate the derivative asset by trading  only the stock and the bond.  We will consider a replicating portfolio containing the stock, the bond, and another option.
\\[0.5em]
In the stochastic volatility setting \eqref{eq:vol-process}, we expect the value $V=(V_t)_{0 \leq t \leq T}$ of any option to be of the form
\begin{align}
V_t 
	&= v(t, S_t, Y_t) ,
\end{align}
for some function $v$.  Suppose we have \textit{sold} an option with payoff $h^{(1)}(S_{T_1})$ expiring at time $T_1$.
We will denote the value of this option as $V_t^{(1)} = v^{(1)}(t,S_t,Y_t)$. 
We will attempt to replicate this option with a portfolio consisting of the underlying stock $S$, the bond $B$, and an option expiring at time $T_2 > T_1$ with payoff $h^{(2)}(S_{T_2})$.
We will denote the value of this option as $V_t^{(2)} = v^{(2)}(t,S_t,Y_t)$.
As usual, we will assume $r=0$ for simplicity.
\\[0.5em]
We set up our hedging portfolio $X$ so that we have $\Del$ shares of $S$ at time $t$ and $\Gamma$ shares of $V^{(2)}$.  The dynamics of $X$ are then given by
\begin{align}
\dd X_t 
	&= \Delta_t \dd S_t + \Gamma_t \dd V_t^{(2)} + (X_t - \Del_t S_t - \Gam_t V_t^{(2)}) \frac{1}{B_t} \dd B_t \\
	&= \Delta_t \dd S_t + \Gamma_t \d_s v^{(2)} \dd S_t + \Gamma_t \d_y v^{(2)} \dd Y_t + \Gamma_t (\d_t + \mathscr{M}) v^{(2)} \dd t, \label{eq:dX.sv}
\end{align}
where we have used the shorthand  $v^{(2)}=v^{(2)}(t,S_t,Y_t)$ for notational clarity and we have introduced an operator $\Mc$, which is defined as follows
\begin{align}
\Mc
	&:= \frac{1}{2} \sigma^2(y) s^2 \d_{s}^2 + \frac{1}{2} \beta^2(y) \d_y^2 + \rho \sigma(y) \beta(y) s \d_s \d_y .
\end{align}
We would like the change in the value of our portfolio $\dd X_t$ to equal the change in the value of the option $\dd V_t^{(1)}$.  Thus, we compute
\begin{align}
\dd V_t^{(1)}
	&=	\dd v_t^{(1)} 
	=		\d_s v^{(1)} \dd S_t + \d_y v^{(1)} \dd Y_t + (\d_t + \Mc ) v^{(1)} \dd t . \label{eq:dV1}
\end{align}
In order to have $\dd X_t = \dd V_t^{(1)}$ we must match the $\dd S_t$, the $\dd Y_t$ and the $\dd t$ terms in \eqref{eq:dX.sv} and \eqref{eq:dV1}.  From
\begin{align}
\Delta_t + \Gamma_t \d_s v^{(2)} 
		&= \d_s v^{(1)} , & 
		&\text{(matching the $\dd S_t$ terms)} \\
\Gamma_t \d_y v^{(2)} 
		&= \d_y v^{(1)} , &
		&\text{(matching the $\dd Y_t$ terms)} 
\end{align}
we deduce that $\Gam$ and $\Del$ must satisfy
\begin{align}
\Gamma_t 
	&= \frac{ \d_y v^{(1)} }{ \d_y v^{(2)} } , &
\Delta_t 
	&= \d_s v^{(1)} - \Gamma_t \d_s v^{(2)} . \label{eq:Gamma-Delta}
\end{align}
We are left with $\dd t$ terms.  Comparing \eqref{eq:dX.sv} and \eqref{eq:dV1}, and using the expression for $\Gam_t$ in \eqref{eq:Gamma-Delta}, we must have
\begin{align} \label{eq:stoch_vol}
	\underbrace{\frac{1}{\d_y v^{(1)}} (\d_t + \Mc) v^{(1)}}_{\text{independent of } T_2} 
		&= \underbrace{\frac{1}{ \d_y v^{(2)} } (\d_t + \Mc) v^{(2)}}_{\text{independent of } T_1} =: f(t,s,y).
\end{align}
As the left-hand side is independent of $T_2$ and the right-hand side is independent of $T_1$ we deduce that both sides must be equal to a function $f$ that is independent of both $T_1$ and $T_2$. For reasons that will become apparent later, we define a function $\Lam$ as follows
\begin{align} 
\Lambda (t,s,y)
	&:=	\frac{\alpha(y) - f(t,s,y)}{\beta(y)} &
	&\Rightarrow&
f(t,s,y) 
	&= \alpha(y) - \beta(y) \Lambda (t,s,y) . \label{eq:stoch_vol2}
\end{align}
Substituting \eqref{eq:stoch_vol2} into \eqref{eq:stoch_vol} we find that $v^{(1)}$ and $v^{(2)}$ must satisfy
\begin{align}	
(\d_t + \Act ) v^{(i)} 
	&= 0 , &
v^{(i)}(T_i,s,y)
		&=	h^{(i)}(s) , \label{eq:pde.sv}
\end{align}
where the operator $\Act$ is defined as follows
\begin{align}
\Act 
	:= \frac{1}{2} \sigma^2(y) s^2 \d_s^2 + \rho \sigma(y) \beta(y) s \d_s \d_y
			+ \frac{1}{2} \beta^2(y) \d_y^2 + \[ \alpha(y) - \beta(y) \Lambda(t,s,y) \] \d_y , \label{eq:A-tilde}
\end{align}
and we have added the obvious terminal condition $v^{(i)}(T_i,s,y)=h^{(i)}(s)$, which is needed to ensure that $V_{T_i}^{(i)} = v^{(i)}(T_i,S_{T_i},Y_{T_i}) = h^{(i)}(S_{T_i})$.  Thus, we have derived a pricing PDE \eqref{eq:pde.sv} for European options in a stochastic volatility setting.
\\[0.5em]
The function $\Lambda$ is called the \textit{market price of volatility risk}.  Note that we have \textit{not} uniquely determined $\Lambda$ yet!
A good student will be asking him/herself: \textit{what should $\Lambda$ be}?  The situation is similar to the one we encountered in the trinomial model. Recall that, in the trinomial model, there was no unique risk-neutral probability measure $\Pbt=(\pt, \qt, \rt)$ to price options in a no-arbitrage fashion.  
However, we could parametrize $\Pbt(\rt) = (\pt(\rt),\qt(\rt),\rt)$ in terms of $\rt$.
Given the price of a single option, we could determine $\rt$, and therefore $(\pt(\rt), \qt(\rt), \rt)$ uniquely.  And then we could price other options in a unique way.
\\[0.5em]
In the stochastic volatility setting the role of $\rt$ is played by the function $\Lam$.  In other words, different choices of $\Lam$ result in different risk-neutral probability measures $\Pbt(\Lam)$.  
We generally assume that the market has chosen a unique $\Lam$, under which all options are priced.
Unlike the trinomial model, we cannot determine $\Lam$ from a single option price.
To find $\Lambda$ we specify a particular form (e.g., $\Lambda(t,s,y) = \lambda \sqrt{y}$).  We then \textit{calibrate} our model to the observed option prices (meaning we search for the parameters that result in options prices that most closely match the option prices on the market in some sense).
We will discuss calibration in detail in Section \ref{sec:calibration}.

\section{Risk neutral pricing}

The 1st fundamental theorem of asset pricing (Theorem \ref{thm:1st}) states that a market defined under a probability measure $\Pb$ is free of arbitrage if and only if there exists a probability measure $\Pbt$, equivalent to $\Pb$, under which all portfolios $X$, denominated in units of a common num\'eraire $N$ are martingales.  
In the Stochastic volatility setting \eqref{eq:vol-process} this leads to a very simple procedure for finding the value $V = (V_t)_{0 \leq t \leq T}$ of an option that pays $\phi(S_T)$ at time $T$.

\begin{enumerate}
\item Choose a num\'eraire.  We will choose the bond $B=(B_t)_{0 \leq t \leq T}$, as this choice leads to the easiest computations.
\item Search for a probability measure $\Pbt$, under which $(X/B)$ is a martingale for all portfolios $X$.  That is, find $\Pbt$ such that
\begin{align}
\frac{X_t}{B_t}
	&=	\Ebt \Big( \frac{X_T}{B_T} \Big| \Fc_t \Big) , 
\end{align} 
where $X$ is given by \eqref{eq:BS-dX} and $\Fc$ is a filtration for the Brownian motions $(W^S,W^Y)$.
\item The value $V$ of the option can then be computed using
\begin{align}
\frac{V_t}{B_t}
	&=	\Ebt \Big( \frac{V_T}{B_T} \Big| \Fc_t \Big) 
	 = 	\Ebt \Big( \frac{\phi(S_T)}{B_T} \Big| \Fc_t \Big) . 
\end{align}
\end{enumerate}

\noindent
Let us carry this procedure out.  As usual, we will assume for simplicity that $r=0$, which implies that $\dd B_t = 0$.  Because $B$ is a constant, $X/B$ will be a martingale if and only if $X$ is a martingale.  Thus, we look for a probability measure $\Pbt$ under which $X$ is a martingale.  To begin, let us define $\Wt^S = (\Wt_t^S)_{0 \leq t \leq T}$ and $\Wt = (\Wt_t)_{0 \leq t \leq T}$ as follows
\begin{align}
\Wt_t^S
	&=	\int_0^t \gam_s \dd s  + W_t^S , &
\Wt_t
	&=	\int_0^t \lam_s \dd s + W_t ,
\end{align}
where $\gam = (\gam_t)_{0 \leq t \leq T}$ and $\lam = (\lam_t)_{0 \leq t \leq T}$ are, for the moment, unspecified stochastic processes.
We know from Girsanov's Theorem \ref{thm:girsanov} that there exists a probability measure $\Pbt$ under which $\Wt^S$ and $\Wt$ are Brownian motions.  Let us write the dynamics of $X$ under this probability measure.  We have
\begin{align}
\dd X_t
	&=	\Del_t \dd S_t + (X_t - \Del_t S_t) \frac{1}{B_t} \dd B_t \\
	&=	\Del_t \mu(Y_t) S_t \dd t + \Del_t \sig(Y_t) S_t \dd W_t^S \\
	&=	\Del_t \mu(Y_t) S_t \dd t + \Del_t \sig(Y_t) S_t ( \dd \Wt_t^S - \gam_t \dd t ) \\
	&=	\Del_t \big( \mu(Y_t) - \sig(Y_t) \gam_t \big) S_t \dd t + \Del_t \sig(Y_t) S_t \dd \Wt_t^S . \label{eq:dX-sv}
\end{align}
Examining \eqref{eq:dX-sv} we see that $X$ will be a martingale under $\Pbt$ if we choose
\begin{align}
\gam_t
	&=	\frac{ \mu(Y_t) }{\sig(Y_t)} . \label{eq:gam-sv}
\end{align}
Thus, the dynamics of $S$ under $\Pbt$ must be given by
\begin{align}
\dd S_t 
	&= \mu(Y_t) S_t \dd t + \sigma(Y_t) S_t \dd W_t^S \\
	&=	\mu(Y_t) S_t \dd t + \sigma(Y_t) S_t \big( \dd \Wt_t^S - \gam_t \dd t \big) \\
	&=	\sigma(Y_t) S_t \dd \Wt_t^S , \label{eq:dS-rn}
\end{align}
where we have used \eqref{eq:gam-sv}.  What about the dynamics of $Y$?  Well, they must be of the form
\begin{align}
\dd Y_t
	&=	\alpha(Y_t) \dd t + \beta(Y_t) \dd W_t^Y \\
	&=	\alpha(Y_t) \dd t + \beta(Y_t) ( \rho \dd W_t^S + \rhob \dd W_t ) \\
	&=	\alpha(Y_t) \dd t + \beta(Y_t) \rho ( \dd \Wt_t^S - \gam_t \dd t)  + \beta(Y_t) \rhob ( \dd \Wt_t  - \lam_t \dd t) \\
	&=	\big( \alpha(Y_t) - \rho \beta(Y_t) \gam_t - \rhob \beta(Y_t) \lam_t \big) \dd t + \beta(Y_t) ( \rho \dd \Wt_t^S + \rhob \dd \Wt_t )  \\
	&=	\big( \alpha(Y_t) - \rho \beta(Y_t) \frac{ \mu(Y_t) }{\sig(Y_t)} - \rhob \beta(Y_t) \lam_t \big) \dd t + \beta(Y_t) \dd \Wt_t^Y ,
\end{align}
where we have introduced $\Wt^Y = (\Wt_t^Y)_{0 \leq t \leq T}$, which is defined as follows
\begin{align}
\Wt_t^Y
	&:=	\rho  \Wt_t^S + \rhob \Wt_t .
\end{align}
One can check that $\Wt^Y$ is a Brownian motion under $\Pbt$ and that $\dd \Wt_t^S \dd \Wt_t^Y = \rho \dd t$.  As $Y$ is not a traded asset, there is no reason why it should be a martingale.  As such, we, have the freedom to choose $\lam$ as we like.  It is reasonable to assume that $\lam$ is of the form $\lam_t = \lam(t,S_t,Y_t)$.  Defining
\begin{align}
\Lam(t,s,y)
	&:=	\rho \frac{ \mu(y) }{\sig(y)} + \rhob  \lam(t,s,y) ,
\end{align}
we can express the dynamics of $Y$ under $\Pbt$ as follows
\begin{align}
\dd Y_t
	&=	( \alpha(Y_t) - \beta(Y_t) \Lam(t,S_t,Y_t) ) \dd t + \beta(Y_t) \dd \Wt_t^Y . \label{eq:dY-rn}
\end{align}
Finally, as the bond price $B$ is a constant, the price of the option $V$ is given by
\begin{align}
V_t
	&=	\Ebt[ \phi(S_T) | \Fc_t ]
	=		\Ebt[ \phi(S_T) | S_t, Y_t ] =: v(t,S_t,Y_t) , \label{eq:V-sv}
\end{align}
where the dynamics of $(S,Y)$ under $\Pbt$ are given by \eqref{eq:dS-rn} and \eqref{eq:dY-rn}, respectively.
Note that, in the second equality above, we have used the fact that $(S,Y)$, as the solution of an SDE, is a Markov process.
If we knew the conditional density of $S_T$ given $(S_t,Y_t)$ under $\Pbt$, we could compute the option price $V_t$ using \eqref{eq:V-sv}.
Unfortunately, it is typically not possible to find $\Pbt( S_T \in \dd s | S_t, Y_t)$ in a stochastic volatility setting.
Nevertheless, we can derive a PDE and terminal condition for the option price $V_t = v(t,S_t,Y_t)$.
\\[0.5em]
The option price $V$ must be a martingale under $\Pbt$ (otherwise there would be an arbitrage).  Thus, if we take the differential of $\dd V_t$, the $\dd t$-term must equal zero.  By It\^o's Lemma, we have from \eqref{eq:dS-rn} and \eqref{eq:dY-rn} that
\begin{align}
\dd v(t,S_t,Y_t)
	&=	\d_t v \dd t + \d_s v \dd S_t + \d_y v \dd Y_t + \tfrac{1}{2} \d_s^2 v (\dd S_t )^2 + \tfrac{1}{2} \d_y^2 v (\dd Y_t )^2 + \d_s \d_y v \dd S_t \dd Y_t \\
	&=	( \d_t + \Act ) v(t,S_t,Y_t) \dd t + ( \ldots ) \dd \Wt_t^S + ( \ldots ) \dd \Wt_t^Y ,
\end{align}
where the operator $\Act$ is given by \eqref{eq:A-tilde}  and the particular form of the $( \ldots )$ terms is not particularly important.  Setting the $\dd t$-term equal to zero, we find that $v$ must satisfy
\begin{align}
0
	&=	(\d_t + \Act) v , &
v(T,s,y)
	&=	\phi(s) , \label{eq:pde.sv-2}
\end{align}
where the terminal condition comes from the fact that $v(T,S_T,Y_T) = \phi(S_T)$.
Comparing \eqref{eq:pde.sv-2} with \eqref{eq:pde.sv}, we see that we have obtained the same PDE for the option price using risk-neutral pricing as we did using a replication argument.

\section{The Heston model}
In this section, we will show how to find the solution $v$ of the pricing PDE \eqref{eq:pde.sv-2} when the underlying stock price $S$ has Heston dynamics.
Under the physical measure $\Pb$, the dynamics of $S$ are described in Example \ref{ex:heston}.
The operator $\Act$ defined in \eqref{eq:A-tilde}, is given explicitly by
\begin{align}
\Act 
	&= \frac{1}{2} y s^2 \d_s^2 + \rho \del s y \d_{sy}^2 + \frac{1}{2} \del^2 y \d_{yy}^2 + \[ \kappa(\theta - y) - \delta \sqrt{y} \Lambda(t,s,y) \] \d_y .
\end{align}
One typically guesses that $\Lam$ is of the form $\Lam(t,s,y) = \lam \sqrt{y}$.  Inserting this form into the above expression for $\Act$ and performing some algebra, we obtain
\begin{align}
\Act
	&= \frac{1}{2} y s^2 \d_s^2 + \rho \del y s \d_{sy}^2 + \frac{1}{2} \delta^2 y \d_{yy}^2 + \kappa^* (\theta^* - y) \d_y, \label{eq:heston.A}
\end{align}
where $\kappa^* = \kappa + \delta \lambda$, $\theta^* = \kappa \theta / (\kappa + \delta \lambda)$.
To ease notation, we will make the notational change $\kappa^* \to \kappa$ and $\theta^* \to \theta$.
With this change, pricing PDE \eqref{eq:pde.sv-2} then becomes
\begin{align}
0
	&=	\Big (\d_t + \frac{1}{2} y s^2 \d_s^2 + \rho \del y s \d_s \d_y + \frac{1}{2} \delta^2 y \d_y^2 + \kappa (\theta - y) \d_y \Big) v(t,s,y) , &
v(T,s,y)
	&=	\phi(s) . \label{eq:heston.pde}
\end{align}
In order to solve this PDE, let us define
\begin{align}
v(t,s,y)
	&:=	u(t,x(s),y) , &
x(s)
	&:=	\log s . \label{eq:variable.change}
\end{align}
Inserting \eqref{eq:variable.change} into \eqref{eq:heston.pde} and using
\begin{align}
\d_s v
	&=	\frac{1}{s} \d_x u , &
\d_s^2 v
	&=	\frac{1}{s^2} ( \d_x^2 - \d_x ) u , &
\d_y v
	&= \d_y u , \\
\d_y^2 v
	&=	\d_y^2 u , &
\d_t v
	&=	\d_t u , &
\d_s \d_y v
	&=	\frac{1}{s} \d_x \d_y u ,
\end{align}
we find that $u$ satisfies
\begin{align}
0
	&=	\Big (\d_t + \tfrac{1}{2} y (\d_x^2 - \d_x )+ \rho \del y \d_x \d_y + \tfrac{1}{2} \delta^2 y \d_y^2 + \kappa (\theta - y) \d_y \Big) u(t,x,y) ,
	\label{eq:heston.pde.2} \\
u(T,x,y)
	&=	 \phi(\ee^x) =: \varphi(x) . \label{eq:heston.tc.2}
\end{align}
Now, let us introduce the \textit{Fourier Transform}.  For any function $f$ we define its Fourier transform $\fh$ by
\begin{align}
\fh(\xi)
	&:=	\int_\Rb \dd x \, \ee^{- \ii \xi x} f(x) . \label{eq:ft}
\end{align}
It is well known that one can recover the function $f$ from its Fourier transform $\fh$ using the \textit{Inverse Fourier transform}
\begin{align}
f(x)
	&= \frac{1}{2 \pi} \int_\Rb \dd \xi \, \ee^{\ii \xi x} \fh(\xi) . \label{eq:ift}
\end{align}
Strictly speaking, we should place some conditions on the function $f$ above (e.g., that it is integrable and continuous).  But, for our purposes, we do not need to get into too much technical detail here.  We will use Fourier transforms to solve PDE \eqref{eq:heston.pde.2}.  To begin, we multiply \eqref{eq:heston.pde.2} and \eqref{eq:heston.tc.2} by $\ee^{- \ii \xi x}$ and integrate with respect to $x$.  Defining the Fourier transform of $u$
\begin{align}
\uh(t,\xi,y)
	&:=	\int_\Rb \dd x \, \ee^{- \ii \xi x} u(t,x,y) ,
\end{align}
and noting that
\begin{align}
\int_\Rb \dd x \, \ee^{- \ii \xi x} \d_t u(t,x,y)
	&=	\d_t  \int_\Rb \dd x \, \ee^{- \ii \xi x} u(t,x,y)
	=		\d_t \uh(t,\xi,y) , \\
\int_\Rb \dd x \, \ee^{- \ii \xi x} \d_x u(t,x,y)
	&=	- \int_\Rb \dd x \, u(t,x,y) \d_x \ee^{- \ii \xi x} 
	=		\ii \xi \uh(t,\xi,y) , \\
\int_\Rb \dd x \, \ee^{- \ii \xi x} \d_x^2 u(t,x,y)
	&=	\int_\Rb \dd x \, u(t,x,y) \d_x^2 \ee^{- \ii \xi x} 
	=		- \xi^2 \uh(t,\xi,y) ,
\end{align}
we obtain the following PDE for $\uh$
\begin{align}
0
	&=	\Big (\d_t + \tfrac{1}{2} y (- \xi^2 - \ii \xi ) + \rho \ii \xi \del y \d_y + \tfrac{1}{2} \delta^2 y \d_y^2 + \kappa (\theta - y) \d_y \Big) \uh(t,\xi,y) ,
	\label{eq:heston.pde.3} \\
\uh(T,\xi,y)
	&=	\widehat{\varphi}(\xi) := \int_\Rb \dd x \, \ee^{-\ii \xi x }\varphi(x) . \label{eq:heston.tc.3}
\end{align}
To solve PDE \eqref{eq:heston.pde.3} we guess that
\begin{align}
\uh(t,\xi,y)
	&=	\widehat{\varphi}(\xi) \ee^{C(t,\xi) + y D(t,\xi)} . \label{eq:guess}
\end{align}
where we impose that
\begin{align}
C(T,\xi) 
	&=	0 , &
D(T,\xi)
	&=	0 , &
	&\text{so that}& 
\uh(T,\xi,y) 
	&= \widehat{\varphi}(\xi) .
\end{align}
Inserting our guess \eqref{eq:guess} into \eqref{eq:heston.pde.3}, using
\begin{align}
\d_t \uh(t,\xi,y)
	&=	\big(\d_t C(t,\xi) + y \d_t D(t,\xi) \big) \uh(t,\xi,y) , \\
\d_y \uh(t,\xi,y)
	&=	D(t,\xi) \uh(t,\xi,y) , \\
\d_y^2 \uh(t,\xi,y)
	&=	D^2(t,\xi) \uh(t,\xi,y) , 
\end{align}
and then dividing the result by $\uh(t,\xi,y)$, we obtain
\begin{align}
0
	&=	\d_t C(t,\xi) + y \d_t D(t,\xi) + \tfrac{1}{2} y (- \xi^2 - \ii \xi ) + \rho \ii \xi \del y D(t,\xi) 
			+ \tfrac{1}{2} \delta^2 y D^2(t,\xi) + \kappa (\theta - y) D(t,\xi) .
\end{align}
The above equation must hold for \textit{every} possible value of $y$.    The only way for this to hold is for the terms that contain $y$ and the terms that do not contain $y$ to sum to zero.  That is, we must have
\begin{align}
0
	&=	\d_t C(t,\xi)  + \kappa \theta D(t,\xi), \\
0
	&=	\d_t D(t,\xi) + \tfrac{1}{2} (- \xi^2 - \ii \xi ) + \rho \ii \xi \del D(t,\xi)  + \tfrac{1}{2} \delta^2 D^2(t,\xi) - \kappa D(t,\xi) .
\end{align}
The second equation above is an Ordinary Differential Equation (ODE) in $t$ for the function $D(t,\xi)$.  Specifically, it is a \textit{Riccati} ODE, which can be solved explicitly.  Once we have found the function $D(t,\xi)$ we can obtain $C(t,\xi)$ by integrating.  From 
$ -\dd C(t,\xi)  = \kappa \theta D(t,\xi) \dd t $
we obtain
\begin{align}
- \int_t^T \dd C(t,\xi)
	&=	- \Big( C(T,\xi) - C(t,\xi) \Big)
	=		C(t,\xi) 
	=	 \kappa \theta \int_t^T D(s,\xi) \dd s .
\end{align}
We will not give the expressions for $C$ and $D$ here, as the formulas are rather complicated and we would not learn much by staring at them.  
A curious reader can find expressions for $C$ and $D$ in any paper written on the Heston model (of which there are \textit{hundreds}).
To obtain $u(t,x,y)$ from $\uh(t,\xi,y)$ we make use of the inverse Fourier transform \eqref{eq:ift}.  We have
\begin{align}
u(t,x,y)
	&=	\frac{1}{2 \pi} \int_\Rb \dd \xi \, \ee^{\ii \xi x} \uh(t,\xi,y) \\
	&=	\frac{1}{2 \pi} \int_\Rb \dd \xi \, \ee^{\ii \xi x} \widehat{\varphi}(\xi)\ee^{C(t,\xi) + y D(t,\xi)} .
\end{align}
Finally, to obtain the option price $v(t,s,y)$, simply use $v(t,s,y) = u(t,x(s),y) =  u(t,\log s,y)$.

\section{Calibration of the Heston model}
\label{sec:calibration}
Let us denote the price of a Call option in the Heston model with Strike $K$ and maturity $T$ as follows
\begin{align}
	&	C^\text{H}(t,S_t;T,K,\Xi) , &
\Xi
	&:= (Y_t,\kappa, \theta, \delta, \rho) .
\end{align}
Note that, while $(t,S_t,T,K)$ are all observable, the parameters $\Xi:=(Y_t, \kappa, \theta, \delta, \rho)$ are \textit{not} directly observable and must be found via calibration.  Roughly speaking, \textit{calibration} is the process of finding the unobservable parameters $\Xi$ that minimize the difference (in some sense) between model-induced Call prices and market quotes.  Because Call prices with differing strikes and maturities have wildly different prices, it is common to perform a calibration procedure with respect to implied volatilities, which, as unitless quantities, do not vary too much across strikes and maturities.  In this way, Call options with different strikes and maturities are more equally weighted in the calibration procedure.
\\[0.5em]
With the above in mind, we define the \textit{Heston implied volatility} $\sigma^\text{H}(T,K;\Xi)$ and the \textit{Market implied volatility} $\sigma^\text{M}(T,K)$ as the unique positive solutions of
\begin{align}
C^\text{BS}(t,S_t;\sigma^\text{H}(T,K;\Xi),T,K) 
	&= C^\text{H} (t,S_t;T,K,\Xi) , \\
C^\text{BS} (t,S_t;\sigma^\text{M}(T,K),T,K) 
	&= C_t^\text{M}(T,K) .
\end{align}
where, as usual, $C^{BS}(t,S_t;\sig,T,K)$ denotes the price of a call option in the Black-Scholes setting with maturity $T$, strike $K$ and volatility $\sig$.
In a typical calibration procedure, the optimal unobservable parameters $\Xi^*$ are defined as follows
\begin{align}
	\Xi^* 
		&:= {\arg \min}_\Xi \sum_{(T,K)} \( \sigma^\text{M} (T,K) - \sigma^\text{H}(T,K,\Xi) \)^2.
\end{align}
If the Heston model truly captured the risk-neutral dynamics of the underlying $S$, once the optimal parameters $\Xi^*=(Y_t^*,\kappa^*, \theta^*, \delta^*, \rho^*)$ are obtained, all of them except $Y_t^*$ (which fluctuates randomly over time) would remain fixed forever.  In practice, however, banks will re-calibrate on a daily basis.  Of course, this means that the Heston model does not truly capture the risk-neutral dynamics of the underlyer.  But, recalibration is a standard industry practice.
\\[0.5em]
At this point, one might ask the following: \textit{if we used market prices to calibrate a model, why do we even need a model?  Would it not be better to just buy and sell options at the market price?}  There are various reasons to need a model for an underlying.  First, a model is required in order to derive replication strategies for options.  It is entirely possible that model A and model B result in identical European option prices but result in different replication strategies.  Second, a model is needed to price path-dependent (i.e., not European) derivatives.


\section{Exercises}
\label{sec:exer-9}

\begin{exercise}
\label{exer:9-1}
Consider the following stochastic volatility model, written under the market's chosen pricing measure $\Pbt$
\begin{align}
\dd S_t
	&=	\sig(Y_t) S_t \dd \Wt_t^S , \\
\dd Y_t
	&=	\alpha(Y_t) \dd t + \beta(Y_t) \dd \Wt_t^Y ,
\end{align}
where $\Wt^S$ and $\Wt^Y$ are \textit{independent} Brownian motions.  As in most of this chapter, we will suppose that the risk free rate is zero: $r=0$, so that the time $t$ value $V_t$ of an option with payoff $\phi(S_T)$ at time $T$ is given by 
\begin{align}
V_t
	&=	\Ebt [ \phi(S_T) | \Fc_t ] .
\end{align}
(a) Define $X_t = \log S_t$.  Write the dynamics of $X$.\\
(b) Let us denote by $\Fc_t^Y$ the filtration generated by observing $Y$ (or, alternatively, by observing $\Wt^Y$) up to time $t$.  Show that $X_t | \Fc_t^Y$ is normally distributed with a mean $m_t$ and variance $v_t^2$ that depend on the path of $Y$.  What are the mean $m_t$ and variance $v_t^2$?\\
(c) Show that
\begin{align}
V_t
	&=	\Ebt[ v^\BS(t,S_t; \overline{\sig}_{t,T} ) | \Fc_t ] , &
\overline{\sig}_{t,T}^2
	&:=	\frac{1}{T-t} \int_t^T \sig^2(Y_s) \dd s , \label{eq:hull-white}
\end{align}
where $v^\BS(t,S_t;\overline{\sig})$ is the Black-Scholes price of a derivative with payoff $\phi(S_T)$ and volatility $\overline{\sig}$.\\
(d) Suppose $\Wt^S$ and $\Wt^Y$ are \textit{not} independent.  Does \eqref{eq:hull-white} still hold?  Why or why not?
\end{exercise}




\chapter{Barrier Options}
So far, we have focused mainly on European derivatives.  These are the simplest type of derivatives because their final values depend only on the final value of the underlying asset.  There are, however, many derivatives whose values depend on the entire path of the underlying from inception to maturity.  
In this chapter we will learn how to price and replicate a variety of \textit{Barrier options}, which are \textit{path-dependent} derivatives.
\\[0.5em]
Let $\tau_m$ be the first time $S$ hits a level $m$.  Mathematically, we can write this as
\begin{align}
\tau_m 
	&:= \inf \{ t \geq 0 : S_t = m \} .
\end{align}
Note that $\tau_m$ is a random time.  
The payoff of a variety of {Barrier options} can be expressed in terms of $\tau_m$. 
\begin{align}
\text{Knock-out Payoff} 
		&= \phi(S_T) \Ib_{\{ \tau_m > T \}} , \\
\text{Knock-in Payoff} 
		&= \phi(S_T) \Ib_{\{ \tau_m \leq T \}} .
\end{align}
Recall that an \textit{indicator function} $\Ib$ is defined as follows
\begin{align}
\Ib_A
	&=	\left\{ \begin{aligned}
			&1 && \text{if event A occurs}, \\
			&0	&& \text{if event A does not occur} .
			\end{aligned} \right. 
\end{align}
Thus, a knock-out option pays the option holder $\phi(S_T)$ only if $S$ does not hit a level $m$ prior to the maturity date $T$.
A knock-in option, on the other hand, pays the option holder $\phi(S_T)$ only if $S$ hits a level $m$ prior to the maturity date $T$.

\section{Parity Relations} 
Let $V^\text{E}=(V_t^\text{E})_{0 \leq t \leq T}$, $V^\text{KO}=(V_t^\text{KO})_{0 \leq t \leq T}$ and $V^\text{KI}=(V_t^\text{KI})_{0 \leq t \leq T}$ denote, respectively, the values of a European, Knock-out and Knock-in option, each with same payoff function $\phi$.  Noting that $1 = \Ib_{\{ \tau_m > T \}} + \Ib_{\{ \tau_m \leq T \}}$, we have
\begin{align}
\phi(S_T) &= \phi(S_T) \Ib_{\{ \tau_m > T \}} + \phi(S_T) \Ib_{\{ \tau_m \leq T \}} .
\end{align}
Dividing both sides of the above equation by $B_T$, and taking conditional expectations under the risk-neutral pricing measure $\Pbt$, we obtain
\begin{align}
	& &
\Ebt \[ \frac{\phi(S_T)}{B_T} \Big| \Fc_t^S \] 
	&= \Ebt \[ \frac{ \phi(S_T) \Ib_{\{ \tau_m > T \}} }{B_T} \Big| \Fc_t^S \] + \Ebt \[ \frac{ \phi(S_T) \Ib_{\{ \tau_m \leq T \}} }{B_T} \Big| \Fc_t^S \]  \\
	&\Rightarrow& 
\frac{V_t^\text{E} }{B_t}
	&= \frac{V_t^\text{KO}}{B_t} + \frac{V_t^\text{KI}}{B_t} \\
	&\Rightarrow& 
V_t^\text{E} 
	&= V_t^\text{KO} + V_t^\text{KI} . \label{eq:KI-KO-parity}
\end{align}
Equation \eqref{eq:KI-KO-parity} is a \textit{knock-in knock-out parity} relation.  
If we can compute or observe the values of a European and knock-out options, then we can use \eqref{eq:KI-KO-parity} to deduce the value of a knock-in option.
Note that the above result is a model-free result, in the sense that we have not had to assume anything about the dynamics of the underlying asset $S$.
 
\section{Static Hedging of Barrier options}
In this section, we will show how to price and replicate a knock-out barrier option under the assumption that $r = 0$ and that $S$ follows a geometric Brownian motion (i.e., the Black-Scholes model).  The price and replication strategy for a knock-in option can be deduced from \eqref{eq:KI-KO-parity}.  Under the risk-neutral pricing measure, the dynamics of $S$ are given by
\begin{align}
\dd S_t 
	&= \sigma S_t \dd \Wt_t , \label{eq:S.gbm}
\end{align}
where $\Wt$ is a Brownian motion under $\Pbt$.

\begin{lemma}[Put-Call Symmetry]
Let $S$ be given by \eqref{eq:S.gbm}.
Then, for any function $G: \Rb_+ \to \Rb$ we have
\begin{align}
\Ebt[ G(S_T) | S_t ]
	&=	\Ebt \[ \frac{S_T}{S_t} G \Big( \frac{S_t^2}{S_T} \Big) \Big| S_t \] . \label{eq:pcs}
\end{align}
\end{lemma}

\begin{proof}
Let $X = (X_t)_{0 \leq t \leq T}$ be defined as follows $X_t = \log S_t$.  Then we have
\begin{align}
X_T | X_t
	&=	X_t - \tfrac{1}{2}\sig^2(T-t) + \sig ( \Wt_T - \Wt_t )
	\sim N \Big( X_t - \tfrac{1}{2}\sig^2(T-t) , \sig^2(T-t) \Big).
\end{align}
We compute
\begin{align}
&\Ebt[ G(S_T) | S_t ] 
	 =	\Ebt[ G(\ee^{X_T}) | X_t ] \\
	&=	\int_\Rb \dd x \, G(\ee^x) \frac{1}{\sqrt{2\pi\sig^2(T-t)}} \exp \Big( \frac{-(x - X_t + \tfrac{1}{2} \sig^2(T-t))^2}{2 \sig^2(T-t)} \Big) \\
	&=	\int_\Rb \dd y \, G(\ee^{2 X_t - y}) \frac{1}{\sqrt{2\pi\sig^2(T-t)}} \exp \Big( \frac{-(-y + X_t + \tfrac{1}{2} \sig^2(T-t))^2}{2 \sig^2(T-t)} \Big) && (x = 2X_t - y)\\
	&=	\int_\Rb \dd y \, \ee^{y - X_t} G(\ee^{2 X_t - y}) \frac{1}{\sqrt{2\pi\sig^2(T-t)}} \exp \Big( \frac{-(y - X_t + \tfrac{1}{2} \sig^2(T-t))^2}{2 \sig^2(T-t)} \Big) \\
	&=	\Ebt[ \ee^{X_T - X_t} G(\ee^{2X_t - X_T}) | X_t ] \\
	&=	\Ebt \[ \frac{S_T}{S_t} G \Big( \frac{S_t^2}{S_T} \Big) \Big| S_t \] ,
\end{align}
which establishes \eqref{eq:pcs}.
\end{proof}

\noindent
We can use Put-Call symmetry in order to statically replicate a knock-out barrier option.

\begin{theorem}
\label{thm:static-hedge}
Let $S$ be given by \eqref{eq:S.gbm}.  Suppose $L < S_0$ and consider a knock-out option with maturity $T$ and payoff $\phi(S_T) \Ib_{\{\tau_L > T \}}$.  
To replicate this claim, purchase at time $t=0$ a European claim with payoff
\begin{align}
\phi^\textup{KO}(S_T)
	&=	\Ib_{\{ S_T > L \}} \phi(S_T) - \Ib_{\{ S_T < L \}} \frac{S_T}{L} \phi \Big(\frac{L^2}{S_T} \Big) .
\end{align}
If and when the barrier claim knocks out, clear the position in $\phi^\textup{KO}(S_T)$ at no cost.
\end{theorem}

\begin{proof}
Suppose that $\tau_L > T$.  This implies that $S_T > L$ and thus
\begin{align}
\phi^\text{KO}(S_T)
	&=	\phi(S_T) , &
	&\text{(when $\tau_L > T$)}
\end{align}
The knock-out claim also pays $\phi(S_T)$ when $\tau_L > T$.  Suppose, on the other hand, that $\tau_L \leq T$.  Then, at time $\tau_L$ we have
\begin{align}
\Ebt[ \Ib_{\{ S_T > L \}} \phi(S_T) | S_{\tau_L} ]
	&=	\Ebt\Big[ \Ib_{\{ S_{\tau_L}^2/S_T > L \}} \frac{S_T}{S_{\tau_L}} \phi \Big( \frac{S_{\tau_L}^2}{S_T} \Big) \Big| S_{\tau_L} \Big] &
	&\text{(by \eqref{eq:pcs})} \\
	&=	\Ebt\Big[ \Ib_{\{ L^2/S_T > L \}} \frac{S_T}{L} \phi \Big( \frac{L^2}{S_T} \Big) \Big| S_{\tau_L} \Big] &
	&\text{(because $S_{\tau_L} = L$)} \\
	&=	\Ebt\Big[ \Ib_{\{ L > S_T \}} \frac{S_T}{L} \phi \Big( \frac{L^2}{S_T} \Big) \Big| S_{\tau_L} \Big] ,
\end{align}
and thus $\Ebt[ \phi^\text{KO}(S_T) | S_{\tau_L} ] = 0$.
\end{proof}

\begin{example}
Consider a knock-out call option with payoff
\begin{align}
(S_T - K)^+ \Ib_{\{\tau_L > T\}} , && 0 < L < S_0 < K . \label{eq:KO-call}
\end{align}
Let us find the European option that statically replicates this claim.  First we identify
\begin{align}
\phi(S_T)
	&=	(S_T - K)^+ .
\end{align}
Then, according to Theorem \ref{thm:static-hedge}, at time zero we should buy a European claim that pays
\begin{align}
\phi^\textup{KO}(S_T)
	&=	\Ib_{\{ S_T > L \}} (S_T-K)^+ - \Ib_{\{ S_T < L \}} \frac{S_T}{L} \Big(\frac{L^2}{S_T} - K\Big)^+ \\
	&=	(S_T-K)^+ - \Ib_{\{ S_T < L \}} \frac{K}{L} \Big( \frac{ L^2}{K} - S_T \Big)^+ \\
	&=	(S_T-K)^+ -  \frac{K}{L} \Big( \frac{ L^2}{K} - S_T \Big)^+ ,
\end{align}
where, in the last line we have used that $S_T < L^2/K$ implies $S_T < L$.  Thus, to statically replicate the knock-out call option \eqref{eq:KO-call}, at time $t=0$ one should buy a single European call option with Strike $K$ and sell $K/L$ European put options with strike $L^2/K$.  Now, consider a knock-in call option with payoff
\begin{align}
(S_T - K)^+ \Ib_{\{\tau_L \leq T\}} , && 0 < L < S_0 < K . \label{eq:KI-call}
\end{align}
How would you statically replicate this claim?  \textit{Hint}: use \eqref{eq:KI-KO-parity}.
\end{example}


\section{Exercises}
\label{sec:exer-10}

\begin{exercise}
\label{exer:10-1}
Assume that the risk-free rate of interest is zero $r=0$ and the dynamics of $S$ under the risk-neutral pricing measure $\Pbt$ are given by $\dd S_t = \sig S_t \dd \Wt_t$.   Recall the put-call symmetry (PCS) result
\begin{align}
\Ebt [ G(S_T) | S_t ]
	&=	\Ebt [ (S_T/S_t) G(S_t^2/S_T) | S_t ] .
\end{align}
Let $\tau_L := \inf \{ t \geq 0 : S_t = L \}$ be the first hitting time of $S$ to level $L < S_0$.\\
\textbf{(a)} Consider a knock-out option with a payoff $\Ib_{\{ \tau_L > T\}} \theta(S_T - K)$ where $\theta$ is a Heaviside function (i.e. $\theta(x)=1$ if $x \geq 0$ and $\theta(x) = 0$ if $x<0$).  Assume $K > L$.  Find the payoff function $\phi(S_T)$ of a European claim that has the same value of the knock-out option at any time $t \leq T \wedge \tau_L$ (i.e., find the payoff function of a European claim that statically hedges the knock-out claim).\\
\textbf{(b)} Consider a knock-in option with a payoff $\Ib_{\{ \tau_L \leq T\}} \theta(S_T - K)$.  Assume $K > L$.  Find the payoff function $\psi(S_T)$ of a European claim that has the same value of the knock-in option at any time $t \leq T \wedge \tau_L$ (i.e., find the payoff function of a European claim that statically hedges the knock-in claim).\\
\textbf{(c)} Plot the functions $\phi(S_T)$ and $\psi(S_T)$ obtained in parts (a) and (b) for $L = 1$ and $K=2$.
\end{exercise}

%
%

%
%

%
%

\appendix

\chapter{Solutions to Exercises}


\section{Exercises from Section \ref{sec:exer-A}}

\subsection*{Solution to Exercise \ref{exer:A-1}}
Recall from Example \ref{ex:die.roll.rv} that $f_X(0) = f_X(1) = 1/2$, $\ft_X(0) = 12/21 = 4/7$ and $\ft_X(1) = 9/21 = 3/7$.  Using \eqref{eq:E.disc} with $g(x) = x$ and $g(x) = \ee^x$, we have
\begin{align}
\text{(a)} \quad \Eb X
	&=	f_X(0) \cdot 0 + f_X(1) \cdot 1
	=		\fbox{$\tfrac{1}{2}$} , &
\text{(b)} \quad \Eb \ee^X
	&=	f_X(0) \ee^0 + f_X(1) \ee^1
	=		\fbox{$\tfrac{1}{2} + \tfrac{1}{2} \ee$} , \\
\text{(c)} \quad \Ebt X
	&=	\ft_X(0) \cdot 0 + \ft_X(1) \cdot 1
	=		\fbox{$\tfrac{3}{7}$} , &
\text{(d)} \quad \Ebt \ee^X
	&=	\ft_X(0) \ee^0 + \ft_X(1) \ee^1
	=		\fbox{$\tfrac{4}{7} + \tfrac{3}{7} \ee$} .
\end{align}


\subsection*{Solution to Exercise \ref{exer:A-2}}
The state space of $S_1$ is $\{ u S_0, d S_0 \}$ and the state space of $S_2$ is $\{ u^2 S_0, u d S_0, d^2 S_0 \}$.  Note that $\Pb(S_2 = u d S_0) = \Pb(HT) + \Pb(TH) = 2pq$.
\begin{enumerate}
\item The joint PMF $f_{S_1,S_2}(s_1,s_2) = \Pb(S_1 = s_1, S_2 = s_2)$ takes the following six values
\begin{align}
f_{S_1,S_2}(uS_0,u^2S_0) &= \Pb(\om_1 = H, \om_2 = H) = p^2 , &
f_{S_1,S_2}(dS_0,u^2S_0) &= 0 , \\
f_{S_1,S_2}(uS_0,udS_0) &= \Pb(\om_1 = H, \om_2 = T) = pq , &
f_{S_1,S_2}(dS_0,udS_0) &= \Pb(\om_1 = T, \om_2 = H) = pq , \\
f_{S_1,S_2}(uS_0,d^2S_0) &= 0 , &
f_{S_1,S_2}(dS_0,d^2S_0) &= \Pb(\om_1 = T, \om_2 = T) = q^2 .
\end{align}
\item Using $f_{S_1|S_2}(s_1,s_2) = f_{S_1,S_2}(s_1,s_2) / f_{S_2}(s_2)$, we obtain
\begin{align}
f_{S_1|S_2}(uS_0,u^2S_0) &= \frac{p^2}{p^2} = 1 , &
f_{S_1|S_2}(uS_0,udS_0) &= \frac{pq}{2pq} = \frac{1}{2} , &
f_{S_1|S_2}(uS_0,d^2S_0) &= 0 , \\
f_{S_1|S_2}(dS_0,u^2S_0) &= 0 , &
f_{S_1|S_2}(dS_0,udS_0) &= \frac{pq}{2pq} = \frac{1}{2} , &
f_{S_1|S_2}(dS_0,d^2S_0) &= \frac{q^2}{q^2} = 1 .
\end{align}
\item Using $f_{S_2|S_1}(s_2,s_1) = f_{S_1,S_2}(s_1,s_2) / f_{S_1}(s_1)$, we obtain
\begin{align}
f_{S_2|S_1}(u^2S_0,uS_0) &= \frac{p^2}{p} = p , &
f_{S_2|S_1}(udS_0,uS_0) &= \frac{pq}{p} = q , &
f_{S_2|S_1}(d^2S_0,uS_0) &= 0 , \\
f_{S_2|S_1}(u^2S_0,dS_0) &= 0 , &
f_{S_2|S_1}(udS_0,dS_0) &= \frac{pq}{q} = p , &
f_{S_2|S_1}(d^2S_0,dS_0) &= \frac{q^2}{q} = q .
\end{align}
\item Using the joint PMF from part (a), we have
\begin{align}
\Eb(S_2/S_1)
	&=	u \, p^2 + d \, pq + u \, pq + d \, q^2
	=		(up + dq)(p + q)
	=		\fbox{$up + dq$} .
\end{align}
\item Using $\Eb[S_2/S_1 | S_1 = s_1] = \sum_{s_2} f_{S_2|S_1}(s_2,s_1) \, s_2/s_1$, where the sum runs over the state space of $S_2$, we have
\begin{align}
\Eb[S_2/S_1|S_1 = uS_0]
	&=	p \frac{u^2S_0}{uS_0} + q \frac{udS_0}{uS_0}
	=		up + dq , &
\Eb[S_2/S_1|S_1 = dS_0]
	&=	p \frac{udS_0}{dS_0} + q \frac{d^2S_0}{dS_0}
	=		up + dq .
\end{align}
Thus, $\Eb[S_2/S_1 | S_1] = \fbox{$up + dq$}$, which is a constant because $S_2/S_1$ is independent of $S_1$.
\end{enumerate}


\subsection*{Solution to Exercise \ref{exer:A-3}}
\textbf{(a)} Let $f_{X,Y}$ be the joint PMF of $(X,Y)$ and recall that $\sum_j f_{X,Y}(x_i,y_j) = f_X(x_i)$ and $\sum_i f_{X,Y}(x_i,y_j) = f_Y(y_j)$.  Then
\begin{align}
\Eb[ a X + b Y ]
	&=	\sum_{i,j} f_{X,Y}(x_i,y_j) (a x_i + b y_j)
	=		a \sum_i x_i \sum_j f_{X,Y}(x_i,y_j) + b \sum_j y_j \sum_i f_{X,Y}(x_i,y_j) \\
	&=	a \sum_i x_i f_X(x_i) + b \sum_j y_j f_Y(y_j)
	=		a \Eb X + b \Eb Y .
\end{align}
\textbf{(b)} Let $f_{X,Y}$ be the joint PDF of $(X,Y)$ and recall that $\int f_{X,Y}(x,y) \dd y = f_X(x)$ and $\int f_{X,Y}(x,y) \dd x = f_Y(y)$.  Then
\begin{align}
\Eb[ a X + b Y ]
	&=	\int \int f_{X,Y}(x,y) (a x + b y) \dd x \dd y
	=		a \int x \Big( \int f_{X,Y}(x,y) \dd y \Big) \dd x + b \int y \Big( \int f_{X,Y}(x,y) \dd x \Big) \dd y \\
	&=	a \int x f_X(x) \dd x + b \int y f_Y(y) \dd y
	=		a \Eb X + b \Eb Y .
\end{align}


\subsection*{Solution to Exercise \ref{exer:A-4}}
\textbf{(a)} Note that $\Eb X$ is a constant, so that $\Eb[ X \Eb X ] = (\Eb X)^2$ and $\Eb[ (\Eb X)^2 ] = (\Eb X)^2$.  Using \eqref{eq:E.linear}, we have
\begin{align}
\Vb X
	&=	\Eb[ X^2 - 2 X \Eb X + (\Eb X)^2 ]
	=		\Eb X^2 - 2 (\Eb X)^2 + (\Eb X)^2
	=		\Eb X^2 - (\Eb X)^2 .
\end{align}
\textbf{(b)} Similarly, as $\Eb X$ and $\Eb Y$ are constants,
\begin{align}
\text{Co}\Vb[X,Y]
	&=	\Eb[ XY - X \Eb Y - Y \Eb X + \Eb X \Eb Y ]
	=		\Eb XY - \Eb X \Eb Y - \Eb Y \Eb X + \Eb X \Eb Y
	=		\Eb XY - \Eb X \Eb Y .
\end{align}
\textbf{(c)} If $X \ind Y$ then, by Section \ref{sec:independence}, $\Eb XY = \Eb X \cdot \Eb Y$, and thus $\text{Co}\Vb[X,Y] = \fbox{0}$.


\subsection*{Solution to Exercise \ref{exer:A-5}}
\textbf{(a)} Suppose $X$ and $Y$ are discrete.\\[0.5em]
\textit{Tower property \eqref{eq:tower}.}  Using \eqref{eq:cond.gen}, \eqref{eq:cond.disc} and \eqref{eq:fXgivenY},
\begin{align}
\Eb \Eb[ g(X,Y) | Y ]
	&=	\Eb \psi(Y)
	=		\sum_j f_Y(y_j) \psi(y_j)
	=		\sum_j f_Y(y_j) \sum_i f_{X|Y}(x_i,y_j) g(x_i,y_j) \\
	&=	\sum_j f_Y(y_j) \sum_i \frac{f_{X,Y}(x_i,y_j)}{f_Y(y_j)} g(x_i,y_j)
	=		\sum_{i,j} f_{X,Y}(x_i,y_j) g(x_i,y_j)
	=		\Eb g(X,Y) .
\end{align}
\textit{Taking out what is known \eqref{eq:known}.}  Using \eqref{eq:cond.disc},
\begin{align}
\Eb[ g(X) h(Y) | Y = y_j ]
	&=	\sum_i f_{X|Y}(x_i,y_j) g(x_i) h(y_j)
	=		h(y_j) \sum_i f_{X|Y}(x_i,y_j) g(x_i)
	=		h(y_j) \Eb[ g(X) | Y = y_j ] .
\end{align}
As this holds for every $y_j$, we have $\Eb[ g(X) h(Y) | Y ] = h(Y) \Eb[ g(X) | Y ]$ by \eqref{eq:cond.gen}.\\[0.5em]
\textit{Independence \eqref{eq:E.ind}.}  If $X \ind Y$ then $f_{X,Y}(x_i,y_j) = f_X(x_i) f_Y(y_j)$, so that
\begin{align}
\Eb[ g(X) | Y = y_j ]
	&=	\sum_i \frac{f_{X,Y}(x_i,y_j)}{f_Y(y_j)} g(x_i)
	=		\sum_i \frac{f_X(x_i) f_Y(y_j)}{f_Y(y_j)} g(x_i)
	=		\sum_i f_X(x_i) g(x_i)
	=		\Eb g(X) .
\end{align}
\textbf{(b)} Suppose $X$ and $Y$ are continuous.  The proofs are identical, with sums replaced by integrals and \eqref{eq:cond.disc}, \eqref{eq:fXgivenY} replaced by \eqref{eq:cond.cont}, \eqref{eq:fXgivenY2}:
\begin{align}
\Eb \Eb[ g(X,Y) | Y ]
	&=	\int f_Y(y) \int \frac{f_{X,Y}(x,y)}{f_Y(y)} g(x,y) \dd x \dd y
	=		\int \int f_{X,Y}(x,y) g(x,y) \dd x \dd y
	=		\Eb g(X,Y) , \\
\Eb[ g(X) h(Y) | Y = y ]
	&=	\int f_{X|Y}(x,y) g(x) h(y) \dd x
	=		h(y) \int f_{X|Y}(x,y) g(x) \dd x
	=		h(y) \Eb[ g(X) | Y = y ] , \\
\Eb[ g(X) | Y = y ]
	&=	\int \frac{f_X(x) f_Y(y)}{f_Y(y)} g(x) \dd x
	=		\int f_X(x) g(x) \dd x
	=		\Eb g(X) , \qquad \text{if $X \ind Y$.}
\end{align}

\subsection*{Solution to Exercise \ref{exer:A-6}}

\begin{enumerate}
    \item We can find
    \begin{align}
        \Eb[1/X] & = \int_{a}^b \frac{1}{x} \frac{1}{b-a} \dd x \\
        {} & = \fbox{$\frac{\log b - \log a}{b - a}$}
    \end{align}
    \item Similarly, $\Eb[1/Y] = \fbox{$\frac{\log d - \log c}{d - c}$} $
    \item Since $X,Y$ are independent, $X,1/Y$ are also independent. This means that $\Eb[X/Y] = \Eb[X]\Eb[1/Y] = \fbox{$\frac{a+b}{2} \frac{\log d - \log c}{d - c}$}$
\end{enumerate}


\clearpage
\section{Exercises from Section \ref{sec:exer-1}}


\subsection*{Solution to Exercise \ref{exer:1-2}}
\textbf{Part (a).}  
We will show that the fair value of the forward contract written on a stock $S$ paying discrete dividends $D_i$ at corresponding times $t_i$ is 
\begin{align}
F_0^T
	= K
	= S_0 \ee^{rT} - \sum_{i=1}^n D_i \ee^{r (T - t_i)},
\end{align}
where $r$ is the risk free rate of interest and $T$ is the expiration time. Suppose that $K > S_0 \ee^{rT} - \sum_{i=1}^n D_i \ee^{r (T - t_i)}$. Then we employ the following strategy:
\begin{enumerate}
	\item At time $t = 0$, enter into the short side of a forward contract on $S$, borrow $S_0$ dollars from the bank at interest rate $r$ and buy $S$. 
	\item At time $t = t_i$, for $i = 1,2,3,\cdots,n$, collect the dividend payment $D_i$ from $S$ and deposit it into the money market account with interest rate $r$. 
	\item At time $t = T$, sell $S$ for $K$ dollars, pay the bank $S_0 \ee^{rT}$ for our loan and withdraw our deposits of the dividend payments.
\end{enumerate}
Our total payoff is thus
\begin{align}
K + \sum_{i=1}^n D_i \ee^{r(T-t_i)} - S_0e^{rT}
	= K - \(   S_0e^{rT} - \sum_{i=1}^n D_i \ee^{r(T-t_i)}   \) > 0.
\end{align}
Since our payoff is deterministic and positive, we have made a riskless profit. 

Now suppose that $K < S_0 \ee^{rT} - \sum_{i=1}^n D_i \ee^{r(T-t_i)}$. We employ the following investment strategy:
\begin{enumerate}
\item At time $t = 0$, enter into the long side of a forward contract on $S$, short-sell $S$ for $S_0$ dollars and invest $S_0$ dollars into a money market account with interest rate $r$. 
\item At times $t = t_i$, for $i = 1,2,3,\cdots, n$, borrow $D_i$ from the bank to pay dividends.
\item At time $t = T$, buy the stock for price $K$, pay back the loans we took from the bank (corresponding to the dividend payments we made) and withdraw our initial investment.
\end{enumerate}
Our payoff is
\begin{align}
S_0 \ee^{rT} - \sum_{i=1}^n D_ie^{r(T-t_i)} - K 
	> 0.
\end{align}
Since our payoff is deterministic and positive, we have made a riskless profit. Thus, the only fair value of the forward contract written on $S$ is
\begin{align}
F_0^T
	= K
	= S_0 \ee^{rT} - \sum_{i=1}^n D_i \ee^{r (T - t_i)}.
\end{align}
Note that $t=0$ is not a special starting point. Therefore, letting $k_t = \min \{ i \mid t \leq t_i \leq T \}$,
\begin{align}
F_t^T
	= S_t \ee^{r(T-t)} - \sum_{i = k_t}^n D_i \ee^{r (T - t_i)}.
\end{align}

\noindent
\textbf{Part (b)}
Assume that we enter the long side of a forward contract on $S$ at $t = 0$ which expires at time $T$. We determined the fair strike price $F_0^T$ in the previous part. At time $t$, we enter the short side of a forward contract expiring at time $T$. The fair strike price of this contract $F_t^T$ was also determined in the last part. Our payoff is thus
\begin{align}
(F_t^T - S_T) + (S_T - F_0^T)
	= F_t^T - F_0^T. 
\end{align}
This is a certain cashflow at time $T$. We know that the present value of this cash flow is
\begin{align}
\(  F_t^T - F_0^T\)e ^{-r(T-t)}.
\end{align}
If time $t$ value $f_t^T$ of a forward contract on $S$ beginning at time $t = 0$ is other than this, an arbitrage opportunity exists. Therefore,
\begin{align}
f_t^T
	&= \( F_t^T - F_0^T \) \ee^{-r (T -t)} \\
	&= \( S_t \ee^{r(T-t)} - \sum_{i = k_t}^n D_i \ee^{r (T - t_i)} - \(S_0 \ee^{r T} - \sum_{i = 1}^n D_i \ee^{r (T - t_i)}\)\) \ee^{-r(T-t)} \\
	&= \( S_t \ee^{r (T-t)} - S_0 \ee^{ r T} + \sum_{i=1}^{k_t-1} D_i \ee^{r(T-t_i)} \) \ee^{-r (T-t)} \\
	&= S_t - S_0 \ee^{r t} + \sum_{i=1}^{k_t - 1} D_i \ee^{ r(t-t_i)}.
\end{align}


\subsection*{Solution to Exercise \ref{exer:1-3}}
\textbf{Part (a).}  The present values at time zero of the two certain cash flows are $\cf_1 \ee^{-r t_1}$ and $\cf_2 \ee^{-r t_2}$.  You receive $\cf_1$ at time $t_1$ and $V_0$ today, and you pay $\cf_2$ at time $t_2$.  The trade is fair (it creates no arbitrage) only if what you receive and what you pay have the same present value
\begin{align}
\cf_1 \ee^{-r t_1} + V_0
	= \cf_2 \ee^{-r t_2} ,
\end{align}
i.e.
\begin{align}
\cf_2
	= \ee^{r t_2} \( \cf_1 \ee^{-r t_1} + V_0 \). \label{cashflow2}
\end{align}

\noindent
\textbf{Part (b)}
Rearranging \eqref{cashflow2} gives
\begin{align}
V_0
	&= \cf_2 \( \ee^{-r t_2} - \frac{\cf_1}{\cf_2} \ee^{-r t_1}\)
	= \cf_2 \( \ee^{-r t_2} - R \ee^{-r t_1}\) .
\end{align}
Since $\cf_2 > 0$, $V_0 > 0$ is equivalent to
\begin{align}
\ee^{-r t_2} - R \ee^{-r t_1}
	> 0
	\qquad \Leftrightarrow \qquad
R < \ee^{-r (t_2 - t_1)}.
\end{align}
\noindent
\textbf{Part (c)}
If $\cf_1 < 0$ and $\cf_2 > 0$, then $R < 0$ and
\begin{align}
V_0
	=\cf_2 \( \ee^{-r t_2} - R \ee^{-r t_1}\)
	> 0.
\end{align}
There is no need for a condition on $R$.


\subsection*{Solution to Exercise \ref{exer:1-4}}
Let us denote the above contract as $\widetilde C_t(T,K_1,K_2)$. The above payoff can be written as
\begin{align}
\text{Payoff}
	&= \( S_T- K_1\)^+ - \( S_T - K_2\)^+.
\end{align}
Therefore, this payoff can be replicated by going long a European option $C_t(T,K_1)$ and going short a European option $C_t(T,K_2)$. Therefore,
\begin{align}
\widetilde C_t(T,K_1,K_2)
	= C_t(T,K_1) - C_t(T,K_2).
\end{align}
Since we know that 
\begin{align}
S_t - K B_t^T 
	\leq C_t(T,K)
	\leq S_t,
\end{align}
we have that 
\begin{align}
-K_1 B_t^T
	&= \( S_t - K_1 B_t^T\) - S_t \\
	&\leq C_t(T,K_1) - C_t(T,K_2) \\
	&\leq S_t - \( S_t - K_2 B_t^T\) \\
	&= K_2 B_t^T.
\end{align}
Therefore,
\begin{align}
-K_1 B_t^T
	\leq \widetilde C_t(T,K_1,K_2)
	\leq K_2 B_t^T.
\end{align}
However, we note that since the payoff of $\widetilde C_t(T,K_1,K_2)$ is non-negative, the value of the option is never negative. Therefore, 
\begin{align}
0
	\leq \widetilde C_t(T,K_1,K_2)
	\leq K_2 B_t^T
\end{align}
is a tighter bound.


\subsection*{Solution to Exercise \ref{exer:1-5}}
At time $t$, sell portfolio $B$ for $X_t^B$, buy portfolio $A$ for $X_t^A$ and put $X_t^B - X_t^A$ in a savings account.  This trading strategy does not cost anything at time $t$ to execute.  At time $T$, the value of the trading strategy is $(X_T^A - X_T^B) + \ee^{r(T-t)} (X_t^B - X_t^A) > 0$.


\subsection*{Solution to Exercise \ref{exer:1-6}}
\textbf{(a)} Observe that
\begin{align}
\begin{pmatrix} h(S_T) \\ g(S_T) \end{pmatrix}
	&=	\begin{pmatrix} a & b \\ c & d \end{pmatrix}
		\begin{pmatrix} (S_T - K)^+ \\ (K - S_T)^+ \end{pmatrix} .
\end{align}
Thus, we have
\begin{align}
\begin{pmatrix} (S_T-K)^+ \\ (K-S_T)^+ \end{pmatrix}
	&=	\frac{1}{ad-bc}
		\begin{pmatrix} d & -b \\ -c & a \end{pmatrix}
		\begin{pmatrix} h(S_T) \\ g(S_T) \end{pmatrix} .
\end{align}
Taking time-zero prices (using linearity of pricing), it follows that
\begin{align}
\begin{pmatrix} C_0(T,K) \\ P_0(T,K) \end{pmatrix}
	&=	\frac{1}{ad-bc}
		\begin{pmatrix} d & -b \\ -c & a \end{pmatrix}
		\begin{pmatrix} H_0 \\ G_0 \end{pmatrix} .
\end{align}
That is,
\begin{align}
C_0(T,K)
	&=	\frac{d H_0 - b G_0}{ad - bc} , &
P_0(T,K)
	&=	\frac{-c H_0 + a G_0}{ad - bc} .
\end{align}
\textbf{(b)} From put-call parity we have
\begin{align}
S_0 - K B_0^T
	&=	C_0(T,K) - P_0(T,K) .
\end{align}
Solving for $B_0^T$ gives
\begin{align}
B_0^T
	&=	\frac{1}{K} \Big( S_0 - C_0(T,K) + P_0(T,K) \Big) .
\end{align}


\subsection*{Solution to Exercise \ref{exer:1-7}}
\textbf{(a)} Denote by $g(S_T)$ the left-hand side of the inequality. We consider four cases.
\begin{itemize}
\item If $S_T \leq K_1$, all three terms are zero, so $g(S_T) = 0$.
\item If $K_1 < S_T \leq K_2$, only the first term is nonzero, so $g(S_T) = \lam (S_T - K_1) > 0$.
\item If $K_2 < S_T \leq K_3$, using $K_2 - \lam K_1 = (1-\lam) K_3$, we have
\begin{align}
g(S_T)
	&=	\lam (S_T - K_1) - (S_T - K_2)
	=		K_2 - \lam K_1 - (1 - \lam) S_T
	=		(1 - \lam)(K_3 - S_T)
	\geq	0 .
\end{align}
\item If $S_T > K_3$, we have
\begin{align}
g(S_T)
	&=	\lam (S_T - K_1) + (1 - \lam)(S_T - K_3) - (S_T - K_2)
	=		K_2 - \lam K_1 - (1-\lam) K_3
	=		0 .
\end{align}
\end{itemize}
Thus $g(S_T) \geq 0$ for every value of $S_T$. The function $g$ is zero outside $(K_1, K_3)$, increases linearly on $(K_1, K_2]$ and decreases linearly on $(K_2, K_3]$.
\\[0.5em]
\textbf{(b)} An arbitrage (see Definition \ref{def:arb}) is a portfolio $X$ that satisfies the following three conditions for some $T > 0$:
\begin{align}
&X_0 = 0 , &
&X_T \geq 0 , &
&\Pb(X_T > 0) > 0 .
\end{align}
Here we start the portfolio at time $t$ rather than at time $0$, which changes nothing in the definition. Suppose that
\begin{align}
\del
	&:=	C_t(T,K_2) - \lam \, C_t(T,K_1) - (1 - \lam) \, C_t(T,K_3)
	>		0 .
\end{align}
At time $t$, consider the following strategy:
\begin{itemize}
\item sell one call with strike $K_2$ (receive $C_t(T,K_2)$),
\item buy $\lam$ calls with strike $K_1$ (pay $\lam \, C_t(T,K_1)$),
\item buy $1-\lam$ calls with strike $K_3$ (pay $(1-\lam) \, C_t(T,K_3)$),
\item invest the remaining amount $\del$ in the bond, i.e., buy $\del / B_t^T$ bonds.
\end{itemize}
The initial value of this portfolio is $X_t = \lam \, C_t(T,K_1) + (1-\lam) \, C_t(T,K_3) - C_t(T,K_2) + \del = 0$. At time $T$, using part (a), we have
\begin{align}
X_T
	&=	\lam (S_T - K_1)^+ + (1 - \lam)(S_T - K_3)^+ - (S_T - K_2)^+ + \frac{\del}{B_t^T}
	=		g(S_T) + \del \, \ee^{r(T-t)}
	\geq	\del \, \ee^{r(T-t)}
	>		0 .
\end{align}
Thus $\Pb(X_T \geq 0) = 1$ and $\Pb(X_T > 0) = 1 > 0$, so $X$ is an arbitrage. We conclude that, in a market with no arbitrage,
\begin{align}
C_t(T,K_2)
	&\leq	\lam \, C_t(T,K_1) + (1 - \lam) \, C_t(T,K_3) ,
\end{align}
i.e., call prices are convex functions of the strike $K$.
\\[0.5em]
\textbf{(c)} By put-call parity, $P_t(T,K) = C_t(T,K) - S_t + K B_t^T$ for every strike $K$. Therefore
\begin{align}
&\lam \, P_t(T,K_1) + (1 - \lam) \, P_t(T,K_3) - P_t(T,K_2) \\
	&\qquad =	\lam \, C_t(T,K_1) + (1 - \lam) \, C_t(T,K_3) - C_t(T,K_2)
				- \big( \lam + (1-\lam) - 1 \big) S_t
				+ \big( \lam K_1 + (1-\lam) K_3 - K_2 \big) B_t^T \\
	&\qquad =	\lam \, C_t(T,K_1) + (1 - \lam) \, C_t(T,K_3) - C_t(T,K_2)
	\geq	0 ,
\end{align}
where we used $\lam + (1-\lam) = 1$, $\lam K_1 + (1-\lam) K_3 = K_2$ and part (b). Hence $P_t(T,K_2) \leq \lam \, P_t(T,K_1) + (1 - \lam) \, P_t(T,K_3)$.


\clearpage
\section{Exercises from Section \ref{sec:exer-2}}

\subsection*{Solution to Exercise \ref{exer:2-1}}
\textit{Case 1: $0 < d < u < 1+r < \infty$:} Consider the following investment strategy. At $n = 0$ we short-sell $S$ for $S_0$ dollars. We then put that money in the bank. At $n = 1$ there are two cases: $S_1 = u S_0$ or $S_1 = d S_0$. In either case, at time $n = 1$ we buy back the asset at its time $n = 1$ price and withdraw the money from the bank. If $S_1 = u S_0$, then our payoff is $S_0 (1+r) - u S_0 > 0$, and if $S_1 = d S_0$, then our payoff is $S_0 (1+r) - d S_0 > 0$. In either case we have made riskless profit i.e. arbitrage has occurred.  \\
\\[0.5em]
\textit{Case 2: $0 < 1+r < d < u < \infty$:} Consider the following investment strategy. At $n=0$ we borrow $S_0$ dollars from the bank and buy the asset $S$.  At $n = 1$ there are two cases: $S_1 = u S_0$ or $S_1 = d S_0$. In either case, we pay back the loan i.e we pay the bank $S_0 (1+r)$ dollars. In the case $S_1 = u S_0$, our payoff is $u S_0 - S_0 (1+r) > 0$. In the case $S_1 = d S_0$, our payoff is $d S_0 - S_0 (1+r) > 0$. In either case, we have made riskless profit i.e. arbitrage has occurred. 


\subsection*{Solution to Exercise \ref{exer:2-2}}
\noindent
\textbf{Part (a):} Let us construct a portfolio $X$ such that 
\begin{align}
X_0
	= \Del_0 S_0 + \frac{ X_0 - \Del_0 S_0}{B_0} B_0,
\end{align}
where $\Del_0$ is the number of shares of $S$ we initially own. At time $n = 1$, our portfolio has the following value
\begin{align}
X_1
	= 
	\left\{ 
	\begin{array}{ll}
	A:= \Del_0 u S_0 + ( X_0 - \Del_0 S_0) (1+r) + a \Del_0 S_0, & \text{if the stock price increased} \\
	B:= \Del_0 d S_0 + ( X_0 - \Del_0 S_0) (1+r) + a \Del_0 S_0, & \text{if the stock price decreased} \\
	\end{array}
	\right. . \label{div.xT}
\end{align}
Now, the value of the option $V$ written on $S$ with expiration date $n = 1$ expires with the value
\begin{align}
V_1
	= 
	\left\{ 
	\begin{array}{lr}
	C:= \phi(u S_0), & \text{if the stock price increased} \\
	D:= \phi(d S_0), & \text{if the stock price decreased} \\
	\end{array}
	\right. . 
\end{align}
Setting $A = C$ and $B = D$, and solving the system of equations for $X_0$ and $\Del_0$ gives
\begin{align}
X_0
	&= \frac{1}{1+r} \Big( \frac{ \phi (d S_0) \left(u + a-(1+r)\right)}{u-d}+\frac{\phi (u S_0) \left((1+r)-d-a\right)}{u-d}\Big), \label{div.x0}\\
\Del_0
	&= \frac{\phi (d S_0)}{S_0 (d-u)}+\frac{\phi (u S_0)}{S_0 (u-d)}. \label{div.del0}
\end{align}

\noindent
\textbf{Part (b).} We set
\begin{align}
\Ebt\[ \frac{X_1}{B_1}\]	
	= \pt A + \qt B
	= \pt A + (1 - \pt) B 
	= \frac{ X_0}{ B_0},
\end{align}
where $A$ and $B$ were defined in \eqref{div.xT}. Solving the above equation for $\pt$ yields 
\begin{align}
\pt
	= \frac{B-X_0 (1+r)}{B-A}.
\end{align}
Inserting $A$ and $B$ from \eqref{div.xT} and $X_0$ and $\Del_0$ from \eqref{div.x0} and \eqref{div.del0} into the above equation gives
\begin{align}
&\pt
	= \frac{1+r-d-a}{u-d},
&\qt
	= 1 - \pt. \label{n.probs}
\end{align}

\noindent
\textbf{Part (c).} Note that in our new probability measure, we have assigned $\Pbt( S_1 = u S_0) = \pt$ and $\Pbt(S_1 = d S_0) = \qt$. Therefore, 
\begin{align}
B_0 \Ebt\[ \frac{\phi(S_1)}{B_1}\]
	= \frac{1}{1+r} \( \pt \phi( u S_0) + \qt \phi(d S_0) \).
\end{align}
Inserting the probabilities from \eqref{n.probs} and simplifying gives
\begin{align}
\frac{1}{1+r}  \( \pt \phi( u S_0) + \qt \phi(d S_0) \)
	&= \frac{1}{1+r} \Big( \frac{ \left(u+a-1-r\right)\phi (d S_0)}{u-d}+\frac{ \left(1+r-d-a\right)\phi (u S_0) }{u-d}\Big).
\end{align}
This value is exactly the value for $X_0$ given in \eqref{div.x0}.

\noindent
\textbf{Part (d).} We have
\begin{align}
\Ebt\[ \frac{S_1}{B_1}\]
	= \frac{ \pt \(u S_0 \) + \qt \( d S_0 \) }{B_0(1+r)} .
\end{align}
Inserting the values of $\pt$ and $\qt$ from \eqref{n.probs} gives
\begin{align}
\frac{ \pt \( u S_0 \) + \qt \( d S_0 \) }{B_0 (1+r)}
	&= \frac{S_0}{B_0} \left(1-\frac{a}{1+r}\right).
\end{align}
Clearly, $\Ebt [ S_1 / B_1] \neq S_0/ B_0$ when $a \neq 0$. 
\\[0.5em]
This can be explained intuitively. We have chosen $\pt$ and $\qt$ such that $ B_0 \Ebt X_1 = X_0$ i.e. we have chosen our probability measure such that there is equal weight on upward and downward movement in our portfolio. The value of our portfolio at $n = 1$ includes a dividend payment $a S_0$. This puts a bias on the movement of the discounted asset price. For example, if $a > 0$, then our selection of $\pt$ and $\qt $ is such that on average, our discounted portfolio $X_1 / B_1$ ends where it started at $n =0$. Since we have a positive dividend payment included, then \textit{on average} the asset price should decrease. 



\subsection*{Solution to Exercise \ref{exer:2-3}}
\textbf{Part (a):} Let us construct a portfolio $X$ such that 
\begin{align}
X_0
	= \Del_0 S_0 + \frac{ X_0 - \Del_0 S_0}{B_0} B_0,
\end{align}
where $\Del_0$ is the number of shares of $S$ we initially own. At time $n = 1$, our portfolio has the following value
\begin{align}
X_1
	= 
	\left\{ 
	\begin{array}{ll}
	A:= \Del_0 u S_0 + ( X_0 - \Del_0 S_0) (1+r), & \text{if the stock price increased} \\
	B:= \Del_0 d S_0 + ( X_0 - \Del_0 S_0) (1+r), & \text{if the stock price decreased} \\
	\end{array}
	\right.
\end{align}
Now, the value of the option $V$ written on $S$ with expiration date $n = 1$ expires with the value
\begin{align}
V_1
	= 
	\left\{ 
	\begin{array}{lr}
	C:= \phi(u S_0), & \text{if the stock price increased} \\
	D:= \phi(d S_0), & \text{if the stock price decreased} \\
	\end{array}
	\right. . 
\end{align}
Setting $A = C$ and $B = D$, and solving the system of equations for $X_0$ and $\Del_0$ gives
\begin{align}
X_0
	&= \frac{1}{1+r} \Big( \frac{ \phi (d S_0) \left(u -(1+r)\right)}{u-d}+\frac{\phi (u S_0) \left((1+r)-d\right)}{u-d}\Big), \\
\Del_0
	&= \frac{\phi (d S_0)}{S_0 (d-u)}+\frac{\phi (u S_0)}{S_0 (u-d)}.
\end{align}

\begin{itemize}
\item[Step 1:] By definition of $\Ebt$,
\begin{align}
\Ebt\[ \frac{ X_1}{B_1} \]
	&= \pt A + \qt B 
	= \pt A + (1 - \pt) B = \frac{X_0}{B_0}.
\end{align}
Since $B_1/B_0 = 1+r$, we set 
\begin{align}
\pt A + (1 - \pt) B
	&= (1+r) X_0
&&\Rightarrow
&&\pt
	= \frac{B-X_0(1+r)}{B-A},\qquad
\qt
	= 1 - \pt. 
\end{align}
Inserting $A$ and $B$ we obtain
\begin{align}
&\pt
	= \frac{1+r-d}{u-d},
&\qt
	= 1 - \pt. \label{n.probs-2}
\end{align}

\item[Step 2:] We then have
\begin{align}
V_0
	= B_0 \Ebt\[ \frac{ V_1}{B_1}\] 
	&= B_0 \Ebt\[ \frac{ \phi(S_1)}{B_1}\] \\
	&= \frac{1}{1+r} \( \pt \phi(u S_0)+ \qt \phi(d S_0) \) \\
	&= \frac{1}{1+r} \Big(\frac{\left(1+r-d \right) \phi (u S_0)}{u-d} + \frac{\phi (d S_0) \left(u - 1-r\right)}{u-d}\Big).\label{v0.parta}
\end{align}
\end{itemize}

\noindent
\textbf{Part (b).} By definition, we have
\begin{align}
\Ebh \[ \frac{ X_1}{S_1}\]
	&= \ph\cdot \frac{A}{u S_0} + \qh\cdot \frac B{d S_0}
	= \ph\cdot \frac{A}{u S_0} + (1-\ph)\cdot \frac B{d S_0}.
\end{align}
Using $B_1/B_0 = 1+r$, and inserting $A$ and $B$ we obtain
\begin{align}\ph\cdot \frac{A}{u S_0} + (1-\ph)\cdot \frac B{d S_0}
	&= \frac{X_0}{S_0}
&&\Rightarrow
&&\ph
	= \frac{u(1+r-d)}{(1+r)(u - d)},\qquad
\qh
	= 1 - \ph.
\end{align}
Computing $V_0$, we have
\begin{align}
V_0
	&= S_0 \Ebh\[ \frac{ V_1}{S_1}\] \\
	&= S_0 \Ebh\[ \frac{ \phi(S_1)}{S_1}\] \\
	&= S_0 \( \ph \frac{ \phi(u S_0)}{u S_0} + \qh \frac{ \phi(d S_0)}{d S_0}\) \\
	&= S_0 \( \[ \frac{u(1+r-d)}{(1+r)(u - d)}\] \frac{ \phi(u S_0)}{u S_0} + \[ 1 - \frac{u(1+r-d)}{(1+r)(u - d)}\] \frac{ \phi(d S_0)}{d S_0}\) \\
	&=  \frac{1}{1+r} \Big(\frac{\left(1+r-d \right) \phi (u S_0)}{u-d} + \frac{\phi (d S_0) \left(u - 1-r\right)}{u-d}\Big). \label{v0.partb}
\end{align}
Note that our expression for $V_0$ \eqref{v0.parta} and \eqref{v0.partb} are equal. 


\subsection*{Solution to Exercise \ref{exer:2-4}}
The key idea is dynamic programming (backwards recursion).  We initialize the option values at the terminal nodes $n=N$ using the put payoff $\phi(s)=(K-s)^+$, and then sweep backwards through the tree using
\begin{align}
X_n^{(i)}
  &= \frac{1}{1+r} \bigl(\pt\, X_{n+1}^{(i+1)} + \qt\, X_{n+1}^{(i)}\bigr),
  \qquad
  \Del_n^{(i)} = \frac{X_{n+1}^{(i+1)}-X_{n+1}^{(i)}}{S_{n+1}^{(i+1)}-S_{n+1}^{(i)}},
\end{align}
where $u=e^{\sig\sqrt{\del t}}$, $d=e^{-\sig\sqrt{\del t}}$, $\pt=\frac{e^{r\del t}-d}{u-d}$, and $\qt=1-\pt$.
The program below reads the six parameters from \texttt{input.txt}, runs the backward recursion using a one-dimensional array (overwriting it in place), and writes $X_0$ and $\Del_0$ to \texttt{output.txt}.

\begin{lstlisting}[caption={European put -- binomial model (\texttt{european\_put.py})}]
"""
european_put.py
Prices and hedges a European put in the N-step binomial model.

Input  (input.txt):  S0, T, K, r, sig, N  (comma-separated)
Output (output.txt): V0, Delta0            (comma-separated)
"""

import math

def price_european_put(S0, T, K, r, sig, N):
    dt   = T / N
    u    = math.exp( sig * math.sqrt(dt))
    d    = math.exp(-sig * math.sqrt(dt))
    disc = math.exp(-r * dt)          # one-step discount factor
    p    = (math.exp(r * dt) - d) / (u - d)   # risk-neutral prob of up
    q    = 1.0 - p

    # --- terminal payoffs ------------------------------------------------
    # node i at time N has stock price u^(N-i) * d^i * S0
    V = [max(K - u**(N - i) * d**i * S0, 0.0) for i in range(N + 1)]

    # --- backward recursion down to time 1 ------------------------------
    for n in range(N - 1, 0, -1):
        for i in range(n + 1):
            V[i] = disc * (p * V[i] + q * V[i + 1])

    # V[0] (up) and V[1] (down) are now the time-1 values
    V0     = disc * (p * V[0] + q * V[1])
    Delta0 = (V[0] - V[1]) / ((u - d) * S0)
    return V0, Delta0


if __name__ == "__main__":
    with open("input.txt", "r") as f:
        S0, T, K, r, sig, N = [
            int(x) if i == 5 else float(x)
            for i, x in enumerate(f.read().split(","))
        ]

    V0, Delta0 = price_european_put(S0, T, K, r, sig, N)

    print(f"V0 = {V0:.6f},  Delta0 = {Delta0:.6f}")

    with open("output.txt", "w") as f:
        f.write(f"{V0},{Delta0}\n")
\end{lstlisting}

\noindent
\textbf{Example.}  With the input file \texttt{50.0, 2.0, 55.0, 0.05, 0.1, 50} the program produces $V_0 \approx 2.6955$ and $\Del_0 \approx -0.4623$ (the Black--Scholes price is $\approx 2.69$).


\subsection*{Solution to Exercise \ref{exer:2-5}}
The only change relative to the European put is that at each interior node we take the maximum of the hold value and the immediate exercise value:
\begin{align}
X_n^{(i)}
  &= \max\!\Bigl(\phi(S_n^{(i)}),\;
      \frac{1}{1+r}\bigl(\pt\, X_{n+1}^{(i+1)} + \qt\, X_{n+1}^{(i)}\bigr)\Bigr).
\end{align}

\begin{lstlisting}[caption={American put -- binomial model (\texttt{american\_put.py})}]
"""
american_put.py
Prices and hedges an American put in the N-step binomial model.

Input  (input.txt):  S0, T, K, r, sig, N  (comma-separated)
Output (output.txt): V0, Delta0            (comma-separated)
"""

import math

def price_american_put(S0, T, K, r, sig, N):
    dt   = T / N
    u    = math.exp( sig * math.sqrt(dt))
    d    = math.exp(-sig * math.sqrt(dt))
    disc = math.exp(-r * dt)
    p    = (math.exp(r * dt) - d) / (u - d)
    q    = 1.0 - p

    def payoff(s):
        return max(K - s, 0.0)

    # --- terminal payoffs ------------------------------------------------
    V = [payoff(u**(N - i) * d**i * S0) for i in range(N + 1)]

    # --- backward recursion with early exercise, down to time 1 ---------
    for n in range(N - 1, 0, -1):
        for i in range(n + 1):
            hold     = disc * (p * V[i] + q * V[i + 1])
            exercise = payoff(u**(n - i) * d**i * S0)
            V[i]     = max(hold, exercise)

    V0     = max(disc * (p * V[0] + q * V[1]), payoff(S0))
    Delta0 = (V[0] - V[1]) / ((u - d) * S0)
    return V0, Delta0


if __name__ == "__main__":
    with open("input.txt", "r") as f:
        S0, T, K, r, sig, N = [
            int(x) if i == 5 else float(x)
            for i, x in enumerate(f.read().split(","))
        ]

    V0, Delta0 = price_american_put(S0, T, K, r, sig, N)

    print(f"V0 = {V0:.6f},  Delta0 = {Delta0:.6f}")

    with open("output.txt", "w") as f:
        f.write(f"{V0},{Delta0}\n")
\end{lstlisting}

\noindent
\textbf{Example.}  With input \texttt{50.0, 2.0, 55.0, 0.05, 0.1, 50} the American put gives $V_0 = 5.0000 = K - S_0$: with these parameters it is optimal to exercise immediately, so the hedge is irrelevant (the holder receives $K - S_0$ now).  The American price exceeds the European price of $\approx 2.6955$, consistent with the early-exercise premium.


\subsection*{Solution to Exercise \ref{exer:2-6}}
Let $\Pbt$ denote the unique risk-neutral measure.  Defining $\pt := \Pbt(S_1 = u S_0)$ and $\qt := \Pbt(S_1 = d S_0)$, we have
\begin{align}
\pt = \frac{1-d}{u-d} , \qquad \qt = \frac{u-1}{u-d} .
\end{align}
Note that $\Ebt[S_1] = S_0$.  Applying Jensen's Inequality under $\Pbt$,
\begin{align}
\Ebt[\phi(S_1)] \geq \phi(\Ebt[S_1]) = \phi(S_0) , \qquad
\Ebt[\theta(S_1)] \leq \theta(\Ebt[S_1]) = \theta(S_0) .
\end{align}
\textbf{(a)} The time-zero value of the American option paying $\phi(S_n)$ if exercised at time $n$ is
\begin{align}
\max\{ \phi(S_0) , \Ebt[\phi(S_1)] \} = \Ebt[\phi(S_1)] = \pt \phi(uS_0) + \qt \phi(dS_0) .
\end{align}
The maximum is achieved by waiting, since $\Ebt[\phi(S_1)] \geq \phi(S_0)$.\\[0.5em]
\textbf{(b)} The time-zero value of the American option paying $\theta(S_n)$ if exercised at time $n$ is
\begin{align}
\max\{ \theta(S_0) , \Ebt[\theta(S_1)] \} = \theta(S_0) .
\end{align}
The maximum is achieved by exercising immediately, since $\Ebt[\theta(S_1)] \leq \theta(S_0)$.


\subsection*{Solution to Exercise \ref{exer:2-7}}
\textbf{(a)} At time $n=0$ we can express $X_0$ as
\begin{align}
X_0
	&=	\Del_0 Z_0 S_0 + \Gam_0 Z_0 B_0 + \Big( X_0 - \Del_0 Z_0 S_0 - \Gam_0 Z_0 B_0 \Big) \frac{1}{A_0} A_0 .
\end{align}
Thus, at time $n=1$,
\begin{align}
X_1
	&=	\Del_0 Z_1 S_1 + \Gam_0 Z_1 B_1 + \Big( X_0 - \Del_0 Z_0 S_0 - \Gam_0 Z_0 B_0 \Big) \frac{A_1}{A_0} .
\end{align}
\textbf{(b)} If the investor puts all his wealth in an American bond, then $X_1 = (1+a)X_0$.  If instead he puts all his wealth in a British bond, then
\begin{align}
X_1 = \frac{X_0}{Z_0 B_0} Z_1 B_1 = (1+b)(1+z) X_0 .
\end{align}
To avoid arbitrage, these must agree, so
\begin{align}
1+a = (1+b)(1+z) \quad \Rightarrow \quad z = \frac{1+a}{1+b} - 1 .
\end{align}
\textbf{(c)} We seek $\pt$ and $\qt$ such that $X_0/A_0 = \Ebt[X_1/A_1]$.  Working through the algebra (using $Z_1 = (1+z)Z_0$ and $A_1 = (1+a)A_0$), the condition reduces to
\begin{align}
S_0 = \frac{\Ebt[S_1]}{1+b} = \frac{S_0(\pt u + \qt d)}{1+b} ,
\end{align}
from which
\begin{align}
\pt = \frac{(1+b)-d}{u-d} , \qquad \qt = 1 - \pt = \frac{u-(1+b)}{u-d} .
\end{align}
\textbf{(d)} Using risk-neutral pricing with the American bond as numeraire,
\begin{align}
\frac{V_0}{A_0}
	&=	\Ebt \Big[ \frac{Z_1 \phi(S_1)}{A_1} \Big]
	=	\frac{(1+z)Z_0}{(1+a)A_0} \Big( \pt \phi(uS_0) + \qt \phi(dS_0) \Big) ,
\end{align}
so
\begin{align}
V_0
	&=	\frac{(1+z)Z_0}{1+a} \Big( \pt \phi(uS_0) + \qt \phi(dS_0) \Big)
	=	\frac{Z_0}{1+b} \Big( \pt \phi(uS_0) + \qt \phi(dS_0) \Big) ,
\end{align}
where we used $1+a = (1+b)(1+z)$.


\subsection*{Solution to Exercise \ref{exer:2-8}}
\textbf{(a)} At time $n=0$ we have
\begin{align}
L_0
	&=	\frac{\beta L_0}{S_0} S_0 + (L_0 - \beta L_0) \frac{1}{B_0} B_0 .
\end{align}
Thus, at time $n=1$,
\begin{align}
L_1
	&=	\frac{\beta L_0}{S_0} S_1 + (L_0 - \beta L_0) \frac{B_1}{B_0} .
\end{align}
\textbf{(b)} If the portfolio has initial value $X_0$ and $\Del_0$ shares of $S$, then
\begin{align}
X_1
	&=	\Del_0 S_1 + (X_0 - \Del_0 S_0)(1+r) .
\end{align}
We also have
\begin{align}
\phi(L_1/L_0) = \phi\!\Big(\beta \tfrac{S_1}{S_0} + (1-\beta)(1+r)\Big) .
\end{align}
Setting $X_1 = \phi(L_1/L_0)$ for both outcomes $S_1 = uS_0$ and $S_1 = dS_0$ and solving, we obtain
\begin{align}
\Del_0
	&=	\frac{\phi(\beta u + (1-\beta)(1+r)) - \phi(\beta d + (1-\beta)(1+r))}{(u-d)S_0} , \\
X_0
	&=	\frac{1}{1+r} \Big( \pt \phi(\beta u + (1-\beta)(1+r)) + \qt \phi(\beta d + (1-\beta)(1+r)) \Big) ,
\end{align}
where $\pt = ((1+r)-d)/(u-d)$ and $\qt = 1 - \pt$ are the risk-neutral probabilities.


\subsection*{Solution to Exercise \ref{exer:2-9}}
\textbf{(a)} We need $\pt + \qt = 1$ and $X_0/B_0 = \Ebt[X_1/B_1]$ for all portfolios.  For a portfolio holding $\Del_0$ shares of $S$, the martingale condition reduces to the single asset condition.  If the stock goes up, $B_1 = B_0(1+r_u)$; if it goes down, $B_1 = B_0(1+r_d)$.  Thus, the conditions are
\begin{align}
\pt \frac{u}{1+r_u} + \qt \frac{d}{1+r_d} &= 1 , \qquad \pt + \qt = 1 .
\end{align}
\textbf{(b)} Using risk-neutral pricing,
\begin{align}
V_0
	&=	\Ebt\!\Big[\frac{\phi(S_1)}{B_1}\Big] \cdot B_0
	=	\pt \frac{\phi(uS_0)}{1+r_u} + \qt \frac{\phi(dS_0)}{1+r_d} .
\end{align}


\subsection*{Solution to Exercise \ref{exer:2-10}}
\textbf{(a)} Since either party can terminate the contract at time $n=0$, we must have $V_0 = \phi(S_0)$.  If $V_0 > \phi(S_0)$ the short side terminates immediately and profits $V_0 - \phi(S_0) > 0$; if $V_0 < \phi(S_0)$ the long side terminates immediately and profits $\phi(S_0) - V_0 > 0$.\\[0.5em]
\textbf{(b)} Suppose the short side holds $\Del_n$ shares of $S$ and the rest in bonds.  For $X_n = \phi(S_n)$ we need $X_{n+1} = \phi(S_{n+1})$.  The portfolio dynamics with the cash flow are
\begin{align}
X_{n+1}
	&=	\Del_n S_{n+1} + (X_n - \Del_n S_n)(1+r) + F_n .
\end{align}
Setting $X_{n+1} = \phi(S_{n+1})$ for the up and down outcomes and subtracting gives
\begin{align}
\Del_n
	&=	\frac{\phi(uS_n) - \phi(dS_n)}{(u-d)S_n} .
\end{align}
\textbf{(c)} From the up-state equation (using $X_n = \phi(S_n)$ and $\Del_n$ from part (b)),
\begin{align}
F_n
	&=	\phi(uS_n) - \Del_n uS_n - (\phi(S_n) - \Del_n S_n)(1+r)
	=	\phi(dS_n) - \Del_n dS_n - (\phi(S_n) - \Del_n S_n)(1+r) .
\end{align}
Substituting $\Del_n$ and simplifying yields the risk-neutral pricing formula:
\begin{align}
F_n
	&=	\pt \phi(uS_n) + \qt \phi(dS_n) - \phi(S_n)(1+r) , \qquad
\pt := \frac{(1+r)-d}{u-d} , \quad \qt := 1 - \pt .
\end{align}


\subsection*{Solution to Exercise \ref{exer:2-11}}
Throughout, we use the notation of \eqref{eq:Sni}: $S_1^{(1)} = u S_0 = 200$ and $S_1^{(0)} = d S_0 = 50$ denote the two nodes at time $n=1$, and $V_1^{(1)}$, $V_1^{(0)}$, $X_1^{(1)}$, $X_1^{(0)}$, $\Del_1^{(1)}$, $\Del_1^{(0)}$ denote the value of the option, the value of the portfolio and the number of shares held at these nodes. Because the payoff depends on the path, we denote by $V_2(\om)$ the value of the option at time $n=2$ along the path $\om = \om_1 \om_2$.
\\[0.5em]
\textbf{(a)} A portfolio $X$ that holds $\Del_n$ shares at time $n$ and invests the rest in the bond satisfies
\begin{align}
X_{n+1}
	&=	\Del_n S_{n+1} + (X_n - \Del_n S_n)(1+r) .
\end{align}
We seek $\pt$ and $\qt$ such that $X_n/B_n = \Ebt_n (X_{n+1}/B_{n+1})$. We have
\begin{align}
\Ebt_n \( \frac{X_{n+1}}{B_{n+1}} \)
	&=	\Del_n \Ebt_n \( \frac{S_{n+1}}{B_{n+1}} \) + (X_n - \Del_n S_n) \frac{(1+r)}{B_{n+1}}
	=		\Del_n \( \pt \frac{u S_n}{B_n (1+r)} + \qt \frac{d S_n}{B_n (1+r)} \) + \frac{X_n}{B_n} - \frac{\Del_n S_n}{B_n} .
\end{align}
In order for the above to equal $X_n/B_n$ for every $\Del_n$ we must have $\pt u + \qt d = 1 + r$. Using $\pt + \qt = 1$ we obtain
\begin{align}
\pt
	&=	\frac{(1+r) - d}{u - d}
	=		\frac{\tfrac{5}{4} - \tfrac{1}{2}}{2 - \tfrac{1}{2}}
	=		\frac{1}{2} , &
\qt
	&=	1 - \pt
	=		\frac{1}{2} .
\end{align}
\textbf{(b)} On each of the four paths we have
\begin{align}
\om &= HH : &  S_1 &= 200 , & S_2 &= 400 , & A &= 300 , & V_2(HH) &= (A-K)^+ = 200 , \\
\om &= HT : &  S_1 &= 200 , & S_2 &= 100 , & A &= 150 , & V_2(HT) &= (A-K)^+ = 50 , \\
\om &= TH : &  S_1 &= 50 ,  & S_2 &= 100 , & A &= 75 ,  & V_2(TH) &= (A-K)^+ = 0 , \\
\om &= TT : &  S_1 &= 50 ,  & S_2 &= 25 ,  & A &= 37.5 , & V_2(TT) &= (A-K)^+ = 0 .
\end{align}
Using risk-neutral pricing $V_n/B_n = \Ebt_n (V_{n+1}/B_{n+1})$, i.e., $V_n = \frac{1}{1+r} \Ebt_n V_{n+1}$, we have
\begin{align}
V_1^{(1)}
	&=	\frac{1}{1+r} \Big( \pt \, V_2(HH) + \qt \, V_2(HT) \Big)
	=		\frac{4}{5} \cdot \frac{1}{2} ( 200 + 50 )
	=		100 , \\
V_1^{(0)}
	&=	\frac{1}{1+r} \Big( \pt \, V_2(TH) + \qt \, V_2(TT) \Big)
	=		\frac{4}{5} \cdot \frac{1}{2} ( 0 + 0 )
	=		0 , \\
V_0
	&=	\frac{1}{1+r} \Big( \pt \, V_1^{(1)} + \qt \, V_1^{(0)} \Big)
	=		\frac{4}{5} \cdot \frac{1}{2} ( 100 + 0 )
	=		40 .
\end{align}
Equivalently, $V_0 = \frac{1}{(1+r)^2} \Ebt (A-K)^+ = \frac{16}{25} \cdot \frac{1}{4} ( 200 + 50 + 0 + 0 ) = 40$.
\\[0.5em]
For a European option that pays $\varphi(S_2)$, the value at time $n=2$ is $\varphi(S_2)$, which depends only on the node $S_2$. Here, the node $S_2 = u d S_0 = 100$ is reached along two different paths with different payoffs: $V_2(HT) = 50$ and $V_2(TH) = 0$. The value of the Asian call depends on the whole path $(S_1, S_2)$, not only on $S_2$. Thus, we must keep track of four values at time $n=2$ (one for each path) rather than three (one for each node), and the formula $V_0 = (1+r)^{-2} \sum_{k=0}^{2} \binom{2}{k} \pt^{\, k} \qt^{\, 2-k} \varphi(u^k d^{2-k} S_0)$ cannot be used.
\\[0.5em]
\textbf{(c)} Because the portfolio replicates the option, we must have $X_n = V_n$ at every node. In particular, $X_0 = V_0 = 40$, $X_1^{(1)} = V_1^{(1)} = 100$ and $X_1^{(0)} = V_1^{(0)} = 0$.
\\[0.5em]
\textit{Time $n=1$, node $S_1 = uS_0 = 200$.} We need $\Del_1^{(1)} S_2 + (X_1^{(1)} - \Del_1^{(1)} S_1^{(1)})(1+r) = V_2(\om)$ for $\om \in \{HH, HT\}$. Subtracting the two equations, we obtain
\begin{align}
\Del_1^{(1)}
	&=	\frac{V_2(HH) - V_2(HT)}{u^2 S_0 - u d S_0}
	=		\frac{200 - 50}{400 - 100}
	=		\frac{1}{2} , &
X_1^{(1)} - \Del_1^{(1)} S_1^{(1)}
	&=	100 - \frac{1}{2} \cdot 200
	=		0 .
\end{align}
Check: $\tfrac{1}{2} \cdot 400 + 0 = 200 = V_2(HH)$ and $\tfrac{1}{2} \cdot 100 + 0 = 50 = V_2(HT)$.
\\[0.5em]
\textit{Time $n=1$, node $S_1 = dS_0 = 50$.} Since $V_2(TH) = V_2(TT) = 0$, we have
\begin{align}
\Del_1^{(0)}
	&=	\frac{V_2(TH) - V_2(TT)}{u d S_0 - d^2 S_0}
	=		\frac{0 - 0}{100 - 25}
	=		0 , &
X_1^{(0)} - \Del_1^{(0)} S_1^{(0)}
	&=	0 - 0
	=		0 .
\end{align}
\textit{Time $n=0$.} We need $\Del_0 S_1 + (X_0 - \Del_0 S_0)(1+r) = V_1$ at both nodes. Subtracting the two equations, we obtain
\begin{align}
\Del_0
	&=	\frac{V_1^{(1)} - V_1^{(0)}}{u S_0 - d S_0}
	=		\frac{100 - 0}{200 - 50}
	=		\frac{2}{3} , &
X_0 - \Del_0 S_0
	&=	40 - \frac{2}{3} \cdot 100
	=		- \frac{80}{3} .
\end{align}
That is, at time $n=0$ we buy $\tfrac{2}{3}$ of a share and borrow $\tfrac{80}{3}$ from the bank. Check:
\begin{align}
\tfrac{2}{3} \cdot 200 - \tfrac{80}{3} \cdot \tfrac{5}{4}
	&=	\tfrac{400}{3} - \tfrac{100}{3}
	=		100
	=		V_1^{(1)} , &
\tfrac{2}{3} \cdot 50 - \tfrac{80}{3} \cdot \tfrac{5}{4}
	&=	\tfrac{100}{3} - \tfrac{100}{3}
	=		0
	=		V_1^{(0)} .
\end{align}


\clearpage
 \section{Exercises from Section \ref{sec:exer-3}}

 \subsection*{Solution to Exercise \ref{exer:3-2}}
 \textbf{Part (a)} Since $\pt + \qt = 1$, we have that $\qt = 1 - \pt$. By the definition of expected value, under the $\Pbt$ measure
 \begin{align}
 \tilde\Eb\[ 
 		\frac{ S_T}{B_T}
          \]
 	&= \pt \frac{u S_0}{R B_0} + \qt \frac{d S_0}{r B_0} \\
 	&= \pt \frac{u S_0}{R B_0} + (1 - \pt) \frac{d S_0}{r B_0}.\label{evalp2}
 \end{align}
 Setting \eqref{evalp2} equal to $S_0 / B_0$, we get
 \begin{align}
 &\pt \frac{u S_0}{R B_0} + (1 - \pt) \frac{d S_0}{r B_0}
 	= \frac{S_0}{B_0}
 &&\Rightarrow
 &&\pt 
 	= \frac{R (d-r)}{d R-r u},\quad
 \qt
 	= \frac{r (R-u)}{d R-r u}.
 \end{align}

 \noindent
 \textbf{Part (b)} We were able to solve for a unique risk-neutral pricing measure under which all assets with numeraire $B$ are martingales. By the Second Fundamental Theorem of Asset Pricing, the market $(B,S)$ is complete.

 \noindent
 \textbf{Part (c)} Let $h(x) = (x-100)^+$. Using the risk neutral measure, we get
 \begin{align}
 V_0 
 	&= B_0 \Ebt \[ \frac{h(S_T)}{B_T}\] \\
 	&= \pt \frac{ (u S_0 - 100)^+}{R} + (1 - \pt) \frac{ (d S_0 - 100)^+}{r} \\
 	&= \frac{25}{2}.	
 \end{align}
 Since we can price the option exactly, $25/2$ is a sharp upper and lower bound.

 \noindent
 \textbf{Part (d)} A portfolio $X$ is an arbitrage if $X_0 = 0$ and $\Pb (X_T \geq 0) =1$ and $\Pb( X_T > 0) > 0$ for some $T > 0$. 
 \\[0.5em]
 Suppose we would like to replicate the call option $V$ with payoff $h(x) = (x - 100)^+$ by trading in $S$ and $B$. We set up the portfolio
 \begin{align}
 X_0
 	= \Del_0 S_0 + \frac{ X_0 - \Del_0 S_0}{B_0} B_0.
 \end{align}
 At time $T$, we have
 \begin{align}
 X_T
 	= \Del_0 S_T + \frac{ X_0 - \Del_0 S_0}{B_0} B_T.
 \end{align}
 Since we want to replicate $V$ with payoff $h$, we have the system
 \begin{align}
 \Del_0 u S_0 + \( X_0 - \Del_0 S_0\) R
 	&= (u S_0 - 100)^+ \\
 \Del_0 d S_0 + \( X_0 - \Del_0 S_0\) r
 	&= (d S_0 - 100)^+,
 \end{align}
 Plugging in our problem parameters (given in part c) and solving yields
 \begin{align}
 X_0
 	&= \frac{25}{2},
 &\Del_0
 	&= \frac 58.
 \end{align}
 Suppose that $V_0 > 25/2$. Then at time 0, we sell the option for $V_0$. Now, there exists a $\delta > 0$ such that $V_0 = X_0 + \delta = 25/2 + \delta$. We buy $5/8$ shares of $S$, borrow $X_0 - \Del_0 S_0 = 25/2 - 125/2 = 50$ dollars from the bank (i.e., hold $-50$ bonds, as $B_0 = 1$), and invest the remaining $\delta$ in bonds. Then at time $T$, the replicating part of our portfolio is worth $X_T = h(S_T)$. We pay the payoff $h(S_T)$ which we owe for selling the option at $t=0$. We are left with $\delta B_T/B_0 >0$ dollars. Since we took zero risk, this amounts to arbitrage. 


\subsection*{Solution to Exercise \ref{exer:3-3}}
(a)
We seek a probability measure $\Pbt$ under which $X/M$ is a martingale for all portfolios $X$.  The existence of {$\Pbt$} will {guarantee} no arbitrage. We have
\begin{align}
\frac{X_n}{M_n}
	&=	\Ebt_n \frac{X_{n+1}}{M_{n+1}} \\
	&=	\Del_n \Ebt_n \frac{B_{n+1}}{M_{n+1}} + ( X_n - \Del_n B_n) \frac{1}{M_n} \frac{M_{n+1}}{M_{n+1}} \\
	&=	\Del_n \pt \frac{(1+R) B_n}{(1+r)M_n} + \frac{X_n}{M_n} - \frac{\Del_n B_n}{M_n} .
\end{align}
The above equality holds true for all $B_n$ only when 
\begin{align}
\pt 
	&= \frac{1+r}{1+R} , &
\pt
	&\in (0,1) &
	&\Leftrightarrow&
r 
	&< R ,
\end{align}
which is the no-arbitrage condition.\\
(b) Because the risk-neutral measure $\Pbt$ is unique, the market is complete and all derivative assets can be replicated.\\
(c) First, consider the case $B_n = 0$.  In this case $B_N = 0$.  It follows that
\begin{align}
\frac{V_n}{M_n}
	&=	\Ebt_n \frac{V_N}{M_N}
	=		\frac{\varphi(0)}{M_n(1+r)^{N-n}} . \\
V_n
	&=	\frac{\varphi(0)}{(1+r)^{N-n}} .
\end{align}
Next, consider the case $B_n > 0$.  In this case we have
\begin{align}
\Pbt \big( B_N = B_0 (1+R)^N \big)
	&=	\pt^{N-n} , &
\Pbt ( B_N = 0 )
	&=	1 - \pt^{N-n} .
\end{align}
Thus, using risk-neutral pricing, we have
\begin{align}
\frac{V_n}{M_n}
	&=	\Ebt_n \frac{V_N}{M_N}
	=		\pt^{N-n} \frac{\varphi(B_0(1+R)^{N})}{M_n(1+r)^{N-n}} + (1 - \pt^{N-n} ) \frac{\varphi(0)}{M_n(1+r)^{N-n}} ,
\end{align}
from which it follows
\begin{align}
V_n
	&=	\frac{1}{(1+r)^{N-n}} \Big( \pt^{N-n} \varphi(B_0(1+R)^{N}) + (1 - \pt^{N-n} ) \varphi(0) \Big) .
\end{align}
where $\pt = (1+r)/(1+R)$.\\
(d) The replicating portfolio $X$ must have the same initial value as the derivative
\begin{align}
X_0
	&=	V_0
	=		\frac{1}{(1+r)^{N}} \Big( \pt^{N} \varphi(B_0(1+R)^{N}) + (1 - \pt^{N} ) \varphi(0) \Big).
\end{align}
When $B_n = 0$ it does not matter what $\Del_n$ is, because the entire portfolio value will automatically be invested in the money market account $M$.
When $B_n > 0$ we need to set up a replicating portfolio: $X_n = V_n$ for all $n$.  We have
\begin{align}
V_{n+1}
	&=	\Del_n B_{n+1} + ( V_n - \Del_n B_n ) (1 + r) . \label{eq:del-n}
\end{align}
Note that when $B_n > 0$ we have
\begin{align}
V_n
	&=	\frac{1}{(1+r)^{N-n}} \Big( \pt^{N-n} \varphi(B_0(1+R)^{N}) + (1 - \pt^{N-n} ) \varphi(0) \Big) . \label{eq:Vn}
\end{align}
It is easiest to solve \eqref{eq:del-n} for $\Del_n$ assuming 
\begin{align}
B_{n+1} 
	&= 0  &
	&\Rightarrow& 
V_{n+1} 
	&= \varphi(0) / (1+r)^{N-n-1} .  \label{eq:Vn1}
\end{align}
Thus, solving \eqref{eq:del-n} for $\Del_n$, and using $B_n = (1+R)^n B_0$, as well as expressions \eqref{eq:Vn} and \eqref{eq:Vn1} for $V_n$ and $V_{n+1}$, respectively, we obtain 
\begin{align}
\Del_n
	&=	\frac{V_n - V_{n+1}/(1+r)}{B_0(1+R)^n} \\
	&=	\frac{\pt^{N-n}}{B_0(1+R)^n (1+r)^{N-n}} \Big(  \varphi(B_0(1+R)^{N})  - \varphi(0) \Big) \\
	&=	\frac{1}{B_0(1+R)^N} \Big(  \varphi(B_0(1+R)^{N})  - \varphi(0) \Big) .
\end{align}


\subsection*{Solution to Exercise \ref{exer:3-4}}
(a) 
Take $B$ as numeraire.  There will be no arbitrage if there exists a probability measure $\Pbt \sim \Pb$ under which $V/B$ is a martingale.  We compute
\begin{align}
\frac{V_0}{B_0}
	&=	\Ebt \frac{g(T_1)}{B_1}
	=		\pt \frac{g(u T_0)}{(1+r)B_0} + \qt \frac{g(d T_0)}{(1+r)B_0} , \\
V_0
	&=	\frac{1}{1+r}\Big(\pt g(u T_0) + \qt g(d T_0) \Big) .
\end{align}
We know that $0 < \pt= 1 - \qt < 1$.  Assume without loss of generality that $g(d T_0) < g( u T_0 )$.  Then $V_0$ is largest when $\pt$ nears 1 and is smallest when $\qt$ nears 1.  It follows that
\begin{align}
\frac{1}{1+r} g(d T_0 )
	&< V_0 < \frac{1}{1+r} g(u T_0 ) .
\end{align}
(b) Now, if we can observe $V_0$ then we can find $\pt$ and $\qt$ using
\begin{align}
V_0
	&=	\frac{1}{1+r}\Big(\pt g(u T_0) + \qt g(d T_0) \Big) , &
1
	&=	\pt + \qt &
	&\Rightarrow&
\pt
	&=	\frac{(1+r) V_0 - g(d T_0)}{g (u T_0) - g(d T_0)} , &
\qt
	&= 1 - \pt .  
\end{align}
Then, we can then find $P_0$ by computing
\begin{align}
P_0
	&=	\frac{1}{1+r}\Big(\pt h(u T_0) + \qt h(d T_0) \Big) .
\end{align}
where $\pt$ and $\qt$ are as given above.


\subsection*{Solution to Exercise \ref{exer:3-5}}
\textbf{(a)} Let $\Del_0$ be shares of $S$, $\Gam_0$ shares of the option, and let $X_0$ be the initial portfolio value.  The portfolio value at time $n=1$ satisfies
\begin{align}
X_1 = \Del_0 S_1 + \Gam_0 h(S_1) + (X_0 - \Del_0 S_0 - \Gam_0 V_0)(1+r) .
\end{align}
For $X_1 = g(S_1)$ we must have
\begin{align}
g(uS_0) &= \Del_0 uS_0 + \Gam_0 h(uS_0) + (X_0 - \Del_0 S_0 - \Gam_0 V_0)(1+r) , \\
g(mS_0) &= \Del_0 mS_0 + \Gam_0 h(mS_0) + (X_0 - \Del_0 S_0 - \Gam_0 V_0)(1+r) , \\
g(dS_0) &= \Del_0 dS_0 + \Gam_0 h(dS_0) + (X_0 - \Del_0 S_0 - \Gam_0 V_0)(1+r) .
\end{align}
The unknowns are $(X_0, \Del_0, \Gam_0)$.\\[0.5em]
\textbf{(b)} Under the assumption of no arbitrage, there exists a probability measure $\Pbt$ under which $\Ebt[A_1/B_1] = A_0/B_0$ for all traded assets $A$.  Thus,
\begin{align}
\frac{V_0}{B_0} = \Ebt\!\Big[\frac{h(S_1)}{B_1}\Big] = \alpha \Ebt\!\Big[\frac{S_1}{B_1}\Big] = \alpha \frac{S_0}{B_0} ,
\end{align}
so $V_0 = \alpha S_0$.\\[0.5em]
\textbf{(c)} When $h(S_1) = \alpha S_1$, we have $V_0 = \alpha S_0$.  The system becomes
\begin{align}
g(uS_0) &= \Del_0 uS_0 + \Gam_0 \alpha uS_0 + (X_0 - \Del_0 S_0 - \Gam_0 \alpha S_0)(1+r) , \\
g(mS_0) &= \Del_0 mS_0 + \Gam_0 \alpha mS_0 + (X_0 - \Del_0 S_0 - \Gam_0 \alpha S_0)(1+r) , \\
g(dS_0) &= \Del_0 dS_0 + \Gam_0 \alpha dS_0 + (X_0 - \Del_0 S_0 - \Gam_0 \alpha S_0)(1+r) .
\end{align}
In matrix form,
\begin{align}
\begin{pmatrix}
(1+r) & (u-(1+r))S_0 & \alpha(u-(1+r))S_0 \\
(1+r) & (m-(1+r))S_0 & \alpha(m-(1+r))S_0 \\
(1+r) & (d-(1+r))S_0 & \alpha(d-(1+r))S_0
\end{pmatrix}
\begin{pmatrix} X_0 \\ \Del_0 \\ \Gam_0 \end{pmatrix}
=
\begin{pmatrix} g(uS_0) \\ g(mS_0) \\ g(dS_0) \end{pmatrix} .
\end{align}
The $3\times 3$ matrix is \emph{not} invertible because the third column is a scalar multiple of the second column.  Therefore the system cannot in general be solved.


\subsection*{Solution to Exercise \ref{exer:3-6}}
\textbf{(a)} We have
\begin{align}
X_1
	&=	\Del_0 S_1 + \Gam_0 Z_1 + (X_0 - \Del_0 S_0 - \Gam_0 Z_0)(1+r) .
\end{align}
\textbf{(b)} In order for $X_1 = g(S_1,Z_1)$ we must have
\begin{align}
g(uS_0,aZ_0) &= \Del_0 uS_0 + \Gam_0 aZ_0 + (X_0-\Del_0 S_0-\Gam_0 Z_0)(1+r) , \\
g(dS_0,aZ_0) &= \Del_0 dS_0 + \Gam_0 aZ_0 + (X_0-\Del_0 S_0-\Gam_0 Z_0)(1+r) , \\
g(uS_0,bZ_0) &= \Del_0 uS_0 + \Gam_0 bZ_0 + (X_0-\Del_0 S_0-\Gam_0 Z_0)(1+r) , \\
g(dS_0,bZ_0) &= \Del_0 dS_0 + \Gam_0 bZ_0 + (X_0-\Del_0 S_0-\Gam_0 Z_0)(1+r) .
\end{align}
\textbf{(c)} In order for $X_0/B_0 = \Ebt(X_1/B_1)$ to hold for all $(\Del_0,\Gam_0)$, we must have
\begin{align}
(\pt_{ua}+\pt_{ub})u + (\pt_{da}+\pt_{db})d &= 1+r , \\
(\pt_{ua}+\pt_{da})a + (\pt_{ub}+\pt_{db})b &= 1+r , \\
\pt_{ua}+\pt_{ub}+\pt_{da}+\pt_{db} &= 1 .
\end{align}
\textbf{(d)} No.  The market $(S,Z,B)$ is not complete because there is no unique choice of $(\pt_{ua},\pt_{ub},\pt_{da},\pt_{db})$: the above is a system of 3 equations in 4 unknowns, so infinitely many risk-neutral measures exist.  Equivalently, there is no $(X_0,\Del_0,\Gam_0)$ replicating an arbitrary $g$.\\[0.5em]
\textbf{(e)} Consider the portfolio that at time $n=0$ holds $\Del_0 = \alpha$ shares of $S$, $\Gam_0 = \beta$ shares of $Z$, and $\gam/((1+r)B_0)$ shares of $B$ (i.e., $\gam/(1+r)$ dollars in the bond).  Then $X_1 = \alpha S_1 + \beta Z_1 + \gam = g(S_1,Z_1)$, and the unique price is
\begin{align}
X_0 = \alpha S_0 + \beta Z_0 + \frac{\gam}{1+r} .
\end{align}
\textbf{(f)} We have
\begin{align}
Y_1
	&=	\Del_0 S_1 + \Gam_0 Z_1 + \Phi_0 h(S_1,Z_1) + (Y_0-\Del_0 S_0-\Gam_0 Z_0-\Phi_0 V_0)(1+r) .
\end{align}
\textbf{(g)} In order for $Y_1 = f(S_1,Z_1)$ we must have
\begin{align}
f(uS_0,aZ_0) &= \Del_0 uS_0 + \Gam_0 aZ_0 + \Phi_0 h(uS_0,aZ_0) + (Y_0-\Del_0 S_0-\Gam_0 Z_0-\Phi_0 V_0)(1+r) , \\
f(dS_0,aZ_0) &= \Del_0 dS_0 + \Gam_0 aZ_0 + \Phi_0 h(dS_0,aZ_0) + (Y_0-\Del_0 S_0-\Gam_0 Z_0-\Phi_0 V_0)(1+r) , \\
f(uS_0,bZ_0) &= \Del_0 uS_0 + \Gam_0 bZ_0 + \Phi_0 h(uS_0,bZ_0) + (Y_0-\Del_0 S_0-\Gam_0 Z_0-\Phi_0 V_0)(1+r) , \\
f(dS_0,bZ_0) &= \Del_0 dS_0 + \Gam_0 bZ_0 + \Phi_0 h(dS_0,bZ_0) + (Y_0-\Del_0 S_0-\Gam_0 Z_0-\Phi_0 V_0)(1+r) .
\end{align}
\textbf{(h)} For $Y_0/B_0 = \Ebt(Y_1/B_1)$ to hold for all $(\Del_0,\Gam_0,\Phi_0)$,
\begin{align}
(\pt_{ua}+\pt_{ub})u + (\pt_{da}+\pt_{db})d &= 1+r , \\
(\pt_{ua}+\pt_{da})a + (\pt_{ub}+\pt_{db})b &= 1+r , \\
\pt_{ua}h(uS_0,aZ_0) + \pt_{ub}h(uS_0,bZ_0) + \pt_{da}h(dS_0,aZ_0) + \pt_{db}h(dS_0,bZ_0) &= (1+r)V_0 , \\
\pt_{ua}+\pt_{ub}+\pt_{da}+\pt_{db} &= 1 .
\end{align}
\textbf{(i)} Yes.  The market $(S,Z,V,B)$ is complete.  The above is now a system of 4 equations in 4 unknowns and generically has a unique solution, yielding a unique risk-neutral measure.  Equivalently, for any $f$, there exists a unique $(Y_0,\Del_0,\Gam_0,\Phi_0)$ such that $Y_1 = f(S_1,Z_1)$.


\subsection*{Solution to Exercise \ref{exer:3-7}}
\textbf{(a)(i)} For any portfolio $X$ we have $X_1 = \Del_0 S_1 + (X_0 - \Del_0 S_0)(1+r)$. Thus
\begin{align}
\Ebt \( \frac{X_1}{B_1} \)
	&=	\Del_0 \Ebt \( \frac{S_1}{B_1} \) + (X_0 - \Del_0 S_0) \frac{1+r}{B_1}
	=		\Del_0 \frac{(\pt u + \qt m + \rt d) S_0}{B_0 (1+r)} + \frac{X_0}{B_0} - \frac{\Del_0 S_0}{B_0} .
\end{align}
In order for the above to equal $X_0/B_0$ for every $\Del_0$, we must have
\begin{align}
\pt u + \qt m + \rt d
	&=	1 + r , &
\pt + \qt + \rt
	&=	1 ,
\end{align}
that is, with the given parameters, $1.2 \, \pt + \qt + 0.7 \, \rt = 1$ and $\pt + \qt + \rt = 1$. Subtracting the second equation from the first, we obtain $0.2 \, \pt - 0.3 \, \rt = 0$. Therefore
\begin{align}
\pt
	&=	\frac{3}{2} \, \rt , &
\qt
	&=	1 - \pt - \rt
	=		1 - \frac{5}{2} \, \rt ,
\end{align}
in agreement with \eqref{eq:pt.rt}. $\Pbt$ is a probability measure if its probabilities are non-negative and sum to one. $\Pbt$ is equivalent to $\Pb$ if $\Pb(A) = 0 \Leftrightarrow \Pbt(A) = 0$ for every event $A$. Since $\Pb$ assigns strictly positive probability to each of the three outcomes, $\Pbt$ must do the same, i.e., $\pt > 0$, $\qt > 0$ and $\rt > 0$. This means $\rt > 0$ and $1 - \frac{5}{2} \rt > 0$, so
\begin{align}
\rt
	&\in	\( 0 , \tfrac{2}{5} \) .
\end{align}
Each probability is then automatically smaller than one. The endpoints must be excluded: if $\rt = 0$ then $\pt = \rt = 0$, so $\Pbt(S_1 = u S_0) = 0$ while $\Pb(S_1 = u S_0) = p_u > 0$; if $\rt = \frac{2}{5}$ then $\qt = 0$, so $\Pbt(S_1 = m S_0) = 0$ while $\Pb(S_1 = m S_0) = p_m > 0$. In both cases $\Pbt$ is not equivalent to $\Pb$.
\\[0.5em]
\textbf{(a)(ii)} There exists a probability measure $\Pbt$ equivalent to $\Pb$ under which $X/B$ is a martingale for all portfolios $X$, so the market is free of arbitrage by the 1st Fundamental Theorem of Asset Pricing (Theorem \ref{thm:1st}). However, there are infinitely many such measures (one for each $\rt \in (0, \frac{2}{5})$), so the market is incomplete by the 2nd Fundamental Theorem of Asset Pricing (Theorem \ref{thm:second}). Concretely, replicating a general payoff with $S$ and $B$ requires solving three equations (one per state) in two unknowns $(X_0, \Del_0)$.
\\[0.5em]
\textbf{(b)(i)} The call pays $(S_1 - 100)^+ = 20, 0, 0$ in the states $u S_0 = 120$, $m S_0 = 100$, $d S_0 = 70$, respectively, and the put pays $(100 - S_1)^+ = 0, 0, 30$. Since $B_0 = B_1 = 1$, any no-arbitrage price is of the form
\begin{align}
C_0(\rt)
	&=	\Ebt (S_1 - 100)^+
	=		20 \, \pt
	=		30 \, \rt , &
P_0(\rt)
	&=	\Ebt (100 - S_1)^+
	=		30 \, \rt ,
\end{align}
for some $\rt \in (0, \frac{2}{5})$. Hence the range of no-arbitrage prices is $(0, 12)$ for both the call and the put.
\\[0.5em]
\textbf{(b)(ii)} By put-call parity, for every pricing measure $\Pbt$ we have $C_0(\rt) - P_0(\rt) = S_0 - K B_0/B_1 = 100 - 100 = 0$. Hence, as $\rt$ ranges over $(0,\frac{2}{5})$, the call and the put take exactly the same values, and their ranges coincide.
\\[0.5em]
\textbf{(b)(iii)} By Theorem \ref{thm:superhedge}, the minimal super-hedging price is the supremum of the no-arbitrage prices:
\begin{align}
\sup_{\rt \in (0, 2/5)} 30 \, \rt
	&=	12 .
\end{align}
\textbf{(c)(i)} If there is no arbitrage, then by part (b) the call price is $C_0 = 30 \, \rt$ for the pricing measure chosen by the market. Setting $30 \, \rt = 6$ gives
\begin{align}
\rt
	&=	\frac{1}{5} , &
\pt
	&=	\frac{3}{2} \cdot \frac{1}{5}
	=		\frac{3}{10} , &
\qt
	&=	1 - \frac{5}{2} \cdot \frac{1}{5}
	=		\frac{1}{2} .
\end{align}
These are strictly positive and $\rt \in (0, \frac{2}{5})$, so this $\Pbt$ is equivalent to $\Pb$. The derivative pays $|S_1 - 100| = 20, 0, 30$ in the three states. Hence its price is
\begin{align}
\Ebt |S_1 - 100|
	&=	\frac{3}{10} \cdot 20 + \frac{1}{2} \cdot 0 + \frac{1}{5} \cdot 30
	=		6 + 6
	=		12 .
\end{align}
\textbf{(c)(ii)} The portfolio satisfies $X_1 = \Del_0 S_1 + \Gam_0 (S_1 - 100)^+ + (X_0 - \Del_0 S_0 - \Gam_0 C_0)(1+r)$. Since $(K - x)^+ = (x - K)^+ - (x - K)$, we have
\begin{align}
|S_1 - 100|
	&=	(S_1 - 100)^+ + (100 - S_1)^+
	=		2 (S_1 - 100)^+ - S_1 + 100 .
\end{align}
Thus the payoff $|S_1 - 100|$ is replicated by $\Gam_0 = 2$ calls, $\Del_0 = -1$ share and $100$ in the bond, i.e., $X_0 - \Del_0 S_0 - \Gam_0 C_0 = 100$. Its initial value is
\begin{align}
X_0
	&=	2 \cdot 6 - 100 + 100
	=		12 ,
\end{align}
which equals the price found in (c)(i). Check: the portfolio is worth $2 \cdot 20 - 120 + 100 = 20$ if $S_1 = 120$, $0 - 100 + 100 = 0$ if $S_1 = 100$ and $0 - 70 + 100 = 30$ if $S_1 = 70$.
\\[0.5em]
Alternatively, one can write the three equations $X_1 = |S_1 - 100|$, one per state,
\begin{align}
120 \, \Del_0 + 20 \, \Gam_0 + (X_0 - 100 \, \Del_0 - 6 \, \Gam_0)
	&=	20 , \\
100 \, \Del_0 + 0 \cdot \Gam_0 + (X_0 - 100 \, \Del_0 - 6 \, \Gam_0)
	&=	0 , \\
70 \, \Del_0 + 0 \cdot \Gam_0 + (X_0 - 100 \, \Del_0 - 6 \, \Gam_0)
	&=	30 .
\end{align}
Subtracting the third equation from the second gives $30 \, \Del_0 = -30$, so $\Del_0 = -1$. Subtracting the second from the first gives $20 \, \Del_0 + 20 \, \Gam_0 = 20$, so $\Gam_0 = 2$. Finally, the second equation gives $X_0 = 6 \, \Gam_0 = 12$.
\\[0.5em]
\textbf{(c)(iii)} No. Under every pricing measure $\Pbt$ with $\rt \in (0, \frac{2}{5})$ we have
\begin{align}
\frac{V_0}{B_0}
	&=	\Ebt \( \frac{2 S_1 + 10}{B_1} \)
	=		2 \, \frac{S_0}{B_0} + \frac{10}{B_1} ,
	&&\Rightarrow &
V_0
	&=	2 S_0 + 10
	=		210 ,
\end{align}
whatever the value of $\rt$ (any other quote would be an arbitrage). Therefore the quote $V_0$ carries no information about $\rt$ and does not pin down the pricing measure. Equivalently, the payoff $2 S_1 + 10$ can already be replicated with $2$ shares of $S$ and $10$ in the bond, so adding this derivative to the market does not allow us to replicate any new payoffs. The market $(S, B, V)$ remains incomplete, and the price of the derivative that pays $|S_1 - 100|$ can be anywhere in its no-arbitrage range $\Ebt |S_1 - 100| = 20 \, \pt + 30 \, \rt = 60 \, \rt \in (0, 24)$.


\clearpage
\section{Exercises from Section \ref{sec:exer-4}}

\subsection*{Solution to Exercise \ref{exer:4-1}}
Let $f(x) = x^{-1}$. Then we have that $X_t = S_t f(M_t)$. So, by the chain rule
\begin{align}
\dd X_t
	= \dd S_t f(M_t) + S_t  \dd f(M_t) + \dd S_t  \dd f(M_t). 
\end{align}
Using Ito's Lemma,
\begin{align}
\dd f(M_t)
	&= f' (M_t) \dd M_t + \frac 12 f''(M_t) \dd M_t \dd M_t \\ 
	&= - \frac 1{M_t} r\dd t.
\end{align}
Therefore, 
\begin{align}
\dd X_t
	&= \dd S_t f(M_t) + S_t  \dd f(M_t) + \dd S_t  \dd f(M_t) \\
	&= \( \mu S_t \dd t + \sig S_t \dd W_t \) \frac{1}{M_t} + S_t \( - \frac 1{M_t} r \dd t\) 
		+ \( \mu S_t \dd t + \sig S_t \dd W_t \)\( - \frac 1{M_t} r \dd t\) \\
	&=  \mu X_t \dd t + \sig X_t \dd W_t  - r X_t \dd t 
		+0 \\
	&= \( \mu - r\) X_t \dd t + \sig X_t \dd W_t. 
\end{align}


\subsection*{Solution to Exercise \ref{exer:4-2}}
Let $f(x) = x^{-1}$. Then we can write $Z_t = S_t f(Y_t)$. Using the chain rule we have
\begin{align}
\dd Z_t
	= \dd S_t\, f(Y_t) + S_t\, \dd f(Y_t) + \dd S_t\, \dd f(Y_t).
\end{align}
Let $\bar \rho = \sqrt{1 - \rho^2}$ and note that $\rho^2 + \bar \rho ^2 = 1$. Using Ito's Lemma,
\begin{align}
\dd f(Y_t)
	&= - \frac 1{Y_t^2} \dd Y_t + \frac 1{Y_t^3} \dd Y_t\, \dd Y_t \\
	&= - \frac 1{Y_t^2}\( \alpha Y_t \dd t + \rho \beta Y_t \dd W_t + \bar \rho \beta Y_t \dd B_t\) 
	  + \frac{1}{Y_t^3} \( \rho^2 \beta^2 Y_t^2 \dd t + \bar \rho^2 \beta^2 Y_t^2 \dd t\) \\
	&= \frac 1{Y_t} \[ ( \beta^2 - \alpha) \dd t - \rho \beta \dd W_t - \bar \rho \beta \dd B_t\].
\end{align}
Therefore,
\begin{align}
\dd Z_t
	&= \dd S_t\, f(Y_t) + S_t\, \dd f(Y_t) + \dd S_t\, \dd f(Y_t) \\
	&= \frac 1{Y_t}\( \mu S_t \dd t + \sig S_t \dd W_t\) + S_t\cdot \frac 1{Y_t} \[ ( \beta^2 - \alpha) \dd t - \rho \beta \dd W_t - \bar \rho \beta \dd B_t\] \\ &\qquad
		+ \( \mu S_t \dd t + \sig S_t \dd W_t\) \frac 1{Y_t} \[ ( \beta^2 - \alpha) \dd t - \rho \beta \dd W_t - \bar \rho \beta \dd B_t\] \\
	&= \mu Z_t \dd t + \sig Z_t \dd W_t + \(\beta^2 - \alpha\)  Z_t\dd t 
		- \rho \beta Z_t \dd W_t- \bar \rho \beta Z_t \dd B_t - \rho \beta \sig Z_t \dd t \\
	&= \( \mu + \beta^2 - \alpha - \rho \beta \sig\) Z_t \dd t + ( \sig - \rho \beta) Z_t \dd W_t- \bar \rho \beta Z_t \dd B_t
\end{align}



\subsection*{Solution to Exercise \ref{exer:4-3}}
Let $f(x) = x^\beta$. Then we can write $P_t = f(S_t)$. Using Ito's Lemma,
\begin{align}
\dd P_t
	&= f'(S_t) \dd S_t + \frac 12 f''(S_t)\dd S_t \dd S_t \\
	&= \beta S_t^{\beta - 1} \( \mu S_t \dd t + \sig S_t \dd W_t\) 
		+ \frac 12 \beta (\beta -1)S_t^{\beta -2} \( \sig^2 S_t^2 \dd t\) \\
	&= \mu \beta P_t \dd t + \sig \beta P_t \dd W_t + \frac {\sig^2}{ 2} \beta(\beta-1) P_t \dd t \\
	&= \( \frac 12 \sig^2 \beta(\beta -1 ) + \mu \beta\) P_t \dd t + \sig \beta P_t \dd W_t. 
\end{align}
Now, $P$ will be a martingale if, for all values of $P_t$, $( \sig^2 \beta(\beta -1 )/2 + \mu \beta) P_t \dd t = 0$. We see that
\begin{align}
&\frac 12 \sig^2 \beta(\beta -1 ) + \mu \beta
	= 0 
&&\Rightarrow
&& \beta
	= 0, \frac{\sigma^2 -2 \mu }{\sigma^2 }.
\end{align}


\subsection*{Solution to Exercise \ref{exer:4-4}}
\textbf{Part (a)} By Ito's Lemma
\begin{align}
\dd (X_t^2)
	&= 2 X_t \dd X_t + \dd X_t \dd X_t \\
	&= 2 X_t \( - \frac 12 X_t \dd t - Y_t \dd W_t \) 
		+ Y_t^2 \dd t \\
	&= -\( X_t^2 - Y_t^2\) \dd t - 2 X_t Y_t \dd W_t
\end{align}
and 
\begin{align}
\dd (Y_t^2)
	&= 2 Y_t \dd Y_t + \dd Y_t \dd Y_t \\
	&= 2 Y_t \( - \tfrac{1}{2} Y_t \dd t + X_t \dd W_t \) + X_t^2 \dd t \\
	&= \( X_t^2 - Y_t^2 \) \dd t + 2 X_t Y_t \dd W_t.
\end{align}
Therefore, 
\begin{align} 
\dd \( X_t^2 + Y_t^2\)
	= \dd X_t^2 + \dd Y_t^2
	= 0. 
\end{align}

\noindent
\textbf{Part (b)} Let $(X_t, Y_t) = (\cos (W_t), \sin(W_t))$. By Ito's Lemma
\begin{align}
\dd X_t
	= \dd \cos (W_t)
	&= - \sin ( W_t) \dd W_t - \frac 12 \cos(W_t) \dd W_t \dd W_t \\
	&= - \frac 12 X_t \dd t - Y_t \dd W_t,
\end{align}
and 
\begin{align}
\dd Y_t
	=\dd \sin(W_t)
	&= \cos(W_t) \dd W_t - \frac 12 \sin(W_t) \dd W_t \dd W_t \\
	&= - \frac 12 Y_t \dd t + X_t \dd W_t.
\end{align}
Therefore $(X_t,Y_t) = (\cos (W_t), \sin(W_t))$ is a solution of the SDE \eqref{fivepointfoursde}.

\noindent
\textbf{Part (c)} Note that
\begin{align}
X_t^2 + Y_t^2
	= \cos^2(W_t) + \sin^2(W_t)
	= 1.
\end{align}
Since $X_t^2 + Y_t^2$ is constant, we expect $\dd ( X_t^2 + Y_t^2) = 0$. 



\subsection*{Solution to Exercise \ref{exer:4-5}}
By It\^o's Lemma,
\begin{align}
\dd M_t
	&=	\dd \big( \ee^{-\lam t}\phi(X_t) \big)
	=	\ee^{-\lam t} \Big( -\lam \phi(X_t) + \mu(X_t) \phi'(X_t) + \tfrac{1}{2}\sig^2(X_t) \phi''(X_t) \Big) \dd t + \ee^{-\lam t} \sig(X_t) \phi'(X_t) \dd W_t .
\end{align}
For $M$ to be a martingale, the $\dd t$-term must vanish.  Thus, the ODE that $\phi$ must satisfy is
\begin{align}
\tfrac{1}{2}\sig^2(x) \phi''(x) + \mu(x) \phi'(x) - \lam \phi(x)
	&= 0 .
\end{align}


 \subsection*{Solution to Exercise \ref{exer:4-6}}
Differentiating $Y_t = \lam \int_{-\infty}^t \ee^{-\lam(t-s)} X_s \dd s = \lam \ee^{-\lam t} \int_{-\infty}^t \ee^{\lam s} X_s \dd s$ with respect to $t$ (there is no It\^o correction, because $Y$ has no $\dd W_t$-term), we obtain
\begin{align}
\dd Y_t
	&=	-\lam Y_t \dd t + \lam \ee^{-\lam t} \ee^{\lam t} X_t \dd t
	=		\lam (X_t - Y_t) \dd t
	=		\lam Z_t \dd t .
\end{align}
Therefore,
\begin{align}
\dd Z_t
	&=	\dd X_t - \dd Y_t
	=		\mu \dd t + \sig \dd W_t - \lam Z_t \dd t
	=		\lam \Big( \frac{\mu}{\lam} - Z_t \Big) \dd t + \sig \dd W_t .
\end{align}
That is, $Z$ is an Ornstein--Uhlenbeck process (see Example \ref{ex:OU}) that mean-reverts to $\mu/\lam$ at rate $\lam$.


\subsection*{Solution to Exercise \ref{exer:4-7}}
Define $X_t := - \tfrac{1}{2}\int_0^t \Del_s^2 \dd s + \int_0^t \Del_s \dd W_s$, so that $Z_t = \ee^{X_t}$ and
\begin{align}
\dd X_t
	&= -\tfrac{1}{2}\Del_t^2 \dd t + \Del_t \dd W_t .
\end{align}
By It\^o's Lemma,
\begin{align}
\dd Z_t
	&=	\ee^{X_t} \dd X_t + \tfrac{1}{2} \ee^{X_t} (\dd X_t)^2
	=		\ee^{X_t} \Big( -\tfrac{1}{2}\Del_t^2 \dd t + \Del_t \dd W_t \Big) + \tfrac{1}{2} \ee^{X_t} \Del_t^2 \dd t
	=		Z_t \Del_t \dd W_t .
\end{align}
As $\dd Z_t$ has no $\dd t$-term (and $\Del$ is bounded), $Z$ is a martingale and $\Eb Z_t = Z_0 = 1$.


\clearpage
\section{Exercises from Section \ref{sec:exer-5}}

\subsection*{Solution to Exercise \ref{exer:5-1}}
\noindent
\textbf{Part (a).} In our model, 
\begin{align}
\dd S_t
	&=	\mu S_t \dd t + \sig S_t \dd W_t , &
\dd B_t
	&=	r B_t \dd t .
\end{align}
Under a risk neutral measure $\Pbt$ (see the risk-neutral pricing section of Chapter 6), we have the dynamics
\begin{align}
\dd S_t
	&=	r S_t \dd t + \sig S_t \dd \Wt_t , &
\dd B_t
	&=	r B_t \dd t,
\end{align}
where $\Wt_t$ is a standard Brownian motion. Under $\Pbt$, $S_t$ has the solution
\begin{align}
S_t
	= S_0 \exp\[ \( r - \frac 12 \sig^2\) t + \sig \Wt_t\].
\end{align}
We may write 
\begin{align}
S_T 
	= S_t \exp\[ \( r - \frac 12 \sig^2\) (T-t) + \sig ( \Wt_T - \Wt_t) \].
\end{align}
Therefore, 
\begin{align}
S_T^p
	&= S_t^p \exp \[ p \( r - \frac 12 \sig^2 \) (T-t) + p \sig ( \Wt_T - \Wt_t)\].
\end{align}
Let $V_t$ denote the price of the claim on $S_t$ under consideration. Let $\Fc_t$ be the $\sig$-algebra generated by $\Wt_t$. Then under $\Pbt$ we have
\begin{align}
V_t
	&= B_t \Ebt \[ \frac{V_T}{B_T} \mid \Fc_t \] \\
	&= B_t \Ebt \[ \frac{V_T}{B_T} \mid S_t \] \\
	&= \exp\[ -r(T-t)\] \Ebt \[ S_T^p \mid S_t \] \\
	&= \exp\[ - r(T-t)\] \Ebt \[  S_t^p \exp \[ p \( r - \frac 12 \sig^2 \) (T-t) + p \sig ( \Wt_T - \Wt_t)\] \mid S_t \] \\
	&= \exp\[ - r(T-t)\] S_t^p \exp\[ p \( r - \frac 12 \sig^2 \)(T- t) \] \Ebt\[ \exp\[ p \sig (\Wt_T - \Wt_t)\] \mid S_t \].\label{vtone}
\end{align}
Noting that $\exp\[ p \sig (\Wt_T - \Wt_t)\]$ is \textit{Log-normally distributed} with variance $p^2 \sig^2 (T-t)$, we have that 
\begin{align}
\Ebt\[ \exp\[ p \sig (\Wt_T - \Wt_t)\] \mid S_t \]
	= \exp\[ \frac 12 p^2 \sig^2 (T-t)\].
\end{align}
Inserting the above equation into \eqref{vtone} gives
\begin{align}
V_t
	= S_t^p \exp\[ \( \frac 12 \sig^2 p+r\)(p-1)(T-t)\].
\end{align}

\noindent
\textbf{Part (b)}. Note that under the risk-neutral measure $\Pbt$, because 
\begin{align}
\dd X_t 
	= \Delta_t \dd S_t + \( X_t - \Delta_t S_t\) \frac 1{B_t} \dd B_t
\end{align}
and $\dd B_t / B_t = r \dd t$, we have
\begin{align}
\dd \[ \frac{X_t}{B_t} \]
	&= \dd X_t \frac1{B_t} + X_t \dd \(\frac{1}{B_t}\) + \dd X_t\ \dd \(\frac 1{B_t}\) \\
	&= \frac{1}{B_t}\(\Delta_t \dd S_t + \( X_t - \Delta_t S_t\) \frac 1{B_t} \dd B_t\)  - r \frac{1}{B_t} X_t \dd t \\
	&= \frac{1}{B_t}\( \Delta_t r S_t \dd t + \Delta_t \sig S_t \dd \Wt_t + r X_t \dd t - \Delta r S_t \dd t\) - r \frac 1{B_t} X_t \dd t \\
	&= \frac{1}{B_t} \Delta_t \sig S_t \dd \Wt_t.
\end{align}
Furthermore, if 
\begin{align}
v(t,s)
	= \ee^{ -r (T-t)} \Ebt\[ \phi(S_T) \mid S_t = s\],
\end{align}
the $v$ satisfies the Black-Scholes PDE. Since $V_t = v(t,S_t)$, then under $\Pbt$,
\begin{align}
\dd \[ \frac{ V_t}{B_t} \]
	=& \dd \[ \frac{v(t,S_t)}{B_t}\] \\
	&= \dd v(t,S_t) \frac 1{B_t} + v(t,S_t) \dd \( \frac{1}{B_t}\) + \dd v(t,S_t) \dd \( \frac{1}{B_t}\) \\
	&= \( v_t(t,S_t) \dd t + v_s(t,S_t) \dd S_t + \frac 12 v_{ss} (t,S_t)\dd S_t \dd S_t\) \frac{1}{B_t} - r \frac{1}{B_t} v(t,S_t) \dd t \\
	&= \( v_t (t,S_t) \dd t + r S_t v_s(t,S_t) \dd t + \sig S_t v_s(t,S_t) \dd \Wt_t + \frac 12 \sig^2 S_t^2 v_{ss}(t,S_t) \dd t\) \frac{1}{B_t} \\ &\quad
			- r \frac{1}{B_t} v(t,S_t) \dd t \\
	&= \frac{1}{B_t}\( v_t (t,S_t) + r S_t v_s (t,S_t) + \frac 12 \sig^2 S_t^2 v_{ss}(t,S_t) - r v(t,S_t)\) \dd t + \sig \frac{1}{B_t} v_s(t,S_t) \dd \Wt_t \\
	&= \sig \frac{1}{B_t} v_s(t,S_t) \dd \Wt_t,
\end{align}
where the last equality comes from the fact that $v$ satisfies the Black-Scholes PDE. Setting 
\begin{align}
X_0
	&= V_0, &
\dd \[ \frac{X_t}{B_t} \]
	&= \dd \[ \frac{ V_t}{B_t} \]
\end{align}
guarantees $X$ replicates the option. The above implies that $\Delta_t = v_s(t,S_t)$ i.e. 
\begin{align}
X_0
	&=  S_0^p \exp\[ \( \frac 12 \sig^2 p+r\)(p-1)T\], &
\Delta_t
	&= p S_t^{p-1} \exp\[ \( \frac 12 \sig^2 p+r\)(p-1)(T-t)\].
\end{align}

\noindent
\textbf{Part (c).} The key insight is to simulate $Z_t = \log S_t$ and $Y_t = \log B_t$ (both of which satisfy linear SDEs with constant coefficients) and then exponentiate. The Euler--Maruyama discretization of $\dd Z_t = (\mu - \tfrac{1}{2}\sig^2)\dd t + \sig \dd W_t$ gives $Z_{t+\del t} \approx Z_t + (\mu - \tfrac{1}{2}\sig^2)\del t + \sig \sqrt{\del t}\, \varepsilon$ where $\varepsilon \sim N(0,1)$.

\begin{lstlisting}[caption={Monte Carlo paths of $S$ and $B$ (\texttt{problem6c.py})}]
"""
problem6c.py  --  Plot one sample path of S and B.
Input  (input.txt):  S0, B0, mu, sig, r, T, p, N, M
"""

import math, numpy as np, matplotlib
matplotlib.use("Agg")
import matplotlib.pyplot as plt

with open("input.txt") as f:
    vals = f.read().split(",")
S0, B0, mu, sig, r, T = [float(vals[i]) for i in range(6)]
p, N, M = float(vals[6]), int(vals[7]), int(vals[8])

dt   = T / N
grid = np.linspace(0.0, T, N + 1)

# Simulate log S using Euler--Maruyama
logS = np.empty(N + 1)
logS[0] = math.log(S0)
increments = (mu - 0.5*sig**2)*dt + sig*math.sqrt(dt)*np.random.randn(N)
logS[1:] = logS[0] + np.cumsum(increments)
S_path = np.exp(logS)

# Bond path is deterministic
B_path = B0 * np.exp(r * grid)

fig, ax = plt.subplots()
ax.plot(grid, S_path, label="Stock S")
ax.plot(grid, B_path, label="Bond B")
ax.set_xlabel("time")
ax.legend()
fig.savefig("plot.6c.png")
plt.close(fig)
print("Saved plot.6c.png")
\end{lstlisting}

\noindent
\textbf{Part (d).} The exact option value is $V_t = S_t^p \ee^{(\frac{1}{2}\sig^2 p + r)(p-1)(T-t)}$ and the replicating portfolio satisfies $\Del_t = p S_t^{p-1}\ee^{(\frac{1}{2}\sig^2 p + r)(p-1)(T-t)}$.  The self-financing condition gives the portfolio update:
\begin{align}
X_{t+\del t} \approx X_t + \Del_t (S_{t+\del t} - S_t) + (X_t - \Del_t S_t)\frac{B_{t+\del t}-B_t}{B_t}.
\end{align}
If the simulation is correct, the paths of $X$ and $V$ will be nearly indistinguishable.

\begin{lstlisting}[caption={Replicating portfolio vs.\ option value (\texttt{problem6d.py})}]
"""
problem6d.py  --  Plot one path of V_t and X_t for the power option.
Input  (input.txt):  S0, B0, mu, sig, r, T, p, N, M
"""

import math, numpy as np, matplotlib
matplotlib.use("Agg")
import matplotlib.pyplot as plt

with open("input.txt") as f:
    vals = f.read().split(",")
S0, B0, mu, sig, r, T = [float(vals[i]) for i in range(6)]
p, N, M = float(vals[6]), int(vals[7]), int(vals[8])

dt   = T / N
grid = np.linspace(0.0, T, N + 1)

# Simulate S
logS = np.empty(N + 1)
logS[0] = math.log(S0)
increments = (mu - 0.5*sig**2)*dt + sig*math.sqrt(dt)*np.random.randn(N)
logS[1:] = logS[0] + np.cumsum(increments)
S_path = np.exp(logS)

# Bond path
B_path = B0 * np.exp(r * grid)

def option_value(s, t):
    """Exact price V_t = s^p * exp((sig^2*p/2 + r)*(p-1)*(T-t))"""
    return s**p * math.exp((sig**2*p/2 + r)*(p - 1)*(T - t))

def delta(s, t):
    """Replicating delta Delta_t = p * s^(p-1) * exp(...)"""
    return p * s**(p - 1) * math.exp((sig**2*p/2 + r)*(p - 1)*(T - t))

# Initialize replicating portfolio
X = option_value(S_path[0], grid[0])
X_path = [X]
for n in range(1, N + 1):
    D  = delta(S_path[n-1], grid[n-1])
    dS = S_path[n] - S_path[n-1]
    dB = B_path[n] - B_path[n-1]
    X  = X + D * dS + (X - D * S_path[n-1]) * dB / B_path[n-1]
    X_path.append(X)

V_path = [option_value(S_path[n], grid[n]) for n in range(N + 1)]

fig, ax = plt.subplots()
ax.plot(grid, V_path, label="$V_t$ (exact)", lw=2)
ax.plot(grid, X_path, label="$X_t$ (hedged)", lw=1, linestyle="--")
ax.set_xlabel("time")
ax.legend()
fig.savefig("plot.6d.png")
plt.close(fig)
print("Saved plot.6d.png")
\end{lstlisting}

\noindent
\textbf{Part (e).} The paths of $V$ and $X$ do not overlap exactly because the hedging portfolio is rebalanced only at discrete time steps $\del t = T/N$, whereas the theoretical derivation assumes \emph{continuous} rebalancing.  The replication error is of order $\Oc(\del t)$ in variance; as $N \to \infty$ (i.e.\ $\del t \to 0$) the error shrinks to zero.

\noindent
\textbf{Part (f).}  The histogram of $X_T - V_T$ over $M$ Monte Carlo paths should be centered near zero (no systematic bias) with spread that decreases as $N$ increases, confirming the convergence of the discrete hedge to the continuous one.

\begin{lstlisting}[caption={Histogram of hedging error $X_T - V_T$ (\texttt{problem6f.py})}]
"""
problem6f.py  --  Histogram of X_T - V_T over M paths.
Input  (input.txt):  S0, B0, mu, sig, r, T, p, N, M
"""

import math, numpy as np, matplotlib
matplotlib.use("Agg")
import matplotlib.pyplot as plt

with open("input.txt") as f:
    vals = f.read().split(",")
S0, B0, mu, sig, r, T = [float(vals[i]) for i in range(6)]
p, N, M = float(vals[6]), int(vals[7]), int(vals[8])

dt   = T / N
grid = np.linspace(0.0, T, N + 1)
B_path = B0 * np.exp(r * grid)

def option_value(s, t):
    return s**p * math.exp((sig**2*p/2 + r)*(p - 1)*(T - t))

def delta(s, t):
    return p * s**(p - 1) * math.exp((sig**2*p/2 + r)*(p - 1)*(T - t))

errors = []
for _ in range(M):
    # Simulate a stock path
    logS = np.empty(N + 1)
    logS[0] = math.log(S0)
    incs = (mu - 0.5*sig**2)*dt + sig*math.sqrt(dt)*np.random.randn(N)
    logS[1:] = logS[0] + np.cumsum(incs)
    S_path = np.exp(logS)

    # Run discrete hedge
    X = option_value(S_path[0], grid[0])
    for n in range(1, N + 1):
        D  = delta(S_path[n-1], grid[n-1])
        dS = S_path[n] - S_path[n-1]
        dB = B_path[n] - B_path[n-1]
        X  = X + D * dS + (X - D * S_path[n-1]) * dB / B_path[n-1]

    errors.append(X - option_value(S_path[-1], T))

fig, ax = plt.subplots()
ax.hist(errors, bins=50, edgecolor="black")
ax.set_xlabel("$X_T - V_T$")
ax.set_title(f"Hedging error over {M} paths  (N={N})")
fig.savefig("plot.6f.png")
plt.close(fig)
print(f"Mean error: {sum(errors)/M:.6f},  Std: {float(np.std(errors)):.6f}")
print("Saved plot.6f.png")
\end{lstlisting}



\subsection*{Solution to Exercise \ref{exer:5-2}}
\begin{enumerate}
    \item 
    
    The portfolio value increases by amount $\Delta_t \times q  S_t \dd t$ because our portfolio holds $\Delta_t$ stock at time $t$ so we get
    
    \begin{align*}
        \dd X_t & = \Delta_t \dd S_t + q \Delta_t S_t \dd t+ \frac{X_t - \Delta_t S_t}{B_t} \dd B_t
    \end{align*}
    
    which we simplify further using above value of $\dd S_t$ and the fact that $\dd B_t = r B_t \dd t$ to get
    
    \begin{align*}
        \dd X_t & = \Delta_t (\mu S_t \dd t + \sigma S_t \dd W_t) + q \Delta_t S_t \dd t + r(X_t - \Delta_t S_t) \dd t \\
        {} & = (\Delta_t S_t (\mu+q-r) + r X_t) \dd t + \sigma \Delta_t S_t \dd W_t
    \end{align*}
    
    \item First we calculate $\dd (\frac{X_t}{B_t})$ and $\dd (\frac{V_t}{B_t})$ We can write
    
    \begin{align*}
        \dd (\frac{X_t}{B_t}) & = \frac{1}{B_t} \dd X_t - \frac{X_t}{B^2_t} \dd B_t \\
        {} & = \frac{1}{B_t}((\Delta_t S_t (\mu+q-r) + r X_t) \dd t + \sigma \Delta_t S_t \dd W_t) - \frac{r X_t \dd t}{B_t} \\
        {} & = \frac{\Delta_t S_t}{B_t}(\mu+q-r) \dd t + \sigma \frac{\Delta_t S_t}{B_t} \dd W_t
    \end{align*}
    \begin{align*}
        \dd V_t & = \dd v(t,S_t) \\
        {} & = \Big( v_t(t,S_t) + \mu S_t v_s(t,S_t) + \frac{1}{2}\sigma^2 S^2_t v_{ss}(t,S_t) \Big) \dd t + \sigma v_s(t,S_t)  S_t \dd W_t
    \end{align*}
    
    so that
    
    \begin{align*}
        \dd (\frac{V_t}{B_t}) & = \frac{1}{B_t}[v_t(t,S_t) + \mu S_t v_s(t,S_t) + \frac{1}{2}\sigma^2 S^2_t v_{ss}(t,S_t)-rv(t,S_t)] \dd t + \frac{1}{B_t}\sigma v_s(t,S_t)  S_t \dd W_t
    \end{align*}
    
    Since $X$ replicates the option price $V$ we want
    
    \begin{align*}
        0 & = \dd (\frac{V_t}{B_t}) - \dd (\frac{X_t}{B_t}) \\
        {} & = \frac{1}{B_t}[v_t(t,S_t) + \mu S_t v_s(t,S_t) + \frac{1}{2}\sigma^2 S^2_t v_{ss}(t,S_t)-rv(t,S_t)] \dd t + \frac{1}{B_t}\sigma v_s(t,S_t)  S_t \dd W_t \\
        {} & - \frac{\Delta_t S_t}{B_t}(\mu+q-r) \dd t - \sigma \frac{\Delta_t S_t}{B_t} \dd W_t \\
        {} & = \frac{1}{B_t}[v_t(t,S_t) + \mu S_t v_s(t,S_t) + \frac{1}{2}\sigma^2 S^2_t v_{ss}(t,S_t)-rv(t,S_t) - \Delta_t S_t(\mu+q-r) ] \dd t \\
        {} & + \frac{\sigma S_t}{B_t}(v_s(t,S_t) - \Delta_t) \dd W_t
    \end{align*}
    
    Both coefficients of $\dd t$ and $\dd W_t$ has to be zero. For $\dd W_t$ term to be zero we need $\Delta_t = v_s(t,S_t)$ so we can get from the fact that coefficients of $\dd t$ has to be zero
    
    \begin{align*}
        0 & = v_t(t,S_t) + \mu S_t v_s(t,S_t) + \frac{1}{2}\sigma^2 S^2_t v_{ss}(t,S_t)-rv(t,S_t) - \Delta_t S_t(\mu+q-r) \\
        {} & = v_t(t,S_t) + \mu S_t v_s(t,S_t) + \frac{1}{2}\sigma^2 S^2_t v_{ss}(t,S_t)-rv(t,S_t) - v_s(t,S_t) S_t(\mu+q-r) \\
        {} & = v_t(t,S_t) + (r-q) S_t v_s(t,S_t) + \frac{1}{2}\sigma^2 S^2_t v_{ss}(t,S_t)-rv(t,S_t)
    \end{align*}
    
    \item Using Girsanov's theorem on a risk neutral measure $\widetilde{\mathds{P}}$
    
    \begin{align*}
        \dd (\frac{X_t}{B_t}) & = \frac{\Delta_t S_t}{B_t}(\mu+q-r) \dd t + \sigma \frac{\Delta_t S_t}{B_t} \dd W_t \\
        {} & = \frac{\Delta_t S_t}{B_t}(\mu+q-r - \sigma \gamma_t) \dd t + \sigma \frac{\Delta_t S_t}{B_t} \dd \widetilde{W}_t
    \end{align*}
    
    For $X/B$ to be a martingale we need $\gamma_t = \frac{\mu+q-r}{\sigma}$ which means that we can write dynamics of $S$ in $\widetilde{\mathds{P}}$
    
    \begin{align*}
        \dd S_t & = \mu S_t \dd t + \sigma S_t \dd W_t \\
        {} & = (\mu-\sigma \gamma_t)S_t \dd t + \sigma S_t \dd \widetilde{W}_t \\
        {} & = (r-q) S_t \dd t + \sigma S_t \dd \widetilde{W}_t
    \end{align*}
    
    So that $\widetilde{\mu} = r-q$

\end{enumerate}


\subsection*{Solution to Exercise \ref{exer:5-3}}
Define $S_t := Q_t C_t$.  By the two-dimensional It\^o formula (product rule) we have
\begin{align}
\dd S_t
  &= Q_t \dd C_t + C_t \dd Q_t + \dd Q_t \dd C_t \\
  &= Q_t (R C_t \dd t) + C_t (\mu Q_t \dd t + \sig Q_t \dd W_t) + 0 \\
  &= (R + \mu) S_t \dd t + \sig S_t \dd W_t .
\end{align}
Thus $S_t = Q_t C_t$ is a geometric Brownian motion with drift $R+\mu$ and volatility $\sig$.  The payoff in Dollars is $\phi(Q_T C_T) = \phi(S_T)$, so this is exactly the Black-Scholes problem for a stock $S$ with drift $R+\mu$, volatility $\sig$, and risk-free rate $r$.

Following the usual risk-neutral pricing argument (see Chapter 6), the risk-neutral dynamics of $S$ are
\begin{align}
\dd S_t
  &= r S_t \dd t + \sig S_t \dd \Wt_t ,
\end{align}
and the value at time $t$ of the option is
\begin{align}
V_t
  &= \ee^{-r(T-t)} \Ebt\!\bigl[\phi(S_T) \bigm| S_t\bigr]
   = \int_\Rb \frac{\ee^{-r(T-t)}}{\sqrt{2\pi\sig^2(T-t)}}
       \exp\!\Bigl(-\frac{y^2}{2\sig^2(T-t)}\Bigr)
       \phi\!\Bigl(S_t \ee^{(r-\frac{1}{2}\sig^2)(T-t)+y}\Bigr)\dd y .
\end{align}
The function $v(t,s)$ satisfies the Black-Scholes PDE:
\begin{align}
0
  &= \frac{\d v}{\d t} + r s \frac{\d v}{\d s} + \frac{1}{2}\sig^2 s^2 \frac{\d^2 v}{\d s^2} - r v , \qquad v(T,s) = \phi(s) .
\end{align}
Note that the drift parameters $\mu$ and $R$ do not appear in the pricing formula or the PDE, consistent with the general principle that real-world drifts are irrelevant for option pricing.


\subsection*{Solution to Exercise \ref{exer:5-4}}
Consider a portfolio $X$ that holds $\Del_t$ shares of $S$ and invests the remainder in bonds.  Because owning one share of $S$ generates a dividend cash flow of $q S_t \dd t$, the self-financing dynamics become
\begin{align}
\dd X_t
  &= \Del_t \dd S_t + \Del_t q S_t \dd t + (X_t - \Del_t S_t) r \dd t .
\end{align}
Using the standard replication argument we write $V_t = v(t,S_t)$ and compute $\dd(V_t/B_t)$ and $\dd(X_t/B_t)$.  Setting the $\dd W_t$ terms equal gives $\Del_t = \d_s v(t,S_t)$, and then setting the $\dd t$ terms equal yields the dividend-adjusted Black-Scholes PDE:
\begin{align}
0
  &= \frac{\d v}{\d t} + (r-q) s \frac{\d v}{\d s} + \frac{1}{2}\sig^2 s^2 \frac{\d^2 v}{\d s^2} - r v , \qquad v(T,s) = (s-K)^+ .
\end{align}
Under the risk-neutral measure $\Pbt$ the stock dynamics are $\dd S_t = (r-q) S_t \dd t + \sig S_t \dd \Wt_t$, so
\begin{align}
S_T
  &= S_t \exp\!\Bigl[(r - q - \tfrac{1}{2}\sig^2)(T-t) + \sig(\Wt_T - \Wt_t)\Bigr] .
\end{align}
Risk-neutral pricing then gives
\begin{align}
V_t
  &= \ee^{-r(T-t)} \Ebt\!\bigl[(S_T - K)^+\bigm| S_t\bigr]
   = S_t \ee^{-q(T-t)} \Phi(d_+) - K \ee^{-r(T-t)} \Phi(d_-) ,
\end{align}
where
\begin{align}
d_\pm
  &= \frac{\log(S_t/K) + (r - q \pm \tfrac{1}{2}\sig^2)(T-t)}{\sig\sqrt{T-t}} .
\end{align}
This is the Merton (1973) dividend-adjusted Black-Scholes formula: the only change from the standard formula is the replacement of $S_t$ by $S_t \ee^{-q(T-t)}$, i.e.\ the stock price is discounted at the dividend yield $q$.

\subsection*{Solution to Exercise \ref{exer:5-5}}
(a) We have
\begin{align}
\dd X_t
	&=	\Del_t \dd S_t + (X_t - \Del_t S_t) \frac{1}{M_t} \dd M_t + C_t \dd t \\
	&=	\Del_t ( \mu_t S_t \dd t + \sig_t S_t \dd W_t ) + (X_t - \Del_t S_t) r_t \dd t + C_t \dd t . 
\end{align}
(b) As $V_t = \phi(S_t)$ for all $t$, we have
\begin{align}
\dd V_t
	&=	\dd \phi(S_t)
	=		\phi'(S_t) \dd S_t + \tfrac{1}{2} \phi''(S_t) (\dd S_t)^2 \\
	&=	\phi'(S_t) ( \mu_t S_t \dd t + \sig_t S_t \dd W_t ) + \tfrac{1}{2} \phi''(S_t) \sig_t^2 S_t^2 \dd t .
\end{align}
(c) In order for $X_t = V_t$ for all $t$ we must have $X_0 = V_0$ and $\dd X_t = \dd V_t$.  Comparing the $\dd W_t$ terms in the expressions for $\dd X_t$ and $\dd V_t$ we find
\begin{align}
\Del_t
	&=	\phi'(S_t) .
\end{align}
Next, comparing the $\dd t$ terms in the expressions for $\dd X_t$ and $\dd V_t$ we find that
\begin{align}
C_t
	&=	\tfrac{1}{2} \phi''(S_t) \sig_t^2 S_t^2 - (X_t - \Del_t S_t) r_t \\
	&=	\tfrac{1}{2} \phi''(S_t) \sig_t^2 S_t^2 - (\phi(S_t) - \phi'(S_t) S_t) r_t ,
\end{align}
where we have used $\Del_t = \phi'(S_t)$ and $X_t = \phi(S_t)$.  Thus, we have obtained $\Del_t$ and $C_t$.  Obviously, $X_0 = V_0 = \phi(S_0)$.


\subsection*{Solution to Exercise \ref{exer:5-6}}
\textbf{(a)} Under $\Pbt$, we have $\dd S_t = r S_t \dd t + \sig S_t \dd \Wt_t$, so $\log S_T | \Fc_t \sim N(\log S_t + (r - \tfrac{1}{2}\sig^2)(T-t),\, \sig^2(T-t))$.  Thus,
\begin{align}
v(t,S_t)
	&=	\ee^{-r(T-t)} \Ebt[S_T^p | \Fc_t]
	=	S_t^p \, \ee^{(p-1)(r + \frac{1}{2}\sig^2 p)(T-t)} .
\end{align}
\textbf{(b)} The numeraire for $\Pbh$ is $V_t = v(t,S_t)$.  We compute
\begin{align}
\dd \Big(\frac{X_t}{V_t}\Big)
	&= (\ldots)\dd t + \frac{\Del_t S_t \sig}{V_t} \Big(\dd \Wt_t - \frac{\d_s v(t,S_t) S_t \sig}{v(t,S_t)} \dd t\Big) .
\end{align}
For $X/V$ to be a $\Pbh$-martingale, we set $\dd \Wh_t = \dd \Wt_t - p \sig \dd t$, which gives
\begin{align}
\dd S_t
	&=	(r + p\sig^2) S_t \dd t + \sig S_t \dd \Wh_t .
\end{align}
\textbf{(c)} Under $\Pbh$, $S_T/S_t = \exp\!\Big((r + p\sig^2 - \tfrac{1}{2}\sig^2)(T-t) + \sig(\Wh_T - \Wh_t)\Big)$, so
\begin{align}
U_t
	&=	V_t \Ebh\!\Big[\frac{S_T^q}{V_T}\Big|\Fc_t\Big]
	=	V_t \Ebh\!\Big[ S_T^{q-p} \Big|\Fc_t\Big]
	=	S_t^q\, \ee^{(q-1)(r + \frac{1}{2}\sig^2 q)(T-t)} ,
\end{align}
where we used $V_T = S_T^p$ and $\Ebh[S_T^{k}|\Fc_t] = S_t^{k} \ee^{k(r + p\sig^2 - \frac{1}{2}\sig^2)(T-t) + \frac{1}{2}k^2\sig^2(T-t)}$ with $k = q-p$.\\
\textbf{(d)} Setting $q = p$ in the expression for $U_t$ gives $U_t = S_t^p \ee^{(p-1)(r + \frac{1}{2}\sig^2 p)(T-t)} = v(t,S_t) = V_t$, confirming the result.


\subsection*{Solution to Exercise \ref{exer:5-7}}
\textbf{(a)} In order to preclude arbitrage, $S/B$ must be a martingale under $\Pbt$.  Choosing $\dd S_t = r S_t \dd t + \sig S_t \dd \Wt_t$ achieves this.  By It\^o's Lemma,
\begin{align}
\dd (\log S_t)
	&=	\Big(r - \tfrac{1}{2}\sig^2\Big)\dd t + \sig \dd \Wt_t .
\end{align}
\textbf{(b)} Integrating from $t$ to $T$:
\begin{align}
\log S_T
	&=	\log S_t + \Big(r - \tfrac{1}{2}\sig^2\Big)(T-t) + \sig(\Wt_T - \Wt_t) .
\end{align}
Let $m = (r - \tfrac{1}{2}\sig^2)(T-t)$ for brevity.  Then
\begin{align}
(\log S_T)^2
	&=	(\log S_t)^2 + m^2 + \sig^2(\Wt_T-\Wt_t)^2
		+ 2(\log S_t) m
		+ 2 m \sig(\Wt_T-\Wt_t)
		+ 2 (\log S_t)\sig(\Wt_T-\Wt_t) .
\end{align}
Taking conditional expectation under $\Pbt$ (using $\Ebt[\Wt_T - \Wt_t | \Fc_t] = 0$ and $\Ebt[(\Wt_T-\Wt_t)^2 | \Fc_t] = T-t$),
\begin{align}
\Ebt[(\log S_T)^2 | \Fc_t]
	&=	(\log S_t)^2 + m^2 + \sig^2(T-t) + 2(\log S_t) m .
\end{align}
By risk-neutral pricing,
\begin{align}
u(t,S_t)
	&=	\ee^{-r(T-t)}\Ebt[(\log S_T)^2|\Fc_t] \\
	&=	\ee^{-r(T-t)}\Big[(\log S_t)^2 + \Big(r-\tfrac{1}{2}\sig^2\Big)^2(T-t)^2 + \sig^2(T-t) + 2(\log S_t)\Big(r-\tfrac{1}{2}\sig^2\Big)(T-t)\Big] .
\end{align}
\textbf{(c)} We must set $X_0 = u(0,S_0)$ and $\Del_t = \d_s u(t,S_t)$.  Differentiating with respect to $s$:
\begin{align}
X_0
	&=	\ee^{-rT}\Big[(\log S_0)^2 + \Big(r-\tfrac{1}{2}\sig^2\Big)^2 T^2 + \sig^2 T + 2(\log S_0)\Big(r-\tfrac{1}{2}\sig^2\Big)T\Big] , \\
\Del_t
	&=	\ee^{-r(T-t)}\Big[\frac{2\log S_t}{S_t} + \frac{2}{S_t}\Big(r-\tfrac{1}{2}\sig^2\Big)(T-t)\Big] .
\end{align}


\subsection*{Solution to Exercise \ref{exer:5-8}}
\textbf{(a)} By It\^o's Lemma,
\begin{align}
\dd C(t,S_t)
	&=	\Big[\frac{\d C}{\d t} + \mu(S_t)\frac{\d C}{\d s} + \tfrac{1}{2}\sig^2(S_t)\frac{\d^2 C}{\d s^2}\Big]\dd t
		+ \sig(S_t)\frac{\d C}{\d s}\dd W_t . \label{eq:dC58}
\end{align}
\textbf{(b)} Holding $\Del_t$ put options and investing the remainder in bonds,
\begin{align}
\dd X_t
	&=	\Del_t \dd P(t,S_t) + (X_t - \Del_t P(t,S_t)) r \dd t \\
	&=	\Del_t\Big[\frac{\d P}{\d t} + \mu(S_t)\frac{\d P}{\d s} + \tfrac{1}{2}\sig^2(S_t)\frac{\d^2 P}{\d s^2} - r P\Big]\dd t
		+ r X_t \dd t
		+ \Del_t \sig(S_t)\frac{\d P}{\d s}\dd W_t . \label{eq:dX58}
\end{align}
\textbf{(c)} Since $X_0 = C(0,S_0)$, if $\dd X_t = \dd C(t,S_t)$ then $X_{T_1} = C(T_1,S_{T_1}) = (S_{T_1}-K_1)^+$.  Equating the $\dd W_t$ coefficients in \eqref{eq:dC58} and \eqref{eq:dX58},
\begin{align}
\Del_t
	&=	\frac{\tfrac{\d C}{\d s}(t,S_t)}{\tfrac{\d P}{\d s}(t,S_t)} .
\end{align}


\subsection*{Solution to Exercise \ref{exer:5-9}}
\textbf{(a)} Using the product rule,
\begin{align}
\dd(Q_t B_t)
	&=	Q_t \dd B_t + B_t \dd Q_t
	=	Q_t B_t (R + \mu)\dd t + \sig Q_t B_t \dd W_t .
\end{align}
Thus,
\begin{align}
\dd X_t
	&=	\Del_t\Big(Q_t B_t(R+\mu-r)\dd t + \sig Q_t B_t \dd W_t\Big) + r X_t \dd t .
\end{align}
\textbf{(b)} Computing $\dd(X_t/M_t)$ and requiring it to have no $\dd t$ term, we choose the Girsanov shift $\gam_t = (R+\mu-r)/\sig$, defining $\Wt_t = W_t + \int_0^t \gam_s \dd s$.  Under $\Pbt$,
\begin{align}
\dd Q_t
	&=	(r-R) Q_t \dd t + \sig Q_t \dd \Wt_t .
\end{align}
\textbf{(c)} Under $\Pbt$, $\dd\log(Q_t B_t) = \dd \log Q_t + R \dd t = (r - \tfrac{1}{2}\sig^2)\dd t + \sig \dd \Wt_t$.  By risk-neutral pricing,
\begin{align}
\frac{V_t}{M_t}
	&=	\Ebt\!\Big[\frac{\log(Q_T B_T)}{M_T}\Big|\Fc_t\Big]
	=	\frac{1}{M_T}\Big(\log(Q_t B_t) + (r-\tfrac{1}{2}\sig^2)(T-t)\Big) ,
\end{align}
so
\begin{align}
V_t
	&=	\ee^{-r(T-t)}\Big(\log(Q_t B_t) + (r-\tfrac{1}{2}\sig^2)(T-t)\Big) .
\end{align}


\clearpage
\section{Exercises from Section \ref{sec:exer-6}}

\subsection*{Solution to Exercise \ref{exer:6-1}}

\begin{enumerate}
    \item See Figure \ref{fig:prob6a}
    \begin{figure}
        \centering
        \begin{tikzpicture}
\begin{axis}[greek, ylabel={Call price}, title={Call price vs.\ $s$}, legend pos=north west]
  \addplot[blue] {x*normcdf(dplus(x,1,0.2,1)) - normcdf(dminus(x,1,0.2,1))};
  \addlegendentry{$t=0$}
  \addplot[orange] {x*normcdf(dplus(x,0.5,0.2,1)) - normcdf(dminus(x,0.5,0.2,1))};
  \addlegendentry{$t=T/2$}
  \addplot[teal] {x*normcdf(dplus(x,0.25,0.2,1)) - normcdf(dminus(x,0.25,0.2,1))};
  \addlegendentry{$t=3T/4$}
  \addplot[red] coordinates {(0.7,0) (1,0) (1.3,0.3)};
  \addlegendentry{$(s-K)^+$}
\end{axis}
\end{tikzpicture}
        \caption{Call price vs. s}
        \label{fig:prob6a}
    \end{figure}
    \item We can calculate $\frac{\d}{\d s} d_{\pm} = \frac{1}{s\sigma \sqrt{T-t}}$ and $\Phi'(d_{\pm}) = \frac{e^{-\frac{d^2_{\pm}}{2}}}{\sqrt{2\pi}}$
    
    \begin{align*}
        \Delta^{BS}(t,s;T,K)&= \frac{\d }{\d s}  C^{BS}(t,s;T,K) \\
        {} & = \frac{\d }{\d s} (s \Phi ( d_+ ) - K \Phi( d_- )) \\
        {} & = s \frac{\d}{\d s}(d_+)\Phi'(d_{+}) + \Phi(d_+) - K \frac{\d}{\d s}(d_-)\Phi'(d_{-}) \\
        {} & = \frac{1}{\sigma\sqrt{T-t}} (\Phi'(d_+) - \frac{K}{s} \Phi'(d_-)) + \Phi(d_+)
    \end{align*}
    
    For the plot see Figure \ref{fig:prob6b}
    \begin{figure}
        \centering
        \begin{tikzpicture}
\begin{axis}[greek, ylabel={Delta}, title={Delta vs.\ $s$}, legend pos=north west]
  \addplot[blue] {normcdf(dplus(x,1,0.2,1))};
  \addlegendentry{$t=0$}
  \addplot[orange] {normcdf(dplus(x,0.5,0.2,1))};
  \addlegendentry{$t=T/2$}
  \addplot[teal] {normcdf(dplus(x,0.25,0.2,1))};
  \addlegendentry{$t=3T/4$}
\end{axis}
\end{tikzpicture}
        \caption{Delta vs. s}
        \label{fig:prob6b}
    \end{figure}
    
    \item We can find 
    
    \begin{align*}
        \Gamma^{BS}(t,s;T,K)& = \frac{\d}{\d s} \Delta^{BS}(t,s;T,K) \\
        {} & = \frac{\d}{\d s} (\frac{1}{\sigma\sqrt{T-t}} (\Phi'(d_+) - \frac{K}{s} \Phi'(d_-)) + \Phi(d_+)) \\
        {} & = \frac{1}{\sigma \sqrt{T-t}} \Big( -\frac{d_+}{s\sigma \sqrt{T-t}}\Phi'(d_+) + \frac{K d_-}{s^2\sigma \sqrt{T-t}} \Phi'(d_-) + \frac{K}{s^2} \Phi'(d_-) \Big) + \frac{1}{s\sigma \sqrt{T-t}}\Phi'(d_+)\\
        {} & = \frac{1}{s^2\sigma(T-t)} (-\frac{sd_+}{\sigma}\Phi'(d_+) +  \frac{Kd_-}{\sigma} \Phi'(d_-) + K\sqrt{T-t}\Phi'(d_-)+s\sqrt{T-t}\Phi'(d_+))
    \end{align*}
    
    For the plot see Figure \ref{fig:prob6c}
    \begin{figure}
        \centering
        \begin{tikzpicture}
\begin{axis}[greek, ylabel={Gamma}, title={Gamma vs.\ $s$}, legend pos=north east]
  \addplot[blue] {normpdf(dplus(x,1,0.2,1))/(x*0.2*sqrt(1))};
  \addlegendentry{$t=0$}
  \addplot[orange] {normpdf(dplus(x,0.5,0.2,1))/(x*0.2*sqrt(0.5))};
  \addlegendentry{$t=T/2$}
  \addplot[teal] {normpdf(dplus(x,0.25,0.2,1))/(x*0.2*sqrt(0.25))};
  \addlegendentry{$t=3T/4$}
\end{axis}
\end{tikzpicture}
        \caption{Gamma vs. s}
        \label{fig:prob6c}
    \end{figure}

    \item For the vega we can first note that
    
    \begin{align*}
        \frac{\d d_{\pm}}{ \d \sigma} & = - \frac{1}{\sigma^2\sqrt{T-t}} \log \frac{S}{K} \pm \frac{1}{2}\sqrt{T-t} \\
        {} & = -\frac{1}{\sigma} d_{\mp}
    \end{align*}
    
    and that
    
    \begin{align*}
        d_+ - d_- = \sigma \sqrt{T-t} 
    \end{align*}
    
    we can get that
    
    \begin{align}
        \Vc^\BS(t,s;T,K) & = \frac{\d}{\d \sigma} C^{BS}(t,s;T,K) \\
        {} & = \frac{\d }{\d \sigma} (s \Phi ( d_+ ) - K \Phi( d_- )) \\
        {} & = s \frac{\d}{\d \sigma}(d_+)\Phi'(d_{+}) - K \frac{\d}{\d \sigma}(d_-)\Phi'(d_{-}) \\
        {} & = -\frac{1}{\sigma} (s d_- \Phi'(d_+) - Kd_+ \Phi'(d_-)) \\
        {} & = -\frac{1}{\sigma} (s (d_+ - \sigma \sqrt{T-t}) \Phi'(d_+) - K (d_- + \sigma\sqrt{T-t}) \Phi'(d_-)) \\
        {} & = -\frac{sd_+}{\sigma}\Phi'(d_+) +  \frac{Kd_-}{\sigma} \Phi'(d_-) + K\sqrt{T-t}\Phi'(d_-)+s\sqrt{T-t}\Phi'(d_+)
    \end{align}
    
    From the form of $\Gamma$ we can immediately see that $ \Vc^\BS(t,s;T,K) = (T-t)\sigma s^2 \Gamma^{BS}(t,s;T,K)$ as desired.
    
\end{enumerate}


\subsection*{Solution to Exercise \ref{exer:6-2}}
\textbf{(a)}
We have from the definition of implied volatility that
\begin{align}
C(t,S_t;T,K)
	&=	C^\BS(t,S_t;T,K,\sig(T,K)) \\
	&=	C^\BS(t,S_t;T,K,I(T-t,\log(K/S_t)) ,
\end{align}
Thus we have
\begin{align}
\Del(t,S_t;T,K)
	&=	\frac{\d C}{\d s}C(t,S_t;T,K) \\
	&=	\frac{\d C^\BS}{\d s}(t,S_t;T,K,I(T-t,\log(K/S_t)) \\
	&=	\Del^\BS(t,S_t;T,K,I(T-t,\log(K/S_t)) \\ &\quad 
			- \Vc^\BS(t,S_t;T,K,I(T-t,\log(K/S_t)) \frac{ \d I}{\d x}(T-t,\log(K/S_t)) \frac{1}{S_t} .
\end{align}
\textbf{(b)}
We have
\begin{align}
\Theta(t,S_t;T,K)
	&=	\frac{\d C}{\d t}C(t,S_t;T,K) \\
	&=	\frac{\d C^\BS}{\d t}(t,S_t;T,K,I(T-t,\log(K/S_t)) \\
	&=	\Theta^\BS(t,S_t;T,K,I(T-t,\log(K/S_t)) \\ &\quad 
			- \Vc^\BS(t,S_t;T,K,I(T-t,\log(K/S_t)) \frac{ \d I}{\d \tau}(T-t,\log(K/S_t)) .
\end{align}


\subsection*{Solution to Exercise \ref{exer:6-3}}
\textbf{(a)} 
We can write the value of our portfolio $X_t$ at time $t$ as
\begin{align}
X_t
	&=	\alpha S_t + \beta C^\BS(t,S_t;T,K,\sig) + \gam P^\BS(t,S_t;T,K,\sig) + \del B_t .
\end{align}
In order for our portfolio to be $\Gam$-neutral, we must have
\begin{align}
0
	&=	\d_s^2 X_t \\
	&=	\beta \Gam^\BS(t,S_t;T,K,\sig) + \gam \Gam^\BS(t,S_t;T,K,\sig) , \label{eq:gamma-sol}
\end{align}
where we are using the  fact that $\Gam^\BS(t,S_t;T,K,\sig)$ is the same for both a call and a put.
In order for our portfolio to be $\Vc$-neutral we must have
\begin{align}
0
	&=	\d_\sig X_t \\
	&=	\beta \Vc^\BS(t,S_t;T,K,\sig) + \gam \Vc^\BS(t,S_t;T,K,\sig) , 
\end{align}
where we are using the  fact that $\Vc^\BS(t,S_t;T,K,\sig)$ is the same for both a call and a put.  In order for the portfolio to be $\Gamma$ neutral, we must have $\beta = - \gam$.  But, this would automatically ensure that the portfolio is $\Vc$-neutral.  So, the answer to part (a) is `no'.\\
\textbf{(b)} We can write the value of our portfolio $(X_t + Y_t)$ at time $t$ as
\begin{align}
X_t + Y_t
	&=	\alpha S_t + \beta C^\BS(t,S_t;T_1,K_1,\sig) + \gam P^\BS(t,S_t;T_2,K_2,\sig) + \del B_t + Y_t .
\end{align}
In order for our portfolio to be $\Del$-neutral, we must have
\begin{align}
0
	&=	\d_s X_t \\
	&=	\alpha + \beta \d_s C^\BS(t,S_t;T_1,K_1,\sig) + \gam \d_s P^\BS(t,S_t;T_2,K_2,\sig) + \Del^Y. \label{eq:del2}
\end{align}
In order for our portfolio to be $\Gam$-neutral, we must have
\begin{align}
0
	&=	\d_s^2 X_t \\
	&=	\beta \d_s^2 C^\BS(t,S_t;T_1,K_1,\sig) + \gam \d_s^2 P^\BS(t,S_t;T_2,K_2,\sig) + \Gam^Y. \label{eq:gamma2}
\end{align}
In order for our portfolio to be $\Vc$-neutral we must have
\begin{align}
0
	&=	\d_\sig X_t \\
	&=	\beta \d_\sig C^\BS(t,S_t;T_1,K_1,\sig) + \gam \d_\sig P^\BS(t,S_t;T_2,K_2,\sig) + \Vc^Y . \label{eq:vega2}
\end{align}
In order for our portfolio to be $\Theta$-neutral we must have
\begin{align}
0
	&=	\d_t X_t \\
	&=	\beta \d_t C^\BS(t,S_t;T_1,K_1,\sig) + \gam \d_t P^\BS(t,S_t;T_2,K_2,\sig) + \del r B_t + \Theta^Y. \label{eq:theta2}
\end{align}
These four linear equations can be solved explicitly for $\alpha$, $\beta$, $\gamma$ and $\del$.


\subsection*{Solution to Exercise \ref{exer:6-4}}
\textbf{(a)} Since $r = 0$, the risk-neutral dynamics are $\dd S_t = \sig S_t \dd \Wt_t$.  Using the moment generating function of a Gaussian,
\begin{align}
V_t^{(p)}
	&=	\Ebt[S_T^p|\Fc_t]
	=	S_t^p \, \ee^{\frac{1}{2}\sig^2(T-t)(p^2-p)} .
\end{align}
\textbf{(b)} Differentiating with respect to $S_t$,
\begin{align}
\Del_t^{(p)}
	&=	\frac{\d V_t^{(p)}}{\d S_t}
	=	p S_t^{p-1} \ee^{\frac{1}{2}\sig^2(T-t)(p^2-p)} .
\end{align}
\textbf{(c)} In order for the portfolio to have zero value and be Delta-neutral, we need
\begin{align}
0 &= V_t^{(p)} + \alpha V_t^{(q)} + \beta V_t^{(k)} , \\
0 &= \Del_t^{(p)} + \alpha \Del_t^{(q)} + \beta \Del_t^{(k)} .
\end{align}
In matrix form,
\begin{align}
\begin{pmatrix}
S_t^q \ee^{\frac{1}{2}\sig^2(T-t)(q^2-q)} & S_t^k \ee^{\frac{1}{2}\sig^2(T-t)(k^2-k)} \\
q S_t^{q-1} \ee^{\frac{1}{2}\sig^2(T-t)(q^2-q)} & k S_t^{k-1} \ee^{\frac{1}{2}\sig^2(T-t)(k^2-k)}
\end{pmatrix}
\begin{pmatrix} \alpha \\ \beta \end{pmatrix}
=
\begin{pmatrix}
-S_t^p \ee^{\frac{1}{2}\sig^2(T-t)(p^2-p)} \\
-p S_t^{p-1} \ee^{\frac{1}{2}\sig^2(T-t)(p^2-p)}
\end{pmatrix} .
\end{align}


\clearpage
\section{Exercises from Section \ref{sec:exer-7}}

\subsection*{Solution to Exercise \ref{exer:7-1}}
The central observation is that, although the \emph{true} stock volatility is $\sig_1$, the trader prices and hedges using $\sig_2 > \sig_1$.  The P\&L of the delta-hedged short call position is
\begin{align}
X_T - (S_T - K)^+
  &= \frac{1}{2}\int_0^T \Gam^\BS(t,S_t;\sig_2)(\sig_2^2 - \sig_1^2) S_t^2\, \dd t \;\geq\; 0,
\end{align}
because $\sig_2 > \sig_1$ and $\Gam^\BS \geq 0$.  The Monte Carlo simulation below verifies this: the histogram of terminal hedging errors should be concentrated on the positive half-line.

\begin{lstlisting}[caption={Delta-hedging with mis-specified volatility (\texttt{hw8.py})}]
"""
hw8.py  --  Simulate delta-hedging a short call with wrong volatility.

Input  (input.txt):  S0, K, T, sig1, sig2, mu, N, M
Output: plot of one hedged path, histogram of X_T - (S_T - K)+
"""

import math, numpy as np, matplotlib
matplotlib.use("Agg")
import matplotlib.pyplot as plt
from scipy.stats import norm

# ---- Black-Scholes helpers ------------------------------------------------

def bs_call(t, S, T, K, sig):
    """Black-Scholes call price (r=0)."""
    if T - t < 1e-12 or S <= 0:
        return max(S - K, 0.0)
    sq  = sig * math.sqrt(T - t)
    d1  = math.log(S / K) / sq + 0.5 * sq
    d2  = d1 - sq
    return S * norm.cdf(d1) - K * norm.cdf(d2)

def bs_delta(t, S, T, K, sig):
    """Black-Scholes call delta (r=0)."""
    if T - t < 1e-12 or S <= 0:
        return 1.0 if S > K else 0.0
    sq  = sig * math.sqrt(T - t)
    d1  = math.log(S / K) / sq + 0.5 * sq
    return norm.cdf(d1)

# ---- Read inputs -----------------------------------------------------------

with open("input.txt") as f:
    vals = f.read().split(",")
S0, K, T, sig1, sig2, mu = [float(vals[i]) for i in range(6)]
N, M = int(vals[6]), int(vals[7])

dt   = T / N
grid = np.linspace(0.0, T, N + 1)

# ---- Single-path illustration ---------------------------------------------

np.random.seed(42)
logS = math.log(S0)
S_path = [S0]
for _ in range(N):
    logS += (mu - 0.5*sig1**2)*dt + sig1*math.sqrt(dt)*np.random.randn()
    S_path.append(math.exp(logS))
S_path = np.array(S_path)

# Run the delta hedge (sell call at sig2, hedge with sig2-delta)
X = bs_call(0, S0, T, K, sig2)   # proceeds from selling the call
X_path = [X]
for n in range(N):
    t  = grid[n]
    D  = bs_delta(t, S_path[n], T, K, sig2)
    X  = X + D * (S_path[n+1] - S_path[n])   # r=0, so no bond term
    X_path.append(X)

C_path = [bs_call(grid[n], S_path[n], T, K, sig2) for n in range(N+1)]

fig, ax = plt.subplots()
ax.plot(grid, X_path, label="$X_t$ (hedge portfolio)")
ax.plot(grid, C_path, label="$C^{BS}(t,S_t;\\sigma_2)$", linestyle="--")
ax.set_xlabel("time");  ax.legend()
fig.savefig("plot.hw8.path.png");  plt.close(fig)

# ---- Histogram over M paths -----------------------------------------------

errors = []
for _ in range(M):
    logS = math.log(S0)
    S_sim = [S0]
    for _ in range(N):
        logS += (mu - 0.5*sig1**2)*dt + sig1*math.sqrt(dt)*np.random.randn()
        S_sim.append(math.exp(logS))
    X = bs_call(0, S0, T, K, sig2)
    for n in range(N):
        D = bs_delta(grid[n], S_sim[n], T, K, sig2)
        X = X + D * (S_sim[n+1] - S_sim[n])
    errors.append(X - max(S_sim[-1] - K, 0.0))

fig, ax = plt.subplots()
ax.hist(errors, bins=50, edgecolor="black")
ax.set_xlabel("$X_T - (S_T-K)^+$")
ax.set_title(f"Hedging P&L over {M} paths  "
             f"($\\sigma_1={sig1}$, $\\sigma_2={sig2}$)")
fig.savefig("plot.hw8.hist.png");  plt.close(fig)
print(f"Mean P&L: {np.mean(errors):.4f},  Min: {np.min(errors):.4f}")
print("Saved plot.hw8.path.png and plot.hw8.hist.png")
\end{lstlisting}

\noindent
Because $\sig_2 > \sig_1$, the hedge always collects a non-negative P\&L: the histogram should be supported entirely (or almost entirely) on $[0,\infty)$.


\subsection*{Solution to Exercise \ref{exer:7-2}}
As the value of the replicating portfolio at time $t$ is equal to the expected payoff of the claim, we have
\begin{align}
\Ebt \Big[ \int_0^T \sig_t^2 \dd t \big| \Fc_t \Big]
	&=	\frac{2}{S_t} S_t + \Ebt \Big[ -2 \log \Big( \frac{S_T}{S_0} \Big)  \big| \Fc_t \Big] + \Gam_t B_t . \label{eq1}
\end{align}
where $2/S_t$ is the number of shares of $S$ held at time $t$ and $\Ebt \Big[ -2 \log ( S_T/S_0 )| \Fc_t]$ is the value at time $t$ of the European contract with payoff $-2 \log (S_T/S_0)$.  From \eqref{eq:RV} we have
\begin{align}
\Ebt \Big[ \int_0^T \sig_t^2 \dd t \big| \Fc_t \Big]
	&=	\Ebt \Big[ \int_0^T \frac{2}{S_u} \dd S_u \big| \Fc_t \Big] + \Ebt \Big[ - 2 \log \( \frac{S_T}{S_0} \) \big| \Fc_t \Big] \\
	&=	\int_0^t \frac{2}{S_u} \dd S_u+ \Ebt \Big[ \int_t^T \frac{2}{S_u} \dd S_u \big| \Fc_t \Big] + \Ebt \Big[ - 2 \log \( \frac{S_T}{S_0} \) \big| \Fc_t \Big] . \label{eq2}
\end{align}
Inserting \eqref{eq2} into \eqref{eq1} and solving for $\Gam_t$ we obtain
\begin{align}
\Gam_t
	&=	-2 + \int_0^t \frac{2}{S_u} \dd S_u .
\end{align}
where we have used $B_t = 1$.


\clearpage
\section{Exercises from Section \ref{sec:exer-8}}

\subsection*{Solution to Exercise \ref{exer:8-1}}
\textbf{(a)}
By the definition of implied volatility we have
\begin{align}
C(t,s;T,K)
	&=	C^\BS(t,s;T,K,\sig(T)) .
\end{align}
Thus, we have
\begin{align}
\d_T C(t,s;T,K)
	&=	\d_T C^\BS(t,s;T,K,\sig(T)) + \d_\sig C^\BS(t,s;T,K,\sig(T)) \sig'(T) \\
	&=	- \Theta^\BS(t,s;T,K,\sig(T)) + \Vc^\BS(t,s;T,K,\sig(T)) \sig'(T) \\
	&=	\tfrac{1}{2} \sig^2(T) s^2 \Gam^\BS(t,s;T,K,\sig(T)) + (T-t) \sig(T) s^2 \Gam^\BS(t,s;T,K,\sig(T)) \sig'(T) \\
	&=	\sig^2(T) s^2 \Gam^\BS(t,s;T,K,\sig(T)) \Big( \tfrac{1}{2} + (T-t) \sig'(T) /\sig(T) \Big) \\
	&=	\frac{1}{\sqrt{T-t}}\sig(T) s \Phi'(d_+) \Big( \tfrac{1}{2} + (T-t) \sig'(T) /\sig(T) \Big) , \\
\d_K^2 C(t,s;T,K)
	&=	\d_K^2 C^\BS(t,s;T,K,\sig(T)) \\
	&=	\pt^\BS(t,s;T,K,\sig(T)) \\
	&=	\frac{1}{K \sig(T) \sqrt{T-t}}\Phi'(d_-) .
\end{align}
Inserting the above expressions into the Dupire formula, we find
\begin{align}
f^2(T,K)
	&=	\frac{\d_T C(t,s,T,K)}{\tfrac{1}{2} K^2 \d_K^2 C(t,s,T,K)} \\
	&=	\frac{\frac{1}{\sqrt{T-t}}\sig(T) s \Phi'(d_+) \Big( \tfrac{1}{2} + (T-t) \sig'(T) / \sig(T)\Big)}{\tfrac{1}{2} K^2\frac{1}{K \sig(T) \sqrt{T-t}}\Phi'(d_-)} \\
	&=	2 \sig^2(T) \frac{s}{K} \frac{\Phi'(d_+)}{\Phi'(d_-)} \Big( \tfrac{1}{2} + (T-t) \sig'(T)/ \sig(T) \Big) \\
	&=	2 \sig^2(T)\Big( \tfrac{1}{2} + (T-t) \sig'(T)/ \sig(T) \Big) \\
	&=	\sig^2(T) + 2 (T-t) \sig(T) \sig'(T) . \label{eq:sigma.ode} 
\end{align}
\textbf{(b)}
Suppose that, as claimed, we have
\begin{align}
\sig^2(T)
	&=	\frac{1}{T-t}\int_t^T f^2(u,K) \dd u . \label{eq:sigma.explicit}
\end{align}
Then, differentiating both sides with respect to $T$ we find
\begin{align}
2 \sig(T) \sig'(T)
	&=	-\frac{1}{(T-t)^2} \int_t^T f^2(u,K) \dd u + \frac{1}{T-t} f^2(T,K) \\
	&=	-\frac{1}{T-t} \sig^2(T) + \frac{1}{T-t} f^2(T,K) .
\end{align}
Thus we have
\begin{align}
\sig^2(T) + 2 (T-t) \sig(T) \sig'(T)
	&=	\sig^2(T) + (T-t) \Big( -\frac{1}{T-t} \sig^2(T) + \frac{1}{T-t} f^2(T,K) \Big) \\
	&=	\sig^2(T) - \sig^2(T) + f^2(T,K) \\
	&=	f^2(T,K) .
\end{align}
Thus, if $\sig$ is given by \eqref{eq:sigma.explicit}, then $\sig$ satisfies \eqref{eq:sigma.ode}.


\clearpage
\section{Exercises from Section \ref{sec:exer-9}}

\subsection*{Solution to Exercise \ref{exer:9-1}}
\textbf{(a)} By It\^o's Lemma applied to $X_t = \log S_t$,
\begin{align}
\dd X_t
  &= -\tfrac{1}{2}\sig^2(Y_t)\,\dd t + \sig(Y_t)\,\dd \Wt_t^S .
\end{align}
\textbf{(b)} Integrating from $0$ to $t$ gives
\begin{align}
X_t
  &= X_0 - \tfrac{1}{2}\int_0^t \sig^2(Y_s)\,\dd s
        + \int_0^t \sig(Y_s)\,\dd \Wt_s^S .
\end{align}
Because $\Wt^S$ and $\Wt^Y$ are independent, for any fixed realization of the path $(Y_s)_{0 \le s \le t}$ the stochastic integral $\int_0^t \sig(Y_s)\,\dd \Wt_s^S$ is a Gaussian random variable with mean zero and variance $\int_0^t \sig^2(Y_s)\,\dd s$.  Therefore, conditional on $\Fc_t^Y$,
\begin{align}
X_t \,\Big|\, \Fc_t^Y
  &\sim N\!\Bigl(m_t,\; v_t^2\Bigr) ,
\end{align}
where
\begin{align}
m_t
  &= X_0 - \tfrac{1}{2}\int_0^t \sig^2(Y_s)\,\dd s , &
v_t^2
  &= \int_0^t \sig^2(Y_s)\,\dd s .
\end{align}
\textbf{(c)} Write the option value by conditioning first on $\Fc_T^Y$ and then using the Black-Scholes formula.  We have $r=0$ so
\begin{align}
V_0
  &= \Ebt[\phi(S_T)]
   = \Ebt\!\Bigl[\Ebt\!\bigl[\phi(S_T)\bigm|\Fc_T^Y\bigr]\Bigr] .
\end{align}
Given $\Fc_T^Y$, the log-stock $X_T = \log S_T$ is normally distributed with mean $m_T = \log S_0 - \tfrac{1}{2}\int_0^T\sig^2(Y_s)\,\dd s$ and variance $v_T^2 = \int_0^T \sig^2(Y_s)\,\dd s$.  But this is exactly the conditional distribution of $\log S_T$ under a Black-Scholes model with constant volatility $\bar{\sig}_{0,T}$ defined by $\bar{\sig}_{0,T}^2 = v_T^2/T$.  Hence
\begin{align}
\Ebt\!\bigl[\phi(S_T)\bigm|\Fc_T^Y\bigr]
  &= v^\BS\!\bigl(0,S_0;\bar{\sig}_{0,T}\bigr) ,
\end{align}
and taking outer expectations gives
\begin{align}
V_0
  &= \Ebt\!\Bigl[v^\BS\!\bigl(0,S_0;\bar{\sig}_{0,T}\bigr)\Bigr] ,
\end{align}
which is \eqref{eq:hull-white} at $t=0$.  The argument at general $t$ is identical, conditioning on $\Fc_t^Y$ and using the Markov property of $(S,Y)$.  This result is known as the \textit{Hull-White formula} (Hull and White, 1987).\\[0.5em]
\textbf{(d)} No, \eqref{eq:hull-white} does \emph{not} hold in general when $\Wt^S$ and $\Wt^Y$ are correlated.  The derivation in part (b) relied crucially on the independence of $\Wt^S$ and $\Wt^Y$: once we condition on the path of $Y$ (equivalently, on $\Fc_T^Y$), the stochastic integral $\int_0^T \sig(Y_s)\,\dd \Wt_s^S$ must still be Gaussian with zero mean.  When $\Wt^S$ and $\Wt^Y$ are correlated, conditioning on $\Fc_T^Y$ changes the distribution of $\Wt^S$ (it is no longer independent of $\Wt^Y$), so the conditional distribution of $X_T$ is no longer the same as in a Black-Scholes model with a deterministic volatility.  As a result, the option price depends on the full joint distribution of $(S,Y)$ and cannot be written purely as an average of Black-Scholes prices.


\clearpage
\section{Exercises from Section \ref{sec:exer-10}}

\subsection*{Solution to Exercise \ref{exer:10-1}}
\begin{figure}
\centering
\begin{tikzpicture}
\begin{axis}[width=0.45\textwidth, height=5cm, axis lines=middle, xmin=0, xmax=3.1, ymin=-1, ymax=1.2,
             xtick={0.5,1,...,3}, ytick={-1,-0.5,0.5,1}, title={knock-out: $\phi(s)$},
             every axis title/.style={above, at={(0.5,1.05)}},
             tick label style={font=\footnotesize}, xlabel={$s$}]
  \addplot[blue, thick] coordinates {(0,0) (0.5,-0.5)};
  \addplot[blue, thick] coordinates {(0.5,0) (2,0)};
  \addplot[blue, thick] coordinates {(2,1) (3,1)};
  \addplot[blue, only marks, mark=*, mark size=1.5pt] coordinates {(0.5,-0.5) (2,1)};
  \addplot[blue, only marks, mark=o, mark size=1.5pt, fill=white] coordinates {(0.5,0) (2,0)};
\end{axis}
\end{tikzpicture}
\hfill
\begin{tikzpicture}
\begin{axis}[width=0.45\textwidth, height=5cm, axis lines=middle, xmin=0, xmax=3.1, ymin=-1, ymax=1.2,
             xtick={0.5,1,...,3}, ytick={-1,-0.5,0.5,1}, title={knock-in: $\psi(s)$},
             every axis title/.style={above, at={(0.5,1.05)}},
             tick label style={font=\footnotesize}, xlabel={$s$}]
  \addplot[blue, thick] coordinates {(0,0) (0.5,0.5)};
  \addplot[blue, thick] coordinates {(0.5,0) (3,0)};
  \addplot[blue, only marks, mark=*, mark size=1.5pt] coordinates {(0.5,0.5)};
  \addplot[blue, only marks, mark=o, mark size=1.5pt, fill=white] coordinates {(0.5,0)};
\end{axis}
\end{tikzpicture}
\caption{We plot the payoff functions $\phi$ and $\psi$ of European options that statically hedge the knock-in and knock-out options, respectively.}
\label{fig:phi.psi}
\end{figure}
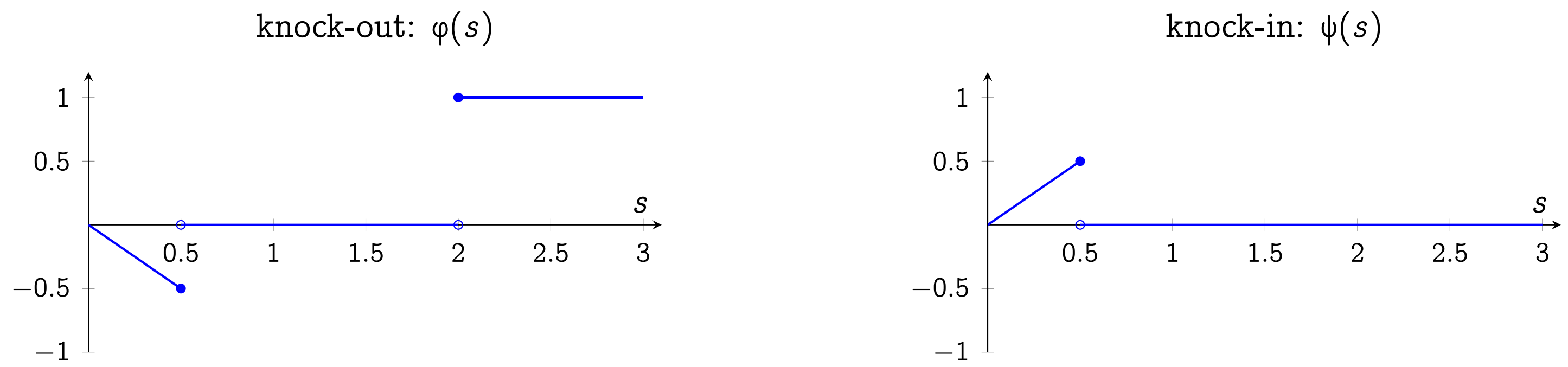

\textbf{(a)}  
If $\tau_L \leq T$ we have by PCS that
\begin{align}
\Ebt [ \theta(S_T-K) | S_{\tau_L} ]
	&=	\Ebt [ (S_T/S_{\tau_L}) \theta(S_{\tau_L}^2/S_T-K) | S_{\tau_L} ] 
	=	\Ebt [ (S_T/L) \theta(L^2/S_T-K) | S_{\tau_L} ] ,
\end{align}
where we have used $S_{\tau_L} = L$.
Thus, if $\tau_L \leq T$ we have
\begin{align}
\Ebt[ \theta(S_T-K) -  (S_T/L) \theta(L^2/S_T-K) | S_{\tau_L} ] 
	&= 0 .
\end{align}
In other words, the value of a European claim with payoff $\phi(S_T) = \theta(S_T-K) -  (S_T/L)  \theta(L^2/S_T-K)$ at time $\tau_L$ is equal to zero at time $\tau_L$ if $\tau_L \leq T$.
Moreover, if $\tau_L > T$ then, at time $T$ we have $S_T > L$ and thus $\theta(L^2/S_T-K)=0$.  It follows that, if $\tau_L > T$, we have
\begin{align}
\phi(S_T) 
	&=	\theta(S_T-K) - (S_T/L)  \theta(L^2/S_T-K)
	=	\theta(S_T-K) .
\end{align}
Thus, a European claim with payoff $\phi(S_T) = \theta(S_T-K) -  (S_T/L)  \theta(L^2/S_T-K)$ will statically hedge the knock-out claim prior to time $\tau_L \wedge T$.\\
\textbf{(b)}
We know from knock-in knock-out parity that the value $V_t^\text{KI} + V_t^\text{KO} = V_t^\text{Euro}$.  Prior to time $\tau_L \wedge T$ we have
\begin{align}
V_t^\text{Euro}
	&=	\Ebt[ \theta(S_T-K) | \Fc_t ] , &
V_t^\text{KO}
	&=	\Ebt[ \phi(S_T) | \Fc_t ] .
\end{align}
Thus, we have
\begin{align}
V_t^\text{KI}
	&=	V_t^\text{Euro} - V_t^\text{KO} \\
	&=	\Ebt[ \theta(S_T-K) | \Fc_t ] - \Ebt[ \phi(S_T) | \Fc_t ] \\
	&=	\Ebt[ \psi(S_T) | \Fc_t ] , \\
\psi(S_T)
	&:=	\theta(S_T-K) - \phi(S_T) = (S_T/L) \theta(L^2/S_T-K) .
\end{align}
In other words, the value of a European claim with payoff $\psi(S_T)$, prior to time $\tau_L$ is equal to the value of the knock-in claim, if $\tau_L \leq T$.  On the other hand, if $\tau_L > T$ we have $S_T > L$ and thus $\psi(S_T) = 0$.  Thus the payoff function $\psi(S_T)$ statically hedges the knock-in claim up to time $\tau_L \wedge T$.\\
\textbf{(c)}  We plot $\phi(s)$ and $\psi(s)$ in Figure \ref{fig:phi.psi}.

\end{document}